%% file: main.tex
\documentclass[10pt]{article}

\input{pkgs.tex}
\input{cmds.tex}

\begin{document}

\pagenumbering{arabic}
\setcounter{page}{1}

\begin{center}
    \vspace*{4mm}
    \LARGE \textbf{
    On spatially irregular ordinary\\
    differential equations and a pathwise\\
    volatility modelling framework\\
    }
    \vspace*{10mm}
    \large A thesis presented for the degree\\
    \vspace*{4mm}
    
    Doctor of Philosophy\\
    \emph{from}\\
    Department of Mathematics, Imperial College London\\
    \emph{by}\\
    \href{mailto:rmccrickerd@chathamfinancial.com}{{\color{main}Ryan McCrickerd}}\\

    \vspace*{6mm}
    \href{mailto:rmccrickerd@chathamfinancial.com}{rmccrickerd@chathamfinancial.com}
\end{center}

\input{frontmatter.tex}

\input{prologue}

\input{chapters/1_introduction}
\input{chapters/2_wellposed}
\input{chapters/3_solutions}
\input{chapters/4_framework}
\input{chapters/5_conclusion}

\input{epilogue.tex}

\input{notation.tex}

\clearpage
\phantomsection
\bibliographystyle{apa}
\renewcommand\refname{Bibliography}
\addcontentsline{toc}{section}{\bibname}

\pagestyle{fancy}
\fancyhf{}
\renewcommand{\headrulewidth}{0pt}
% \rhead{\thepage}
\cfoot{\thepage}
\chead{\nouppercase{\sc Bibliography}}

{\small
\bibliography{phd}
}

\input{appendix.tex}

\end{document}

%% file: pkgs.tex
\usepackage[utf8]{inputenc}

\usepackage{geometry}
\renewcommand{\baselinestretch}{1.5} % required by Imperial

\usepackage{float}
\usepackage[justification=centering]{caption}
\usepackage{xcolor}
\definecolor{main}{RGB}{0,25,51}   % text colour
\definecolor{alt1}{RGB}{0,153,153} % internal links
\definecolor{alt2}{RGB}{76,0,153}  % external links
\definecolor{alt3}{RGB}{0,128,255} % all links
\color{main} % set to main colour

\usepackage[pdftex,
    bookmarks=true,         % show bookmarks bar?
    unicode=true,          % non-Latin characters in Acrobat’s bookmarks
    pdftoolbar=false,        % show Acrobat’s toolbar?
    pdfmenubar=false,        % show Acrobat’s menu?
    pdffitwindow=true,     % window fit to page when opened
    pdfstartview={FitV},    % fits the width of the page to the window
    pdftitle={Irregular ODEs \& pathwise volatility},    % title
    pdfauthor={Ryan McCrickerd},     % author
    pdfsubject={},   % subject of the document
    pdfcreator={},   % creator of the document
    pdfproducer={}, % producer of the document
    pdfkeywords={}, % list of keywords
    pdfnewwindow=true,      % links in new PDF window
    colorlinks=true,       % false: boxed links; true: colored links
    linkcolor=alt3,          % color of internal links (change box color with linkbordercolor)
    citecolor=alt3,        % color of links to bibliography
    filecolor=alt3,      % color of file links
    urlcolor=alt3         % color of external links
]{hyperref}

\usepackage{aliascnt}  
\usepackage{titlesec}

\titlespacing\section{0pt}{12pt plus 0pt minus 0pt}{12pt plus 0pt minus 0pt}
\titlespacing\subsection{0pt}{8pt plus 0pt minus 0pt}{8pt plus 0pt minus 0pt}
\titlespacing\subsubsection{0pt}{8pt plus 0pt minus 0pt}{8pt plus 0pt minus 0pt}

\titleformat{\section}[hang]{\Large\bfseries\filcenter}{\thesection}{1em}{}
\titleformat{\subsection}[hang]{\large\bfseries\filcenter}{\thesubsection}{1em}{}

\usepackage{fancyhdr}

\usepackage{amsmath,amssymb,amsthm,wasysym,stmaryrd,mathrsfs}
\usepackage[UKenglish]{babel}
\newcommand{\mynewtheorem}[2]{
  \newaliascnt{#1}{dummy}
  \newtheorem{#1}[#1]{#2}
  \aliascntresetthe{#1}
  \expandafter\def\csname #1autorefname\endcsname{#2}
}

\newtheoremstyle{mythm}%
    {6pt plus 0pt minus 0pt}% Space above
    {2pt plus 0pt minus 0pt}% Space below
    {\itshape}% Body font 
    {}% Indent amount
    {\bfseries}% ⟨Theorem head font⟩
    {\bfseries.}% ⟨Punctuation after theorem head ⟩
    {.5em}% Space after theorem head 
    {}%
    
\newtheoremstyle{myrem}%
    {6pt plus 0pt minus 0pt}% Space above
    {2pt plus 0pt minus 0pt}% Space below
    {}% Body font 
    {}% Indent amount
    {\bfseries}% ⟨Theorem head font⟩
    {\bfseries.}% ⟨Punctuation after theorem head ⟩
    {.5em}% Space after theorem head 
    {}%

\theoremstyle{mythm}
  \mynewtheorem{theorem}{Theorem}
  \mynewtheorem{lemma}{Lemma}
  \mynewtheorem{corollary}{Corollary}
  \mynewtheorem{question}{Question}

\theoremstyle{myrem}
  \mynewtheorem{example}{Example}
  \mynewtheorem{problem}{Problem}
  \mynewtheorem{definition}{Definition}
  \mynewtheorem{remark}{Remark}
  \mynewtheorem{algorithm}{Algorithm}
  \mynewtheorem{model}{Model}
  \mynewtheorem{framework}{Framework}
  \mynewtheorem{assumption}{Assumption}
  \mynewtheorem{project}{Project}
  \mynewtheorem{conjecture}{Conjecture}
  \mynewtheorem{proof2}{Proof}

\renewenvironment{proof}
{\textbf{Proof.}\begin{mdseries}}{\qed\end{mdseries}}

\usepackage[longnamesfirst]{natbib}

\usepackage{graphicx}

\usepackage{datetime}
\newdateformat{myformat}{\monthname[\THEMONTH]\ \THEYEAR}

\usepackage{bbm} % for indicator function

\usepackage{listings}

\usepackage{mathtools}

\usepackage{changepage}
\usepackage[all]{nowidow}
\usepackage{lineno}

\usepackage{centernot}

%% file: cmds.tex
\newcommand{\dd}{\mathrm{d}}
\newcommand{\id}{\mathrm{e}}
\newcommand{\RR}{\mathbb{R}}
\newcommand{\II}{\mathbb{I}}
\newcommand{\QQ}{\mathbb{Q}}
\newcommand{\PP}{\mathbb{P}}
\newcommand{\EE}{\mathbb{E}}

\newcommand{\WW}{\mathbb{W}}

\newcommand{\BB}{\mathbb{B}}

\newcommand{\bS}{\mathbb{S}}

\newcommand{\NN}{\mathbb{N}}

\newcommand{\cX}{\mathcal{X}}
\newcommand{\cE}{\mathcal{E}}
\newcommand{\cF}{\mathcal{F}}

\newcommand{\cG}{\mathcal{G}}
\newcommand{\cD}{\mathcal{D}}

\newcommand{\cB}{\mathcal{B}}
\newcommand{\uW}{\mathrm{W}}
\newcommand{\uF}{\mathrm{F}}
\newcommand{\uFe}{\mathrm{F_{\:\!\! e}}}

\newcommand{\uG}{\mathrm{G}}
\newcommand{\uD}{\mathrm{D}}

\newcommand{\uS}{\mathrm{S}}
\newcommand{\uC}{\mathrm{C}}

\newcommand{\uN}{\mathrm{N}}
\newcommand{\uH}{\mathrm{H}}

\newcommand{\uAC}{\mathrm{AC}}
\newcommand{\uE}{\mathrm{E}}
\newcommand{\uM}{\mathrm{M}}

\newcommand{\uL}{\mathrm{L}}
\newcommand{\uJ}{\mathrm{J}}

\newcommand{\tX}{\tilde{X}}

\newcommand{\tW}{\tilde{W}}
\newcommand{\tS}{\tilde{S}}

\newcommand{\cR}{\mathcal{R}}

\DeclareMathOperator*{\Leb}{Leb}

\newcommand{\ve}{\varepsilon}
\newcommand{\vt}{\vartheta}

\newcommand{\vp}{\varphi}
\newcommand{\hvp}{\hat{\varphi}}

\newcommand{\bvp}{\overline{\varphi}}
\newcommand{\ubvp}{\underline{\varphi}}

\newcommand{\bX}{{\overline{X}}}

\newcommand{\bRR}{{\overline{\mathbb{R}}}}

\newcommand{\bPhi}{\overline{\Phi}}

\newcommand{\bXX}{{\overline{X}}}

\newcommand{\tY}{{\tilde{Y}}}

\newcommand{\hY}{{\hat{Y}}}
\newcommand{\hW}{{\hat{W}}}
\newcommand{\hX}{{\hat{X}}}
\newcommand{\hS}{{\hat{S}}}

\newcommand{\cd}{\xrightarrow{\dd}}
\newcommand{\cfd}{\xrightarrow{\mathrm{f.d.}}}
\newcommand{\cp}{\xrightarrow{\mathrm{p}}}
\newcommand{\asc}{\xrightarrow{\mathrm{a.s.}}}
\newcommand{\ed}{\stackrel{\dd}{=}}

\newcommand{\cw}{\xRightarrow{n\to\infty}}

\numberwithin{equation}{section}

\newcommand{\ep}{\epsilon}

\newcommand{\sgn}{\mathrm{sgn}}

\newcommand{\tZ}{\tilde{Z}}
\newcommand{\tvt}{\tilde{\vt}}

\usepackage[T1]{fontenc}

%% file: frontmatter.tex
\clearpage
% \pagenumbering{arabic}
% \setcounter{page}{2}
\setlength{\parskip}{2mm}

\pagestyle{fancy}
\fancyhf{}
\renewcommand{\headrulewidth}{0pt}
\cfoot{\thepage}

\input{abstract}

\clearpage
\section*{Copyright Declaration}

The copyright of this thesis rests with the author. Unless otherwise
indicated, its contents are licensed under a Creative Commons
Attribution-Non Commercial 4.0 International Licence (CC BY-NC).

Under this licence, you may copy and redistribute the material in any
medium or format. You may also create and distribute modified
versions of the work. This is on the condition that: you credit the
author and do not use it, or any derivative works, for a commercial
purpose.

When reusing or sharing this work, ensure you make the licence
terms clear to others by naming the licence and linking to the licence
text. Where a work has been adapted, you should indicate that the
work has been changed and describe those changes.

Please seek permission from the copyright holder for uses of this
work that are not included in this licence or permitted under UK
Copyright Law.

\clearpage
\section*{Statement of Originality}

The parts of this thesis which are presented as my own work are my own work. The parts of this thesis which are not my own work are not presented as such and are appropriately referenced.

\clearpage
\section*{Acknowledgements}

My primary thanks go to those without whom this thesis would not exist. These are foremost my parents, brother and sister, who provided the initial conditions for me to think independently, impartially and creatively. My early interests in sports, video games and cartoons can explain my intrigue with reality and how abstract models relate to it.

Without encouragement from Peter O'Grady at school, I may not have started a mathematics degree. The quality of his teaching additionally enabled me to continue sports both in reality and virtually during my studies. Outclassing friends on video games proved important:~intoxicating booby prizes are avoided; a clearer mind for mathematics is maintained.

It took several years working at JCRA to develop my appreciation of the models utilised in finance, as opposed to the sexier ones in physics. The level of support which my manager Ivan Harkins gave to me over those years was always surprising, and his enabling of PhD research alongside work will be forever valued tremendously. The wider culture set up by the late John Rathbone at JCRA was remarkable, like him, in my eyes and many friends'.

The patience and advice of my supervisors Mikko Pakkanen and Martin Rasmussen at Imperial College, especially as my research took unconventional and risky directions, has been vital to its completion. Hindsight clarifies that I have at times been guided too much by mathematical aesthetics, rather than the more valuable consequences of my research.

I have been delighted to grow even closer to my partner-in-crime Lavinia Singer and her family, and am excited to start our own. Watching Lavinia grow, magnified by Covid-19, has been inspirational, and I could not imagine feeling luckier than I do today. Witnessing a poetry editor endure my thoughts on function topologies has provided further inspiration.

Finally I thank the authors of the texts \cite{Coddington_1955}, \cite{Agarwal_1993} and \cite{Whitt_2002}, whose shoulders I feel I have stood on most.

\vspace{4mm}
This thesis was funded directly by the EPSRC Centre for Doctoral Training in Financial Computing and Analytics, and indirectly by my employers JCRA and Chatham Financial.

\setlength{\parskip}{0mm}
\renewcommand{\baselinestretch}{1.4}
\clearpage
\renewcommand{\contentsname}{Table of Contents}
{\tableofcontents}
% \clearpage

\setlength{\parskip}{2mm}
\renewcommand{\baselinestretch}{1.5} % required by imperial
% \pagenumbering{arabic}

\renewcommand{\listfigurename}{Figures}
\renewcommand{\listtablename}{Tables}

%% file: abstract.tex
\section*{\hypertarget{abstract}{Abstract}}

This thesis develops a new framework for modelling price processes in finance, such as an equity price or foreign exchange rate. This can be related to the conventional It\^o calculus-based framework through the time integral of a price's squared volatility, or `cumulative variance'. In the new framework, corresponding processes are strictly increasing, solve random ordinary differential equations (ODEs), and are composed with geometric Brownian motion. The new framework has no dependence on stochastic calculus, so processes can be studied on a pathwise basis using probability-free ODE techniques and functional analysis.

The ODEs considered depend on continuous driving functions which are `spatially irregular', meaning they need not have any spatial regularity properties such as H\"older continuity. They are however strictly increasing in time, thus temporally asymmetric. When sensible initial values are chosen, initial value problem (IVP) solutions are also strictly increasing, and the solution set of such IVPs is shown to contain all differentiable bijections on the non-negative reals. This enables the modelling of any non-negative volatility path which is not zero over intervals, via the time derivative of solutions. Despite this generality, new well-posedness results establish the uniqueness of solutions going forwards in time. A condition is provided which prohibits explosions, and then the IVPs' solution map is shown to be continuous with respect to uniform convergence over compacts. 

Motivation to explore this framework comes from its connection with a time-changed Heston volatility model. The framework shows how Heston price processes can converge to a generalisation of the normal-inverse Gaussian (NIG) L\'evy process, and reveals a deeper relationship between integrated Cox-Ingersoll-Ross (CIR) processes and the inverse Gaussian (IG) process. Within this framework, a `Riemann-Liouville-Heston' (RLH) martingale model is defined which generalises these relationships to fractional counterparts. This model's implied volatilities are simulated, and exhibit features characteristic of leading volatility models.

%% file: prologue.tex
\clearpage

\pagestyle{fancy}
\fancyhf{}
\renewcommand{\headrulewidth}{0pt}
\cfoot{\thepage}
\chead{\nouppercase{\sc Prologue}}

\phantomsection\section*{\hypertarget{prologue}{Prologue: Heston-NIG motivating relationships}}
\addcontentsline{toc}{section}{Prologue: Heston-NIG motivating relationships}

This prologue presents the author's personal account of the preliminary motivations behind this thesis. These originate primarily from a desire to validate, strengthen and generalise the main result of \cite{Mechkov_2015}, after experiencing modelling benefits of this in financial risk management. Although not strictly required to appreciate the thesis's mathematical contributions, results and goals presented here will be referred to throughout the remainder.

\vspace{3mm}\textbf{Personal modelling experience.} I have worked as a quantitative analyst in risk management for nine years thus far, mostly at an advisory firm called JCRA, set up by John Rathbone in 1989 and bought by Chatham Financial in 2019. In 2015, a year before starting to work towards a PhD at Imperial College, I was reviewing models of financial variables (e.g.~interest rates, foreign exchange rates, stock prices) for the purpose of various simulation-based computations related to the possible future values of clients' derivative portfolios. 

For this purpose, we had at our disposal Numerix's model library (Numerix is a trading and risk management technology provider, see e.g.~\href{https://www.numerix.com/}{Numerix.com}). This is how I came across Numerix's `fast-reversion Heston' (FRH) model, specifically when comparing calibration accuracies and stabilities of various foreign exchange (FX) models. The details of this are available in \cite{Mechkov_2015}, and an implementation at \href{https://github.com/ryanmccrickerd/frh-fx}{github.com/ryanmccrickerd/frh-fx}.

It became clear that this FRH extension of the classical stochastic volatility model of \cite{Heston_1993} was excellent for our purposes. In simple terms, this model can, like alternative extensions of the Heston model (e.g.~local-stochastic volatility or jump-diffusion), replicate the 100 or so quotes in an FX implied volatility surface with near-perfect accuracy, and it can also be easily simulated accurately, unlike its classical namesake. In my experience, it achieves this with relatively few, stable and physically meaningful parameters, the effects of which one stands a chance of explaining to non-mathematical colleagues and clients.

I was therefore surprised to find that this FRH model is not really a new model, but, at least in its basic form, a repackaged old one. Specifically, it is a reparameterised normal inverse-Gaussian (NIG) model from \cite{Barndorff-Nielsen_1997}, wherein the variable of concern, e.g.~FX rate or equity price, is modelled by an exponentiated NIG (exp-NIG) L\'evy process. Valuably, the influences of Heston's parameters on this FRH version are preserved, for which MSc graduates and quantitative analysts in general usually have strong intuition. 

My immediate surprise originated from the fact that these apparently related Heston and NIG models are two of the most popular in finance, and so vast sums of companies' wealth, and their related decision making, depend on their properties. In my line of work, this dependence primarily manifests from what such models say about derivative values.

Mathematically, the models being related here exist in different frameworks. One depends inseparably on It\^o calculus, accommodating continuous sample paths like Heston's, and the other on non-Brownian L\'evy processes, with discontinuous paths like those of the NIG model. Popularity aside, these models are exemplars as good as any for these contrasting theoretical frameworks. Supplementing practical experiences, such theoretical considerations reinforce the value of developing a deeper understanding of the relationship between these two models.

\vspace{3mm}\textbf{Mechkov's Heston-NIG relationship.} The Heston and NIG relationship from \cite{Mechkov_2015} is now summarised, presented in notation consistent with the core of this thesis. This is a relationship manifesting through limits of parameters, so consider a general family of classical Heston price processes $S^n=\{S^n_t\}_{t\in\RR_+}$, with $n>0$, exactly as in \cite{Heston_1993},~
\begin{equation}\label{eq:heston_prologue}
    \dd V^n_t = \sigma_n\sqrt{V^n_t}\dd W^1_t + \kappa_n(\theta_n - V^n_t)\dd t,\quad \dd S^n_t = \sqrt{V^n_t}S^n_t\dd W^{\rho_n}_t,\quad (V^n_0,S^n_0) := (v_n,1),
\end{equation}
where $W^0,W^1$ are independent 1d Brownian motions starting at 0, and $W^{\rho_n}:=\sqrt{1-\rho_n^2}W^0+\rho_n W^1$. In finance, the stochastic differential equations (SDEs) which the variance processes $V^n=\{V^n_t\}_{t\in\RR_+}$ verify are called CIR SDEs, because of \cite{Cox_1985}. 

The parameters $\sigma_n,\kappa_n,\theta_n, v_n >0$ are respectively known as volatility of volatility, reversion speed, reversion level, and starting variance, and $\rho_n\in[-1,1]$ correlation. For some fixed $\sigma,\theta,v>0$ and $\rho\in[-1,1]$, now set $(\sigma_n,\kappa_n,\theta_n,\rho_n,v_n):=(n\sigma,n,\theta,\rho,v)$, so that $n$ indexes the reversion speeds of this family $\{S^n\}_{n>0}$, and \autoref{eq:heston_prologue} more simply reads
\begin{equation}\label{eq:heston_prologue2}
    \dd V^n_t = n\sigma\sqrt{V^n_t}\dd W^1_t + n(\theta - V^n_t)\dd t,\quad \dd S^n_t = \sqrt{V^n_t}S^n_t\dd W^{\rho}_t,\quad (V^n_0,S^n_0) := (v,1).
\end{equation}
The fact that \emph{both} the diffusive and reversionary components of the CIR SDEs for each $V^n$ here scale linearly with $n$, so grow at the same rate as $n\to\infty$, is critical to the novelty of this parameterisation, and the resulting NIG relationship. This differs from that considered extensively in \cite{Fouque_2011} and preceding articles by the same authors where, in the same notation here, one would instead set $\sigma_n:=\sqrt{n}\sigma$.

When defining L\'evy processes, it is sufficient and common to state their marginal characteristic function, and the NIG one can be found, alongside Heston's, in \cite{Mechkov_2015}. But following \cite{Applebaum_2009}, it is possible and informative to construct an exp-NIG process $S^0=\{S^0_t\}_{t\in\RR_+}$ from the same Brownian motions $W^0,W^1$ as in \autoref{eq:heston_prologue2}, by
\begin{equation}\label{eq:nig_prologue}
    S^0_t := \exp\left(\sqrt{1-\rho^2}W^0_{X^0_t} + \frac{2\rho-\sigma}{2\sigma}X^0_t -\frac{\rho\theta}{\sigma} t\right),\quad X^0_t:=\inf\bigg\{x>0:x-\sigma W^1_x > \theta t\bigg\}.
\end{equation}
The process $X^0=\{X^0_t\}_{t\in\RR_+}$ is an inverse-Gaussian (IG) subordinator, which is a non-decreasing L\'evy process. The main result of \cite{Mechkov_2015} can now be stated as follows.
\begin{theorem}[Mechkov's Heston-NIG relationship]\label{thm:mech_pro}
    Let $\{S^n\}_{n>0}$ be the family of Heston price processes from \autoref{eq:heston_prologue2}, and $S^0$ the exponentiated NIG process from \autoref{eq:nig_prologue}. Then for each fixed $t\in\RR_+$, the convergence in distribution $S^n_t\cd S^0_t$ takes place as $n\to\infty$.
\end{theorem}

At first this result can seem related to the relationship known earlier between the distribution of a \emph{fixed} Heston process at large times and the NIG distribution, established independently in \cite{Keller_2011} and \cite{Forde_2011}. However, when one tries to map this large-time result onto parameters of a \emph{family} of Heston models (through scaling properties of Brownian motion), the resulting family is not quite like those in \autoref{eq:heston_prologue2}, but those obtained when instead setting $(\sigma_n,\kappa_n,\theta_n,\rho_n,v_n):=(n\sigma,n,n\theta,\rho,nv)$ in \autoref{eq:heston_prologue}. 

With these starting variances and reversion levels additionally exploding as $n\to\infty$, the resulting distribution of the family $S^n$ at any fixed time grows in a manner which cannot be reconciled with those of a fixed exp-NIG process. So these earlier large-time results for a fixed Heston model cannot be adapted to a relationship with a fixed limiting model. Indeed, the authors concluded this at the time, which now emphasises the novelty of \autoref{thm:mech_pro}.

\vspace{3mm}\textbf{The volatility skew paradox.} With \autoref{thm:mech_pro} stated, it is worth highlighting a paradox of sorts which it raises. Around the time when I became aware of and had computationally verified Mechkov's Heston-NIG relationship, I also became aware of the preprint of \cite{Gatheral_2018} and the growing popularity of `rough' volatility models. These models usually extend a classical counterpart like Heston's, and are distinguished by the depending volatility or variance process having a comparably low H\"older regularity. So these processes look rough, like fractional Brownian motion with a low Hurst parameter. 

In my area of derivative-related work, such models are supported by their ability to reproduce observed `implied volatility skews' in equity markets. First demonstrated in \cite{Bayer_2015}, this is backed up by the theory of \cite{Alos_2007} and \cite{Fukasawa_2011}. For a price process $S=\{S_t\}_{t\in\RR_+}$ and future time, this skew is loosely related, via some transformations, to the third moment of $S_t$. See \cite{Bergomi_2016} or \autoref{eq:skew_curv}.

\begin{figure}[ht]
    \centering
    \includegraphics[width=0.50\linewidth]{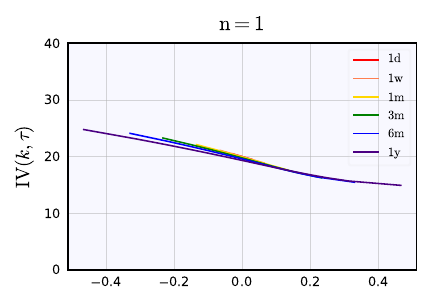}
    \includegraphics[width=0.47\linewidth]{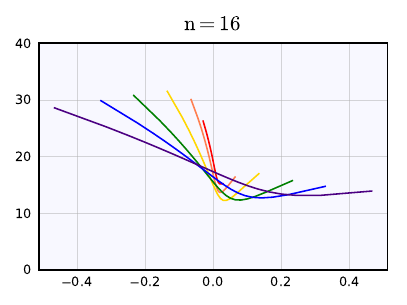}
    
    \includegraphics[width=0.50\linewidth]{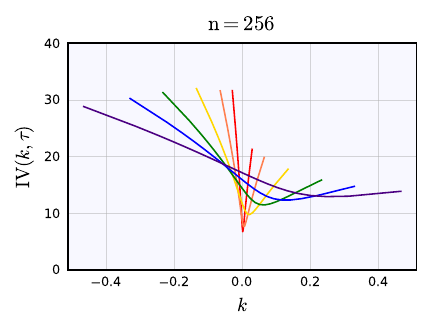}
    \includegraphics[width=0.47\linewidth]{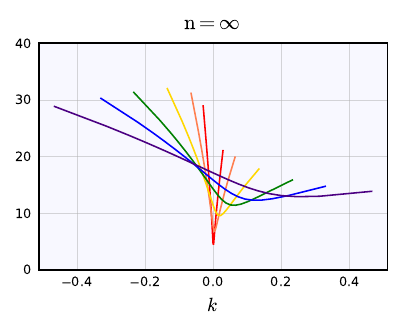}
    \caption{Like in \cite{Mechkov_2015}, implied volatilities IV$(k,\tau)$ from the Heston model in \autoref{eq:heston_prologue2}, with $n=1,16,256$, are shown to converge across logstrikes $k$ to those of the NIG model ($n=\infty$) from \autoref{eq:nig_prologue}. This is a consequence of \autoref{thm:mech_pro}. Maturities $\tau$ range from a day ($\tau=1/256$) to a year ($\tau=1$), and $\sigma\!=\!0.2$, $\theta\!=\!v\!=\!0.04$, $\rho\!=\!-0.7$.}
    \label{fig:vols_converge}
\end{figure}

The theory says that the classical Heston and NIG models featuring in \autoref{thm:mech_pro} will respectively \emph{under} and \emph{over}-emphasise how a skew consistent with rough volatility evolves backwards in time, towards $t=0$. The theoretical justification of this for the NIG model is in \cite{Gerhold_2016}. But a consequence of \autoref{thm:mech_pro} is that the implied volatilities from these models, from which skews derive, will \emph{converge} as $n\to\infty$. This is demonstrated graphically in \cite{Mechkov_2015}, and we reproduce something similar in \autoref{fig:vols_converge}. So how can this under and over-emphasis be explained, in light of this convergence? 

% 20201210-Heston-NIG-surface-convergence-no-sim

Despite few being aware of \autoref{thm:mech_pro} and its consequences, this paradox is actually understood, especially by practitioners, many of whom do not see a problem. A resolution is provided in \cite{Bergomi_2016}, and the fact there is one is because these theoretical under and over-emphasis statements sometimes only apply with a meaningful degree of accuracy over an impractically short timescale, possibly even where no observable data exists. 

As in \cite{Bergomi_2016}, practitioners like this author have long been successfully bypassing needs for rough volatility models by foremost employing fast reversion speeds in classical models. A theoretical relationship between rough and fast reverting models was later revealed by certain representations of rough processes which show they depend implicitly on \emph{arbitrarily} large reversion speeds, see e.g.~\cite{Muravlev_2011} and \cite{Abi_Jaber_2019c}. This dependence is just carefully controlled, in a manner which does not produce jumps like those present in the NIG limit of \autoref{thm:mech_pro}, but instead just reduces paths' H\"older regularity. Following private discussions, the presentation \cite{Abi_Jaber_2019} was the first to expose this link between rough volatility and jumps, via different reversionary properties. 

The point here is \emph{not} about precise reproductions of skews, which I sincerely believe is done best by rough volatility models, at least in the equity markets. It is about where we collectively place our research focus, and the value of it. Having co-written \cite{McCrickerd_2018} on derivative pricing for a particular rough volatility model, I felt like I had spent as much time as anyone treating practical difficulties associated with rough volatility models, especially regarding \emph{simulation}, and as a result was ready to consider alternatives. 

Although simulation of the Heston price process $S^n$ from \autoref{eq:heston_prologue2} becomes more difficult as the reversion speed $n$ is raised, justifying various approximation techniques like those of \cite{Andersen_2007}, the exp-NIG limit $S^0$ as $n\to\infty$ in \autoref{thm:mech_pro} can be simulated exactly. As already noted, fast reversion speeds in classical models are being relied upon in practice, as an alternative to rough volatility. See e.g.~\cite{Bergomi_2016} and \cite{De_Col_2013} for calibrated values of the order of 1,000\% deriving respectively from equity and FX derivative price data, but note \cite{Fouque_2011} obtain values as high as 10,000\% from realised data, still corresponding to a plausible reversionary timescale of 2--3 trading days. 

Given positive personal experiences with the FRH model, and that the Heston-NIG relationship from \autoref{thm:mech_pro} can help to alleviate \emph{both} of these volatility skew and simulation problems, it seemed clear that I, if not others also, should first spend more time trying to better understand this surprising new fast reversion relationship between these relatively simple existing models, before seriously reconsidering rough volatility models again.

This version of Occam's razor is particularly salient in finance, because regulators often neglect the complexity of output requested from companies. The requirement to compute valuation adjustments (XVAs) demonstrates this. These depend on the future values of derivative portfolios mentioned earlier, and are what led me to Numerix's FRH model. At JCRA, we used this model for FX XVAs for five years, because it consistently calibrates well to FX implied volatility surfaces, and can be simulated efficiently and exactly thereafter.

\vspace{3mm}\textbf{The general preliminary goal.} In the knowledge of \autoref{thm:mech_pro} and some of its consequences, the distant goal was to strengthen and generalise this, to widen the applicability from practitioners relying on the Heston and NIG models like me to those relying on others.

It was not immediately clear how to generalise \autoref{thm:mech_pro}, so strengthening it appeared to be the better starting point, with the hope that a stronger understanding of it would later reveal how to generalise. Towards this strengthening, notice that while Mechkov's relationship in \autoref{thm:mech_pro} references Heston and NIG \emph{models}, unlike the earlier large-time connection, it is still not a relationship between them. Rather, it should be considered a family of relationships, between the random variables $\{S^n_t\}_{n\ge0}$ relating to each fixed time. 

Nevertheless the convergence $S^n_t\cd S^0_t$ for a fixed time is valuable in practice, because from it we obtain $\EE[\#(S^n_t)]\to\EE[\#(S^0_t)]$ for sufficiently-behaved functions $\#:\RR\to\RR$. As covered in \autoref{sec:martingale}, such a function $\#$ and values like $\EE[\#(S^0_t)]$ can be related to derivative payoffs and prices respectively, under some sensible assumptions. So this convergence tells us how a class of `European' derivative prices will behave in the limit of \autoref{thm:mech_pro}. Indeed, letting $\#$ correspond to a particular (put option) derivative, this confirms the convergence of Heston's implied volatilities to those of the NIG model, as demonstrated in \autoref{fig:vols_converge}.

To widen the applicability to other common, path-dependent, derivatives, we require $\EE[\#(S^n)]\to\EE[\#(S^0)]$ with $\#$ now generalising to a suitably-behaved function from a set $\cX$ of \emph{paths} containing those of $S^n$ and $S^0$. This is provided, almost by definition, by the weak convergence $S^n\cw S^0$ on a metric space $(\cX,d_\cX)$, where $\#:(\cX,d_\cX)\to(\RR,d_\RR)$ must in general be bounded and continuous, and $d_\RR$ can be taken as the usual Euclidean metric on $\RR$. 

Taking this deeper, if we want to understand how derivative payoffs $\#(S^n)$ relate to those of $\#(S^0)$, not just resulting prices hidden behind expectations (i.e.~integrals), we require stronger notions of convergence $S^n\xrightarrow{n\to\infty} S^0$ still, say those of convergence in probability or almost sure (a.s.). Understanding whether this is actually possible is intimately connected to whether the Brownian motions $W^0,W^1$ in both \autoref{eq:heston_prologue2} and \autoref{eq:nig_prologue} are related.

To emphasise difficulties in obtaining these goals, it is worth providing a spoiler. Despite the relevant set $\cX$ of \emph{c\`adl\`ag} paths containing those of both Heston and NIG models, it turns out even weak convergence $S^n\cw S^0$ is violated on all Skorokhod metric spaces. From \cite{Skorokhod_1956}, these spaces appear often in financial stochastic process limit theorems.

There is thankfully a recipe for establishing weak convergence, sometimes called `Prokhorov's approach' after \cite{Prokhorov_1956}, summarised well in \cite{Jacod_2003} and depending on something called `tightness'. But this does not say what to do when things go wrong; when tightness is violated. Unfortunately this is the setting we are in here, despite working with some of the most popular, and relatively simple, models. Indeed, much of my research originated from a need for a different approach to stochastic process limit theorems.

As the title of this thesis suggests, the focus has shifted away from these preliminary goals. It has become about the mathematics developed to obtain them, about a change of perspective on the Heston model, moving away from a dependence on It\^o calculus in favour of random ODEs, and about a resulting robust modelling framework which accommodates generalisations of this Heston and NIG relationship, and eventually rough volatility as well. 

Answers to specific questions here relating to classical Heston models are found in \autoref{sec:price_limits}, where a surprising \emph{interval-valued} generalisation $S^\bullet$ of the NIG process must be introduced,~
\begin{equation}\label{eq:ex_nig_pro}
    X^0_t:=\inf\bigg\{x>0:x-\sigma W^1_x > \theta t\bigg\},\quad S^\bullet_t := \bigg\{\exp\left(W^\rho_x - \frac12 x\right) : x\in[X^0_{t_-},X^0_t] \bigg\}.
\end{equation}
This has a beautifully intimate relationship with the standard NIG process from \autoref{eq:nig_prologue}, which can actually be expressed more compactly as $S^0_t = \exp(W^\rho_{X^0_t} - \frac12 X^0_t)$. The \hyperlink{epilogue}{Epilogue} focuses in on the \emph{origin} of the resulting weak convergence, and is presented from the accessible perspective of It\^o SDEs. There, relationships between the CIR process and several L\'evy processes connected with $X^0$ are established on a new `exit-time' metric space.

%% file: chapters/1_introduction.tex
\clearpage

\setlength{\headsep}{10mm}

\pagestyle{fancy}
\fancyhf{}
\renewcommand{\headrulewidth}{0pt}
% \rhead{\thepage}
\cfoot{\thepage}
% \lhead{\nouppercase\leftmark}
\chead{\textsc{\nouppercase{\leftmark}}}

\section{Introduction}\label{ch:intro}

With primary motivations covered in the \hyperlink{prologue}{Prologue}, this introduction provides an overview of this thesis by chapter and section, highlighting more specific motivations behind these and main results within them. When it is clear to do so, background mathematics and related literature is also provided in this overview, rather than in the following core chapters' context. 

The only background needed at the moment is notational: following convention, $\uC$ and $\uD$ will be used to denote sets of continuous and c\`adl\`ag functions, e.g.~like in \cite{Billingsley_1999}. Extending this, $\uC_0^1$ denotes first-order differentiability and that all functions start from zero.

\vspace{3mm}\textbf{From the Heston model to ODEs.} This thesis is generally presented from foundations to applications. With the preliminary Heston-related goals discussed in the \hyperlink{prologue}{Prologue} constituting one of the final applications, specifically covered in \autoref{sec:price_limits}, the connection of both \autoref{chap:wellposed} and \autoref{chap:solutions} with conventional volatility modelling, let alone the specific Heston model, may not be clear without this introductory explanation. Indeed, given that the volatility of a price is conventionally a probabilistic and model-dependent object, it is not until \autoref{sec:base_frame} that this process is defined within our framework, and not until \autoref{sec:heston_frame} that the earlier probability-free ODE analysis is precisely related to the Heston model. 

We now forgo some precision in order to help develop readers' intuition for how classical stochastic volatility models like Heston's can be related on a pathwise basis to the ODEs treated in this thesis, and therefore how these ODEs can be related to volatility. Towards this, fix a probability space $(\Omega,\cF,\PP)$ which supports a standard 2d Brownian motion $W=(W^0,W^1)$ over $\RR_+$, and let $S=\{S_t\}_{t\in\RR_+}$ be the Heston price process constructed from $W$ and parameters $\sigma,\kappa,\theta,v>0$, $\rho\in[-1,1]$ as in \autoref{eq:heston_prologue}, i.e.~verifying the It\^o SDEs
\begin{equation}\label{eq:heston_intro}
    \dd V_t = \sigma\sqrt{V_t}\dd W^1_t + \kappa(\theta - V_t)\dd t,\quad \dd S_t = \sqrt{V_t}S_t\dd W^{\rho}_t,\quad (V_0,S_0) := (v,1).
\end{equation}
We may explicitly write this CIR SDE for $V$ in its integrated form starting from time zero, and solve the SDE for $S$ in terms of $V$ to obtain the equivalent model representation
\begin{equation}\label{eq:heston_intro2}
    V_t = \sigma\int_0^t\sqrt{V_s}\dd W^1_s + \kappa\int_0^t(\theta - V_s)\dd s + v,\quad S_t = \exp\left(\int_0^t\sqrt{V_s}\dd W^{\rho}_s-\frac12\int_0^t V_s\dd s\right).
\end{equation}
Now using a result due to \cite{Dambis_1965} and \cite{Dubins_1965}, stated precisely in \autoref{thm:DDS} and applied in \autoref{thm:heston_ode}, we may again write this model equivalently as
\begin{equation}\label{eq:tce}
    V_t = \sigma B^1_{\int_0^t V_s\dd s} + \kappa\left(\theta t - \int_0^t V_s\dd s\right) + v,\quad S_t = \exp\left(B^\rho_{\int_0^t V_s\dd s} - \frac12\int_0^t V_s\dd s\right)
\end{equation}
where $B=(B^0,B^1)$ is another standard 2d Brownian motion over $\RR_+$ on $(\Omega,\cF,\PP)$, connected with $W$ according to \autoref{lem:heston_timechange}, and $B^\rho$ is defined like $W^\rho$ by $B^\rho:=\rho B^1+\sqrt{1-\rho^2}B^0$.

The rich history of the `change of time' method which results in the representation of the CIR process $V$ in \autoref{eq:tce} is presented in \cite{Swishchuk_2016}, with the general theory covered concisely in \cite{Barndorff_Nielsen_2010}. Although \cite{Ikeda_1989} makes some important contributions, this line of research originates from the work of Wolfgang Doeblin, in 1940. Before It\^o's calculus, e.g.~the integral of \cite{Ito_1944} and SDEs of \cite{Ito_1951}, Doeblin had shown that diffusions like $V$ admit `time-changed' representations like that in \autoref{eq:tce}, even establishing properties of martingales before the concept existed. This is not widely known because Doeblin unfortunately died in 1940, and his work only discovered in 2000. See \cite{Bru_2002} for the surprising history of Doeblin's work. 

Specifically, in \autoref{eq:tce} we would call the integrated (or cumulative) variance process $\int_0^t V_s\dd s$ a `time-change' of $B^1$. Technically, a time-change must possess certain adaptedness properties related to stopping times, given precisely in \autoref{def:time_change}, but these, and related properties of martingales, are not important for the core analysis of this thesis, especially regarding the Heston and NIG relationship discussed in the \hyperlink{prologue}{Prologue}. So this is where, until \autoref{sec:martingale}, this thesis diverges from research related to changes of time and martingales, because we will more generally consider this process $\int_0^t V_s\dd s$ as a random ODE solution. 

Towards this, there turns out to be nothing special about $B$ in \autoref{eq:tce}, so we can instead construct this model from the arbitrary Brownian motion $W$ on $(\Omega,\cF,\PP)$, and then define the process $X_t:=\int_0^t V_s\dd s$ to obtain another representation of a Heston price process
\begin{equation}\label{eq:tce2}
    X'_t = \sigma W^1_{X_t} + \kappa\left(\theta t - X_t\right) + v,\quad S_t = \exp\left(W^\rho_{X_t} - \tfrac12 X_t\right),
\end{equation}
where $X'_t:=\frac{\dd}{\dd t}X_t = V_t$. Now for each $(t,x)\in\RR^2_+$, define the real random variable $Y_{t,x}:= \sigma W^1_x + \kappa\left(\theta t - x\right) + v$, and the random field $Y=\{Y_{t,x}\}_{(t,x)\in\RR_+^2}$. A random field is just a random element of $\uC(\RR^2_+,\RR)$ which, drawing upon the text \cite{Nielsen_2018}, is defined precisely in \autoref{def:rand_field}. Now write \autoref{eq:tce2} succinctly as
\begin{equation}\label{eq:rand_ode}
    X'_t = Y_{t,X_t},\quad S_t = \exp\left(W^\rho_{X_t} - \tfrac12 X_t\right).
\end{equation}
The process $X$ thus verifies $X'_t=Y_{t,X_t}$ and $X_0=0$, and so we will define it to be a solution of the random IVP $x'=Y_{t,x}$, $x_0=0$, as in \autoref{def:rand_ivp}. It is through the study of such random IVPs that we have been able to strengthen and generalise the Heston and NIG relationship discussed in the \hyperlink{prologue}{Prologue}, and more generally develop a robust volatility modelling framework summarised by \autoref{eq:rand_ode}, but specified more precisely in \autoref{def:price_frame}.

Focusing now just on the process $X$, which we will always call a cumulative variance process even when not explicitly referring to a related price process $S$ like in \autoref{eq:rand_ode}, we may fix an outcome $\omega\in\Omega$ and, defining the fixed function $g:=Y(\omega)\in\uC(\RR^2_+,\RR)$, may analyse the deterministic IVP $x'=g(t,x)$, $x(0)=0$, and in particular look for a path $X(\omega):=\vp\in\uC^1_0(\RR_+,\RR)$ which verifies $\vp'(t)=g(t,\vp(t))$ over $\RR_+$. Then in the Heston case,
\begin{equation}\label{eq:hest_ode}
    g(t,x) := \sigma w(x) + \kappa\left(\theta t - x\right) + v,
\end{equation}
% PART REVISED FOLLOWING VIVA
% where $w:=W^1(\omega)\in\uC_0(\RR_+,\RR)$. Now supposing the path $w(x)$ in \autoref{eq:hest_ode} is replaced by $w(t)$ instead, and for simplicity set $\theta=v$, then we find a unique IVP solution given by
% \begin{equation}
%     \vp(t) = v t + \sigma\int_0^te^{-\kappa(t-s)}w(s)\dd s.
% \end{equation}
% This solution can be interpreted as an integrated Ornstein-Uhlenbeck sample path. This is not immediately helpful for volatility modelling, primarily because $\vp'$ can become negative. New ODE theory is required if we want to understand whether the general Heston case in \autoref{eq:hest_ode} has a unique positive global solution and want to avoid placing impractically restrictive regularity constraints on the path $w\in\uC_0(\RR_+,\RR)$, such as Lipschitz continuity.
where $w:=W^1(\omega)\in\uC_0(\RR_+,\RR)$. At this point, if we were to remove the spatial influence of $w$ on $g$, by replacing $w(x)$ in \autoref{eq:hest_ode} by $w(t)$, then \emph{existing} ODE theory can establish a unique global solution $\vp$, since $g$ becomes spatially Lipschitz. In the reduced case where the reversion level $\theta$ coincides with the starting variance $v$, this unique solution is given by
\begin{equation}
    \vp(t) = v t + \sigma\int_0^te^{-\kappa(t-s)}w(s)\dd s.
\end{equation}
Whether $\theta=v$ or not, this solution can be interpreted as an integrated Ornstein-Uhlenbeck (OU) path, because after this replacement of $w(x)$ by $w(t)$, the counterpart of the CIR SDE in \autoref{eq:heston_intro} for $V$ is the simpler OU SDE which has $\sqrt{V_t}$ removed. It is unfortunate but not surprising that these simplifications, treated by existing theory, are not directly helpful for volatility modelling, because the OU `variance' paths $\vp'=V(\omega)$ can become negative. 

New ODE theory is required if we want to understand whether the general Heston case in \autoref{eq:hest_ode} has a unique positive global solution and want to avoid placing impractically restrictive regularity constraints on the path $w\in\uC_0(\RR_+,\RR)$, such as Lipschitz continuity.

This makes a case for avoiding considerations of the Heston model on this pathwise ODE basis, and sticking to the It\^o SDEs in \autoref{eq:heston_intro}. However, we believe the benefits of this alternative endeavour now speak for themselves. A surprising consequence of our new ODE theory is that, in the Heston case of \autoref{eq:hest_ode}, the IVP $x'=g(t,x)$, $x(0)=0$ actually has a unique maximal solution $\vp$ \emph{for all} $w\in\uC_0(\RR_+,\RR)$. Loosely, a `maximal' solution is any which `reaches the boundary of $\RR_+\!\times\!\RR$', and one which also exists over $\RR_+$ is `global'.

This maximal solution turns out to \emph{always} be strictly increasing, so can always be used to meaningfully model a cumulative variance path $\vp$, with corresponding non-negative volatility $\sqrt{\vp'}$. Given that such functions $g$ need not have any spatial regularity properties, which is clear given the path $w$ in \autoref{eq:hest_ode} may be any in $\uC_0(\RR_+,\RR)$, we call these IVPs spatially irregular, and these can be used to model a counter-intuitively wide set of volatility paths.

In the case of \autoref{eq:hest_ode}, this maximal uniqueness not only extends to a global result if $w$ verifies the condition $\sup_{x\in\RR_+}\kappa x - \sigma w(x) = \infty$, more generally given by $\inf_{x\in\RR_+}g(t,x)<0$ for each $t\in\RR_+$, but the solution $\vp$, which defines a bijection in $\uC^1_0(\RR_+,\RR_+)$, is then bounded above by the strictly increasing c\`adl\`ag path $\bvp\in\uD(\RR_+,\RR_+)$ deriving from $g$ according to 
\begin{equation}\label{eq:cadlag_bound_intro}
    \bvp(t) := \inf\{x>0:g(t,x)<0\}.
\end{equation}
Substituting the representation of $g$ from \autoref{eq:hest_ode}, this c\`adl\`ag path takes the form
\begin{equation}\label{eq:path_bound}
    \bvp(t) := \inf\{x>0: \kappa x - \sigma w(x) > \kappa\theta t + v\}.
\end{equation}
Now replacing the Brownian motion path $w := W^1(\omega)$ with the process $W^1$, and defining the c\`adl\`ag process $\bX$ on a pathwise basis by $\bX(\omega):=\bvp$, then from \autoref{eq:path_bound} we have 
\begin{equation}\label{eq:ig_intro}
    \bX_t := \inf\{x>0: \kappa x - \sigma W^1_x > \kappa\theta t + v\}.
\end{equation}
Consulting \cite{Applebaum_2009}, this process $\bX$ is non other than the IG L\'evy process, also with an IG-distributed random starting point $\bX_0>0$. So relatively succinctly, we have demonstrated how considering the Heston model on a pathwise basis using these spatially irregular ODEs could be helpful in determining how this model is related to the NIG L\'evy process. For clarity, as in \autoref{eq:nig_prologue}, the NIG process depends heavily on the IG process, just like the Heston price process depends on the integrated variance process $X_t:=\int_0^t V_s\dd s$.

The fact that the path $\bvp$ in \autoref{eq:path_bound} is a well-defined element of $\uD(\RR_+,\RR_+)$ for \emph{all} $w\in\uC_0(\RR_+,\RR)$ satisfying this condition of $\sup_{x\in\RR_+}\kappa x - \sigma w(x) = \infty$ provides some more intuition as to how this approach may generalise the Heston and NIG connection. Indeed, we will eventually understand how \emph{any} strictly increasing path in $\uD(\RR_+,\RR_+)$ can be constructed as a limit of our IVP solutions, so also any strictly increasing process $\bX$ with paths in $\uD(\RR_+,\RR_+)$ as a limit of \emph{random} IVP solutions, then any price process of type $S=\exp(W^\rho_\bX-\frac12 \bX)$.

\vspace{2mm}Now that the connection between the spatially irregular ODEs studied in this thesis and pathwise volatility modelling is clear, we move onto a detailed overview of the main results in each chapter and section, starting with the well-posedness foundations of such ODEs.

\vspace{3mm}\textbf{\autoref{chap:wellposed}: Well-posedness for spatially irregular ODEs.} The focus of this chapter is a class of first-order, one-dimensional ODEs $x'=f(t,x)$ where $f$ is a function in $\uC(\RR^2,\RR)$ with some additional simple properties. These properties are captured by the following set.
\begin{definition}[Set $\uF$ of functions]\label{def:driving_func}
    Let the subset $\uF\subset\uC(\RR^2,\RR)$ contain the functions $f$ such that $f(\cdot,x)$ is strictly increasing for every $x\in\RR$, and $f(\tau,\xi)>0$ for some $(\tau,\xi)\in\RR^2$.
\end{definition}

These properties given to the functions in $\uF$ constitute a balance between simplicity and generality when considering various possibilities which we know results in strictly increasing IVP solutions $\vp$. As discussed, this ensures $\vp$ can be used to model a price's cumulative variance path with a meaningful volatility $\sqrt{\vp'}$, given that $\vp'$ is then always non-negative. It is the fact that the functions in $\uF$ need not have any spatial regularity properties, e.g.~$f(t,\cdot)$ need not be Lipschitz or H\"older continuous, which, on the one hand puts these ODEs outside of classical theory, but on the other enables the modelling of a rich set of volatility paths.

Notice that the functions in $\uF\subset\uC(\RR^2,\RR)$ here differ from $g\in\uC(\RR_+^2,\RR)$ defined in \autoref{eq:hest_ode} and related to the Heston model. This is because, at this stage, we do not want to assume that solutions $\vp$ goes through $(0,0)\in\RR^2$, i.e.~verify $\vp(0)=0$, and are thereafter contained in $\RR^2_+$. Treating different initial values $(\tau,\xi)\in\RR^2$ turns out to be delicate, specifically when $f(\tau,\xi)=0$, so this chapter just focuses on IVPs related to $\uF$, with the next chapter simplifying this to IVPs related to functions $g$ in a subset $\uG\subset\uC(\RR_+^2,\RR)$ and with $(\tau,\xi)=(0,0)$, only once $\uF$ and arbitrary initial values $(\tau,\xi)\in\RR^2$ are fully understood.

As with the Heston model in \autoref{eq:heston_intro}, we will always be interested in solutions $\vp$ which go forwards in time from an initial, i.e.~present, state. This state is described by the values $\tau$, $\xi$ and $f(\tau,\xi)$, given the requirements $\vp(\tau)=\xi$ and $\vp'(\tau)=f(\tau,\xi)$ which $\vp$ must verify. Given these values, any `history' of a solution $\vp$ can be considered as being described by parameters like $\sigma,\kappa,\theta,v$ in \autoref{eq:hest_ode}, defining $f$. Only in \autoref{chap:solutions} will we briefly consider such histories over some interval $(T,\tau]$, which must solve a \emph{terminal} value problem (TVP) $x'=f(t,x)$, $x(\tau)=\xi$, to help understand the sensible values of $\tau$, $\xi$ and $f(\tau,\xi)$.

For a given IVP $x'=f(t,x)$, $x(\tau)=\xi$, it will prove vitally important to understand the \emph{maximum} domain $[\tau,T_*)$ over which a solution $\vp$ exists, i.e.~remains finite. This is because, when we move to a probabilistic setting where a stochastic process $X$ solves a random IVP as in \autoref{eq:tce2}, we must prohibit e.g.~having $X_t\xrightarrow{t\to T_*}\infty$ with positive probability for some $T_*<\infty$. This explosion would not only be unnatural, given the possible behaviour of a price process like $S_t=\exp(W^\rho_{X_t} - \frac12 X_t)$ as $t\to T_*$, but mathematically it is then not even clear how to give the sense in which these stochastic processes should actually be considered conventional stochastic processes, i.e.~to provide a function set and $\sigma$-algebra into which these objects define measurable maps from $(\Omega,\cF,\PP)$, so that probability can be conducted.

The problem of this chapter's focus is thus as follows, and given the importance of understanding maximum domains, \emph{maximal} solutions will always be emphasised. These differ from non-maximal solutions only through the final condition on $T_*\vee\sup_{t\in[\tau,T_*)}|\vp(t)|$ here.
\begin{problem}[IVPs of \autoref{chap:wellposed}]\label{prob:ivp}
    For $f\in \uF$ and $(\tau,\xi)\in\RR^2$ where $f(\tau,\xi)>0$, find a maximal solution $\vp\in\uC^1([\tau,T_*),\RR)$ of the IVP $x'=f(t,x)$, $x(\tau)=\xi$. By definition, this means $\vp$ verifies $\vp'(t)=f(t,\vp(t))$ for each $t\in[\tau,T_*)$, $\vp(\tau)=\xi$ and also $T_*\vee\sup_{t\in[\tau,T_*)}|\vp(t)|=\infty$.
\end{problem}

This maximal condition $T_*\vee\sup_{t\in[\tau,T_*)}|\vp(t)|=\infty$ is equivalent to the description that, going forwards in time from $\tau$, the solution $\vp$ `reaches the boundary of $\RR^2$'. Once we know that solutions $\vp$ of \autoref{prob:ivp} are strictly increasing, then we obtain the representation $\sup_{t\in[\tau,T_*)}|\vp(t)|=|\xi|\vee\lim_{t\uparrow T_*}\vp(t)=:|\xi|\vee X_*$, so can start to simply write $T_*\vee X_* = \infty$.

Classical ODE theory dating back to \cite{Peano_1890} establishes that a \emph{maximal} IVP solution as in \autoref{prob:ivp} always exists for any initial conditions $(\tau,\xi)\in\RR^2$, provided $f\in\uC(\RR^2,\RR)$. \cite{Lakshmikantham_1969} can be consulted for this theory, specifically Theorems 1.1.2 and 1.1.3 regarding the existence and `continuation' of solutions respectively. 

In \autoref{sec:main_problem} some important subsets $\uF_\vt\subset \uF$ are introduced, containing functions with a simple additively separable representation $f(t,x)=\vt(t) - w(x)$ for some $\vt,w\in\uC(\RR,\RR)$. The sets $\uF_\vt$ of functions, and the related cases of \autoref{prob:ivp}, are relevant to the entirety of this thesis, so will always be used to help clarify new results. Notice that the Heston function $g$ in \autoref{eq:hest_ode} can be written in a similar additively separable form.

In \autoref{sec:global_bijectivity} we start to build up some properties of maximal solutions $\vp$ of \autoref{prob:ivp}, without assuming that these solutions are unique. But before this, in \autoref{sec:function_zeros}, we focus on just understanding the zeros of any function in $\uF$, i.e.~the points in $\RR^2$ where $f(t,x)=0$, because this understanding helps with many later results. In this section, strictly increasing c\`adl\`ag paths $\bvp$ analogous to that defined in \autoref{eq:cadlag_bound_intro} are introduced, which satisfy $f(t,\bvp(t))=0$ whenever $\bvp(t)<\infty$ and will turn out to bound any solution from above.

Before \autoref{sec:uniqueness} we will understand that, provided we select initial conditions such that $f(\tau,\xi)>0$, then any solution of \autoref{prob:ivp} is indeed strictly increasing, as desired. This ensures that any maximal solution $\vp$ constitutes a bijection in some set $\uC^1([\tau,T_*),[\xi,X_*))$ with $T_*\vee X_*=\infty$, and we provide additional conditions on $f$, consolidated in \autoref{cor:exist_summary}, which ensure that either of $T_*$ or $X_*$ are greater than any chosen value in $\RR$, or are $\infty$.

% PART REVISED FOLLOWING VIVA
% Ensuring that initially $\vp'(\tau)=f(\tau,\xi)>0$ does not mean we find $\vp'(t)>0$ for all $t\in[\tau,T_*)$. This can be clarified with the Heston example in \autoref{eq:hest_ode}, because if we sample $w=W(\omega)$ under the Wiener measure, then it is widely known that the values of $\vp'(t)=V_t(\omega)$ have a noncentral chi-squared distribution, with positive mass at zero whenever $\sigma^2>2\kappa\theta$. 
We can now consider whether the assumption of $f(\tau,\xi)>0$ in \autoref{prob:ivp}, meaning that initially $\vp'(\tau)>0$, leads to such bijections $\vp\in\uC([\tau,T_*),[\xi,X_*))$ satisfying $\vp'(t)>0$ for all $t\in[\tau,T_*)$. We can actually confirm that this is not the case, i.e.~points where $\vp'(t)=0$ may be found, using the Heston example in \autoref{eq:hest_ode}. In this example, the probability of finding $\vp'(t)=0$ when $w=W(\omega)$ is sampled under the Wiener measure coincides with the probability of finding $V_t=0$, where $V$ solves the CIR SDE from \autoref{eq:heston_intro}. However, it is known that this probability is \emph{strictly} positive whenever the CIR SDE's parameters violate the `Feller condition' $\sigma^2\le2\kappa\theta$. See for example \cite{Feller_1968} or \cite{Cox_1985}.

Always working with maximal solutions and therefore dealing with this possibility of finding $\vp'(t)=0$ makes the uniqueness result in \autoref{thm:global_uniqueness} the single most important of this thesis. Being applicable to maximal solutions is what simultaneously takes this result outside of the scope of existing theory and what leads to a robust probabilistic modelling framework. As discussed in \autoref{sec:uniqueness}, the applicable existing theory ends with \cite{Wend_1969}, which applies only where $\vp'(t)>0$ is known. Even in the Heston case, if $\sigma^2>2\kappa\theta$ (as is often required) then there exists no interval $[0,\ep)$ over which $\vp'(t)=V_t(\omega)>0$ a.s., and so no interval over which we have a uniqueness result helpful for probabilistic applications. If we stick to It\^o SDEs, we do have such a result, provided by \cite{Yamada_1971}.

Following this uniqueness section, in this well-posedness chapter we include continuous dependence and simulation convergence results, respectively in \autoref{sec:cont_dep} and \autoref{sec:simulation}. Besides clarifying stability properties of the modelling framework, the former also serves to later define the sense in which our random IVP solutions like $X$ in \autoref{eq:rand_ode} constitute measurable maps from $(\Omega,\cF,\PP)$, so are bona fide stochastic processes. The goal regarding simulation is just to establish that the most basic, easy to implement, forward Euler schemes will always converge to the unique maximal solution, with optimisations left for the future.

\vspace{3mm}\textbf{\autoref{chap:solutions}: The solution space and exit-time limits.} The first goal of this chapter is to provide conditions which preserve the well-posedness properties of the previous chapter while additionally accommodating initial values $(\tau,\xi)\in\RR^2$ where $f(\tau,\xi)=0$. As discussed, we may find $\vp'(t)=f(t,\vp(t))=0$ for some $t>\tau$, so it is reassuring that the conditions of \autoref{thm:include_zero}, applicable to \autoref{prob:ivp}, are ensured if there exists a strictly increasing solution $\vp$ of the ODE $x'=f(t,x)$ over any time interval $(\tau-\ep,\tau]$ which \emph{arrives} at $(\tau,\xi)$, i.e.~$\vp(\tau)=\xi$. This is to say, having $f(\tau,\xi)=0$ is fine provided there exists a meaningful `history' to the present state $(\tau,\xi)$, in which $\vp'(t)\ge0$ and so volatility $\sqrt{\vp'}$ is defined.

The fact that such histories may not be unique, i.e.~\autoref{prob:ivp} always generates a unique \emph{future} solution but this may not have a unique \emph{past}, clarifies that these IVPs are not time-reversible. This may be considered obvious given that the assumption of each $f(\cdot,x)$ being strictly increasing is clearly not time-reversible. Time-related symmetries were famously treated in finance by \cite{Zumbach_2009}, and popularised by \cite{Blanc_2017}. There is now good evidence for processes in finance, like natural physics at large (cf.~second law of thermodynamics), exhibiting time reversal \emph{asymmetry}. Recent accounts of such asymmetries in finance are given in \cite{El_Euch_2018b} and \cite{Cordi_2020}, the reconciliation of which we leave for the future.

The focus in \autoref{sec:refining_problem} becomes the imposition of additional conditions on functions in $\uF$ which ensure solutions of \autoref{prob:ivp} have desirable properties for setting up a probabilistic volatility modelling framework. Primarily, we want unique bijective maximal solutions $\vp\in\uC^1([\tau,T_*),[\xi,X_*))$ to exist for all time and to be spatially unbounded, i.e.~we want $T_*=X_*=\infty$, because the behaviour of a price process $S_t=\exp(W^\rho_{X_t} - \frac12 X_t)$ is undesirable on a path $X(\omega)=\vp$ as $t\uparrow T_*$ otherwise. Having treated the consequences of different initial states, we now w.l.o.g.~fix $(\tau,\xi)=(0,0)$, impose $f(0,0)\ge0$ and define functions only over $\RR^2_+$, like the Heston case in \autoref{eq:hest_ode}. Related to $\uF$, we then arrive at the following set.
\begin{definition}[Set $\uG$ of functions]\label{def:mod_driving_func}
    Let the subset $\uG\subset\uC(\RR_+^2,\RR)$ contain the functions $g$ which are such that: 1.~$g(0,0)\ge0$; 2.~$g(\cdot,x)$ is strictly increasing for each $x\in\RR_+$, and;
    \begin{equation}\label{eq:assumptions_G}
        \text{3.~}\inf_{x\in\RR_+}g(t,x)<0\ \ \forall t\in\RR_+; \quad\text{4.~} \sup_{t\in\RR_+}g(t,x)>0\ \ \forall x\in\RR_+.
    \end{equation}
\end{definition}

Although the set $\uG$ is more complicated to define than $\uF$, the corresponding problem, stated as follows, is simpler to analyse. We now only consider \emph{global} solutions, which are maximal solutions, defined as in \autoref{prob:ivp}, but where the maximum time interval $[\tau,T_*)$ is $\RR_+$.

\begin{problem}[IVPs of \autoref{chap:solutions}]\label{prob:ivp2}
    For $g\in \uG$, find a global solution $\vp\in\uC^1_0(\RR_+,\RR_+)$ of the IVP $x'=g(t,x)$, $x(0)=0$. That is, $\vp$ verifying $\vp'(t)=g(t,\vp(t))$ for $t\in\RR_+$ and $\vp(0)=0$.
\end{problem}

With this problem, the foundations of the remainder of the thesis are in place, i.e.~for a volatility modelling framework in which cumulative variance processes solve the spatially irregular IVPs $x'=g(t,x)$, $x(0)=0$ of \autoref{prob:ivp2} on a pathwise basis. In \autoref{thm:wellposed}, we consolidate important well-posedness results from the previous chapter but applicable to \autoref{prob:ivp2}. It is conditions 3.~and 4.~of $\uG$ which respectively ensure $T_*=\infty$ and $X_*=\infty$, and so any maximal solution of \autoref{prob:ivp2} is automatically global. In \autoref{thm:wellposed} we also clarify that any such global solution $\vp$ is more specifically in the following set of paths.
\begin{definition}[Set $\Phi$ of paths]\label{def:solutionset}
    Let the set $\Phi$ contain the bijective paths in $\uC^1_0(\RR_+,\RR_+)$.
\end{definition}
In \autoref{sec:solution_space} we first focus on the solution set of \autoref{prob:ivp2}, i.e.~on establishing exactly which cumulative variance paths in $\Phi$ can be modelled using these IVPs. This not only turns out to be the entirety of this set, but in \autoref{thm:solutionset} we provide IVP examples depending on subsets $\uG_\vt\subset\uG$ of additively separable functions $g(t,x) := \vt(t) - w(x)$, like the Heston case in \autoref{eq:hest_ode}, which \emph{generates} any specified $\vp\in\Phi$ as the unique global solution of \autoref{prob:ivp2}. Moreover, in \autoref{thm:solutionmap} we show that one can even \emph{fix} a path $\vt$ in $\uC_0(\RR_+,\RR)$ with $\sup_{t\in\RR_+}\vt(t)=\infty$, and still generate any solution $\vp\in\Phi$ which satisfies $\sup_{t\in\RR_+}\vt(t)-\vp'(t)=\infty$, and so any which satisfies the weaker condition $\liminf_{t\to\infty}\vp'(t)<\infty$. This condition is not restrictive for our purposes, given that we would never need to model volatility paths $\sqrt{\vp'}$ for which $\liminf_{t\to\infty}\vp'(t)=\infty$. By this point, we have shown that the IVPs of \autoref{prob:ivp2} are exceedingly well-suited to volatility modelling.

\autoref{sec:exit_topology} is the most important towards answering the preliminary questions in the \hyperlink{prologue}{Prologue}, regarding the Heston and NIG models. Mathematically, this relates to understanding how discontinuous limit points of the set $\Phi$ can arise from simple sequences of solutions of \autoref{prob:ivp2}. The limits of interest are characterised by the following superset of $\Phi$. Uncoincidentally, this set a.s.~contains the paths $\bvp=\bX(\omega)$ of the IG process in \autoref{eq:ig_intro}.
\begin{definition}[Set $\bPhi$ of paths]\label{def:setE}
    Let the superset $\bPhi\supset\Phi$ contain the strictly increasing c\`adl\`ag paths $\bvp$ in $\uD(\RR_+,\RR_+)$ which are also unbounded, i.e.~which verify $\lim_{t\to\infty}\bvp(t)=\infty$.
\end{definition}
For this analysis we specify a new `uniform exit-time' metric $d_{\bPhi}$, in \autoref{def:exit_metric}, on $\bPhi$. Defined via the `exit-time functional' of \autoref{def:exit_functional}, this metric just considers uniform distances in time between the paths in $\bPhi$, rather than in space. As such, it is far simpler to define and work with compared with alternatives from \cite{Skorokhod_1956}, and on $\bPhi$ is stronger than two of the metrics there. We will eventually show that the solution set $\Phi$ of \autoref{prob:ivp2} is dense in $(\bPhi,d_{\bPhi})$, and that this metric space is both separable and complete. 

Because from \autoref{chap:solutions} onwards we will always work over the unbounded domain $\RR_+$ of the unbounded solutions $\vp\in\Phi$ of \autoref{prob:ivp2}, we let our standard metric $d$ on $\uC:=\uC(\RR_+,\RR)$ be defined through the uniform seminorms $\Vert w \Vert_{[0,T]}:=\sup_{t\in[0,T]}|w(t)|$ on $\uC$ according to
\begin{equation}\label{def:uni_metric}
    d(w_1,w_2) := \Vert w_2 - w_1 \Vert_{\RR_+} := \sum_{n\in\NN} 2^{-n}(1\wedge \Vert w_2-w_1 \Vert_{[0,n]}).
\end{equation}
This can be interpreted as a damped uniform norm on the countable product $\!\text{\Large$\times$}_{\!n}\uC([0,n],\RR)$. As such, $(\uC,d)$ is both separable and complete (see appendix M6 of \cite{Billingsley_1999} for succinct proofs), and convergence on $(\uC,d)$ coincides with convergence on all compact restrictions $\uC([0,n],\RR)$. This can be seen by splitting the sum in \autoref{def:uni_metric} to obtain the bounds $\Vert\cdot\Vert_{\RR_+}\le n \Vert \cdot \Vert_{[0,n]} + 2^{-n}$. It is precisely w.r.t.~this uniform convergence over compacts defined through \autoref{def:uni_metric} that the exit-time metric $d_{\bPhi}$ `considers uniform distances in time'. Specifically, for paths $\vp_{1,2}\in\Phi\subset\bPhi$ with the inverses $\vp_{1,2}^{-1}\in\uC$, we have
\begin{equation}
    d_{\bPhi}(\vp_1,\vp_2) = \Vert \vp^{-1}_2 - \vp^{-1}_1 \Vert_{\RR_+}.
\end{equation}
It is also w.r.t.~the topologies of uniform convergence over compacts in $\RR^2_+$ and $\RR_+$, that \autoref{thm:wellposed} establishes the solution map of \autoref{prob:ivp2}, taking each $g$ to the global solution $\vp$ of the IVP $x'=g(t,x)$, $x(0)=0$, to be continuous from $\uG\subset\uC(\RR_+^2,\RR)$ to $\Phi\subset\uC(\RR_+,\RR)$.

In the main limiting result of this chapter, \autoref{thm:exit_limit}, we show how paths $\bvp\in\bPhi$ arise as limits on $(\bPhi,d_{\bPhi})$ of solutions of \autoref{prob:ivp2} of type $x'=ng_n(t,x)$, $x(0)=0$, as $n\to\infty$. Furthermore, in \autoref{cor:construct_limit} we explicitly construct \emph{any} such limit $\bvp\in\bPhi$, which ultimately provides the pathwise foundations of a considerable generalisation of the Heston and NIG limiting relationship from \autoref{thm:mech_pro}. For example, these results explain how Heston cumulative variance paths $\vp_n =: X^n(\omega)$, which solve the IVPs $x'=ng(t,x)$, $x(0)=0$ with $g$ as in \autoref{eq:hest_ode}, converges on $(\bPhi,d_{\bPhi})$ to paths $\bX(\omega) := \bvp$ of the IG L\'evy process from \autoref{eq:path_bound} as $n\to\infty$. Considered as the deepest origin of our findings regarding the motivating questions in the \hyperlink{prologue}{Prologue}, the \hyperlink{epilogue}{Epilogue} clarifies how several L\'evy processes can arise as \emph{weak} limits on $(\bPhi,d_{\bPhi})$ from integrated CIR processes which solve It\^o SDEs like that in \autoref{eq:heston_intro}, rather than solving the related random IVPs like that in \autoref{eq:rand_ode}.

The final goal of this chapter in \autoref{sec:excursionary_limits} is to develop the pathwise theory for understanding the resulting behaviour of price process paths $S(\omega)$ in \autoref{eq:rand_ode} under these exit-time limits $X^n(\omega) := \vp_n\xrightarrow{n\to\infty}\bvp =: \bX(\omega)$ on $(\bPhi,d_{\bPhi})$. Considering that from \autoref{eq:rand_ode} we may write $S_t := \exp(W^\rho_{X_t}-\frac12 X_t) = \Lambda_{X_t}$ where $\Lambda_x:=\exp(W^\rho_x-\frac12 x)$, we must understand the behaviour of composite paths $\{w\circ\vp_n\}_{n\in\NN}$ for some $(w,\vp_n)\in\uC\times\Phi$ as $n\to\infty$.

In general, the pathwise composite convergence $w\circ\vp_n\xrightarrow{n\to\infty}w\circ\bvp$ turns out to be violated on all of Skorokhod's metric spaces, and following \autoref{thm:para_uniform} we show how limits can be understood through the parametric representations $(\vp_n^{-1},w)$, which constitute natural higher-dimensional representations of price process paths. Through \autoref{cor:graph_hausdorff} we then show how graphs of the sequence $\{w\circ\vp_n\}_{n\in\NN}$ can develop instantaneous but finite excursions as $n\to\infty$, and how this sequence converges to a compact interval-valued limit $w\bullet\bvp$, intimately related to $w\circ\bvp$, with respect to a Hausdorff distance between graphs in $\RR_+\times\RR$. 

For general $(w,\bvp)\in\uC\times\bPhi$ these interval-valued limits $w\bullet\bvp$ are defined for each $t\in\RR_+$ by 
\begin{equation}\label{eq:excursion_path}
    (w\bullet\bvp)(t) := \big\{w(x):x\in[\bvp(t_-),\bvp(t)]\big\}
\end{equation}
where as usual $\bvp(t_-):=\lim_{s\uparrow t}\bvp(s)$. Uncoincidentally, these limits are exactly like the paths of the interval-valued generalisation of the exponentiated NIG process $S^0$ from \autoref{eq:ex_nig_pro}, so provide the theoretical foundations to answer and generalise the questions in the \hyperlink{prologue}{Prologue} related to both Heston and NIG price processes \emph{and} derivatives which depend upon these.

\vspace{3mm}\textbf{\autoref{chap:framework}: A pathwise volatility modelling framework.} By this point, all of the pathwise theory is in place to set up a probabilistic volatility modelling framework which can be summarised by the expressions $X'_t = Y_{t,X_t}$ and $S_t = \exp\left(W^\rho_{X_t} - \tfrac12 X_t\right)$ in \autoref{eq:rand_ode}, where $Y=\{Y_{t,x}\}_{(t,x)\in\RR_+^2}$ is a random field a.s.~returning functions in the set $\uG\subset\uC(\RR_+^2,\RR)$. 

Towards this we first specify in \autoref{sec:base_frame} what is meant by a \emph{random} IVP $x'=Y_{t,x}$, $x_0=0$ and a solution. This constitutes a natural generalisation of the deterministic IVP from \autoref{prob:ivp2}, and coincides with the `SP' (sample path) formulation in \cite{Strand_1970}. This general formulation contrasts the focus of applied texts from \cite{Soong_1973} to \cite{Han_2017}, because of the reliance of these on Lipschitz conditions for well-posedness properties. This reliance often reduces the generality of random ODEs considered to cases of type $x'=h(Z_t,x)$, where $h$ is fixed and spatially Lipschitz. This is clearly too restrictive for volatility modelling in general, because even in the Heston case of \autoref{eq:tce2} we instead have $x'=h(t,Z_x)$ where $h(t,Z_\cdot)$ inherits the $\frac12-\ep$ H\"older regularities of Brownian motion.

After stating the random IVPs of our focus in \autoref{prob:random_ivp}, we consolidate consequences of the pathwise results of the previous two chapters applicable to solutions $X=\{X_t\}_{t\in\RR_+}$. For example, \autoref{thm:solutionset_rand} shows that any process $X$ with paths a.s.~in $\Phi$ can be constructed as the unique solution of \autoref{prob:random_ivp}, meaning that we are theoretically able to model any price process of type $S_t = \exp\left(W^\rho_{X_t} - \tfrac12 X_t\right)$. But following \autoref{thm:solutionmap_rand} we make a case for starting with random fields of additively separable type $Y_{t,x} = \vt(t) - Z_x$ for volatility modelling, the solution set of which still contains all $X\in\Phi$ with a.s.~$\liminf_{t\to\infty}X'_t<\infty$.

As an aside, by such statements as $X\in\Phi$ we always mean in the a.s.~sense, i.e.~on the space $(\Omega,\cF,\PP)$ we have $\PP[X\in\Phi]:=\PP[\omega\in\Omega:X(\omega)\in\Phi]=1$. When we start imposing further a.s.~conditions alongside $Y\in\uG$, note that if $\{\Omega_n\}$ are countable subsets of $\Omega$ with full $\PP$-measure, then so is the intersection $\Omega_*:=\cap_{n}\Omega_n$, since, consulting \cite{billingsley_1995},
\begin{equation}\label{eq:full_measure}
    \PP[\cap_n\Omega_n]=1-\PP[(\cap_n\Omega_n)^c]=1-\PP[\cup_n\Omega^c_n]\ge1-\sum_n\PP[\Omega_n^c]=1.
\end{equation}
In this chapter it will always be possible to explicitly define such an intersecting set $\Omega_*$ with full measure for which our analysis and results hold \emph{for every} outcome $\omega\in\Omega_*$. This justifies the description as a pathwise framework, in which all models have probability-free meaning. 

Finally by this point we are ready to fully specify the price process framework summarised by the expression $S_t = \exp\left(W^\rho_{X_t} - \tfrac12 X_t\right)$, in  \autoref{def:price_frame}. Following this we can call the process $\sqrt{X'}$ the volatility of $S$, which of course proves well-defined despite the generality of the cumulative variance process $X\in\Phi$ and the arbitrary relationship of this with $(W^0,W^1)$.

\begin{figure}[H]
    \centering
    \includegraphics[width=0.60\linewidth]{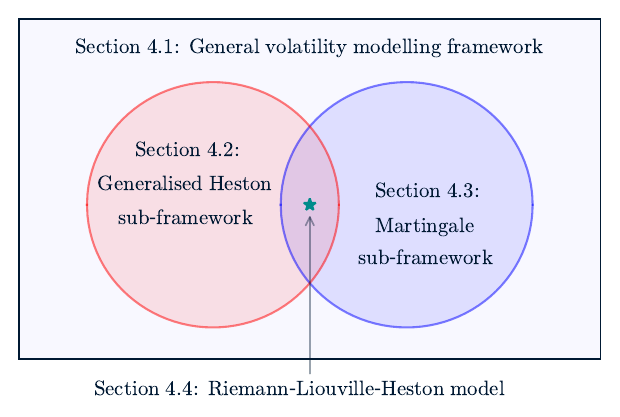}
    \caption{Venn diagram showing the frameworks and model defined in \autoref{chap:framework}.}
    \label{fig:venn}
\end{figure}

The next two sections focus on sub-frameworks of the very general one from \autoref{def:price_frame}, which exhibit certain desirable properties, and \autoref{sec:RLH_model} then introduces a specific model in the intersection of these. This situation is described by the Venn diagram in \autoref{fig:venn}.

\vspace{0mm}
\vspace{-4mm}

More specifically, \autoref{sec:heston_frame} defines a generalised Heston sub-framework in \autoref{def:gen_heston_frame}. In this sub-framework, models for a price $S$ and its cumulative variance $X$ verify equations
\begin{equation}\label{eq:gen_heston_intro}
    X'_t = \sigma Z_{X_t} + \kappa\left(\vt(t) - X_t\right) + v,\quad S_t = \exp\left(W^\rho_{X_t} - \tfrac12 X_t\right),
\end{equation}
where $\sigma,\kappa,v>0$, $\rho\in[-1,1]$ can be interpreted like the usual Heston parameters, $\vt$ is any bijective path in $\uC_0(\RR_+,\RR_+)$ and $Z=\{Z_x\}_{x\in\RR_+}$ a process with paths in $\uC_0(\RR_+,\RR)$. It is clear that this coincides with the classical Heston case in \autoref{eq:tce2} when $\vt(t)=\theta t$ and $Z=W^1$, but, unlike the Heston It\^o SDEs in \autoref{eq:heston_intro}, these models are well-defined over $\RR_+$ for \emph{any} $Z$ a.s.~verifying $\sup_{x\in\RR_+}\kappa x - \sigma Z_x = \infty$. This a.s.~condition simultaneously ensures the implicit random field in \autoref{eq:gen_heston_intro} is a.s.~in $\uG$ and that the c\`adl\`ag process $\bX$ which dominates $X$, analogous to that in \autoref{eq:ig_intro}, exists over the entirety of $\RR_+$.

The power of having this dominating process $\bX$, which derives \emph{directly} from a random field $Y$, is exhibited at the end of this section in \autoref{thm:mgf_existence} and \autoref{cor:gauss_heston_mgf}. These results provide conditions on $Z$ ensuring existence of the moment generating function (MGF) $M_X(p,t):=\EE[e^{pX_t}]$, which is important towards establishing the martingality of $S$ in \autoref{sec:martingale}. The second result focuses on Gaussian processes with sub-linear variance growth, like fractional Brownian motion, which due to \cite{Gatheral_2018} and \cite{Bayer_2015} are gaining prominence in volatility modelling and which we will make use of in \autoref{sec:RLH_model}. 

In \autoref{sec:martingale} the focus is a sub-framework in which all price processes $S$ are martingales. Drawing primarily upon \cite{Cont_2003} and \cite{Guyon_2013}, the importance of martingales for derivative pricing is covered, with a strong emphasis on practicalities. Until this point, no restrictions have been placed on the relationship between a random field $Y$ and the Brownian motion $(W^0,W^1)$, except that both are random elements on $(\Omega,\cF,\PP)$, and so no restrictions on $X$ and $W^\rho$ defining $S$ through $S_t = \exp\left(W^\rho_{X_t} - \tfrac12 X_t\right)$. Culminating with \autoref{them:price_martingale}, we now show how $Y$ should be adapted to the natural filtration $\{\cF_x\}_{x\in\RR_+}$ generated by $(W^0,W^1)$, to ensure that a price $S$ from the framework of \autoref{def:martingale_frame} is a $\cG_t:=\cF_{X_t}$-martingale on the filtered space $(\Omega,\cF,\{\cG_t\}_{t\in\RR_+},\PP)$. 

Consistent with the models introduced thus far, the reality of stochastic interest rates are neglected in this martingale framework. Theoretically, this amounts to the assumption that interest rates are zero, so the usual bank account numeraire $B=\{B_t\}_{t\in\RR_+}$ is constant $B_t=B_0:=1$ and price processes $S$ coincide with their discounted counterpart $B^{-1}S$. See \cite{Brigo_2006} and \cite{Andersen_2010} for backgrounds to numeraires and discounting. In the future, a stochastic interest rate which is \emph{independent} from $W^\rho$ and $X$ may be introduced easily. Otherwise, it would be harmonious to link the interest rate's volatility to the price process's, e.g.~adopting a rate $r_X=\{r_{X_t}\}_{t\in\RR_+}$ adapted to $\{\cG_t\}_{t\in\RR_+}$ and bank account numeraire $B_t:=\exp(\int_0^t r_{X_s}\dd s)$, thus price process $S_t = B_t\exp(W^\rho_{X_t} - \tfrac12 X_t)$.

Returning to some of our motivations summarised by \autoref{eq:tce} and related to Doeblin's work, we clarify in passing that in this martingale sub-framework the random IVP solution $X$ defines a conventional time-change of the Brownian motion $W^\rho$, in the sense of \autoref{def:time_change}, consistent with \cite{Revuz_1999}. A consequence of this is that our general definition $\sqrt{X'}$ of volatility coincides with the most conventional one in It\^o's calculus, depending on quadratic variations $[\cdot]$. Specifically, we can confirm that $[\log S]=X$ a.s.~holds.

At this point all of the probability theory is in place for practitioners to start defining models within our framework, which from the earlier pathwise analysis we know to be both very general and stable when compared with others. In \autoref{sec:RLH_model} the `Riemann-Liouville-Heston' (RLH) model is defined, which showcases both of the generalised Heston and martingale sub-frameworks because it resides in the intersection of these, as per \autoref{fig:venn}.

The idea behind this model is very simple, and is summarised by adapting \autoref{eq:tce} to
\begin{equation}
    X'_t = \sigma W^\alpha_{X_t} + \kappa\left(\theta t - X_t\right) + v,\quad S_t = \exp\left(W^\rho_{X_t} - \tfrac12 X_t\right),
\end{equation}
where we have simply replaced the Brownian motion $W^1$ with its Riemann-Liouville fractional derivative $W^\alpha:=D^\alpha(W^1)$ of some order $\alpha\in(0,\frac12)$, so the classical Heston model is recovered in the boundary case of $\alpha=0$. The fact that this replacement of Brownian motion is possible in our framework, without the need for any additional well-posedness analysis, should not be taken for granted, and is reminiscent of some of the motivations behind rough path theory given in the introductions of \cite{Friz_2010} and \cite{Friz_2014}. Given that, like $W^\alpha$, the variance process $X'$ becomes H\"older regular of orders in $(0,\frac12-\alpha)$, it is clear how to select the fractional derivative $\alpha$ in the RLH model to reproduce the evidence that volatility can exhibit H\"older regularities much lower than Brownian motion.

This model is specified fully in \autoref{def:RLH_model}, but before this a background to Riemann-Liouville fractional derivatives is provided, drawing upon theory from \cite{Hardy_1932} to \cite{Hamadouche_2000}. This theory proves important for establishing the convergence of a simulation scheme for the purpose of derivative pricing, which is the general focus of \autoref{sec:vol_surfaces}. This scheme is used to generate the implied volatilities at the end of \autoref{sec:vol_surfaces}, which are contrasted with those of the classical Heston model, demonstrating desirable features such as power-law scaling of skews (discussed in the \hyperlink{prologue}{Prologue}) and curvatures, as exhibited by leading rough volatility models. Standalone and simplified \texttt{python} code for this simulation scheme is provided in the \hyperlink{appendix}{Appendix}, with seeded output shown in \autoref{fig:simulation_output}.

The final goal is to use the RLH model to illustrate the pathwise limiting results of \autoref{chap:solutions}, and to specialise these to the classical Heston model in order to precisely answer our motivating questions in the \hyperlink{prologue}{Prologue}. The most striking findings in \autoref{sec:price_limits} are as follows. 

First let $S^n$ be the classical Heston process in \autoref{thm:mech_pro}, define $X^0$ to be the IG process in both \autoref{eq:nig_prologue} and \autoref{eq:ex_nig_pro}, and define the c\`adl\`ag and interval-valued processes
\begin{multline}\label{eq:intro_limit_def}
    S^\circ_t := \exp\left(\sqrt{1-\rho^2}W^0_{X^0_t} + \frac{2\rho-\sigma}{2\sigma}X^0_t -\frac{\rho\theta}{\sigma} t\right), \\
    S^\bullet_t := \bigg\{\exp\left(W^\rho_x - \frac12 x\right) : x\in[X^0_{t_-},X^0_t] \bigg\}.
\end{multline}
Then $S^\circ$ is the exp-NIG process in \autoref{thm:mech_pro}, while $S^\bullet$ is a stochastic counterpart of the interval-valued path $w\bullet\bvp$ from \autoref{eq:excursion_path}. The inclusion $S^\circ_t\in S^\bullet_t =:[S^-_t,S^+_t]$ becomes clear following \autoref{lem:NIG_reduc}, which clarifies the counter-intuitive representation $S^\circ_t=\exp(W^\rho_{X^0_t} - \frac12 X^0_t)$, and from which $S^\bullet_t=\{S^\circ_t\}$ almost everywhere (a.e.) follows, meaning $S^-_t=S^\circ_t=S^+_t$. Then, although we show that the convergence in \autoref{thm:mech_pro} can be extended to \emph{either} the convergence of finite-dimensional distributions of $S^n$ to $S^\circ$ \emph{or} the pointwise convergence a.e.~in time on a.e.~path, the graphs of $S^n$ in $\RR_+\times\RR$ actually converge weakly to that of $S^\bullet$ with respect to the Hausdorff distance in \autoref{cor:heston_hausdorff}. So $S^n$ develops compact spatial excursions, of size $S^+_t - S^\circ_t$ upwards and  $S^\circ_t - S^-_t$ downwards, which are almost nowhere but nevertheless dense in time, like the discontinuities of $X^0$. \autoref{fig:hausdorff_composite} helps tremendously to visualise (and validate) this peculiar kind of Hausdorff convergence.

\vspace{3mm}\textbf{\autoref{chap:conclusion}: Conclusion.} Although we did not set out to explore the volatility modelling frameworks of \autoref{chap:framework}, with these instead revealing themselves over several years when considering the questions finally answered in \autoref{sec:price_limits}, we have by this point made a convincing case for the value of our spatially irregular IVPs from \autoref{prob:ivp2} in finance. But actually it is clear that these can theoretically benefit the modelling of any dynamical system, given that the bijective solutions, which we label cumulative variance paths in our context, essentially model time itself, which is by definition central to all dynamical systems. 

Unlike in other fields, the modelling of time itself is a very natural concept in finance, which many authors have exploited, mostly with subordinated L\'evy processes, like \cite{Barndorff_Nielsen_2001}, \cite{Geman_2001}, and \cite{Carr_2004} to name a few. This is because the prices which we aim to model are \emph{fundamentally} observed parametrically, with both the temporal and spatial components of a trade, both appearing random, being indexed by another notion of time captured by deterministic trade identifiers. 

For applications in other fields, the general time-irreversibility of our IVPs, clarified in \autoref{chap:solutions}, is peculiarly consistent with apparent asymmetries in nature between future and past, which is related to the strict increase of entropy and second law of thermodynamics. Of course we leave such exciting general considerations for the future, and in \autoref{chap:conclusion} focus on several more specific ideas for future financial research which this thesis has made possible. This ranges from theoretical `Carath\'eodory' extensions of the ODEs treated here, to the practical implications of the surprising interval-valued limits like $S^\bullet$ in \autoref{eq:intro_limit_def}.

\vspace{3mm}\textbf{\hyperlink{epilogue}{Epilogue}: Integrated CIR-L\'evy relationships.} As discussed briefly already, the \hyperlink{epilogue}{Epilogue} consolidates and generalises what we consider to be the origin of the Heston and NIG limiting relationship, entirely from the perspective of It\^o SDEs. This is presented purely from this more accessible perspective, although our proofs of course require random IVPs. 

To this end, we first drastically over-parameterise Heston's CIR SDE, in accordance with
\begin{equation}
    \dd V^n_t = n^\alpha a \sqrt{V^n_t}\dd W_t + n(b - n^{\beta-1} V^n_t)\dd t,\quad V^n_0=n^\gamma c.
\end{equation}
The exponents $\alpha,\beta,\gamma\in(-\infty,1]$ then control how each term scales as $n\to\infty$ in comparison with the reversionary term $nb$, and specific selections are provided which coincide with the regimes of \cite{Heston_1993}, \cite{Fouque_2011} and \cite{Mechkov_2015}. Depending on the selection of these exponents, \autoref{tab:cir_limits} identifies the eight possible L\'evy processes which arise from the integrated CIR process. Two of these are degenerate, but also two arise with random starting points. So not only does this thesis accommodate continuous, rough and jump models of volatility through the novel application of random IVPs, but here we find randomised models arising as well, as studied in \cite{Mechkov_2016} and \cite{Jacquier_2016}.

%% file: chapters/2_wellposed.tex
\clearpage
\section{Well-posedness for spatially irregular ODEs}\label{chap:wellposed}

The main results of this chapter regard first-order, one-dimensional IVPs $x'=f(t,x)$, $x(\tau)=\xi$, where $f$ belongs to the subset $\uF\subset\uC(\RR^2,\RR)$ from \autoref{ch:intro}, repeated here for convenience. 

\textbf{\autoref{def:driving_func}} (Set $\uF$ of functions)\textbf{.}
    Let the subset $\uF\subset\uC(\RR^2,\RR)$ contain the functions $f$ such that $f(\cdot,x)$ is strictly increasing for every $x\in\RR$, and $f(\tau,\xi)>0$ for some $(\tau,\xi)\in\RR^2$.

The focus of this chapter is well-posedness for these IVPs specifically, which for us means addressing questions related to the existence and uniqueness, continuous dependence and simulation of solutions. Following \cite{Lipschitz_1876} and the extensive line of spatial regularity-based uniqueness theory, e.g.~collected tremendously in \cite{Agarwal_1993}, the well-posedness of \emph{maximal} solutions for such IVPs, i.e.~well-posedness for \autoref{prob:ivp}, has not yet been considered. This is despite ODEs depending implicitly on such functions in $\uF$ appearing in Wolfgang Doeblin's 1940 treatment of diffusions (presented like \autoref{eq:tce} in a `time-changed' form), as discussed in \autoref{ch:intro} and extensively in \cite{Bru_2002}.

Local uniqueness theory for IVPs driven by functions in $\uF$ does exist, and $f(\cdot,x)$ can be relaxed to being non-decreasing even. This line of research can be considered to originate from a simple uniqueness result in \cite{Peano_1890}, but essentially terminates with \cite{Wend_1969}. This is covered in more detail in \autoref{sec:uniqueness}. The most relevant consequence of this terminal article is presented as Theorem 2.6.1 in \cite{Agarwal_1993}, but is omitted from mainstream texts like the classic \cite{Hartman_2002}. The practical problem is \emph{precisely} this locality, which reduces the time interval $[\tau,T)$ of consideration until we know $f(t,\vp(t))>0$ holds for a local solution $\vp\in\uC^1([\tau,T),\RR)$. Given this, a uniqueness proof becomes straightforward, and is accommodated by the general work of \cite{Cid_2009}.

The impracticality of this locality constraint was discussed in \autoref{ch:intro} alongside an IVP deriving from the Heston model in \autoref{eq:heston_intro}, which is important in volatility modelling. Considering \autoref{eq:hest_ode}, fix $\sigma,\kappa,\theta,v>0$ and $w\in\uC_0(\RR,\RR)$, then define $f\in\uF$ using
\begin{equation}\label{eq:cir_ivp_well_posed}
    f(t,x) := \sigma w(x) + \kappa(\theta t - x) + v.
\end{equation}
Then although, for each $w\in\uC_0(\RR,\RR)$, existing theory provides the uniqueness of a local solution $\vp$ to the IVP $x'=f(t,x)$, $x(0)=0$ over some interval $[0,T)$, it is still possible (in fact probable, in volatility modelling) that there is no fixed interval over which this theory can be applied almost surely, when $w$ is drawn under Brownian motion's Wiener measure.

So no fixed interval $[0,T)$ exists over which a coherent volatility model can be defined using existing ODE theory, even in this relatively simple Heston example in \autoref{eq:cir_ivp_well_posed}. This clarifies a shortcoming of local results like \cite{Wend_1969} and \cite{Cid_2009} for probabilistic applications, helping to explain why these are not in the mainstream theory. With this in mind, new results here include statements like the following immediate consequence of \autoref{thm:global_uniqueness}, which does not prohibit finding $\vp'(t)=f(t,\vp(t))=0$ for some $t>\tau$.

\begin{corollary}[Maximal uniqueness]\label{thm:uniqueness_toy}
    Provided $f\in\uF$ and $f(\tau,\xi)>0$, the IVP $x'=f(t,x)$, $x(\tau)=\xi$ has a unique maximal solution. This is to say, \autoref{prob:ivp} has a unique solution.
\end{corollary}

Recall from \autoref{ch:intro} that solutions $\vp$ of such IVPs will model paths of a price process's cumulative variance, with corresponding volatility $\sqrt{\vp'}$. So we are not interested in initial values $(\tau,\xi)\in\RR^2$ where $\vp'(\tau)=f(\tau,\xi)<0$. As in \autoref{thm:uniqueness_toy} above, this chapter further assumes $f(\tau,\xi)>0$, because treating the case of $f(\tau,\xi)=0$ is delicate. This treatment is reserved until \autoref{sec:refining_problem}, where we look closer at the solution set of such IVPs, and confirm that this accommodates all strictly increasing and differentiable paths. The consequences of this for volatility paths $\sqrt{\vp'}$ are deceptively rich. For example, while it is quite clear that $\vp'$ cannot be zero over intervals, \cite{Royden_2010} use pathological examples to show $\vp'$ can still be zero on a set of points arbitrarily close to full Lebesgue measure.

We will shortly provide an `additively separable' class of IVP examples in \autoref{sec:main_problem} which are relevant to the entirety of this thesis, but for now we demonstrate one specific example of the familiar Heston case in \autoref{eq:cir_ivp_well_posed}, to build intuition for the diverse functions in $\uF$.

\begin{figure}[ht]
    \centering
    \includegraphics[height=0.48\linewidth]{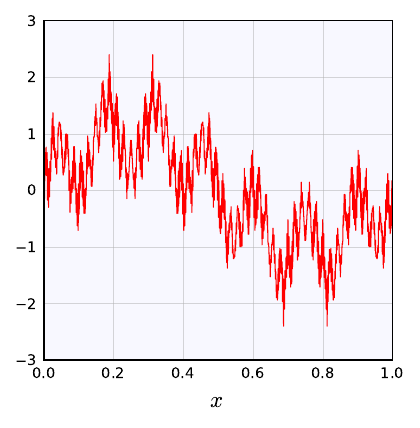}
    \includegraphics[height=0.48\linewidth]{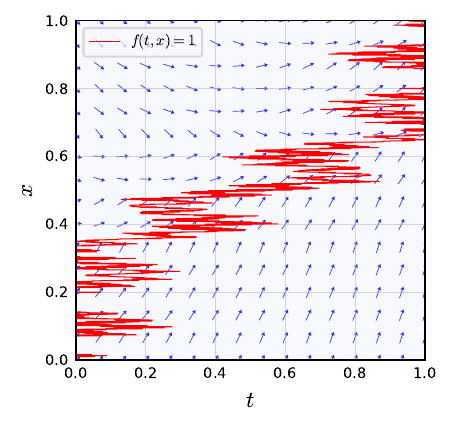}
    \caption{The left panel shows the Weierstrass path $w$ from \autoref{eq:weierstrass_def}, with $a=7$ and $\alpha=0.25$. The right panel demonstrates the corresponding Heston function $f$ (blue arrows) from \autoref{eq:cir_ivp_well_posed}, with $\sigma=0.25$, $\kappa=2$ and $v=\theta=1$. Also shown is the path where $f(t,x)=f(0,0)=1$.}
    \label{fig:weierstrass_example}
\end{figure}

In \autoref{eq:cir_ivp_well_posed} we have the freedom to fix any $w\in\uC_0(\RR,\RR)$, so can choose Karl Weierstrass's pathological function, studied notably in \cite{Hardy_1916}, with its H\"older regularity properties established in \cite{Zygmund_2003}. This admits the Fourier series representation
\begin{equation}\label{eq:weierstrass_def}
    w(x) := \sum_{k\in\NN_0}a^{-\alpha k}\sin(2 a^k\pi x).
\end{equation}
For any $\alpha\in(0,1)$, this series converges provided $a$ is an odd integer greater than 5, by the Weierstrass M-test, and the path $w$ is then nowhere differentiable, but $\alpha$-H\"older continuous. This is demonstrated in \autoref{fig:weierstrass_example}, alongside a corresponding function $f$ from \autoref{eq:cir_ivp_well_posed}. The blue arrows in the right panel, like in the related figures which will follow, provide the direction of the vector $(1,f(t,x))$, to which ODE solutions are necessarily tangential. Notice how the points where $f(t,x)=1$ form a graph over the $x$ axis, which is clearly related to $w$.

% from 20200514-weierstrass-example

Once uniqueness for such highly irregular IVPs is in place, practically relevant properties of the wider modelling framework fall into place. For example, define the truncated functions
\begin{equation}\label{eq:truncated_weierstrass}
    w_n(x):=\sum_{k=0}^n a^{-\alpha k}\sin(2a^k\pi x),\quad f_n(t,x) := \sigma w_n(x) + \kappa(\theta t - x) + v,
\end{equation}
for $n\in\NN$, which unlike $w$ or $f$ can be stored exactly in computer memory. Then, the continuous dependence result of \autoref{thm:continuity_solution_map} establishes that the solutions $\vp_n$ of the IVPs $x'=f_n(t,x)$, $x(0)=0$ will converge to the unique solution of the IVP $x'=f(t,x)$, $x(0)=0$ uniformly over compacts as $n\to\infty$, and the simulation convergence result of \autoref{thm:euler_converge} similarly establishes the convergence of computationally friendly forward Euler polygons.

\vspace{2mm}
Now moving on, the present chapter is structured as follows. \autoref{sec:main_problem} precisely defines the class of spatially irregular IVPs being treated in this chapter, and provides example subsets $\uF_\vt\subset\uF$ of functions containing all of the Heston-related ones referred to thus far. \autoref{sec:function_zeros} takes a step back, analysing the zeros of any function $f\in\uF$. Although not directly related to IVPs, this cannot not be neglected, being important for many of the results which follow. \autoref{sec:global_bijectivity} treats the maximal existence of IVP solutions, but also establishes some basic properties of solutions, like them being strictly increasing, important for volatility modelling. \autoref{sec:uniqueness}, \autoref{sec:cont_dep} and \autoref{sec:simulation} then focus specifically on the uniqueness, continuous dependence and simulation of maximal solutions respectively.

\subsection{The main problem and examples}\label{sec:main_problem}

The programme of this section is to first reiterate the class of IVPs discussed in \autoref{ch:intro} depending on the set $\uF\subset\uC(\RR^2,\RR)$ of functions, to which the main results in this chapter will apply, and to then provide simple subsets $\uF_\vt\subset\uF$ of examples which accommodate the IVPs mentioned thus far, e.g.~those deriving from the Heston case in \autoref{eq:cir_ivp_well_posed}.

Recall that it is \autoref{prob:ivp} to which this chapter applies, repeated here for convenience.

\textbf{\autoref{prob:ivp}} (IVPs of \autoref{chap:wellposed})\textbf{.}
    For $f\in \uF$ and $(\tau,\xi)\in\RR^2$ where $f(\tau,\xi)>0$, find a maximal solution $\vp\in\uC^1([\tau,T_*),\RR)$ of the IVP $x'=f(t,x)$, $x(\tau)=\xi$. By definition, this means $\vp$ verifies $\vp'(t)=f(t,\vp(t))$ for each $t\in[\tau,T_*)$, $\vp(\tau)=\xi$ and also $T_*\vee\sup_{t\in[\tau,T_*)}|\vp(t)|=\infty$.

Some minor points are in order. Firstly, through the statement of \autoref{prob:ivp} it is clear we are only seeking solutions $\vp$ defined over some set $[\tau,T_*)$, i.e.~extending \emph{forwards} in time from initial conditions $(\tau,\xi)$. As discussed in \autoref{ch:intro}, `histories' extending backwards in time will be considered in \autoref{chap:solutions}. As such we should interpret $\vp'(\tau)$ as a right derivative. 

Recall that should we find a solution $\vp$ in a set $\uC^1([\tau,T_*),\RR)$ for some $T_*\in(\tau,\infty]$, then it is only the condition $T_*\vee\sup_{t\in[\tau,T_*)}|\vp(t)|=\infty$ which distinguishes this as a \emph{maximal} solution. This condition means $\vp$ must extend to the boundary of $\RR^2$, i.e.~as far as possible. For $\vp$ to be such a solution, thereby verifying $\vp'(t)=f(t,\vp(t))$ over $[\tau,T_*)$ and $\vp(\tau)=\xi$, is equivalent to $\vp$ verifying the integral equation $\vp(t)=\xi+\int_\tau^t f(s,\vp(s))\dd s$. Although this requires proof, this can be found in any ODE text, e.g.~p.2 of \cite{Coddington_1955}.

Finally, as in the title of this thesis, we refer to the IVP in \autoref{prob:ivp} as `spatially irregular', because we have not imposed regularity conditions, such as Lipschitz or H\"older continuity, on the spatial component of the functions in $\uF$. `Temporally strictly increasing and spatially irregular' certainly provides a more complete description. But, loosely, although our IVPs being strictly increasing in their temporal variable renders them \emph{helpful} for volatility modelling, i.e.~ensures that volatility $\sqrt{\vp'}$ is well-defined given a solution $\vp$, it is the spatial irregularity of these IVPs which thereafter governs \emph{how helpful}, via the IVPs' solution set, and distinguishes the resulting modelling framework from comparatively restrictive others. 

For example, due to this irregularity, the solution set becomes so large that, unlike in the conventional framework depending on It\^o SDEs, the highly irregular paths of volatility observed in reality, with very low, time-varying, or even no apparent  H\"older regularity, can be accommodated. See \cite{Bennedsen_2016} for extensive empirical evidence of such volatility paths, and the recent research into `hyper' or `super' rough volatility models, such as \cite{Jusselin_2020} and \cite{Bayer_2020}. Hence our emphasis on a `spatially irregular' description, given our focus on volatility modelling.

We now look at some examples of \autoref{prob:ivp}, in which the functions in $\uF$ admit the additively separable representation $f(t,x)=\vt(t)-w(x)$ for some $\vt,w\in\uC(\RR,\RR)$. These examples may seem related to the work of \cite{Kaper_1988} at first, wherein the authors consider similarly separable functions. But the assumptions there are actually very different, e.g.~$w$ is assumed monotone and differentiable except at the IVP starting point.

\begin{example}[The subsets $\uF_\vt\subset\uF$]\label{ex:main_example}
    Let the path $\vt\in\uC_0(\RR,\RR)$ be strictly increasing and bijective, so $\lim_{t\to\pm\infty}\vt(t)=\pm\infty$, then let the set $\uF_\vt$ contain functions with representation
    \begin{equation}\label{eq:main_example}
        f(t,x) := \vt(t) - w(x)
    \end{equation}
    for some $w\in\uC(\RR,\RR)$ with $w(0)<0$ and $\sup_{x\in\RR_+}w(x)=\infty$. For any such $\vt$, the inclusion $\uF_\vt\subset\uF\subset\uC(\RR^2,\RR)$ is then quite clear using \autoref{def:driving_func}, given that each $f(\cdot,x)$ is strictly increasing. Moreover, given that $f(0,0)>0$ is ensured, then for any $f\in\uF_\vt$ the IVP $x'=f(t,x)$, $x(0)=0$ provides an example of \autoref{prob:ivp}, specifically with $(\tau,\xi)=(0,0)$.
\end{example}

The assumptions related to $\infty$ in \autoref{ex:main_example} can be ignored for now, but are tremendously helpful later. We will specifically use the assumption  $\sup_{x\in\RR_+}w(x)=\infty$ (which is e.g.~satisfied almost surely by paths of Brownian motion) to guarantee maximal solutions of IVPs $x'=f(t,x)$, $x(0)=0$ are always global, i.e.~remain finite over all of $\RR_+$, for example. The unnecessary minus sign used in \autoref{eq:main_example} will be justified later as well, in \autoref{ex:zeros_example}.

Letting $\Theta$ contain the functions $\vt$ in \autoref{ex:main_example}, then to see that we \emph{do not} have the set equivalence $\cup_{\vt\in\Theta}\uF_\vt=\uF$ we can consider functions of type $f(t,x) = \vt_1(\vt_2(t) - w(x))$, where $\vt_{1,2}\in\Theta$, and also of multiplicative type $f(t,x) = \vt(t)w(x)$ provided $w$ is strictly positive. We will not explore such functions in our applications because we have no reason to believe they would be more helpful for volatility modelling than those in $\uF_\vt$. Notice the Heston function in \autoref{eq:cir_ivp_well_posed} is found in $\uF_\vt$ with $\vt(t):=\kappa\theta t$, provided $\sup_{x\in\RR_+}\kappa x - \sigma w(x)=\infty$.

% 20200514-weierstrass-example
\begin{figure}[H]
    \centering
    \includegraphics[height=0.48\linewidth]{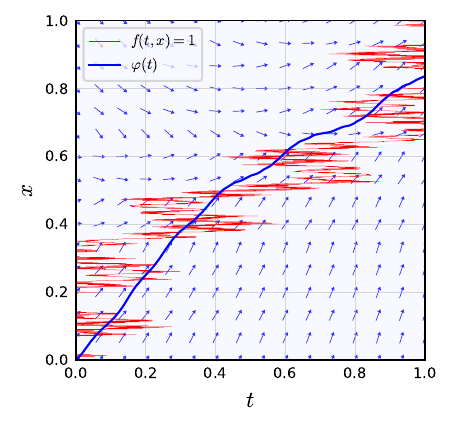}
    \includegraphics[height=0.48\linewidth]{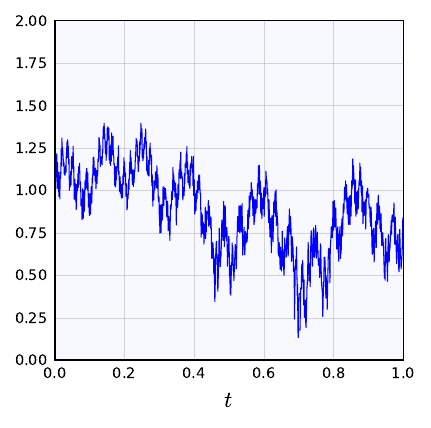}
    \caption{The left panel repeats the right one in \autoref{fig:weierstrass_example} but includes the solution $\vp$ of the IVP $x'=f(t,x)$, $x(0)=0$. The right shows the corresponding `rough' volatility path $\sqrt{\vp'}$.}
    \label{fig:main_examples}
\end{figure}

\vspace{-4mm}
In \autoref{fig:main_examples} we illustrate a \emph{solution} $\vp$ of an IVP $x'=f(t,x)$, $x(0)=0$ where $f\in\uF_\vt$, using the Heston function from the right panel of \autoref{fig:weierstrass_example}, so $\vt(t):=\kappa\theta t$. Also shown in \autoref{fig:main_examples} is the corresponding volatility path $\sqrt{\vp'}$, which clearly inherits properties of the driving Weierstrass function from the left panel of \autoref{fig:weierstrass_example}. Of course both $\vp$ and $\vp'$ must be approximated using a simulation scheme, for which we use that from \autoref{alg:forward_euler}.

% 20200514-weierstrass-example
\begin{figure}[H]
    \centering
    \includegraphics[height=0.48\linewidth]{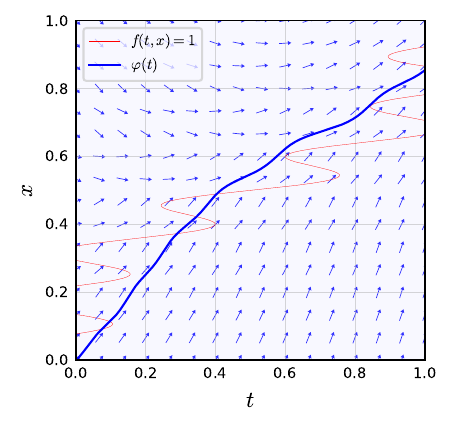}
    \includegraphics[height=0.48\linewidth]{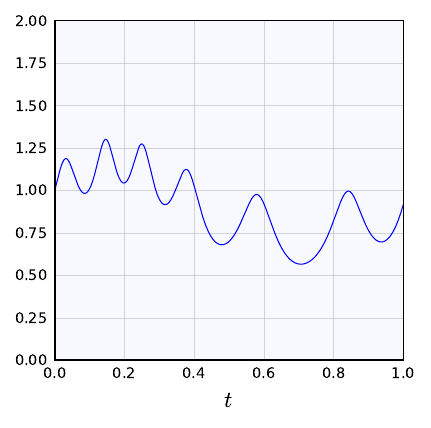}
    \caption{A reproduction of \autoref{fig:main_examples}, but using the truncated differentiable path $w_n$ from \autoref{eq:truncated_weierstrass} with $n=1$, rather than the $\alpha$-H\"older Weierstrass limit as $n\to\infty$.}
    \label{fig:main_examples2}
\end{figure}

\vspace{-3mm}

We now move on to consider the zeros of a function $f\in\uF$, i.e.~the points in $\RR^2$ where $f(t,x)=0$, which are related to the points where $f(t,x)=1$, as shown in \autoref{fig:main_examples}. Notice that for $f\in\uF_\vt$, these zeros verify the simple equation $\vt(t)=w(x)$, justifying the minus sign in \autoref{eq:main_example}. So these zeros also verify $t = \vt^{-1}(w(x))$, given $\vt\in\uC_0(\RR,\RR)$ is bijective.

\subsection{Driving functions' zeros}\label{sec:function_zeros}

For $f\in\uF$, understanding the points in $\RR^2$ where $f(t,x)=0$ turns out to be incredibly fruitful. For example, some basic properties of IVP solutions established in this chapter, the limit theorems of \autoref{chap:solutions} and the martingality result in \autoref{chap:framework} all depend on the c\`adl\`ag path $\bvp$ defined in \autoref{lem:zeros2} here, related to the path defined earlier in \autoref{eq:path_bound}. Because every $f(\cdot,x)$ is strictly increasing, these zero points can be characterised by a single path $\phi$, as covered in \autoref{lem:zeros1}. However, because it is also possible that no point $t\in\RR$ exists where $f(t,x)=0$ for a given $x\in\RR$, we choose to extend the image of $\phi$ beyond $\RR$.

So let $\bRR:=\RR\cup\{\pm\infty\}$ denote the extended real line, and $[x,\infty]:=[x,\infty)\cup\infty$ etc.~for $x\in\RR$. We equip $\bRR$ (and subintervals) with the standard topology (or `two-point compactification') which is homeomorphic, for example, to the Euclidean topology on $[-1,1]$. This can be induced by the metric $d(a,b):=|\tanh(b)-\tanh(a)|$ on $\bRR$, with $\tanh(\pm\infty):=\pm 1$, for example. \cite{Aliprantis_1998} can be consulted for more details. Now denote by $\uC(\RR,\bRR)$ and $\uD(\RR,\bRR)$ the sets of paths which are respectively continuous and c\`adl\`ag given these topologies.

Adopt the conventions $\sup\varnothing:=-\infty$ and $\inf\varnothing:=\infty$. This enables the compact definition of $\phi\in\uC(\RR,\bRR)$ in \autoref{lem:zeros1}, although it is informative to check that this is equivalent to
\begin{equation}\label{eq:define_phi_alt}
\phi(x) := 
\begin{cases}
    -\infty & f(t,x)>0\ \forall t \\
    +\infty & f(t,x)<0\ \forall t \\
    t\in\RR \text{ such that } f(t,x)=0 & \text{otherwise}.
\end{cases}
\end{equation}
This next result \autoref{lem:zeros1} is related to the implicit function theorem e.g.~presented as in Theorem 9.28 of \cite{Rudin_1976}, although we do not assume differentiability of $f$. Our proof relies upon the Bolzano-Weierstrass theorem for sequences (a bounded sequence in $\RR$ has a convergent subsequence), which is given as Theorem 3.4.8 in \cite{Bartle_2000}.

\begin{lemma}[Path of zeros]\label{lem:zeros1}
    For $f\!\in\!\uF\!\subset\!\uC(\RR^2,\RR)$, define the function $\phi=\phi_f\!:\!\RR\to\bRR$ by
    \begin{equation}\label{eq:define_phi}
    \phi(x) := \sup\{t\in\RR: f(t,x) <0 \}.
    \end{equation}
    Then $\phi$ is a well-defined path in $\uC(\RR,\bRR)$, verifying $f(\phi(x),x)=0$ whenever $\phi(x)\in\RR$.
\end{lemma}
\begin{proof}
    It is clear, using \autoref{eq:define_phi}, that $\phi:\RR\to\bRR$ is a well-defined function, and the three cases in the representation of \autoref{eq:define_phi_alt} just follow from $\sup\varnothing:=-\infty$, $\sup\RR=\infty$ and the continuity of each strictly increasing $f(\cdot,x)$ respectively. Whenever $\phi(x)\in\RR$, then we are in the third case in \autoref{eq:define_phi_alt}, so clearly $f(\phi(x),x)=0$, and to establish the claim it just remains to show this function $\phi$ is in $\uC(\RR,\bRR)$. The proof here uses limits of sequences. 
    
    So let $\{x_n\}_{n\in\NN_0}\subset\RR$ be a sequence with $x_n\xrightarrow{n\to\infty}x_0$ in $\RR$, but, for a contradiction, assume that $\phi(x_n)\xrightarrow{n\to\infty}\phi(x_0)$ is violated in $\bRR$. This divergence provides an open ball $\BB\subset\RR$ around $\phi(x_0)$ such that infinitely many $\phi(x_n)$ are in $\bRR\setminus\BB$. By considering a subsequence of $\{x_n\}_{n\in\NN_0}$ if necessary, we can therefore w.l.o.g.~assume that $\phi(x_n)\in\bRR\setminus\BB$ for every $n\in\NN_0$. 
    
    Since the topology on $\bRR$ is homeomorphic to the Euclidean one on $[-1,1]$, the Bolzano-Weierstrass theorem provides a subsequence $\{x_{n_k}\}_{k\in\NN}$ verifying $\phi(x_{n_k})\xrightarrow{k\to\infty}t_0\in\bRR$. Importantly, $t_0\neq\phi(x_0)$ follows from having ensured $\phi(x_n)\in\bRR\setminus\BB$ for every $n$, opposing $\phi(x_0)\in\BB$. By again redefining $\{x_n\}_{n\in\NN_0}$ if necessary, we may assume w.l.o.g.~that $\phi(x_n)\xrightarrow{n\to\infty}t_0\neq\phi(x_0)$ holds. Now assume $t_0<\phi(x_0)$, and fix any $t_*\in(t_0,\phi(x_0))\subseteq\RR$.
    
    Using \autoref{eq:define_phi_alt}, if $\phi(x_0)=\infty$, then we have $f(t_*,x_0)<0$ by definition, and if $\phi(x_0)<\infty$, then $f(t_*,x_0)<f(\phi(x_0),x_0)=0$ follows from $f(\cdot,x_0)$ being strictly increasing. So $f(t_*,x_0)<0$ is ensured. Since $\phi(x_n)\xrightarrow{n\to\infty}t_0<t_*$, we can assume w.l.o.g.~that $\phi(x_n)<t_*$ for each $n$. Like previously, if $\phi(x_n)=-\infty$, then $f(t_*,x_n)>0$ by definition, and if $\phi(x_n)>-\infty$, then $0=f(\phi(x_n),x_n)<f(t_*,x_n)$ follows from each $f(\cdot,x_n)$ being strictly increasing. So $f(t_*,x_n)>0$ is ensured for each $n$. The continuity of $f$ now provides the contradiction
    \begin{equation}
        0<f(t_*,x_n)\xrightarrow{n\to\infty}f(t_*,x_0)<0.
    \end{equation}
    
    The analysis for $t_0>\phi(x_0)$ is practically identical, with these inequalities reversed. Due to these contradictions, the convergence $\phi(x_n)\xrightarrow{n\to\infty}t_0:=\phi(x_0)$ in $\bRR$ must hold. So, for any sequence $\{x_n\}_{n\in\NN_0}$, the convergence $x_n\xrightarrow{n\to\infty}x_0$ in $\RR$ implies $\phi(x_n)\xrightarrow{n\to\infty}\phi(x_0)$ in $\bRR$. This is equivalent to the outstanding claim of $\phi\in\uC(\RR,\bRR)$, so the proof is thus complete.
\end{proof}

The path $\phi\in\uC(\RR,\bRR)$ clearly \emph{characterises} the zeros of $f$, and an entire set $\{\phi_a\}_{a\in\RR}$ of paths could be similarly defined, where $f(\phi_a(x),x)=a$. The red line in \autoref{fig:main_examples} where $f(t,x)=1$ thus coincides with $\phi_1(x)$. With some work, using differential inequalities, each such path $\phi_a\in\uC(\RR,\bRR)$ can be used to construct a bound on any solution $\vp$ of the corresponding IVP $x'=f(t,x)$, $x(\tau)=\xi$. We will just focus on $\phi = \phi_0$, which specifically leads to the path $\bvp$ defined properly in \autoref{lem:zeros2}, which is established later as a solution bound in \autoref{lem:spatial_bounds}.

For $\phi\in\uC(\RR,\bRR)$ and $(\tau,\xi)\in\RR^2$, the next result \autoref{lem:zeros2} utilises `exit-time' notation $E(\phi) = E_{\tau,\xi}(\phi)$, which refers to the function defined over $[\tau,\infty)$, with $\inf\varnothing:=\infty$, through
\begin{equation}\label{eq:exit_def}
    E(\phi)(t) := \inf\{x>\xi:\phi(x)>t\}.
\end{equation}
The exit-time \emph{functional} $E$ is specified properly in \autoref{def:exit_functional}, but these details are superfluous now. It is however very helpful to note some properties of the function $E(\phi)$ defined in \autoref{eq:exit_def}, which are analysed extensively in Section 13.6 of \cite{Whitt_2002}.

To this end, allow $\phi$ to be any path in $\uD(\RR,\RR)$ verifying $\sup_{x\in[\xi,\infty)}\phi(x)=\infty$. Then it is clear that $E(\phi)(\tau)\ge\xi$, and indeed $E(\phi)(t)\in[\xi,\infty)$ for each $t\in[\tau,\infty)$, given that $\sup_{x\in[\xi,\infty)}\phi(x)=\infty$. It is also clear that $E(\phi)$ is non-decreasing. Less clear is that $E(\phi)$ is c\`adl\`ag, so defines a path in $\uD([\tau,\infty),[\xi,\infty))$. Towards this, notice that left limits exist at any $t_*\in[\tau,\infty)$ since, over $[\tau,t_*)$, $E(\phi)$ is monotone and bounded in $[\xi,E(\phi)(t_*)]$. Similarly for right limits. Right continuity is best observed by contradiction: fixing any $x_*:=E(\phi)(t_*)$, then due to the use of `$\phi(x)>t$' in \autoref{eq:exit_def} (opposing `$\phi(x)\ge t$'), the inequality $\sup_{x\in[x_*,x_*+\ep)}\phi(x)>\phi(x_*)$ holds for every $\ep>0$. If a right discontinuity is assumed, namely a point $t_*\in[\tau,\infty)$ where $x_*^+:=\lim_{t\downarrow t_*}E(\phi)(t)>E(\phi)(t_*)=:x_*$, then $\sup_{x\in[x_*,x_*+\ep)}\phi(x)>\phi(x_*)$ is violated for every $\ep\in(0,x_*^+-x_*)$. Finally, there are two situations which can render $E(\phi)$ non-\emph{increasing} over an interval: either $\phi(\xi)>\tau$, then the interval is $[\tau,\phi(\xi))$, or given an upward discontinuity $\lim_{x\uparrow x_*}\phi(x)=:\phi(x_*^-)<\phi(x_*)$, then the interval is $[\phi(x_*^-),\phi(x_*))$. Assuming $\phi(\xi)\le\tau$ and $\phi\in\uC(\RR,\RR)$ therefore precludes such intervals, making $E(\phi)$ \emph{strictly} increasing. Continuous mapping properties of the related functional $E$ are also obtained in \cite{Whitt_1971}, and exploited in \cite{Puhalskii_1997}.

In what follows, minor extensions to these observations will be made, specifically to accommodate the two main differences here, following \autoref{lem:zeros1}, where $\phi$ is in $\uC(\RR,\bRR)$ rather than $\uC(\RR,\RR)$, and where $\sup_{x\in[\xi,\infty)}\phi(x)<\infty$ is possible rather than $\sup_{x\in[\xi,\infty)}\phi(x)=\infty$.

\begin{lemma}[C\`adl\`ag zeros]\label{lem:zeros2}
    Adopt the assumptions of \autoref{lem:zeros1}. Then for any initial value $(\tau,\xi)\in\RR^2$ where $f(\tau,\xi)\ge0$, define the function $\bvp=\bvp_{f,\tau,\xi}:[\tau,\infty)\to[\xi,\infty]$ by
    \begin{equation}\label{eq:up_bound}
        \bvp(t):=\inf\{x>\xi:f(t,x)<0\}.
    \end{equation}
    Then $\bvp$ is a well-defined increasing path in $\uD([\tau,\infty),[\xi,\infty])$, verifying $\bvp = E(\phi) = E_{\tau,\xi}(\phi)$. Moreover, if $\bvp(T)<\infty$, then $\bvp$ is \emph{strictly} increasing and verifies $f(t,\bvp(t))=0$ over $[\tau,T]$.
\end{lemma}
\begin{proof}
    Using \autoref{eq:up_bound}, for $t_*\in[\tau,\infty)$, either $\bvp(t_*)=\inf\varnothing:=\infty$ or, by the continuity of $f$, $\bvp(t_*)$ is the lowest value $x_*\in[\xi,\infty)$ where $f(t_*,x)<0$ for all $x$ in some $(x_*,x_*+\ep)$. This clarifies that $\bvp:[\tau,\infty)\to[\xi,\infty]$ is a well-defined function. If this lowest value $x_*=\bvp(t_*)<\infty$ indeed exists, then $f(t_*,x)\ge0$ for $x$ in $[\xi,x_*]$, while $f(t_*,x)<0$ for $x\in(x_*,x_*+\ep)$, so the continuity of $f$ also ensures $f(t_*,x_*)=f(t_*,\bvp(t_*))=0$. Since $\phi$ from \autoref{lem:zeros1} characterises the zeros of $f$, with $f(\phi(x),x)=0$ when $\phi(x)\in\RR$, then $t_* = \phi(x_*)\in\RR$.
    
    Now the equivalence $\bvp = E(\phi)$ will be established, meaning $\bvp(t)=\inf\{x>\xi:\phi(x)>t\}=:E(\phi)(t)$ over $[\tau,\infty)$. First assume $x_*=\bvp(t_*)<\infty$. If $\phi(x)<\infty$ for $x\in(x_*,x_*+\ep)$, then the ordering $f(t_*,x)<0=f(\phi(x),x)$ holds, and $t_*=\phi(x_*)<\phi(x)$ then follows from $f(\cdot,x)$ being strictly increasing. But clearly $\phi(x_*)<\phi(x)$ also holds if $\phi(x)=\infty$. Now assuming that $x_*\in[\xi,\infty)$ is \emph{not the lowest} value with $\phi(x)>\phi(x_*)$ for $x$ in some $(x_*,x_*+\ep)$, so \emph{not} equal to $E(\phi)(t_*)$, contradicts $x_*=\bvp(t_*)$ being the lowest value in $[\xi,\infty)$ where $f(t_*,x)<0$ for $x\in(x_*,x_*+\ep)$, again using that $f(\cdot,x)$ is strictly increasing. This establishes the equivalence $\bvp(t_*)=E(\phi)(t_*)$ for $t_*\in[\tau,\infty)$ \emph{when} $\bvp(t_*)<\infty$. If instead $\bvp(t_*)=\infty$, so $f(t_*,x)\ge0$ for $x\in[\xi,\infty)$, then $\phi(x)\le t_*$ holds over the same interval $[\xi,\infty)$, and $E(\phi)(t_*)=\infty$. This clarifies that $\bvp=E(\phi)$ holds whether $\infty$ is attained or not.
    
    It just remains to show that $\bvp=E(\phi)$ is in $\uD([\tau,\infty),[\xi,\infty])$ and is increasing or strictly increasing as claimed. The discussion following \autoref{eq:exit_def} clarifies that $E(\phi)$ is strictly increasing and in $\uD([\tau,\infty),[\xi,\infty))$ when $\phi\in\uC(\RR,\RR)$, $\phi(\xi)\le\tau$ and $\sup_{x\in[\xi,\infty)}\phi(x)=\infty$. 
    
    In our case, $\phi(\xi)\le\tau$ clearly holds if $\phi(\xi)=-\infty$, and otherwise is ensured by $f(\phi(\xi),\xi)=0\le f(\tau,\xi)$ and $f(\cdot,\xi)$ being strictly increasing. The consequence of having $\phi\in\uC(\RR,\bRR)\supset\uC(\RR,\RR)$ is just that a point $x_*\in(\xi,\infty)$ could exist where $\lim_{x\uparrow x_*}\phi(x)=\infty$. But assuming $x_*$ to be the lowest such point, then $E(\phi)$ is clearly just found in $\uD([\tau,\infty),[\xi,x_*))$. Likewise, the effect of having $\sup_{x\in[\xi,\infty)}\phi(x)=t_*<\infty$ is simply that $E(\phi)(t)=\infty$ for all $t\in[t_*,\infty)$, using $\inf\varnothing:=\infty$, meaning now $E(\phi)$ is found in $\uD([\tau,\infty),[\xi,\infty])$. In all cases, $\bvp=E(\phi)$ remains in $\uD([\tau,\infty),[\xi,\infty])$, is strictly increasing over any $[\tau,T]\subset[\tau,\infty)$ provided $\bvp(T)<\infty$, and is constant over any $[T,\infty)\subset[\tau,\infty)$ if $\bvp(T)=\infty$, therefore the proof is complete.
\end{proof}

Showing that the path $\phi$ of zeros from \autoref{lem:zeros1} is in $\uC(\RR,\bRR)$ might seem superfluous, but this route appears to be the cleanest towards establishing properties of the path $\bvp$ in \autoref{lem:zeros2}, which are not at all obvious, yet critically important. As an example, if $\phi$ was only continuous into the weaker `one-point compactification' of $\RR$, so e.g.~could jump between $-\infty$ and $\infty$, then the resulting path $\bvp$ is not guaranteed to be strictly increasing.

For our volatility modelling applications, it would be acceptable to assume that any function $f\in\uF\subset\uC(\RR^2,\RR)$ is such that each $f(\cdot,x)$ defines a strictly increasing \emph{bijection} from and to $\RR$, like the Heston case in \autoref{eq:cir_ivp_well_posed} and all functions in the subsets $\uF_\vt\subset\uF$ from \autoref{ex:main_example}. In this case, it is straightforward to show that the zero path $\phi$ from \autoref{eq:define_phi} is in $\uC(\RR,\RR)$. Although tempting to make this bijectivity assumption, the extended reals $\bRR$ are still required to make sense of the more important c\`adl\`ag path $\bvp$ from \autoref{lem:zeros2}, unless $\bvp(t)<\infty$ can be ensured over $[\tau,\infty)$ by further constraints. So at least in this chapter, we stay in the general setting where $f$ simply belongs to $\uF$, and so $\phi\in\uC(\RR,\bRR)$.

The following example clarifies the forms of these two paths, $\phi\in\uC(\RR,\bRR)$ and $\bvp\in\uD(\RR_+,\bRR_+)$, when $f$ is in a subset $\uF_\vt\subset\uF$ from \autoref{ex:main_example}, and when $(\tau,\xi)=(0,0)$. For convenience, let the subsets $\Theta,\uW\subset\uC(\RR,\RR)$ contain the paths $\vt$ and $w$ from \autoref{ex:main_example} respectively.

\begin{example}[The subset $\uF_\vt\subset\uF$]\label{ex:zeros_example}
    Recall functions $f\in\uF_\vt$ admit the representation
    \begin{equation}
        f(t,x) := \vt(t) - w(x),
    \end{equation}
    where $(\vt,w)\in\Theta\times\uW$. Using \autoref{eq:define_phi}, the path $\phi$ from \autoref{lem:zeros1} is then given by
    \begin{equation}
        \phi(x) := \sup\{t\in\RR: f(t,x) <0 \} = \sup\{t\in\RR: \vt(t) < w(x) \} = \vt^{-1}(w(x)),
    \end{equation}
    where the final representation $\phi=\vt^{-1}\circ w$ follows from our assumption that $\vt\in\Theta$ is bijective. The path $\phi$ is thus found in $\uC(\RR,\RR)$, i.e.~we never find $\phi(x)=\pm\infty$, and have $f(\phi(x),x)=0$ for all $x\in\RR$. Now from \autoref{eq:up_bound}, the corresponding path $\bvp$ is given by
    \begin{equation}
        \bvp(t):= \inf\{x>0:f(t,x)<0\} = \inf\{x>0:w(x)>\vt(t)\}.
    \end{equation}
    Given that $w\in\uW$ ensures $\sup_{x\in\RR_+}w(x)=\infty$, then we find $\bvp(t)<\infty$ over $\RR_+$. Combining this with \autoref{lem:zeros2}, then $\bvp$ defines a strictly increasing path in $\uD(\RR_+,\RR_+)$ where $f(t,\bvp(t))=0$ holds, and finally we have the exit-time relationship $\bvp = E(\vt^{-1}\circ w)=E(\phi)$.
\end{example}

\begin{figure}[H]
    \centering
    \includegraphics[height=0.46\linewidth]{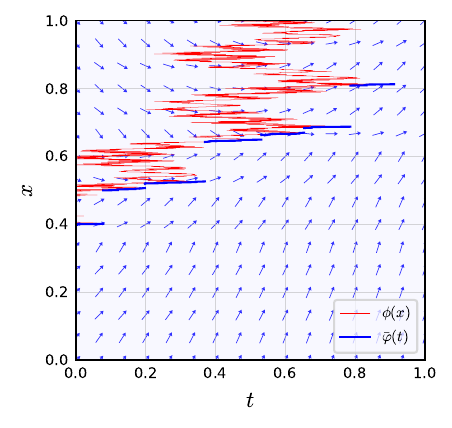}
    \includegraphics[height=0.46\linewidth]{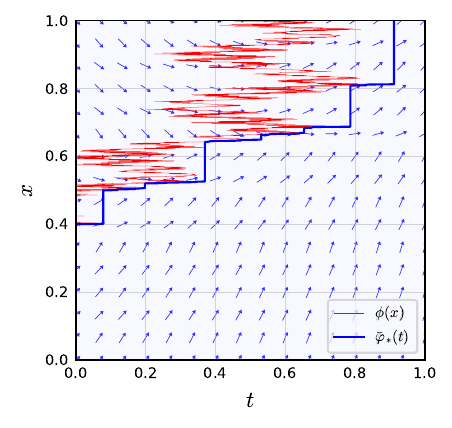}
    \caption{The left panel shows the paths $\phi$ and $\bvp$ from \autoref{eq:zeros_example4}, where the Weierstrass path $w$ and parameters $\sigma,\kappa,\theta,v$ are consistent with those in \autoref{fig:main_examples}. The right panel repeats the left but shows instead $\bvp_*(t):=[\bvp(t_-),\bvp(t)]$.}
    \label{fig:function_zeros}
\end{figure}

Following these examples, now specifically let $f$ take the Heston form in \autoref{eq:cir_ivp_well_posed}, so
\begin{equation}
    f(t,x) := \sigma w(x) + \kappa(\theta t - x) + v,
\end{equation}
which is found in $\uF_\vt$, with $\vt(t):=\kappa\theta t$, provided $\sup_{x\in\RR_+}\kappa x - \sigma w(x)=\infty$. Then we find 
\begin{equation}\label{eq:zeros_example4}
    \phi(x) := (\kappa\theta)^{-1}(\kappa x - \sigma w(x) - v),\quad \bvp(t):=\inf\{x>0: \kappa x - \sigma w(x) > \kappa\theta t + v\}.
\end{equation}
The form of $\bvp$ coincides with that given in \autoref{eq:path_bound} which, as discussed in \autoref{ch:intro}, can be considered as a path of an IG L\'evy process. The left panel of \autoref{fig:function_zeros} demonstrates both paths from \autoref{eq:zeros_example4}, which should be compared with the left panel of \autoref{fig:main_examples}. To help visualise $\bvp$ (the discontinuities of which are technically dense) and the relationship $\bvp=E(\phi)$ of \autoref{lem:zeros2}, the intervals $\bvp_*(t):=[\bvp(t_-),\bvp(t)]$ are shown in the right panel.

% 20200514-weierstrass-example

\subsection{Maximal existence, bijectivity and bounds}\label{sec:global_bijectivity}

The focus henceforth is IVP solutions of \autoref{prob:ivp}, and not just properties of the driving functions $f\in\uF$, like in the previous section. Specifically, the main programme of this section is as follows. First, in \autoref{lem:spatial_bounds}, spatial bounds of solutions are established which, as stated in \autoref{thm:bijectivity_new}, help clarify that maximal solutions $\vp$ are bijective paths in some set $\uC^1([\tau,T_*),[\xi,X_*))$ with $T_*\vee X_*=\infty$. This was discussed following the statement of \autoref{prob:ivp}. \autoref{lem:temporal_unbound} and \autoref{lem:spatial_unbound} then provide simple conditions on $f$ which help control the values of $T_*\in(\tau,\infty]$ and $X_*\in(\xi,\infty]$ respectively, so e.g.~$T_*=X_*=\infty$ can be ensured. As discussed before \autoref{def:mod_driving_func}, this is desirable for volatility modelling. Finally, following \autoref{ex:main_example}, consequences of these results are given in the example of $f\in\uF_\vt$.

\vspace{3mm}\textbf{Solution bounds.} Considering simple geometrical consequences of the next result, as demonstrated in \autoref{fig:function_zeros}, it becomes clear that the spatial solution bounds $\xi\le\vp(t)\le\bvp(t)$ established restrict solutions into a subsets of $[\tau,\infty)\times[\xi,\infty)$ where $f(t,x)\ge0$, leading to the desired strictly increasing solutions. It is important to appreciate that this does \emph{not} mean that the region where instead $f(t,x)<0$ can be neglected, or replaced arbitrarily. On the contrary, this region is required to define the important path $\bvp$ from \autoref{lem:zeros2}.

Proof of \autoref{lem:spatial_bounds} here utilises differential inequalities, as covered extensively in \cite{Lakshmikantham_1969}. Full details are provided here, however, given that we make unconventional use of such inequalities over \emph{c\`adl\`ag} paths related to $\bvp\in\uD([\tau,\infty),[\xi,\infty])$. 

Use of such paths makes this result more complicated that it can seem, and making use of the path $\phi\in\uC(\RR,\bRR)$ from \autoref{lem:zeros1} and mean value theorem (MVT), rather than directly using $\bvp$, can seem superfluous. This is because, although \autoref{eq:spatial_bound} provides $f(t,x)<0$ for $x$ in some interval $(\bvp(t),\bvp(t)+\ep)$, the infimum of such $\ep$ values over any time interval will be zero if the discontinuities of $\bvp$ are dense in $[\tau,T_*)$. This is the situation a.s.~in the Heston example of \autoref{eq:zeros_example4}, so is practically relevant, and means we cannot make use of a set of paths $\bvp_\ep(t):=\bvp(t)+\ep$ above $\bvp$ where $f(t,\bvp_\ep(t))<0$, which would simplify matters.

\begin{lemma}[Spatial solution bounds]\label{lem:spatial_bounds}
    Assume $f\in\uF$ and $f(\tau,\xi)>0$, then define $\bvp$ by
    \begin{equation}\label{eq:spatial_bound}
        \bvp(t) := \inf\{x>\xi:f(t,x)<0\},
    \end{equation}
   which adheres to \autoref{lem:zeros2}. Then any maximal solution $\vp$ of the IVP $x'=f(t,x)$, $x(\tau)=\xi$, which is in some set $\uC^1([\tau,T_*),\RR)$ with $T_*\in(\tau,\infty]$, satisfies $\xi\le\vp(t)\le\bvp(t)$ over $[\tau,T_*)$.
\end{lemma}
\begin{proof}
    The lower bound of $\xi\le\vp(t)$ is simple to establish. Because $\vp'(\tau) = f(\tau,\xi) > 0$, then $\vp(t)>\xi$ is ensured for all $t$ in some $(\tau,\tau+\ep)$, and so $\vp$ enters the quadrant $(\tau,\infty)\times(\xi,\infty)$. A first touching point $t_*>\tau$ where $\vp(t_*)=\xi$ then provides $\vp'(t_*)\le0$, given $\vp(t)$ must approach $\xi$ from above. But $f(\cdot,\xi)$ is strictly increasing, so we run into the contradiction
    \begin{equation}
        0 \ge \vp'(t_*) = f(t_*,\vp(t_*)) = f(t_*,\xi) > f(\tau,\xi) >0,
    \end{equation}
    and instead must conclude that $\vp(t)>\xi$ for all $t\in(\tau,T_*)$. Establishing the upper bound of $\bvp$ is conceptually similar, but complicated by the fact that $f(t,\bvp(t))=0$ whenever $\bvp(t)<\infty$, as shown in \autoref{lem:zeros2}, rather than the more helpful $f(t,\bvp(t))<0$. Moreover, as noted before this result, trying to utilise values $\ep>0$ such that $f(t,x)<0$ for all $x\in(\bvp(t),\bvp(t)+\ep)$ is futile, since the infimum of such $\ep$ values is zero over any interval where $\bvp$ jumps. So we make use of the path $\phi\in\uC(\RR,\bRR)$ of zeros from \autoref{lem:zeros1}, related to $\bvp$ through $\bvp=E(\phi)$.
    
    Since $f(\tau,\xi)>0$, it is clear from \autoref{eq:spatial_bound} and the continuity of $f$ that $\bvp(\tau)>\xi=\vp(\tau)$, so $\bvp$ is a strict bound at this starting point. Likewise, if for $t_*\in[\tau,T_*)$ we find $\bvp(t_*)=\infty$, then clearly the strict bound $\vp(t)<\bvp(t)=\infty$ also holds over $[t_*,T_*)$. The proof is thus trivially complete if $t_*=\tau$, which applies to impractical cases (for us) like $f(t,x):=1 + t$.
    
    Despite $\bvp$ being only c\`adl\`ag, if ever a point where $\vp(t)>\bvp(t)$ is found, a crossing point $t_*<t$ where $\vp(t_*)=\bvp(t_*)=:x_*$ is guaranteed, given $\vp(\tau)<\bvp(\tau)$ and $\bvp$ is strictly increasing. Assume a \emph{first} such crossing point $t_*\in(\tau,T_*)$ exists. Then given the relationship $\bvp=E(\phi)$, $\phi\in\uC(\RR,\bRR)$ \emph{and} that $t_*$ is the first crossing point, there exists $\ep,\delta>0$ such that the parametric path $(t,\vp(t))$ for $t\in(t_*,t_\delta)$ resides \emph{strictly} earlier in time than that of $(\phi(x),x)$ for $x\in(x_*,x_\ep)$, where $t_\delta:=t_*+\delta$ and $\vp(t_\delta)=x_\ep:=x_*+\ep$. For additional clarity, residing `strictly earlier in time' means $\vp(t)=x\implies t<\phi(x)$ whenever $(t,x)\in(t_*,t_\delta)\times(x_*,x_\ep)$.
    
    Given $\phi$ characterises the zeros of $f$ according to $f(\phi(x),x)=0$ by \autoref{lem:zeros1}, and every $f(\cdot,x)$ is strictly increasing, then having $(t,\vp(t))$ positioned earlier in time than $(\phi(x),x)$ provides $f(t,\vp(t))<0$ for $t\in(t_*,t_\delta)$. However, the MVT provides a point $t\in(t_*,t_\delta)$ where $\vp'(t)=(x_\ep-x_*)/(t_\delta-t_*)=\ep/\delta>0$. We have thus established  $f(t,\vp(t))<0<\vp'(t)$ at this point, and $\vp$ therefore cannot solve the ODE $x'=f(t,x)$ over $(t_*,t_\delta)\subset[\tau,T_*)$, if $t_*$ exists.
        
    The assumption of such a point in $[\tau,T_*)$ where $\bvp(t)<\vp(t)$ is therefore absurd, and $\vp(t)\le\bvp(t)$ thus extends from the initial time $\tau$ to the entirety of $[\tau,T_*)$, completing the proof.
\end{proof}

In the above proof, we saw that the bound of $\vp(t)\ge\xi$ is strict for all $t>\tau$, and with the geometry of this proof in mind it is worth covering conditions which make the upper bound $\vp(t)\le\bvp(t)$ also strict, over $[\tau,T_*)$. Towards this, assume that $\bvp$ is not \emph{just} strictly increasing, but verifies $\bvp(t)-\bvp(s)\ge\ep(t-s)$ for some $\ep>0$ and all $s,t\in[\tau,T_*)$ with $s\le t$. Now, if a first touching time $t_*\in(\tau,T_*)$ where $\vp(t_*)=\bvp(t_*)$ is assumed, then $\vp'(t_*)=0$, given $f(t_*,\bvp(t_*))=0$, but $\bvp(t)-\bvp(t_*)\le-\ep(t_*-t)<0$ for $t\in(\tau,t_*)$. So, a contradictory interval in $(\tau,t_*)$ is found where $\bvp(t)<\vp(t)$. Given the relationship $\bvp=E(\phi)$ from \autoref{lem:zeros2}, where $\phi\in\uC(\RR,\bRR)$ and $f(\phi(x),x)=0$ whenever $\phi(x)\in\RR$, then this property of $\bvp(t)-\bvp(t_*)\le-\ep(t_*-t)<0$ is ensured (by elementary geometrical considerations) given the \emph{one-sided} Lipschitz condition $\phi(x) - \phi(u) \le L (x - u)$ of \autoref{lem:strict_upper}, where $\ep:=L^{-1}$. One-sided Lipschitz properties can be found in \cite{Lakshmikantham_1969} and \cite{Agarwal_1993}, regarding bounds and uniqueness of solutions respectively.

\begin{lemma}[Strict upper bound]\label{lem:strict_upper}
    The upper bound in \autoref{lem:spatial_bounds} is strict, i.e. $\vp(t)<\bvp(t)$ over $[\tau,T_*)$, and $\vp'(t)>0$, if the path $\phi\in\uC(\RR,\bRR)$ from \autoref{lem:zeros1} has the one-sided Lipschitz property $\phi(x) - \phi(u) \le L (x - u)$ for some $L\in\RR_+$ and all $u,x\in[\xi,\infty)$ with $x\ge u$.
\end{lemma}

This result is practically relevant because a solution $\vp$, modelling the cumulative variance of a price path, then has a \emph{strictly} positive corresponding volatility $\sqrt{\vp'}$. This can be helpful, in order to relate abstract risk-neutral derivative pricing measures to a real-world probability measure (both introduced in \autoref{sec:martingale}), although the details of this will not be covered.

\vspace{3mm}\textbf{Bijective maximal solutions.} The main purpose of the bounds in \autoref{lem:spatial_bounds} for now is to help enable the bijectivity statement in the following result. The reader should note that the assumption $f(\tau,\xi)>0$ here cannot in general be relaxed to $f(\tau,\xi)\ge0$, which is treated in \autoref{sec:refining_problem}. Now recall, following the statement of \autoref{prob:ivp}, that a \emph{maximal} solution $\vp\in\uC^1([\tau,T_*),\RR)$ is one which reaches the boundary of $\RR^2$, i.e.~which verifies $T_*\vee\sup_{t\in[\tau,T_*)}|\vp(t)|=\infty$, and classical ODE theory, e.g.~Theorem 1.1.3 of \cite{Lakshmikantham_1969}, establishes the existence of such solutions in our setting where $f\in\uC(\RR^2,\RR)$.

\begin{theorem}[Maximal existence and bijectivity]\label{thm:bijectivity_new}
    Assume $f\in\uF$ and $f(\tau,\xi)>0$. Then there exists a maximal solution $\vp$ of the IVP $x'=f(t,x)$, $x(\tau)=\xi$. Moreover, any such $\vp$ defines a strictly increasing bijection in some set $\uC^1([\tau,T_*),[\xi,X_*))$, where $T_*\vee X_*=\infty$.
\end{theorem}
\begin{proof}
    Classical theory gives the existence statement, since $f\in\uC(\RR^2,\RR)$. This provides a maximal solution $\vp\in\uC^1([\tau,T_*),\RR)$ which, by definition, satisfies $T_*\vee\sup_{[\tau,T_*)}|\vp(t)|=\infty$. Because $f\in\uF$ and $f(\tau,\xi)>0$, the bounds $\xi\le\vp(t)\le\bvp(t)$ hold over $[\tau,T_*)$ by \autoref{lem:spatial_bounds}.
    
    Given that $f(t,\xi)>f(\tau,\xi)>0$ for $t>\tau$, it is clear then from the definition $\bvp(t) := \inf\{x>\xi:f(t,x)<0\}$ that $\vp$ is restricted to a subset of $\RR^2$ where $f(t,x)\ge0$, making $\vp$ non-decreasing. Recall from \autoref{lem:zeros2} that $f(t,\bvp(t))=0$ when $\bvp(t)<\infty$, so a touching point $\vp(t)=\bvp(t)$ provides $\vp'(t)=0$. In the opposite direction, if $\vp'(t)=0$, then we must find either $\vp(t)=\bvp(t)$ or $\vp(t)=\bvp(t_-)<\bvp(t)$, given $\bvp$ is strictly increasing. So now, for a contradiction, assume $\vp'(t)=0$ holds over an interval $[a,b]\subset[\tau,T_*)$. Then we must find
    \begin{equation}
        \bvp(a_-)\le\bvp(a) = \vp(a) = \vp(b) = \bvp(b_-) \le \bvp(b).
    \end{equation}
    But having $\bvp(a)=\bvp(b_-)$ implies that $\bvp$ is constant at least over $[a,b)$, which violates the strictly increasing nature of $\bvp$ from \autoref{lem:zeros2}. So $\vp$ is non-decreasing, and $\vp'(t)=0$ \emph{cannot} hold over intervals. So in any such $[a,b]$ we must find a point where $\vp'(t)>0$, and the continuity of $\vp'$ extends this to ensure that $\vp(b)-\vp(a)=\int_a^b\vp'(s)\dd s>0$ for $a,b\in[\tau,T_*)$.
    
    Therefore, like $\bvp$, we find $\vp\in\uC^1([\tau,T_*),\RR)$ to be strictly increasing, and $\vp$ therefore defines a bijection in $\uC^1([\tau,T_*),[\xi,X_*))$, where $X_*=\lim_{t\uparrow T_*}\vp(t)\in(\xi,\infty]$. This allows the maximality condition $T_*\vee\sup_{[\tau,T_*)}|\vp(t)|=\infty$ to be written as $T_*\vee|\xi|\vee X_*=\infty$. In turn this is equivalent to the claim of $T_*\vee X_*=\infty$, given we know $|\xi|<\infty$, completing the proof.
\end{proof}

\vspace{3mm}\textbf{Existence conditions.} From the condition $T_*\vee X_*=\infty$ in \autoref{thm:bijectivity_new}, we know that either $T_*=\infty$, or $X_*=\infty$, or both. The purpose of the next two results is to provide separate practicable conditions on $f$ which independently ensure $T_*=\infty$ or $X_*=\infty$ respectively, both of which are desirable. Establishing the first of these is quite straightforward, as follows.

\begin{lemma}[Temporal existence]\label{lem:temporal_unbound}
    Let $f,\tau,\xi,\bvp$ be as defined in \autoref{lem:spatial_bounds}, and $\vp\in\uC^1([\tau,T_*),[\xi,X_*))$ be any maximal solution  of the IVP $x'=f(t,x)$, $x(\tau)=\xi$. If $\bvp(T)<\infty$ for some $T\in(\tau,\infty)$, then $T_*>T$, and by extension, if $\bvp(t)<\infty$ over $[\tau,\infty)$, then $T_*=\infty$.
\end{lemma}
\begin{proof}
    From \autoref{thm:bijectivity_new}, $\vp$ is a bijective element of some $\uC^1([\tau,T_*),[\xi,X_*))$ with $T_*\vee X_*=\infty$. For a contradiction, assume $\bvp(T)<\infty$ for some $T\in(\tau,\infty)$ but $T_*\le T<\infty$. Then we have $X_*=\infty$, so that $T_*\vee X_*=\infty$ is verified, and therefore $\vp(t)\xrightarrow{t\to T_*}\infty$. Because $\vp$ is strictly increasing over $[\tau,T_*)$ and $\bvp$ is over $[\tau,T]\supset[\tau,T_*)$, then having $\vp(\tau)=\xi<\bvp(\tau)$ and $\bvp(T)<\lim_{t\uparrow T_*}\vp(t)=\infty$ provides a unique touching point $t_*\in(\tau,T_*)$ where $\vp(t_*)=\bvp(T)$, with the strict inequalities $\bvp(t)<\bvp(T)<\vp(t)$ over $(t_*,T_*)$. This contradicts the relationship $\vp(t)\le\bvp(t)$ over $[\tau,T_*)$ from \autoref{lem:spatial_bounds}. So instead we find $T_*>T$ if $\bvp(T)<\infty$. The extension to $T_*=\infty$ follows this argument when letting $T\to\infty$, establishing the claim.
\end{proof}

The simple condition of $\bvp(T)<\infty$ in \autoref{lem:temporal_unbound} of course only ensures existence of maximal solutions over $[\tau,T]$ when $f\in\uF$, and not more generally when $f\in\uC(\RR^2,\RR)$. Nevertheless this is a condition which supplements classical existence theory, such as the main result of \cite{Wintner_1945}, presented concisely as Theorem 5.1 in \cite{Hartman_2002}. This theorem requires checking a limit $\int^\infty\dd x/U(t,x)=\infty$, for some $U$ with $|f(t,x)|\le U(t,|x|)$ over $[\tau,T]$, and applies to cases like $U(t,x)=x$ and $U(t,x)=x\log x$. But this condition depends on $f$ in an unbounded set like $[\tau,T]\times\RR$, so taking an example like $f(t,x):=t-w(x)$ with $w(x):=x^a\sin(x) - 1$ and any $a>1$, we find $\int^\infty\dd x/U(t,x)<\infty$ when making the natural selection $U(t,x)=1+t+x^a$, yet \autoref{lem:temporal_unbound} immediately provides existence of all maximal solutions over $[\tau,\infty)$, given that the upper bound $\bvp$ is the element of $\uD(\RR_+,\RR)$ given by 
\begin{equation}\label{eq:exit_use1}
    \bvp(t) = \inf\{x>0: x^a\sin(x)-1>t\} = E(w)(t) < \infty.
\end{equation}

On the other hand, consider the reduction of this example to $f(t,x):=1 + t$, i.e.~using instead $w(x)=-1$. Then clearly $\vp(t):=t + \frac12 t^2$ is the global solution of $x'=f(t,x)$, $x(0)=0$. However, since one finds $\bvp(t):=\inf\{x>0:-1>t\}=\inf\varnothing:=\infty$ for all $t\in\RR_+$, \autoref{lem:temporal_unbound} is useless in this elementary example. But, as discussed after \autoref{def:solutionset}, such examples with $\liminf_{t\to\infty}\vp'(t)=\infty$ are not helpful when $\sqrt{\vp'}$ will model a volatility path.

Now we want a condition on $f\in\uF$ which can ensure $X_*=\infty$ in \autoref{thm:bijectivity_new}. Towards this, consider the example of $f(t,x):=2-e^{-t}-2x$. Then it is straightforward to check that $f\in\uF$ and  $\vp(t):=1-e^{-t}$ is a global solution of the IVP $x'=f(t,x)$, $x(0)=0$. In \autoref{thm:bijectivity_new} we therefore have $(T_*,X_*)=(\infty,1)$, i.e.~$\vp\in\uC^1(\RR_+,[0,1))$. The property of this IVP leading to $X_*=1<\infty$ is that $f(t,1)=-e^{-t}<0$ for all $t\in\RR_+$, so $\vp$ cannot pass through the line where $x=1$. Indeed, this is equivalent to having $\lim_{x\uparrow 1}\phi(x)=\infty$ where $\phi\in\uC(\RR,\bRR)$ is from \autoref{lem:zeros1}. The condition in the following result serves to rule out such examples, thereby enforcing $X_*=\infty$, in fact characterising this property when $f\in\uF$.

In the following statement, note that any such maximal solution $\vp$ is a bijective element of some set $\uC^1([\tau,T_*),[\xi,X_*))$ with $T_*\vee X_*=\infty$ by \autoref{thm:bijectivity_new}, so e.g.~$\vp(t)\xrightarrow{t\to T_*}X_*$ holds.

\begin{lemma}[Spatial existence]\label{lem:spatial_unbound}
    Assume $f\in\uF$ and $f(\tau,\xi)>0$. Then any maximal solution $\vp\in\uC^1([\tau,T_*),[\xi,X_*))$ of the IVP $x'=f(t,x)$, $x(\tau)=\xi$ satisfies $X_*>X$ iff $\lim_{t\to\infty}f(t,x)>0$ for all $x\in[\xi,X]$. So $X_*=\infty$ iff $\lim_{t\to\infty}f(t,x)>0$ for all $x\in[\xi,\infty)$.
\end{lemma}
\begin{proof}
    The easier only if direction is dealt with first, so suppose $X_*>X$ for some $X>\xi$. Given that $\vp\in\uC^1([\tau,T_*),[\xi,X_*))$ is bijective by \autoref{thm:bijectivity_new}, then $\vp(t)\xrightarrow{t\to T_*}X_*>X$ and there exists unique $T\in(\tau,T_*)$ where $\vp(T)=X$, with $\vp'(t)\ge0$ over $[\tau,T]$. So we find $f(t,\vp(t))= \vp'(t)\ge0$ over $[\tau,T]$, and given $\{\vp(t):t\in[\tau,T]\}=[\xi,X]$ and $f(\cdot,x)$ is strictly increasing, then clearly $\lim_{t\to\infty}f(t,x)>0$ for every $x\in[\xi,X]$. If $X_*=\infty$, this argument applies to any $X\in(\xi,\infty)$, so $\lim_{t\to\infty}f(t,x)>0$ for $x\in[\xi,\infty)$, completing half of the proof.
    
    In the other direction, we are trying to establish $X_*>X$ for some $X>\xi$. Notice that if $T_*<\infty$, then we must have $X_*=\infty>X$ to verify the requirement $T_*\vee X_*=\infty$ from \autoref{thm:bijectivity_new}. The result is thus obvious, and we should now assume $T_*=\infty$. Further suppose that for some $X\in(\xi,\infty)$, $\lim_{t\to\infty}f(t,x)>0$ holds for every $x\in[\xi,X]$. A time $T\in(\tau,T_*)$ will now be constructed where $\vp(T)>X$ must hold, confirming $X_*>X$. First notice that this condition of $\lim_{t\to\infty}f(t,x)>0$ implies $\phi(x)<\infty$ for $x\in[\xi,X]$, where $\phi$ from \autoref{lem:zeros1} characterises the zeros of $f$ according to $f(\phi(x),x)=0$ when $\phi(x)\in\RR$.
    
    For $\ep>0$, define the shifted path $\phi_\ep\in\uC([\xi,X],\RR)$ by $\phi_\ep(x):=(\tau\vee \phi(x))+\ep$. Then the strict inequality $f(\phi_\ep(x),x)>0$ holds, given every $f(\cdot,x)$ is strictly increasing. Now define
    \begin{equation}
        c:=\min_{x\in[\xi,X]} f(\phi_\ep(x),x)\in(0,\infty),\quad t_*:= \max_{x\in[\xi,X]}\phi_\ep(x) \in (\tau,\infty).
    \end{equation}
    These values ensure the inequality $f(t,x)\ge c$ on the vertical line $(t_*,x)$ for $x\in[\xi,X]$, so also $f(t,x)> c$ in the rectangle $(t_*,\infty)\times[\xi,X]$. Now define the time $T:=t_* + c^{-1}(X-\xi)$ and the line $\underline{\vp}\in\uC([t_*,T],[\xi,X])$, from points $(t_*,\xi)$ to $(T,X)$ with gradient $c>0$, by
    \begin{equation}
        \underline{\vp}(t):= \xi + c(t - t_*).
    \end{equation}
    We have constructed a line $\underline\vp$ where clearly $\underline\vp'(t)=c$, and also $f(t,\underline{\vp}(t))>c$ over $(t_*,T)$. Clearly $\ubvp(t_*)=\xi<\vp(t_*)$ holds, and assuming a first touching point where $\vp(t)=\ubvp(t)$, thus $\vp'(t)\le c$, provides the contradiction $c<f(t,\ubvp(t))=f(t,\vp(t))=\vp'(t)\le c$. So instead we must have $\vp(t)>\underline\vp(t)$ over $[t_*,T]$, and so also $\vp(T)>\underline\vp(T)=X$, confirming that $X_*>X$. By taking $X\to\infty$, the extension to $X_*=\infty$ is straightforward, completing the proof.
\end{proof}

The proof of \autoref{lem:spatial_unbound} is constructive in that it actually establishes a lower bound $\ubvp$ for the point $T\in(\tau,T_*)$, i.e.~verifying $\xi<\ubvp(T)<\vp(T)\le\bvp(T)$, so improves on the lower bound from \autoref{lem:spatial_bounds}. Such bounds, and their derivation via differential inequalities, are central to establishing the limit theorems of \autoref{chap:solutions}, especially \autoref{thm:exit_limit}, where it is natural to interpreted these \emph{not} as lower bounds in space, but rather upper bounds in time.

This next result just bring together those of \autoref{lem:temporal_unbound} and \autoref{lem:spatial_unbound}, in the way that these will be applied when treating \autoref{prob:ivp2} in \autoref{chap:solutions}. For this we present the condition $\bvp(t)<\infty$ from \autoref{lem:temporal_unbound} directly on $f\in\uF$ as $\inf_{x\in[\xi,\infty)}f(t,x)<0$, which is clearly equivalent using \autoref{eq:up_bound}. Similarly, the condition $\lim_{t\to\infty}f(t,x)>0$ of \autoref{lem:spatial_unbound} is presented as $\sup_{t\in[\tau,\infty)}f(t,x)>0$, which is equivalent given $f(\cdot,x)$ is strictly increasing. After the statement of \autoref{cor:exist_summary}, the purpose of the assumptions made in \autoref{eq:assumptions_G}, defining the subset $\uG\subset\uC(\RR^2_+,\RR)$ to which \autoref{chap:solutions} applies, will be clear.

Like for \autoref{lem:spatial_unbound}, by writing $\vp\in\uC^1([\tau,T_*),[\xi,X_*))$ in the following statement we mean the values $T_*,X_*$ are as established in \autoref{thm:bijectivity_new}, so e.g.~the limit $\vp(t)\xrightarrow{t\to T_*}X_*$ holds.

\begin{corollary}[Spatio-temporal existence]\label{cor:exist_summary}
    Assume $f\in\uF$ and $f(\tau,\xi)>0$. Then maximal solutions $\vp\in\uC^1([\tau,T_*),[\xi,X_*))$ of the IVP $x'\!=\!f(t,x)$, $x(\tau)\!=\!\xi$ verify the conditions~
    \begin{equation}\label{eq:exist_summary}
        \inf_{x\in[\xi,\infty)}f(t,x)<0\ \ \forall t\in[\tau,\infty)\implies T_*=\infty,\ \sup_{t\in[\tau,\infty)}f(t,x)>0\ \ \forall x\in[\xi,\infty)\implies X_*=\infty.
    \end{equation}
\end{corollary}

\vspace{2mm}This completes coverage of some basic properties of these spatially irregular IVPs from \autoref{prob:ivp}, and their solutions, and these properties are now clarified with examples depending on the subsets $\uF_\vt\subset\uF$ used in \autoref{ex:zeros_example}. Particularly notable is how the existence condition of $\inf_{x\in[\xi,\infty)}f(t,x)<0$ in \autoref{eq:exist_summary} places no constraint on the positive growth of $f(t,x)$ for $x\in[\xi,\infty)$, which  in \autoref{ex:global_exist} below is controlled by the negative growth of the paths $w\in\uW$. So the conditions of \autoref{eq:exist_summary} indeed supplement classical existence conditions depending on such positive growth constraints, as discussed following \autoref{lem:temporal_unbound}.

\begin{example}[The subset $\uF_\vt\subset\uF$]\label{ex:global_exist}
    Recall functions $f\in\uF_\vt$ admit the representation
    \begin{equation}
        f(t,x) := \vt(t) - w(x),
    \end{equation}
    where $(\vt,w)\in\Theta\times\uW$, so e.g.~$\lim_{t\to\pm\infty}\vt(t)=\pm\infty$ and $\sup_{x\in\RR_+}w(x)=\infty$. In \autoref{ex:zeros_example} we saw that the path $\bvp$ from \autoref{lem:zeros1} for the IVP $x'=f(t,x)$, $x(0)=0$ is then given by 
    \begin{equation}
        \bvp(t)= \inf\{x>0:w(x)>\vt(t)\},
    \end{equation}
    and this is in $\uD(\RR_+,\RR_+)$, given $\sup_{x\in\RR_+}w(x)=\infty$. We can more compactly write this as $\bvp=E(\vt^{-1}\circ w)=E(w)\circ\vt$, using the exit-time notation from \autoref{eq:exit_def}. Now \autoref{thm:bijectivity_new} tells us that any maximal solution $\vp$ of this IVP is a bijective path in some set $\uC^1([0,T_*),[0,X_*))$, where $T_*\vee X_*=\infty$. Given $\bvp(t)<\infty$ for all $t\in\RR_+$, then \autoref{lem:temporal_unbound} provides $T_*=\infty$, and given $\lim_{t\to\infty}f(t,x)=\infty$ for every $x\in\RR$, then \autoref{lem:spatial_unbound} provides $X_*=\infty$ also. \autoref{cor:exist_summary} can equivalently be used to obtain $T_*=X_*=\infty$, that is
    \begin{equation}
        \inf_{x\in\RR_+}\vt(t) - w(x)<0\ \ \forall t\in\RR_+\implies T_*=\infty,\ \sup_{t\in\RR_+}\vt(t) - w(x)>0\ \ \forall x\in\RR_+\implies X_*=\infty.
    \end{equation}
\end{example}

\begin{figure}[H]
    \centering
    % \vspace{60mm}
    \includegraphics[height=0.46\linewidth]{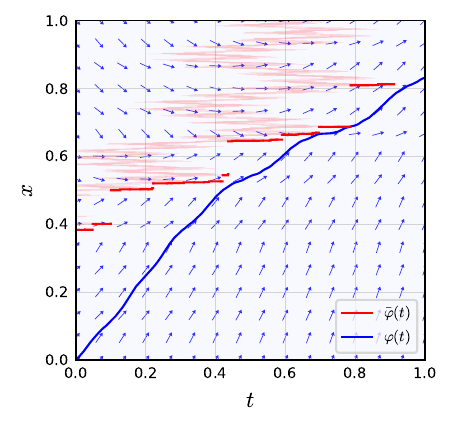}
    \includegraphics[height=0.46\linewidth]{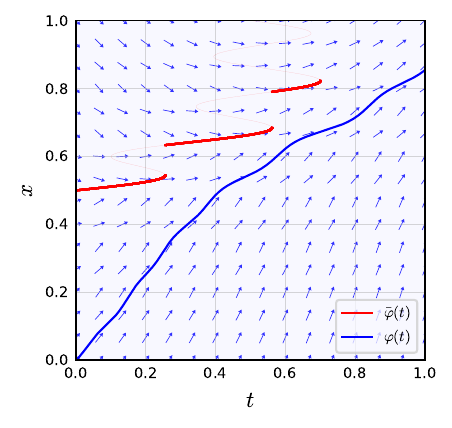}
    % \vspace{-2mm}
    \caption{Solutions $\vp$ of the IVP $x'=f(t,x)$, $x(0)=0$ are shown alongside their upper bounds $\bvp$ from \autoref{lem:spatial_bounds}. The functions $f$ are consistent with \autoref{fig:main_examples} and \autoref{fig:main_examples2}.}
    \label{fig:global_exist_examples}
\end{figure}

\vspace{4mm}Following this example, now specifically let $f$ take the Heston form in \autoref{eq:cir_ivp_well_posed}, so
\begin{equation}
    f(t,x) := \sigma w(x) + \kappa(\theta t - x) + v,
\end{equation}
which is found in $\uF_\vt$, with $\vt(t):=\kappa\theta t$, provided $\sup_{x\in\RR_+}\kappa x - \sigma w(x)=\infty$. Then we have 
\begin{equation}
    \bvp(t):=\inf\{x>0: \kappa x - \sigma w(x) > \kappa\theta t + v\}
\end{equation}
which, given $\sup_{x\in\RR_+}\kappa x - \sigma w(x)=\infty$, defines a path in $\uD(\RR_+,\RR_+)$ and provides an upper bound to any maximal solution $\vp$ of the IVP $x'=f(t,x)$, $x(0)=0$. Such maximal solutions are thus found in $\uC^1_0(\RR_+,\RR_+)$, as we hope when modelling volatility. In \autoref{fig:global_exist_examples} we illustrate solutions $\vp$ and their bounds $\bvp$ in this Heston setting, consistently with earlier figures.

% 20200514-weierstrass-example

\subsection{Maximal uniqueness}\label{sec:uniqueness}

As promised at the beginning of this chapter, this section starts by clarifying what existing ODE uniqueness theory has to say about \autoref{prob:ivp}. An excellent starting point is the text \cite{Agarwal_1993}, which systematically presents 21 directly applicable first-order ODE uniqueness theorems, as well as further corollaries, nonuniqueness theorems and `Carath\'eodory' extensions, which assume only the a.e.~differentiability of solutions.

\vspace{3mm}\textbf{Historical context.} The only existing result which places no constraints on the spatial behaviour of $f\in\uC(\RR^2,\RR)$, so could in principle be applied to \autoref{prob:ivp}, is that of \cite{Wend_1969}, stated as Theorem 2.6.1 in \cite{Agarwal_1993}. Technically, this is not quite true, since, for example, Theorem 1.21.2 due to \cite{Yosie_1925}, which is the only which actually \emph{characterises} uniqueness, of course accommodates Wend's and all others. Although this characterisation is achieved very intuitively, with the result being so general it is hard to imagine establishing its conditions without having to use those of another.

Perhaps unsurprisingly, the situation in Wend's theorem is similar to ours, as covered in \autoref{ch:intro} when discussing the importance of maximal solutions. Specifically, a temporal monotonicity constraint is assumed, and solutions are sought only forwards in time. This result will be introduced properly shortly, but we first consider the earlier result of \cite{Peano_1890}. This instead adopts a \emph{spatial} monotonicity constraint, and applies to examples like $f(t,x)=-\mathrm{sgn}(x)|x|^a$ for $a\in(0,1)$, which generate unique solution $\vp(t)=0$ through $(0,0)$.

\begin{theorem}[Peano's uniqueness]
    Let $f\in\uC(\RR^2,\RR)$ be such that $f(t,\cdot)$ is non-increasing for every $t\in\RR$. Then the IVP $x'=f(t,x)$, $x(\tau)=\xi$ has a unique maximal solution.
\end{theorem}

Proving this result takes just a few lines, and these are presented following Theorem 1.3.1 in \cite{Agarwal_1993}. Now a consequence of the inverse function theorem, e.g.~Theorem 9.24 in \cite{Rudin_1976}, is that if $\vp$ is a solution of an IVP $x'=f(t,x)$, $x(\tau)=\xi$ with a well-defined and differentiable inverse $\vp^{-1}$, then this inverse will solve the inverted IVP $x'=g(t,x)$, $x(\xi)=\tau$ where $g(t,x):=1/f(x,t)$. This is covered and utilised in \cite{Cid_2009}. For now notice that the non-increasing assumption of $f(t,\cdot)$ from Peano's theorem becomes a non-\emph{decreasing} assumption on $g(\cdot,x)$, inverting the constraint between time and space and placing us in the setting which Wend's theorem applies. But also notice that since points where $f(x,t)=\infty$ are precluded in Peano's theorem by the assumption $f\in\uC(\RR^2,\RR)$, then we must preclude points where $g(t,x)=0$ manually in Wend's theorem.

\begin{theorem}[Wend's uniqueness]\label{thm:wend_uniqueness}
    Assume $f\in\uC(\RR^2,\RR)$, $f(\cdot,x)$ is non-decreasing for $x\in\RR$ and $f(t,x)>0$ in $\cX\subset\RR^2$ with $(\tau,\xi)\in\cX$. Then the IVP $x'=f(t,x)$, $x(\tau)=\xi$ has a unique solution in $\cX$, i.e.~some $\vp\in\uC^1([\tau,T],\RR)$ with $(t,\vp(t))\in\cX$ and $(T,\vp(T))\in\partial\cX$. 
\end{theorem}

So Wend's theorem is a local result and not applicable to maximal solutions, unless the assumption of $f(t,x)>0$ in $\cX$ is extended to all of $\RR^2$. This is clearly not acceptable for volatility modelling, ruling out the Heston case at the core of this thesis, in \autoref{eq:cir_ivp_well_posed}, and generally rendering the important path $\bvp$ from \autoref{lem:zeros2} useless, since $\bvp(t)=\inf\varnothing:=\infty$.

This transition from Peano's to Wend's uniqueness results via the inverse function theorem provides a particular example of the idea of \cite{Cid_2009}. Here, the authors map any uniqueness result onto another, using simple spatio-temporal transformations like just shown. Like Wend's theorem, however, regions where $f(t,x)\neq 0$ are always considered, and these can of course be impractically small when considering a general function in $\uC(\RR^2,\RR)$.

So, at least expressed in light of existing results, the main contribution of the \emph{uniqueness} result here, in \autoref{thm:global_uniqueness}, is its applicability to maximal solutions, and so the complete relaxation of these limitations from \cite{Wend_1969} and \cite{Cid_2009}. Of course, the setting here and in Wend's theorem are not otherwise equivalent, given that in \autoref{prob:ivp} we ask that each function $f(\cdot,x)$ is improved from being non-decreasing to strictly increasing. 

We believe this improvement can be relaxed in the future, but see no practical value in doing so now. For example, the solution set $\Phi$ from \autoref{def:solutionset} gets widened to accommodate volatility paths $\sqrt{\vp'}$ which (unrealistically) can be zero over \emph{intervals}, but already (equally unrealistic) paths which can be zero on sets arbitrarily close to full Lebesgue measure are accommodated. The price paid for this widening is solutions' inverses do not always exist, and these can be key to relatively neat proofs, e.g.~\autoref{thm:global_uniqueness} via Lebesgue's calculus.

\vspace{2mm} Moving on, few mathematicians, financial or not, know of these simple monotonicity results due to Peano and Wend. This does seem paradoxical in finance, given such mathematicians' brain-busting knowledge of the comparatively complex It\^o SDEs, from which \autoref{eq:cir_ivp_well_posed} derives. The Lipschitz uniqueness condition, from which Peano's and Wend's results do \emph{not} follow, appears to have encouraged this. As recognised in \cite{Soong_1973}, the Lipschitz condition is often `too restrictive' for applications, and `certainly undesirable from a practical viewpoint', yet such conditions have been prioritised in the teaching of ODE theory, and now researchers do not always appreciate simple alternatives like these just presented. 

The programme of this section is relatively simple as compared with the last, although delves deeper into functional analysis. Having shown in \autoref{thm:bijectivity_new} that solutions $\vp$ of \autoref{prob:ivp} are bijective, they clearly have well-defined inverses, which will be labelled $\hvp$ in this section to avoid having to write $(\vp^{-1})'$. These inverses are key to the proof of \autoref{thm:global_uniqueness}, and so important properties of such inverses are collected first in \autoref{lem:inverse_properties}.

\vspace{3mm}\textbf{The uniqueness result.} As a note of precaution, it is easy to jump to `intuitive' conclusions, regarding the paths $\vp$ and $\hvp:=\vp^{-1}$ here, which turn out to be false in pathological cases. For example, despite \autoref{eq:inv_func_thm}, from the inverse function theorem, one cannot presume $\vp'(t)>0$ a.e.~(with respect to the Lebesgue measure) follows from $\hvp'(x)<\infty$ a.e., even though $\hvp'(x)>0$ indeed follows from $\vp'(t)<\infty$. This is essentially because an inverse $\hvp$, unlike $\vp$, need not map null sets to null sets, which is called the Lusin (N) property, after the thesis \cite{Lusin_1916}. This property is closely related to absolute continuity, but rarely emphasised in modern functional analysis. Section 7.6 of \cite{Saks_1937} is devoted to it. 

Our saviour in a battle against pathology, which local results like Wend's theorem avoid, is Lebesgue's fundamental theorem of calculus from \cite{Lebesgue_1904}. We present this within \autoref{lem:inverse_properties} as in Section 6.2 of \cite{Royden_2010}. Put simply, \autoref{thm:global_uniqueness} is a product of applying Lebesgue's theorem to Wend's, and to highlight the wider importance of Lebesgue's theorem we reproduce an inspiring remark from \cite{Royden_2010}:

\vspace{-6mm}
\begin{adjustwidth}{18pt}{18pt}
\begin{remark}\label{rem:pathology}
    \emph{Frigyes Riesz and Bela Sz.-Nagy remark that Lebesgue's Theorem is `one of the most striking and most important in real variable theory.' Indeed, in 1872 Karl Weierstrass presented mathematics with a continuous function on an open interval which failed to be differentiable at any point. Further pathology was revealed and there followed a period of uncertainty regarding the spread of pathology in mathematical analysis. Lebesgue's Theorem, which was published in 1904, and its consequences, helped restore confidence in the harmony of mathematics analysis.}
\end{remark}
\end{adjustwidth}

\vspace{2mm}Recall that the inequality in \autoref{eq:lebesgue_ftoc}, which is a corollary of Lebesgue's theorem, is strict for the pathological Cantor function from \cite{Cantor_1884}. Simple modifications of the Cantor function are also, counter-intuitively, capable of verifying $\lambda(0)=0$, $\lambda'(x)<0$ for a.e.~$x>0$ yet $\lambda(x)>0$ simultaneously. This is precisely the situation which arises in \autoref{thm:global_uniqueness}, exposing the level of mathematical generality at which this result applies.

As its proof demonstrates, the following collections of properties constitute little more than the inverse function theorem, stated as Theorem 9.24 of \cite{Rudin_1976}, and Lebesgue's theorem and its corollaries, stated as such in Section 6.2 of \cite{Royden_2010}.

\begin{lemma}[Solutions' inverse properties]\label{lem:inverse_properties}
    Assume $f\in\uF$, $f(\tau,\xi)>0$, and let $\vp\in\uC^1([\tau,T_*),[\xi,X_*))$ be a maximal solution of the IVP $x'=f(t,x)$, $x(\tau)=\xi$ as in \autoref{thm:bijectivity_new}, so $X_*=\lim_{t\uparrow T_*}\vp(t)$ and $T_*\vee X_*=\infty$. Then $\vp$ has a well-defined inverse $\hvp:=\vp^{-1}\in\uC([\xi,X_*),[\tau,T_*))$, which is also strictly increasing. For $(t,x)\in[\tau,T_*)\times[\xi,X_*)$,~
    \begin{equation}\label{eq:inv_func_thm}
        \vp'(t)\hvp'(\vp(t)) = \hvp'(x)\vp'(\hvp(x)) = 1
    \end{equation}
    holds whenever $\vp'(t)>0$, or equivalently whenever $\hvp'(x)$ exists, with $x=\vp(t)$. Over any subinterval $[a,b]\subset[\xi,X_*)$, Lebesgue's theorem applies to $\hvp$, i.e.~$\hvp$ is a.e.~differentiable and
    \begin{equation}\label{eq:lebesgue_ftoc}
        \hvp(b) - \hvp(a) \ge \int_{[a,b]}\hvp'(x)\dd x.
    \end{equation}
    Finally, if $\hvp'(x)$ exists everywhere in $[a,b]$ except for a finite number of points, then $\hvp$ is absolutely continuous here, with equality in \autoref{eq:lebesgue_ftoc}, and $\vp'(t)>0$ a.e~then follows.
\end{lemma}
\begin{proof}
    By \autoref{thm:bijectivity_new}, every maximal solution $\vp$ of the IVP $x'=f(t,x)$, $x(\tau)=\xi$ defines a strictly increasing bijection in some set $\uC^1([\tau,T_*),[\xi,X_*))$ with $T_*\vee X_*=\infty$. So its inverse $\hvp$ is clearly also strictly increasing and in $\uC([\xi,X_*),[\tau,T_*))$, and uniquely defined as the function which verifies $\hvp(\vp(t))=t$ and $\vp(\hvp(x))=x$ for any $(t,x)\in[\tau,T_*)\times[\xi,X_*)$. 
    
    In addition, whenever $\vp'(\hvp(x))>0$, the inverse function theorem provides $\hvp'(x)=1/\vp'(\hvp(x))$, so clearly $\hvp'(x)>0$ exists. In reverse, if $\hvp'(\vp(t))>0$ exists, then $\vp'(t)=1/\hvp'(\vp(t))$. These equivalences are more compactly expressed in \autoref{eq:inv_func_thm}, when identifying $x=\vp(t)$.
    
    Although $\hvp$ might not be differentiable, it is strictly increasing over $[\xi,X_*)$, and so by Lebesgue's theorem is a.e.~differentiable and verifies \autoref{eq:lebesgue_ftoc}. If $\hvp$ is improved to being differentiable in $[a,b]$ except for a finite number of points, then it is differentiable in the open subintervals $(a_i,b_i)$ between these points, with $\hvp(b_i) - \hvp(a_i) = \int_{a_i}^{b_i}\hvp'(x)\dd x$. Linearity of the integral with $\Leb[\cup_i(a_i,b_i)]=b-a$ extends this equality to \autoref{eq:lebesgue_ftoc}. 
    
    For a strictly increasing function, having equality in \autoref{eq:lebesgue_ftoc} is equivalent to being absolutely continuous, from Corollary 6.5.12 in \cite{Royden_2010}. Finally, if $\hvp'(x)<\infty$ except at a finite number of points $x_i\in[a,b]$, then clearly $\vp'(t)=1/\hvp'(\vp(t))>0$ except at the finite points $t_i:=\hvp(x_i)$, and therefore $\vp'(t)>0$ a.e.~in $[\hvp(a),\hvp(b)]$ follows.
\end{proof}

\vspace{2mm}We will not utilise the final conclusion $\vp'(t)>0$ a.e.~in $[\hvp(a),\hvp(b)]$ directly, but include it to emphasise that we cannot in general assume it, even though $\vp'(\hvp(x))>0$ a.e.~in $[a,b]$, which follows from $\hvp'(x)<\infty$ a.e.~in $[a,b]$ and $\vp'(t)=1/\hvp'(\vp(t))$. As discussed, this would assume that $\hvp$ has the Lusin (N) property. Equivalently, this \emph{assumes} absolute continuity, using Theorem 7.6.7 of \cite{Saks_1937}, given $\hvp$ is strictly increasing thus of bounded variation.

Now we are in a position to cover arguably the single most important result of this thesis. As with all results thus far, one cannot in general expect this to hold in the extension where $f(\tau,\xi)=f(\tau,\vp(\tau))=0$, despite $f(t,\vp(t))=0$ actually being possible for any $t\in(\tau,T_*)$.
    
\begin{theorem}[Maximal uniqueness]\label{thm:global_uniqueness}
    Assume $f\in\uF$ and $f(\tau,\xi)>0$. Then there exists precisely one maximal solution $\vp$ of the IVP $x'=f(t,x)$, $x(\tau)=\xi$. This unique solution is a strictly increasing and bijective path in some set $\uC^1([\tau,T_*),[\xi,X_*))$, with $T_*\vee X_*=\infty$.
\end{theorem}
\begin{proof}
    Start by letting $\vp_i$ for $i=1,2$ be any two such maximal solutions of this IVP, which by \autoref{thm:bijectivity_new} define strictly increasing and bijective paths in some sets $\uC^1([\tau,T_i),[\xi,X_i))$ respectively, with $T_i\vee X_i=\infty$. This bijectivity ensures $\vp_i(t)\xrightarrow{t\to T_i}X_i$. Now define $T_*:=T_1\wedge T_2$ and $X_*:=X_1\wedge X_2$, so $\vp_i$ both exist in $\cX:=[\tau,T_*)\times[\xi,X_*)$. The task is to now show that $\vp_2=\vp_1$ in $\cX$. From this, one can intuit, and a simple case-by-case analysis clarifies, that $T_2=T_1$ and $X_2=X_1$, so that actually $\vp_i$ are the same maximal solution, in $\uC^1([\tau,T_*),[\xi,X_*))$ with $T_*\vee X_*=\infty$. To cover one such case, let $T_1<\infty$, so that $X_1=\infty$ is given by $T_1\vee X_1=\infty$. Then assume that $X_2<\infty$, so $T_2=\infty$. Now having $\vp_2=\vp_1$ in $\cX$ means in particular that $\vp_2(t)=\vp_1(t)=X_2<\infty$ for some $t\in[\tau,T_1)$, since $\vp_1(t)\xrightarrow{t\to T_1<\infty}X_1=\infty$. For $\vp_2$, this is absurd given that the bijective path $\vp_2$ satisfies $\vp_2(t)\xrightarrow{t\to T_2=\infty}X_2<\infty$ and so cannot satisfy $\vp_2(t)=X_2$ for any $t<\infty$. So now we can just focus on verifying $\vp_2=\vp_1$ in $\cX$, from which the maximal uniqueness claim follows.
    
    The key henceforth is to utilise the inverses $\hvp_i:=\vp_i^{-1}\in\uC([\xi,X_i),[\tau,T_i))$, along with the properties of these paths consolidated in \autoref{lem:inverse_properties}. Notice that $\vp_2=\vp_1$ in $\cX$ will follow from the uniqueness of inverses $\hvp_i$ if we find $\hvp_2=\hvp_1$ in $\hat\cX:=[\xi,X_*)\times[\tau,T_*)$. Towards this, define the function $\lambda\in \uC([\xi,X_*),\RR)$ by $\lambda(x):=\hvp_2(x)-\hvp_1(x)$, which tracks the difference in \emph{time} for each $\vp_i$ to reach the spatial level $x$. This function satisfies $\lambda(\xi)=0$, since $\hvp_2(\xi)=\hvp_1(\xi)=\tau$, and the proof is thus complete if we find $\lambda(x)=0$ over all of $[\xi,X_*)$.
    
    Targeting a contradiction, assume the existence of a point $c\in(\xi,X_*)$ where $\lambda(c)\neq 0$, and assume w.l.o.g.~that $\hvp_2(c)>\hvp_1(c)$, so $\lambda(c)>0$. By the continuity of $\lambda$, and the fact that $\lambda(\xi)=0$, an interval $(a,b)\subset[\xi,X_*)$ exists containing $c$ where $\lambda(x)>0$ and also $\lambda(a)=0$, meaning $\hvp_2(a)=\hvp_1(a)$. Having $a=\xi$ is possible, but should not be assumed. Now having $\lambda(x)>0$ over all of $(a,b)$ means $\hvp_2(x)>\hvp_1(x)$ here, and since every $f(\cdot,x)$ is strictly increasing, then $f(\hvp_2(x),x) > f(\hvp_1(x),x)$. Evaluating the ODEs $\vp'_{1,2}(t)=f(t,\vp_{1,2}(t))$ at times $\hvp_{1,2}(x)$ for $x\in(a,b)$, this inequality regarding $f$ values equivalently provides 
    \begin{equation}\label{eq:uniquenessineq}
        \vp'_2(\hvp_2(x)) - \vp'_1(\hvp_1(x)) > 0.
    \end{equation}
    Now clearly $\vp_1'(\hvp_1(x))\ge0$ over $(a,b)$, and finding $\vp_1'(\hvp_1(x))=0$ is indeed possible. But notice that \autoref{eq:uniquenessineq} instead enforces $\vp_2'(\hvp_2(x))>0$ over $(a,b)$. So \autoref{eq:inv_func_thm} then provides $\hvp_2'(x)=1/\vp_2'(\hvp_2(x))<\infty$, showing $\hvp_2$ is differentiable in $(a,b)$, and thus verifies
    \begin{equation}\label{eq:u1}
        \hvp_2(x) - \hvp_2(a) = \int_{[a,x]} \hvp_2'(u)\dd u
    \end{equation}
    for all $x\in(a,b)$. For completeness, notice that this holds even if $\hvp'(a)$ is not defined, meaning $\vp'(\hvp(a))=0$, using \autoref{lem:inverse_properties} and the finiteness of the singleton $\{a\}\subset[a,b)$.
    
    The equality of \autoref{eq:u1} cannot be assumed to hold for $\hvp_1$, which in general remains just a.e.~differentiable in $(a,b)$, and Lebesgue's theorem in \autoref{eq:lebesgue_ftoc} instead provides
        \begin{equation}\label{eq:u2}
            \hvp_1(x) - \hvp_1(a) \ge \int_{[a,x]} \hvp_1'(u)\dd u.
        \end{equation}
    The inequalities of \autoref{eq:u1} and \autoref{eq:u2} will be invoked momentarily, but first consider again \autoref{eq:uniquenessineq}. By applying \autoref{lem:inverse_properties} to each component here, this can be equivalently expressed as $1/\hvp'_2(x) - 1/\hvp'_1(x) > 0$ a.e.~in $(a,b)$, and therefore also as
    \begin{equation}\label{eq:u3}
        \lambda'(x) = \hvp'_2(x) - \hvp'_1(x) < 0.
    \end{equation}
    Alongside the properties of $\lambda(a)=0$ and $\lambda(x)>0$ in $(a,b)$, a contradiction now seems close, but contrary to intuition is not guaranteed without \autoref{eq:u1} and \autoref{eq:u2}. (As mentioned after \autoref{rem:pathology}, functions exist where $\lambda(a)=0$, $\lambda'(x)<0$ a.e.~and $\lambda(x)>0$ simultaneously). Indeed, only by using \autoref{eq:u1}, \autoref{eq:u2} \emph{and} \autoref{eq:u3} together (in that order) the ordering $\lambda(x)>\lambda(a)=0$ over $(a,b)$ is contradicted as follows:
    \begin{align}
        \lambda(x)-\lambda(a) &= \hvp_2(x) - \hvp_2(a) - (\hvp_1(x) - \hvp_1(a)) \nonumber\\
        &\le \int_{[a,x]} \hvp_2'(u)\dd u - \int_{[a,x]} \hvp_1'(u)\dd u \label{eq:u4} \\
        &= \int_{[a,x]} \hvp_2'(u)- \hvp_1'(u)\dd u \nonumber\\
        &= \int_{[a,x]} \lambda'(u)\dd u\nonumber\\
        &< 0,\label{eq:u5}
    \end{align}
    where \eqref{eq:u4} uses \autoref{eq:u1} and \autoref{eq:u2}, and \eqref{eq:u5} uses \autoref{eq:u3}. This establishes $\lambda(x)<\lambda(a)=0$ over $(a,b)$, so gives the desired contradiction. The existence of a point $c\in(\xi,X_*)$ with $\lambda(c)\neq0$ is therefore absurd, and instead $\lambda(x)=0$ over $[\xi,X_*)$. This provides $\hvp_2(x)=\hvp_1(x)$ over $[\xi,X_*)$, so also $\vp_2(t)=\vp_1(t)$ over $[\tau,T_*)$. As already noted, this implies $T_2=T_1=T_*$, $X_2=X_1=X_*$ and $T_*\vee X_*=\infty$, so both $\vp_i$ are in fact the same unique maximal solution of the IVP $x'=f(t,x)$ $x(\tau)=\xi$, completing the proof.
\end{proof}

This result ends this short section. It should be noted that this constitutes the first known maximal uniqueness result for ODEs $x'=f(t,x)$ where no spatial regularity constraints are imposed on the function $f$. Regarding the IVPs introduced in this chapter and illustrated in several figures, like those depending on the subset $\uF_\vt\subset\uF$ from \autoref{ex:main_example}, there is little to say other than the solutions $\vp\in\uC^1([\tau,T_*),[\xi,X_*))$ referred to are in fact unique.

% \clearpage
\subsection{Continuity of the solution map}\label{sec:cont_dep}

The focus of this section is establishing the continuous dependence result of \autoref{thm:continuity_solution_map}, thereby clarifying certain stability properties of \autoref{prob:ivp} and completing the three conventional requirements of `well-posedness' (existence, uniqueness \& continuous dependence). 

Loosely, the goal is to obtain a statement like those in Section 2.4 of \cite{Coddington_1955}, e.g.~Theorem 2.4.1. However, these results, like most in the literature, regard continuity with respect to initial conditions $(\tau,\xi)\in\RR^2$ and parameters $\mu\in\RR^k$. See also Chapter 5 of \cite{Hartman_2002}, which emphasises the \emph{differentiability} of a solution map, following related assumptions for a function $f\in\uC^{1,1}(\RR^2,\RR)$ which are inapplicable here.

\begin{figure}[H]
    \centering
    \includegraphics[height=0.48\linewidth]{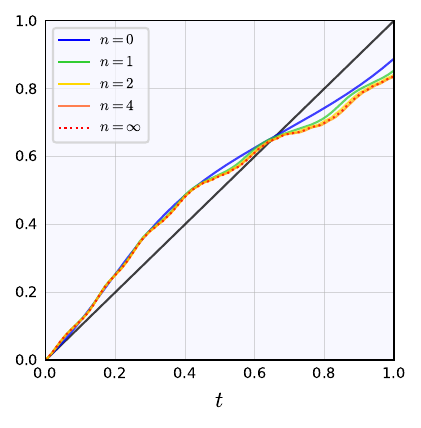}
    \includegraphics[height=0.48\linewidth]{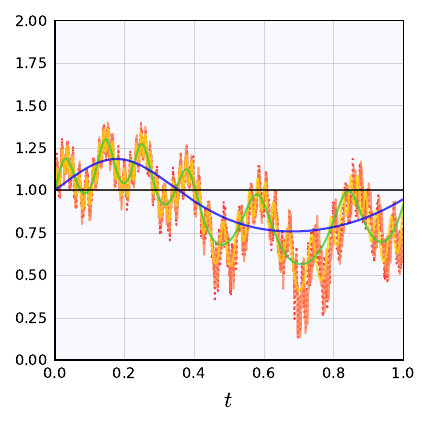}
    % \vspace{-2mm}
    \caption{The left panel shows the solutions $\vp_n$ of the IVPs $x'=f_n(t,x)$, $x(0)=0$ with $f_n$ as in \autoref{eq:cont_ex}. Also shown is the IVP solution when setting $w_n=0$ in \autoref{eq:cont_ex} (black). The right panel shows the corresponding volatility paths $\sqrt{\vp'_n}$.}
    \label{fig:cont_dependence}
\end{figure}

These results related to initial conditions and parameters are simply too restrictive for us. For example, we are interested in knowing whether the cumulative variance solution $\vp$ of the Heston IVP $x'=f(t,x)$, $x(0)=0$ from  
\autoref{eq:cir_ivp_well_posed}, with $f(t,x) := \sigma w(x) + \kappa(\theta t - x) + v$, is continuous with respect to the sample path $W(\omega)=:w\in\uC_0:=\uC_0(\RR,\RR)$ of a process like Brownian motion. Practically, this tells us if and how we may approximate such a path $w$ with a computationally helpful sequence $\{w_n\}_{n\in\NN}\subset\uC_0$, such as linear polygons or the truncated Fourier series of \autoref{eq:truncated_weierstrass}. The latter is used for a demonstration in \autoref{fig:cont_dependence}.

Such considerations are also theoretically helpful, because for the modelling framework of price processes $S_t=\exp(W^\rho_{X_t}-\frac12 X_t)$ outlined in \autoref{ch:intro} to actually be well-defined, we require that the cumulative variance and price processes $X$ and $S$ define \emph{measurable} maps from $(\Omega,\cF,\PP)$ into explicit measurable spaces, so that probability can actually be conducted. Of course, measurability will follow if we can establish a suitably general form of continuity. 

The relevant measurable spaces will always be sets equipped with their Borel $\sigma$-algebra induced by a specified metric or norm. Contrasting results related to initial conditions and parameters, this clarifies why we seek general sequential continuity statements, such as
\begin{equation}\label{eq:cont_summary}
    f_n\xrightarrow{n\to\infty}f_0 \text{ on } (\uF,d_{\uF}) \implies \vp_n\xrightarrow{n\to\infty}\vp_0 \text{ on } (\Phi,d_{\Phi}),
\end{equation}
for solutions $\vp_n$ of the IVPs $x'=f_n(t,x)$, $x(\tau)=\xi$, and $\Phi,d_\uF,d_\Phi$ appropriately defined. The assumption here then reduces to a requirement $w_n\xrightarrow{n\to\infty}w_0$ in the Heston example.

We wait until \autoref{chap:solutions}, specifically point 3.~of \autoref{thm:wellposed}, to give the continuity statement which will be relied upon in the probabilistic framework of \autoref{chap:framework}, and for now settle for that in \autoref{thm:continuity_solution_map}, which is more informative in the probability-free setting here.

To interpret \autoref{eq:continuity_statement} as intended, let uniform seminorms on $\uC(\RR^2,\RR)$ be defined by
\begin{equation}
    \Vert f \Vert_{[\tau,T]\times[\xi,X]}:=\sup\{|f(t,x)|:(t,x)\in[\tau,T]\times[\xi,X]\}
\end{equation}
for any rectangle $[\tau,T]\times[\xi,X]\subset\RR^2$. Similarly, on the sets $\uC([\tau,T_*),\RR)$ with $T_*\in(\tau,\infty]$, define seminorms by $\Vert\vp\Vert_{[\tau,T]}:=\sup_{t\in[\tau,T]}|\vp(t)|$ for any $T>\tau$, where this should be interpreted as $\Vert\vp\Vert_{[\tau,T]}:=\infty$ when $T\ge T_*$. This unusual extension for $T\ge T_*$ is just to accommodate the situation where $\vp_n(t)\xrightarrow{t\to T_n\le T}\infty$ in \autoref{eq:continuity_statement}, so $\Vert \vp_0 - \vp_n \Vert_{[\tau,T]}=\infty$. It is clear from \autoref{thm:continuity_solution_map} that $\Vert \vp_0 - \vp_n \Vert_{[\tau,T]}=\infty$ for at most a finite number of terms.

Finally, proof of \autoref{thm:continuity_solution_map} depends on the Ascoli Lemma, stated as such in Chapter 1 of \cite{Coddington_1955}. This fundamental result says that an \emph{equibounded} and \emph{equicontinuous} sequence $\{\vp_n\}_{n\in\NN_0}\subset\uC([\tau,T],\RR)$ has a uniformly convergent subsequence. For a differentiable sequence, these equi-conditions respectively follow from $\Vert \vp_n\Vert_{[\tau,T]}< X$ and $\Vert \vp'_n\Vert_{[\tau,T]}< M$. In fact the latter with a consistent starting point $\vp_n(\tau)=\xi$ suffices.

\begin{theorem}[Solution map continuity]\label{thm:continuity_solution_map}
    Assume the subset $\{f_n\}_{n\in\NN_0}\subset\uF$ is such that $f_n(\tau,\xi)>0$ for all $n\in\NN_0$, and let $\vp_n\in\uC^1([\tau,T_n),\RR)$ denote the unique maximal solution of each IVP $x'=f_n(t,x)$, $x(\tau)=\xi$. Then for any values $T\in(\tau,T_0)$ and $X\in(\vp_0(T),\infty)$, 
    \begin{equation}\label{eq:continuity_statement}
        \Vert f_0 - f_n \Vert_{[\tau,T]\times[\xi,X]} \xrightarrow{n\to\infty}0 \implies \Vert \vp_0 - \vp_n\Vert_{[\tau,T]} \xrightarrow{n\to\infty}0.
    \end{equation}
\end{theorem}
\begin{proof}
    Recall, following \autoref{thm:bijectivity_new} and \autoref{thm:global_uniqueness}, that such maximal solutions $\vp_n$ exist, are unique, and define strictly increasing bijections in sets $\uC^1([\tau,T_n),[\xi,X_n))$ with $T_n\vee X_n=\infty$. Now fix $T\in(\tau,T_0)$ and $X\in(\vp_0(T),X_0)$, then define $\cX:=[\tau,T]\times[\xi,X]$. Also fix $M$ such that $\Vert f_0\Vert_\cX<M<\infty$, which exists since $f_0\in\uC(\RR^2,\RR)$. Given $\Vert f_0 - f_n \Vert_\cX \xrightarrow{n\to\infty}0$, then we find $\Vert f_n\Vert_\cX<M$ for all $n$ greater than some $N\in\NN$. So we can w.l.o.g.~assume $\Vert f_n\Vert_\cX< M$ for all $n\in\NN_0$ by either redefining $M$ or just by removing the first $N$ terms. 
    
    Having $\Vert f_n\Vert_\cX< M$ ensures $(t,\vp_n(t))\in\cX$ for $t\in[\tau,t_1\wedge T]$ where $t_1=\tau+M^{-1}(X-\xi)$. For consistency later, define instead the earlier time $t_1:=\tau+M^{-1}(X-\vp_0(T))$. To alleviate the use of $t_1\wedge T$, just assume the worst, i.e.~$t_1<T$. Still, this provides $(t,\vp_n(t))\in\cX$ over $[\tau,t_1]$ for all $n$, and now we aim to establish the reduced conclusion $\Vert \vp_0 - \vp_n\Vert_{[\tau,t_1]} \xrightarrow{n\to\infty}0$.
    
    % note this part just adds clarity and could actually be removed
    The reasoning used to obtain this will then be repeated for the times $t_k:=\tau + kM^{-1}(X-\vp_0(T))$, until the convergence over $[\tau,T]$, as in \autoref{eq:continuity_statement}, is obtained with at most 
    \begin{equation}\label{eq:num_its}
        \left\lceil\frac{T-\tau}{t_1-\tau}\right\rceil = \left\lceil\frac{M(T-\tau)}{X - \vp_0(T)}\right\rceil
    \end{equation}
    iterations of this procedure. This procedure towards the convergence $\Vert \vp_0 - \vp_n\Vert_{[\tau,t_1]} \xrightarrow{n\to\infty}0$ just over $[\tau,t_1]$ follows a relatively standard argument via the Ascoli Lemma, used in the related Theorem 2.4.1 of \cite{Coddington_1955} and very loosely summarised by 
    \begin{equation}\label{eq:approach_summary}
        \underbrace{\text{equiboundedness + equicontinuity}}_{\text{$\implies$Ascoli Lemma}}\text{ + uniqueness $\implies$ convergence}.
    \end{equation}
    Here, equiboundedness and equicontinuity regards the set $\{\vp_n\}_{n\in\NN_0}$ of IVP solutions. In our setting, we have already ensured these properties over $[\tau,t_1]$ since we find $\Vert\vp_n\Vert_{[\tau,t_1]}\le|\xi|\vee|X|<\infty$ and $\Vert\vp'_n\Vert_{[\tau,t_1]}\le \Vert f_n \Vert_\cX <M<\infty$ for all $n\in\NN_0$. The Ascoli Lemma then provides the convergence $\Vert\vp_* - \vp_{n_k}\Vert_{[\tau,t_1]}\xrightarrow{k\to\infty}0$ of a subsequence $\{\vp_{n_k}\}_{k\in\NN}$ to a limit $\vp_*$.
    % end of this part which could be removed
    
    Seeking a contradiction, now assume $\Vert \vp_0 - \vp_n\Vert_{[\tau,t_1]} \xrightarrow{n\to\infty}0$ is violated. Then there is an infinite subsequence of $\{\vp_n\}_{n\in\NN_0}$ remaining \emph{outside} some open ball around $\vp_0$, and the Ascoli Lemma provides a subsequence of this with $\Vert \vp_* - \vp_{n_k}\Vert_{[\tau,t_1]} \xrightarrow{k\to\infty}0$, where $\vp_*\neq\vp_0$.
    
    But each $\vp_{n_k}$ verifies $\vp_{n_k}(t)=\xi+\int_\tau^t f_{n_k}(s,\vp_{n_k}(s))\dd s$ over $[\tau,t_1]$, which may be written as
    \begin{equation}\label{eq:cont_dep_main}
        \vp_{n_k}(t) = \xi + \int_\tau^t f_0(s,\vp_{n_k}(s)) + \lambda_{n_k}(s)\dd s,\quad \lambda_n(t) := f_n(t,\vp_n(t)) - f_0(t,\vp_n(t)).
    \end{equation}
    Now $\Vert \lambda_{n_k}\Vert_{[\tau,t_1]}\le \Vert f_0 - f_{n_k} \Vert_\cX \xrightarrow{k\to\infty}0$ by assumption, and having $f_0\in\uC(\RR^2,\RR)$ extends $\Vert \vp_* - \vp_{n_k}\Vert_{[\tau,t_1]} \xrightarrow{k\to\infty}0$ to $\Vert f_0(\cdot,\vp_*(\cdot)) - f_0(\cdot,\vp_{n_k}(\cdot))\Vert_{[\tau,t_1]} \xrightarrow{k\to\infty}0$, so taking $k\to\infty$ in \autoref{eq:cont_dep_main} we find $\vp_*(t)=\xi+\int_\tau^t f_0(s,\vp_*(s))\dd s$ over $[\tau,t_1]$. So $\vp_*$ solves $x'=f_0(t,x)$, $x(\tau)=\xi$ over $[\tau,t_1]$, contradicting the uniqueness result of \autoref{thm:global_uniqueness} for this IVP given that $\vp_*\neq\vp_0$. The assumption that $\Vert \vp_0 - \vp_n\Vert_{[\tau,t_1]} \xrightarrow{n\to\infty}0$ is violated is therefore absurd.
    
    The preceding argument will now be repeated to extend the interval $[\tau,t_1]$. Having $\Vert \vp_0 - \vp_n\Vert_{[\tau,t_1]} \xrightarrow{n\to\infty}0$, there exists $N_1\in\NN$ such that $\vp_n(t_1)<\vp_0(T)$ for $n>N_1$, given $\vp_0(t_1)<\vp_0(T)$. By removing the first $N_1$ terms, we can w.l.o.g.~assume $\vp_n(t_1)<\vp_0(T)$ for $n\in\NN$. But this bound, alongside $\vp_0(T)<X$ and $\Vert f_n\Vert_\cX< M$, lets us extend the previous statement $(t,\vp_n(t))\in\cX$ over $[\tau,t_1]$ to the same over $[\tau,t_2\wedge T]$, where $t_2:=t_1 + M^{-1}(X-\vp_0(T))$. 
    
    So $\{\vp_n\}_{n\in\NN_0}$ is now equibounded and equicontinuous over $[\tau,t_2]$, assuming again that $t_2<T$. A repeat of the procedure summarised in \autoref{eq:approach_summary} then gives $\Vert \vp_0 - \vp_n\Vert_{[\tau,t_2]} \xrightarrow{n\to\infty}0$. Repeating this further, the sequence $t_k:=\tau + kM^{-1}(X-\vp_0(T))$ of times are generated, with the conclusions $\Vert \vp_0 - \vp_n\Vert_{[\tau,t_k\wedge T]} \xrightarrow{n\to\infty}0$. So the claim of $\Vert \vp_0 - \vp_n\Vert_{[\tau, T]} \xrightarrow{n\to\infty}0$ is ensured after a finite number of iterations of this procedure, provided in \autoref{eq:num_its}.
\end{proof}

The dependence of the domain $[\tau,T]\times[\xi,X]$ in \autoref{eq:continuity_statement} on the path $\vp_0$ clarifies that \autoref{thm:continuity_solution_map} cannot yet be extended to a statement like \autoref{eq:cont_summary} on metric spaces, helpful in probability. As discussed, in \autoref{chap:solutions} this dependence will be alleviated once solutions $\vp_n$ are guaranteed to be global, i.e.~exist over $\RR_+$. Then we obtain a statement
\begin{equation}\label{eq:cont_sum}
    \Vert f_0 - f_n \Vert_{\RR_+^2} \xrightarrow{n\to\infty}0 \implies \Vert \vp_0 - \vp_n\Vert_{\RR_+} \xrightarrow{n\to\infty}0
\end{equation}
where these are not literally uniform norms, but consistent with that in  \autoref{def:uni_metric}, inducing the topologies of uniform convergence \emph{over compact subsets} of $\RR_+^2$ and $\RR_+$ respectively. For when we move to a probabilistic setting, note that an analogous pathwise statement to \autoref{eq:cont_sum} \emph{does not} hold for the It\^o SDE map. This is explained in the introductions of \cite{Friz_2010} and \cite{Friz_2014}, and used to motivate rough path theory.

It is worth noting that in \autoref{thm:continuity_solution_map} the assumptions $\{f_n\}_{n\in\NN_0}\subset\uF$ and $f_n(\tau,\xi)>0$, relevant to this chapter, are primarily used for our convenience here, since we then know solutions are strictly increasing, etc. There is little difficulty extending this result to any IVPs known to have unique solutions. The repeated application of the approach in \autoref{eq:approach_summary} over the sequence of intervals $[\tau,t_k]$ can still be utilised, with minor adaptations. 

Finally, the implicit convergence $f_n(\tau,\xi) \xrightarrow{n\to\infty}f_0(\tau,\xi)$ from \autoref{eq:continuity_statement} allows the assumption of $f_n(\tau,\xi)>0$ for all $n\in\NN_0$ to be relaxed to just $f_0(\tau,\xi)>0$, which is the case also in the forthcoming simulation convergence result of \autoref{thm:euler_converge}. However, this is no longer the case when in \autoref{chap:solutions} we introduce the delicate possibility of $f_0(\tau,\xi)=0$, so for a smoother transition between chapters we leave these limiting results as they are.

\vspace{4mm}Now to illustrate \autoref{thm:continuity_solution_map}, let functions $f_n\in\uF$ for $n\ge1$ be defined as in \autoref{eq:truncated_weierstrass},~
\begin{equation}\label{eq:cont_ex}
    f_n(t,x) := \sigma w_n(x) + \kappa(\theta t - x) + v,\quad w_n(x):=\sum_{k=0}^n a^{-\alpha k}\sin(2a^k\pi x),
\end{equation}
with $f_0(t,x) := \sigma w_0(x) + \kappa(\theta t - x) + v$ and $w_0(x):=\sum_{k=0}^\infty a^{-\alpha k}\sin(2a^k\pi x)$. For consistency, fix all values $\sigma,\kappa,\theta,v,a,\alpha$ as in \autoref{fig:weierstrass_example}. Now let $\vp_n$ denote the unique maximal solution of each IVP $x'=f_n(t,x)$, $x(0)=0$. Then using $\Vert w_0-w_n\Vert_{[0,X]}\xrightarrow{n\to\infty}0$ for all $X>0$ (which can be established via pointwise convergence and equicontinuity) we obtain $\Vert f_0-f_n\Vert_{[0,T]\times[0,X]}\xrightarrow{n\to\infty}0$ for all $T,X>0$. The existence conditions of \autoref{cor:exist_summary} can be used to establish $\vp_0\in\uC^1(\RR_+,\RR_+)$, so in particular $T_0=\infty$ in \autoref{thm:continuity_solution_map}. This result then provides $\Vert \vp_0-\vp_n\Vert_{[0,T]}\xrightarrow{n\to\infty}0$ for all $T>0$. An application of the triangle inequality then also provides the convergence $\Vert \vp'_0-\vp'_n\Vert_{[0,T]}\xrightarrow{n\to\infty}0$ for all $T>0$. In summary,
\begin{multline}
    \Vert \vp'_0-\vp'_n\Vert = \Vert f_0(\cdot,\vp_0(\cdot)) - f_n(\cdot,\vp_n(\cdot))\Vert\\
    \le \Vert f_0(\cdot,\vp_0(\cdot)) - f_0(\cdot,\vp_n(\cdot))\Vert + \Vert f_0(\cdot,\vp_n(\cdot)) - f_n(\cdot,\vp_n(\cdot))\Vert \xrightarrow{n\to\infty}0.
\end{multline}
\autoref{fig:cont_dependence} illustrates the uniform convergence $\vp_n\xrightarrow{n\to\infty}\vp_0$ over $[0,1]$, and the corresponding volatility $\sqrt{\vp'_n}\xrightarrow{n\to\infty}\sqrt{\vp'_0}$. Notice this figure contains paths also in \autoref{fig:main_examples} and \autoref{fig:main_examples2}.

% from 20201117-weierstrass-cont-dep

\subsection{Simulation of solutions}\label{sec:simulation}

The focus of this section is \autoref{thm:euler_converge}, which relates to a simple forward Euler simulation scheme for our IVPs $x'=f(t,x)$, $x(\tau)=\xi$ with $f\in\uF$. This result is not standard, however, since we never actually presume that values of this function $f$ can be reproduced \emph{exactly} on a computer, but just those of a \emph{convenient} sequence $\{f_n\}_{n\in\NN}\subset\uF$ converging uniformly (over compacts) to it. By convenient, we mean that (over compacts) each $f_n$ can be stored in computer memory. Beyond simple models, this generalised type of simulation convergence is often required in volatility modelling, and we will rely on it for applications in \autoref{chap:framework}.

Although computational ODE texts such as \cite{Griffiths_2010} and \cite{Han_2017} can be consulted for the Euler method and extensions, the related Cauchy-Peano existence theorem proof and its dependencies, e.g.~as in \cite{Coddington_1955}, will prepare the reader better for \autoref{thm:euler_converge}. This is because we are not targeting a fancy simulation scheme of any sort, but for now a framework-wide one applicable for any $f\in\uF$, which is as simple as possible, notwithstanding the necessary generalisation just mentioned.

At first it can seem that \autoref{thm:euler_converge} here could follow from the continuous dependence result of \autoref{thm:continuity_solution_map}, or vice versa. But here, practicable polygons are being simulated which are not differentiable and can only be considered to solve IVPs driven by discontinuous functions, i.e.~in a Carath\'eordory sense. So \autoref{thm:continuity_solution_map}$\implies$\autoref{thm:euler_converge} would require an extension of \autoref{thm:continuity_solution_map} applicable to discontinuous functions such as $f_n\in\uD(\RR^2,\RR)$. This approach for obtaining \autoref{thm:euler_converge} is circumvented here through a simple application of Lebesgue's calculus to (absolutely continuous) polygons. In the other direction, we only get \autoref{thm:euler_converge}$\implies$\autoref{thm:continuity_solution_map} if in \autoref{thm:euler_converge} we take the mesh limit $\Vert\pi_n\Vert\to0$ \emph{before} taking $f_n\to f_0$. But such an iterated limit cannot actually be realised on a computer.

We now define an IVP's forward Euler polygon. Toward this, call an unbounded set $\pi:=\{t_k\}_{k\in\NN_0}\subset[\tau,\infty)$, with $\tau=:t_0<t_1<\dots$, a partition of $[\tau,\infty)$, and let $\Pi([\tau,\infty))$ be the set of such objects. For $T\in(\tau,\infty)$, define the mesh $\Vert\pi\Vert_{[\tau,T]}:=\max_{k\in\NN_0}\{t_{k+1}\wedge T - t_k\wedge T\}$. Finally, we write $\vp_\pi\in\uAC\subset\uC$ to emphasise \emph{absolute} continuity of the following polygons.

\begin{definition}[Forward Euler polygon]\label{alg:forward_euler}
    Fix $f\in\uC(\RR^2,\RR)$ and $\pi:=\{t_k\}_{k\in\NN_0}\in\Pi([\tau,\infty))$ for some $\tau\in\RR$, e.g.~one could set $t_k:=\tau+k\Delta$ for some $\Delta>0$. Define the polygon $\vp_\pi\in\uAC([\tau,\infty),\RR)$ using $\vp_\pi(\tau):=\xi\in\RR$, then recursively over each interval $(t_k,t_{k+1}]$ set
    \begin{equation}\label{alg:equation}
        \vp_\pi(t) := \vp_\pi(t_k) + (t-t_k) f(t_k,\vp_\pi(t_k))
    \end{equation}
    noticing that indeed $\vp_\pi(t)\xrightarrow{t\downarrow t_k}\vp_\pi(t_k)$ for every $k\in\NN_0$, and $\cup_{k\in\NN_0}(t_k,t_{k+1}]=(\tau,\infty)$. Such a path $\vp_\pi$ will be called the forward Euler $\pi$-polygon for the IVP $x'=f(t,x)$, $x(\tau)=\xi$.
\end{definition}

No harm comes from defining forward Euler polygons over all of $[\tau,\infty)$, even if maximal solutions $\vp\in\uC^1([\tau,T_*),\RR)$ of the related IVP $x'=f(t,x)$, $x(\tau)=\xi$ explode in finite time, e.g.~satisfy $T_*<X_*:=\lim_{t\uparrow T_*}\vp(t)=\infty$. Theoretically, our forward Euler polygons $\vp_\pi$ cannot explode over a compact $[\tau,T]$, as this would demand an infinitude of time points from $\pi$ to be in $[\tau,T]$. Practically, $\vp_\pi$ could of course exceed a computer's largest number.

We now give our convergence result on forward Euler polygons for \autoref{prob:ivp}. Its proof uses the modulus of continuity of a function $f\in\uF$ which, over compact $\cX\subset\RR^2$, is defined
\begin{equation}\label{eq:mod_cont_def}
    w(r) = w_{f,\cX}(r) := \sup\{|f(t_2,x_2)-f(t_1,x_1)|:|(t_2,x_2)-(t_1,x_1)|<r,(t_i,x_i)\in\cX\}.
\end{equation}
For any such $f$ and $\cX$, the existence of $w$ and the limit $w(r)\xrightarrow{r\downarrow0}0$ follow from $f\in\uC(\RR^2,\RR)$.

\begin{theorem}[Forward Euler convergence]\label{thm:euler_converge}
    Assume $\{f_n\}_{n\in\NN_0}\subset\uF$ and $f_n(\tau,\xi)>0$. For partitions $\{\pi_n\}_{n\in\NN}\subset\Pi([\tau,\infty))$, let $\vp_n\in\uAC([\tau,\infty),\RR)$ be the forward Euler $\pi_n$-polygon for the IVP $x'=f_n(t,x)$, $x(\tau)=\xi$, and let $\vp_0\in\uC^1([\tau,T_0),\RR)$ be the unique maximal solution of the IVP $x'=f_0(t,x)$, $x(\tau)=\xi$. Then for any values $T\in(\tau,T_0)$ and $X\in(\vp_0(T),\infty)$,
    \begin{equation}\label{eq:euler_converge}
        \left(\Vert f_0 - f_n \Vert_{[\tau,T]\times[\xi,X]}, \Vert\pi_n\Vert_{[\tau,T]}\right) \xrightarrow{n\to\infty}(0,0) \implies \Vert\vp_0 - \vp_n\Vert_{[\tau,T]}\xrightarrow{n\to\infty}0.
    \end{equation}
\end{theorem}
\begin{proof}
    Unlike the IVP solutions from \autoref{thm:continuity_solution_map}, a given polygon $\vp_n$ need not be strictly increasing like $\vp_0$, and indeed the bounds from \autoref{lem:spatial_bounds} may be violated, e.g.~we could find $\vp_n(t)<\xi$ for some $n$ and $t>\tau$. Nevertheless, as with \autoref{thm:continuity_solution_map}, fix $T\in(\tau,T_0)$ and $X\in(\vp_0(T),X_0)$, then define the rectangle $\cX:=[\tau,T]\times[\xi,X]$. We will now clarify that, over some subinterval $[\tau,t_1]\subset[\tau,T]$, we still find $(t,\vp(t))\in\cX$ for sufficiently large $n$.
    
    Since $f_0(\tau,\xi)>0$ and $f_0(\cdot,\xi)$ is strictly increasing, then $f_0(t,\xi)>0$ over $[\tau,T]$, and the continuity of $f_0$ provides a rectangle $\cX_\ep:=[\tau,T]\times[\xi,\xi+\ep]\subset\cX$ where $f_0(t,x)>0$ also. Using $\Vert f_0 - f_n \Vert_{\cX_\ep}\xrightarrow{n\to\infty}0$, this can be extended w.l.o.g.~to every $n$, i.e.~we can assume $f_n(t,x)>0$ for all $n$ provided $(t,x)\in\cX_\ep$. Precisely as in \autoref{thm:continuity_solution_map}, the wider convergence $\Vert f_0 - f_n \Vert_\cX \xrightarrow{n\to\infty}0$ also allows us to assume w.l.o.g.~a bound $\Vert f_n \Vert_\cX<M$ for all $n$. 
    
    The rectangular sliver $\cX_\ep$ will become a reflecting barrier for the polygons $\vp_n$ for sufficiently large $n$, reestablishing the bound $\vp_n(t)\ge\xi$ as $n\to\infty$. To see this, consider $\vp_n$ over $[\tau,t_1]$ where $t_1:=\tau+M^{-1}(X-\vp_0(T))$ as in \autoref{thm:continuity_solution_map}. Since $\Vert f_n \Vert_\cX<M$, $\vp_n$ cannot reach the top of $\cX$ over $[\tau,t_1]$, but it could reach the bottom. To pull this off, however, $\vp_n$ must first escape through the top of $\cX_\ep$, to find a point where $f_n(t,x)<0$. For $\vp_n$ to \emph{then} reach the bottom of $\cX$ requires the crossing of $\cX_\ep$ in one forward Euler step, since $f_n(t,x)>0$ in $\cX_\ep$. This becomes impossible once the simulation mesh is sufficiently small, specifically once $\Vert\pi_n\Vert_{[\tau,t_1]}<M^{-1}\ep$. Since this is guaranteed by assumption as $n\to\infty$, we can now assume w.l.o.g.~that it holds for all $n$ so that, over $[\tau,t_1]$ at least, all $\vp_n$ are now contained in $\cX$.
    
    Now getting $\Vert \vp_0 - \vp_n\Vert_{[\tau,t_1]} \xrightarrow{n\to\infty}0$ utilises the general approach in \autoref{eq:approach_summary}, but the details are a little different to those of \autoref{thm:continuity_solution_map}, since $\vp_n(t)\neq \xi + \int_\tau^t f_n(s,\vp_n(s))\dd s$, i.e.~$\vp_n$ is not an IVP solution but rather a forward Euler polygon. Given however $\vp_n\in\uAC([\tau,\infty),\RR)$, then we at least have $\vp_n(t) = \xi + \int_{[\tau,t]} \vp'_n(s)\dd s$, which can be written
    \begin{equation}\label{eq:main_sim}
        \vp_n(t) = \xi + \int_{[\tau,t]} f_0(s,\vp_n(s)) + \lambda_n(s)\dd s,\quad \lambda_n(t) := \vp'_n(t) - f_0(t,\vp_n(t)).
    \end{equation}
    Notice the subtle difference between $\lambda_n$ here and in \autoref{eq:cont_dep_main}. For each $t\in[\tau,T]$, let $t_n=t_n(t):=\max\{t_k\in\pi_n:t_k<t\}$ denote the time point in the partition $\pi_n$ which immediately precedes $t$, so that $0<t-t_n\le \Vert\pi_{n}\Vert_{[\tau,T]}$ and $\vp'_n(t)=f_n(t_n,\vp_n(t_n))$ wherever $\vp'_n$ exists. Substituting this equality into $\lambda_n$ from \autoref{eq:main_sim}, for a.e.~$t\in[\tau,t_1]$ we obtain
    \begin{multline}\label{eq:the_ugly_one}
        |\lambda_n(t)| = |f_n(t_n,\vp_n(t_n)) - f_0(t,\vp_n(t))|\\
        \le\underbrace{|f_n(t_n,\vp_n(t_n)) - f_0(t_n,\vp_n(t_n))|}_{\le \Vert f_n-f_0\Vert_\cX\xrightarrow{n\to\infty}0} + \underbrace{|f_0(t_n,\vp_n(t_n)) - f_0(t,\vp_n(t)) |}_{\le w(\sqrt{1+M^2}\Vert\pi_{n}\Vert_{[\tau,T]})\xrightarrow{n\to\infty}0}
    \end{multline}
    which utilises the triangle inequality, the bound $|(t_n,\vp_n(t_n)) - (t,\vp_n(t)) |<\sqrt{1+M^2}\Vert\pi_{n}\Vert_{[\tau,T]}$, and the modulus of continuity $w=w_{f_0,\cX}$ from \autoref{eq:mod_cont_def}, which satisfies $w(r)\xrightarrow{r\downarrow0}0$.
    
    Having ensured $\Vert\vp_n\Vert_{[\tau,t_1]}<|\xi|\vee|X|$ and $\Vert\vp'_n\Vert_{[\tau,t_1]}<M$, the set $\{\vp_n\}_{n\in\NN_0}$ is equibounded and equicontinuous over $[\tau,t_1]$. Like in the proof of \autoref{thm:continuity_solution_map}, assuming the limit $\Vert\vp_0 - \vp_n\Vert_{[\tau,t_1]}\xrightarrow{n\to\infty}0$ is violated leads to a contradiction, by invoking the Ascoli Lemma and the limit $\int_{[\tau,t]} \lambda_n(s)\dd s\xrightarrow{n\to\infty}0$ in \autoref{eq:main_sim} to get a limit $\vp_*:=\lim_{k\to\infty}\vp_{n_k}\neq\vp_0$ which verifies $\vp_*(t)= \xi + \int_\tau^t f_0(s,\vp_*(s))\dd s$ like $\vp_0$. So we must conclude $\Vert\vp_n-\vp_0\Vert_{[\tau,t_1]}\xrightarrow{n\to\infty}0$, and again as in \autoref{thm:continuity_solution_map}, this can be extended in steps to any $[\tau,t_k\wedge T]$ with $t_k:=\tau + kM^{-1}(X-\vp_0(T))$, providing the claim of $\Vert\vp_n-\vp_0\Vert_{[\tau,T]}\xrightarrow{n\to\infty}0$ after $\lceil\frac{T-\tau}{t_1-\tau}\rceil$ steps.
\end{proof}

Given the generality of the partitions $\{\pi_n\}_{n\in\NN}\subset\Pi([\tau,\infty))$ in \autoref{thm:euler_converge}, we are free to decouple the assumption in \autoref{eq:euler_converge} to $\lim_{(n,m)\to(\infty,\infty)}\left(\Vert f_0 - f_n \Vert, \Vert\pi_m\Vert\right)$ $=(0,0)$. By utilising \autoref{thm:euler_converge} as stated and \autoref{thm:continuity_solution_map} from the previous section, it is then straightforward to establish that this joint limit can be replaced with either iterated variant $\lim_{n\to\infty}\lim_{m\to\infty}$ or $\lim_{m\to\infty}\lim_{n\to\infty}$. As already discussed, however, such iterated limits cannot be realised on a computer and so by themselves are not actually practically useful.

\begin{figure}[H]
    \centering
    \includegraphics[height=0.48\linewidth]{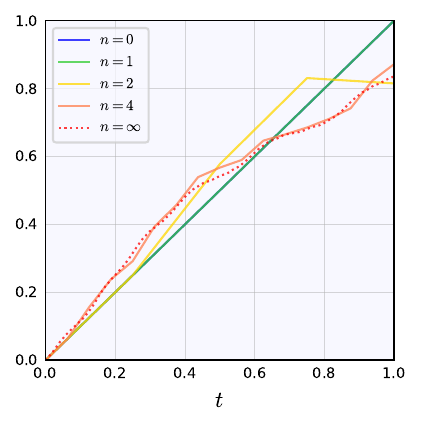}
    \includegraphics[height=0.48\linewidth]{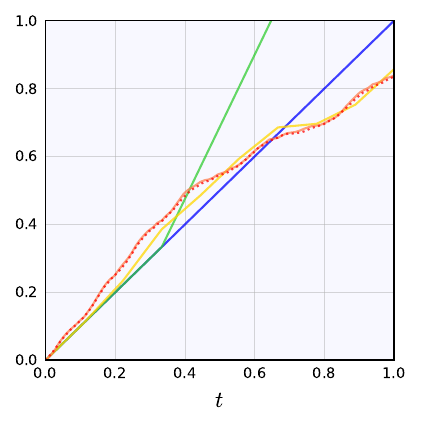}
    % \vspace{-2mm}
    \caption{The left panel shows forward Euler polygons converging to the IVP solution from \autoref{fig:main_examples} when using dyadic partitions. The right panel repeats the left but uses instead triadic partitions, so converges faster.}
    \label{fig:sim_converge1}
\end{figure}

\vspace{4mm}To illustrate \autoref{thm:euler_converge}, in \autoref{fig:sim_converge1} we reproduce \autoref{fig:main_examples} but also show sequences of forward Euler polygons $\vp_n$ converging to the IVP solution $\vp_0$. Dyadic and triadic partitions are used, defined by $\pi_n:=\{k m^{-n}:k\in\NN_0\}$ for $m=2,3$ respectively. For simplicity, $f_n:=f$ for all $n$, where $f(t,x):=\sigma w(x)+\kappa(\theta t - x)+v$ is the Heston function from \autoref{fig:weierstrass_example}.

% from 20201117-weierstrass-cont-dep

\vspace*{-4mm}
In general, we cannot bound the rates of convergence illustrated in \autoref{fig:sim_converge1} without imposing constraints on the spatial regularity of $f(t,x):=\sigma w(x)+\kappa(\theta t - x)+v$ via $w$. Nevertheless when we come to draw comparisons between our new volatility models and existing ones in \autoref{sec:vol_surfaces}, it is still important to gauge the error imparted by our forward Euler scheme on the metrics being compared. In \autoref{fig:heston_comparison_skewed} and \autoref{fig:heston_comparison_symmetric} the metrics being compared are implied volatilities, and we find that 4,096 simulated paths are sufficient to bring all simulated implied volatilities within 0.1 of the analytically generated Heston ones. Each one of these 4,096 paths depends on our forward Euler scheme from \autoref{alg:forward_euler}, and in the \hyperlink{appendix}{Appendix} we provide succinct \texttt{python} code which generates one such path with 4,096 time steps over 10 years, i.e.~using $\Delta:=10/4,096\approx0.0024$ in \autoref{alg:forward_euler}. As explained in the \hyperlink{appendix}{Appendix}, this code takes 75 ms to run on a 2.3 GHz Intel Core i5 MacBook Pro.

This concludes the present chapter, so the treatment of well-posedness for \autoref{prob:ivp}. While it is tempting to include here some minor modifications of this problem, especially regarding the delicate possibility of relaxing the assumption of $f(\tau,\xi)>0$ to $f(\tau,\xi)\ge0$, this seems better placed in the next chapter, which explores the solution space and its limits.

%% file: chapters/3_solutions.tex
\clearpage
\section{The solution space and exit-time limits}\label{chap:solutions}

This chapter has three main goals. Firstly, following the well-posedness analysis of \autoref{chap:wellposed}, \autoref{thm:include_zero} provides a condition applicable to \autoref{prob:ivp} which preserves results thus far and also accommodates initial values where $\vp'(\tau)=f(\tau,\xi) = 0$, rather than $f(\tau,\xi)>0$. Despite having negligible modelling consequences, \autoref{thm:include_zero} both instils harmony in our IVPs' solution space, and provides a tangible interpretation on why the maximal uniqueness result of \autoref{thm:global_uniqueness} holds despite $f(t,\vp(t)) = 0$ being possible for any future time $t>\tau$.

As with \autoref{chap:wellposed}, the subdomain $[\tau,\infty)\times[\xi,\infty)\subset\RR^2$ of a function $f\in\uF\subset\uC(\RR^2,\RR)$ determines the behaviour of solutions, so to avoid repetition of assumptions, the pragmatic conclusion of \autoref{sec:refining_problem} is to simplify the IVPs being considered, to be driven by related functions $g\in\uG\subset\uC(\RR_+^2,\RR)$ like the Heston example in \autoref{eq:hest_ode}. Having treated initial values with due care, only IVPs of type $x'=g(t,x)$, $x(0)=0$ are then w.l.o.g.~considered.

Such IVPs feature in \autoref{prob:ivp2}, and with this the modelling foundations for the remainder of the thesis are set. The second goal of this chapter is to precisely understand the solution set of these IVPs, and how certain discontinuous limit points can arise through simple sequences of them. This analysis takes us to \autoref{sec:bounds_limits} which, via \autoref{thm:exit_limit}, provides foundations for understanding surprising limiting relationship between the time-integrated CIR process and L\'evy subordinators such as the IG process from \autoref{eq:ig_intro}. But more generally a simple recipe is provided to construct \emph{any} strictly increasing and unbounded path in $\uD(\RR_+,\RR_+)$ as a limit of IVP solutions on an intuitive `exit-time' metric space $(\bPhi,d_{\bPhi})$. 

The final goal of this chapter is to expose the consequences of such limits $\vp_n\xrightarrow{n\to\infty}\vp_0$ on $(\bPhi,d_{\bPhi})$ for composite paths $w\circ\vp_n$, for any $w\in\uC(\RR_+,\RR)$. The composite convergence $w\circ\vp_n\xrightarrow{n\to\infty}w\circ\vp_0$ takes place pointwise a.e., but is violated on Skorokhod's topologies because $w\circ\vp_n$ can develop instantaneous `excursions' as $n\to\infty$. So in \autoref{sec:excursionary_limits} a Hausdorff metric space $(\uE,d_\uE)$ is introduced on which convergence of graphs is established. This provides the foundations to answer questions in the \hyperlink{prologue}{Prologue} on the Heston and NIG relationship, eventually extending \autoref{thm:mech_pro} to a practically valuable functional result.

With this pathwise theory in place, we will be ready to move to the probabilistic framework of \autoref{chap:framework} in which the results thus far, relating to IVP solutions $\vp$, will apply a.s.~to cumulative variance and price processes, $X$ and $S$. As outlined in \autoref{ch:intro}, these will be related through a \emph{composition} $S=\exp(W^\rho_X-\frac12 X)$, hence our emphasis on composite paths.

\subsection{Simplifying the problem}\label{sec:refining_problem}

In \autoref{chap:wellposed}, only IVPs $x'=f(t,x)$, $x(\tau)=\xi$ with initial values $(\tau,\xi)\in\RR^2$ where $f(\tau,\xi)>0$ were considered, meaning that a solution $\vp$ verifies $\vp'(\tau)=f(\tau,\xi)>0$. However, the maximal uniqueness result of \autoref{thm:global_uniqueness} accommodates times $t>\tau$ where $\vp'(t)=f(t,\vp(t))=0$, suggesting some naturally occurring conditions where $f(\tau_*,\xi_*)=0$ for some $\tau_*>\tau$ and $\xi_*>\xi$, yet uniqueness of the translated IVP $x'=f(t,x)$, $x(\tau_*)=\xi_*$ still holds.

Indeed, a consequence of \autoref{thm:include_zero} here is that:~provided the point $(\tau,\xi)\in\RR^2$ is attainable by some strictly increasing (`history') $\vp\in\uC^1((T,\tau],\RR)$ which solves the \emph{terminal} value problem $x'=f(t,x)$, $x(\tau)=\xi$, then there exists a unique maximal solution (`future') of the corresponding \emph{initial} value problem, whether $f(\tau,\xi)=0$ or not. This natural stability is peculiar, given the forthcoming counterexamples to uniqueness when $f(\tau,\xi)=0$ in general.

Following these counterexamples and the proof of \autoref{thm:include_zero}, the new set of IVP functions $\uG\subset\uC(\RR_+^2,\RR)$ introduced in \autoref{ch:intro} is properly defined, which will be prioritised in the remainder of this chapter and taken into \autoref{chap:framework}. Although this set $\uG$ is not quite as simple to define as $\uF$, the related maximal solutions $\vp$ are certainly simpler to analyse. This is primarily because these always constitute differentiable bijections from and to $\RR_+$, so maximal solutions are in fact global. As discussed, this is very helpful when moving to a probabilistic setting, given requirements to understand the solution space topologically.

\vspace{3mm}\textbf{Non-uniqueness examples.} Recall the common counterexample to IVP uniqueness given by $x'=f(t,x):=|x|^\alpha$, $x(0)=0$ for some $\alpha\in(0,1)$. It is straightforward to verify the two global solutions $\vp_\infty(t):=0$ and $\vp_0(t):=((1-\alpha)t)^\frac{1}{1-\alpha}$, and to combine these to get others,
\begin{equation}\label{eq:standard_nonunique}
    \vp_T(t) := 
    \begin{cases}
        \vp_\infty(t) & t\in[0,T),\\
        \vp_0(t-T) & t\in[T,\infty),
    \end{cases}
\end{equation}
for any $T\in(0,\infty)$. Indeed, the convergences $\vp_T\xrightarrow{T\downarrow 0}\vp_0$ and  $\vp_T\xrightarrow{T\uparrow\infty}\vp_\infty$ then take place uniformly over compacts. The point of raising this example is to clarify that, in our case where $f\in\uF$, we do not need to worry about this translation of solutions along a line like $x=0$ here. This is because such lines, where $f(\cdot,x)$ is necessarily zero, are precluded by the condition that every $f(\cdot,x)$ is strictly increasing. However, notice that when this IVP is adapted to $x'=\sgn(x)|x|^\alpha$, $x(0)=0$, then we get additional \emph{negative} solutions like $-\vp_0$. These are precisely the kind of non-uniqueness examples which we \emph{do} need to worry about.

For a thorough extension of these cases to our setting, we could consider functions of type 
\begin{equation}
    f(t,x) := I_{a,b}(t)|t|^\alpha + I_{c,d}(x)|x|^\beta,\quad I_{a,b}(t):=a\mathbbm{1}_{t<0} + b\mathbbm{1}_{t\ge0}
\end{equation}
for $a,b,c,d\in\RR$ and $\alpha,\beta\in(0,1)$. Notice that $f\in\uF$ provided that $a<0<b$, given every $f(\cdot,x)$ is then strictly increasing. Checking that $I_{c,d}(x)=\sgn(x)$ when $(c,d)=(-1,1)$, it is interesting to note that we can still get non-uniqueness without this ordering $c<0<d$.

For the sake of clarity, we will just consider cases $f\in\uF$ with the simplified representation
\begin{equation}\label{eq:nonunique_example}
    f(t,x) := t + I_{a,b}(x)\sqrt{|x|}
\end{equation}
for some $a,b\in\RR$. Notice that the IVP $x'=f(t,x)$, $x(\tau)=0$ then provides an example of \autoref{prob:ivp} whenever $\tau>0$ because $f(\tau,0)=\tau>0$, contrasting $\tau=0$ given $f(0,0)=0$.

\begin{figure}[ht]
    \centering
    \includegraphics[width=0.48\linewidth]{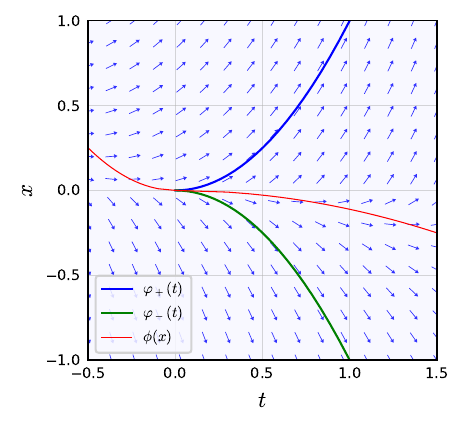}
    \includegraphics[width=0.48\linewidth]{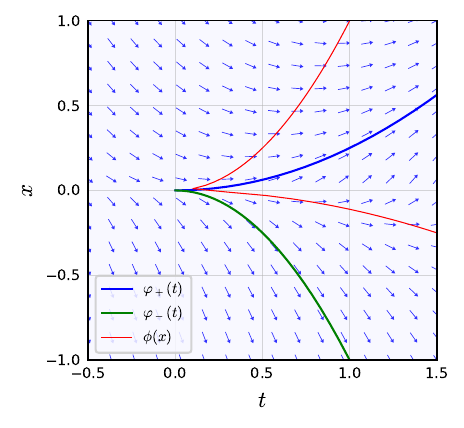}
    \caption{The functions $f(t,x)$ (blue arrows) from \autoref{eq:nonunique_example}, where $(a,b)=(-3,1)$ in the left panel and $(a,b)=(-3,-1)$ in the right panel. The two solutions $\vp_\pm(t)$ of the IVP $x'=f(t,x)$, $x(0)=0$ are shown, with the path $\phi(x) = - I_{a,b}(x)\sqrt{|x|}$.}
    \label{fig:nonuniqueness}
\end{figure}

In the case of $-3=:a<0<b:=1$ and $\tau=0$, we get the two parabolic solutions $\vp_{\pm}(t):=\pm t^2$, which is straightforward to confirm. But even if we set $-3=:a<b:=-1<0$, so the sign of $I_{a,b}$ no longer changes over the line $x=0$, we still get two solutions, namely $\vp_+(t)=\frac14 t^2$ and $\vp_-(t)=-t^2$. The solutions $\vp_\pm$ for these examples are demonstrated in \autoref{fig:nonuniqueness}, also with the path $\phi(x):=-I_{a,b}(x)\sqrt{|x|}$ from \autoref{lem:zeros1}, where $f(\phi(x),x)=0$. 

So while the results of \autoref{chap:wellposed} apply to these examples from \autoref{eq:nonunique_example} when $\tau>0$, few do when $\tau=0$. This clearly demonstrates that we cannot in general relax the condition $f(\tau,\xi)>0$ in the statement of \autoref{prob:ivp} to $f(\tau,\xi)\ge0$, and retain well-posedness qualities. 

Following this breakdown of uniqueness, it is straightforward to construct examples of `discontinuous dependence', violating \autoref{thm:continuity_solution_map}. For example, let the function $f_0$ be defined as in \autoref{eq:nonunique_example} with $(a,b)=(-3,1)$ and take the global solution $\vp_0(t):=-t^2$ of the IVP $x'=f_0(t,x)$, $x(0)=0$. Then define $f_n(t,x):=f_0(t+n^{-1},x)$ which generates unique global parabolic solutions $\vp_n(t):=(t+\frac12 n^{-1})^2$. Now clearly $\Vert f_0-f_n \Vert_{[0,T]\times[0,X]}\xrightarrow{n\to\infty}0$ for all $T,X>0$, as per the conditions of \autoref{thm:continuity_solution_map}, but we find $\Vert\vp_0-\vp_n\Vert_{[0,T]}\xrightarrow{n\to\infty}2T^2$. This distance $2T^2$ is nothing but the difference $\vp_+(T)-\vp_-(T)$ in the left panel of \autoref{fig:nonuniqueness}.

Before proceeding, it is worth making two pragmatic remarks. Firstly, it is practically reassuring that the forward Euler scheme of \autoref{alg:forward_euler} will not only converge for any example like these, but will converge to the desirable strictly increasing solution. This is to say, \autoref{thm:euler_converge} still holds when $f(\tau,\xi)=0$. This is straightforward to see with \autoref{fig:nonuniqueness} in mind; if $f(\tau,\xi)=0$, then any $\pi$-polygon for the IVP $x'=f(t,x)$, $x(\tau)=\xi$ satisfies $\vp_\pi(t_1)=\vp_\pi(t_0) + (t_1-t_0)\times 0=\xi$, and then $\vp_\pi(t_2)=\vp_\pi(t_1) + (t_1-t_0)f(t_1,\vp_\pi(t_1)) = \xi + (t_1-\tau)f(t_1,\xi)>\xi$, given $t_1>\tau$ and $f(\cdot,\xi)$ is strictly increasing. Such polygons thus converge to the \emph{strictly increasing} limit as $\Vert\pi\Vert\to0$, for the same reasons as in \autoref{thm:euler_converge}.

\begin{figure}[ht]
    \centering
    \includegraphics[width=0.48\linewidth]{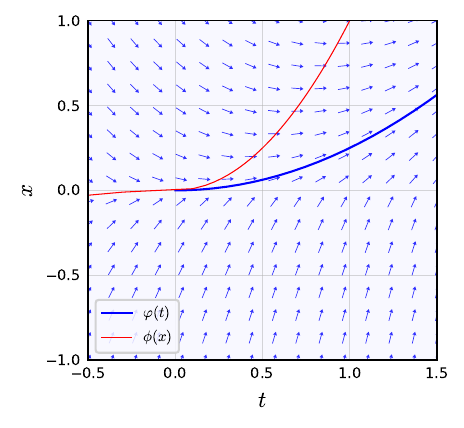}
    \includegraphics[width=0.48\linewidth]{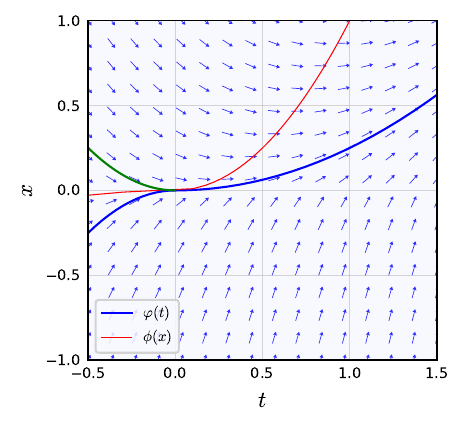}
    \caption{A repeat of the right panel of \autoref{fig:nonuniqueness}, with instead $(a,b)=(3,-1)$. Now a physically meaningful (strictly increasing) TVP solution which arrives at $(0,0)$ exists over $-\RR_+$, and the corresponding IVP has a unique global solution over $\RR_+$.}
    \label{fig:nonuniqueness2}
\end{figure}

Secondly, it is worth at this point noticing in 
\autoref{fig:nonuniqueness} that, given one finds $f(t,x)<0$ in all of $(-\infty,\tau)\times(-\infty,\xi)$, there can clearly be no strictly increasing solution $\vp\in\uC^1((T,\tau],\RR)$ to the TVP $x'=f(t,x)$, $x(\tau)=\xi$ when $(\tau,\xi)=(0,0)$, i.e.~no physically meaningful `history' which arrives at the point $(0,0)$. This renders $(0,0)$ an unnatural initial value to choose, given we cannot make sense of volatility $\sqrt{\vp'}$ in the past. We only have to reverse the sign of $a$ in \autoref{fig:nonuniqueness} to avoid this, as \autoref{fig:nonuniqueness2} illustrates. But the right hand panel of \autoref{fig:nonuniqueness2} shows that, as a result, uniqueness can now be violated going \emph{backwards} in time from $(0,0)$, which justifies earlier remarks related to the time-irreversibility of the IVPs considered here.

\vspace{3mm}\textbf{Conditions enabling $\boldsymbol{f(\tau,\xi)=0}$.} We now provide a condition which ensure we are in the setting of \autoref{fig:nonuniqueness2} rather than \autoref{fig:nonuniqueness} when selecting initial values $(\tau,\xi)$. While there may not be a unique solution arriving at $(\tau,\xi)$ from the past, this will ensure there is always a unique and physically meaningful solution into the future. It is important to recognise that this condition is applicable to the general setting of \autoref{chap:wellposed} where $f\in\uF$, and not just the visually helpful examples just covered. The relatively simple functions $I_{a,b}(x)\sqrt{x}$ from \autoref{eq:nonunique_example} could therefore be replaced by any path $w\in\uC(\RR,\RR)$, like those from \autoref{ex:main_example}, and these problematic points, like $x=0$ in \autoref{fig:nonuniqueness}, could be dense in $\RR$.

The forthcoming condition of \autoref{thm:include_zero} applies to the triple $(f,\tau,\xi)\in\uF\times\RR^2$ and is geometrically intuitive, given \autoref{fig:nonuniqueness2}. Following this, a more abstract interpretation of this condition, in terms of the path $\phi\in\uC(\RR,\bRR)$ of zeros from \autoref{lem:zeros1}, is given. First, for each initial value $(\tau,\xi)\in\RR^2$ and $c\in\RR$, $\ep>0$, let the line $\cX_{c,\ep}(\tau,\xi)\subset\RR^2$ be defined by
\begin{equation}\label{eq:line_def}
    \cX_{c,\ep}(\tau,\xi) := \{(\tau+cx,\xi-x):x\in(0,\ep)\}.
\end{equation}

\begin{theorem}[Additional initial values]\label{thm:include_zero}
    All of the results of \autoref{chap:wellposed}, relating to an IVP $x'=f(t,x)$, $x(\tau)=\xi$ with $f\in\uF$ and $f(\tau,\xi)>0$, hold if the requirement `$f(\tau,\xi)>0$' is replaced by the existence of any line $\cX_{c,\ep}(\tau,\xi)$ such that $f(t,x)>0$ for all $(t,x)\in\cX_{c,\ep}(\tau,\xi)$.
\end{theorem}
\begin{proof}
    There is nothing to prove if $f(\tau,\xi)>0$. The assumption that $f(t,x)>0$ on some such line $\cX_{c,\ep}(\tau,\xi)$ precludes $f(\tau,\xi)<0$ by the continuity of $f$, so now suppose $f(\tau,\xi)=0$.
    
    In \autoref{chap:wellposed} we swiftly concluded that any maximal solution $\vp\in\uC^1([\tau,T_*),\RR)$ is strictly increasing, in \autoref{thm:bijectivity_new}, which of course assumes $f(\tau,\xi)>0$. If the existence of a line $\cX_{c,\ep}(\tau,\xi)$ ensures that any solution going forward in time from $(\tau,\xi)$ is still strictly increasing, even when $f(\tau,\xi)=0$, then the results of \autoref{chap:wellposed} (e.g.~the bounds of \autoref{lem:spatial_bounds}, uniqueness of \autoref{thm:global_uniqueness}, continuous dependence of \autoref{thm:continuity_solution_map}) hold by the proofs given there. Loosely, if a solution here \emph{is} still strictly increasing, this clearly only depends on the subdomain $[\tau,\infty)\times[\xi,\infty)$ of $f$, and so we can simply imagine that a solution from \autoref{chap:wellposed} \emph{starting from an earlier time} has arrived at this point $(\tau,\xi)$ where $f(\tau,\xi)=0$.
    
    So let $\vp\in\uC^1([\tau,T),\RR)$ be a local solution of the IVP $x'=f(t,x)$, $x(\tau)=\xi$ with $f(\tau,\xi)=0$, and assume $f(t,x)>0$ on $\cX_{c,\ep}(\tau,\xi)$ for some $c\in\RR$ and $\ep>0$. It will help to appreciate that if $c=0$, then $\cX_{c,\ep}(\tau,\xi)$ is just the open vertical line of length $\ep$ below the point $(\tau,\xi)$, and otherwise it covers the same distance of $\ep$ downwards, but with gradient $-1/c\in\RR\setminus\{0\}$.
    
    As discussed following the example in \autoref{eq:standard_nonunique}, we cannot find $\vp'(t)=0$ over an interval $[\tau,\tau+\ep)$, given that $f(\cdot,\xi)$ is strictly increasing. So as in \autoref{fig:nonuniqueness}, $\vp$ immediately enters one of the quadrants $\cX_-:=(\tau,\infty)\times(-\infty,\xi)$ or $\cX_+:=(\tau,\infty)\times(\xi,\infty)$, like $\vp_-$ and $\vp_+$ in \autoref{fig:nonuniqueness} respectively. If $\vp$ immediately enters $\cX_+$, then the bounds $\xi\le\vp(t)\le\bvp(t)$ from \autoref{lem:spatial_bounds} are thereafter enforced and $\vp$ is strictly increasing using the proof of \autoref{thm:bijectivity_new}. So now we just have to show that such a line $\cX_{c,\ep}(\tau,\xi)$ precludes immediate entry into $\cX_-$.
    
    By definition, we have $f(t,x)>0$ on the line $\cX_{c,\ep}(\tau,\xi)$, and so also in the entire region
    \begin{equation}
        \cX_{c,\ep}^\to(\tau,\xi):=\{(t,\xi-x):x\in(0,\ep), t\ge\tau+cx\}\subset (\tau,\infty)\times(\xi,\xi-\ep)
    \end{equation}
    on or to the right of $\cX_{c,\ep}(\tau,\xi)$, given each $f(\cdot,x)$ is strictly increasing. But $\vp'(\tau)=f(\tau,\xi)=0$, so $\vp$ cannot immediately enter $\cX_-$ without entering $\cX_{c,\ep}^\to(\tau,\xi)$. But this is now clearly impossible, given $f(t,x)>0$ in $\cX_{c,\ep}^\to(\tau,\xi)$. So any local solution like $\vp\in\uC^1([\tau,T),\RR)$ of the IVP $x'=f(t,x)$, $x(\tau)=\xi$ is instead strictly increasing, and the proof is thus complete.
\end{proof}

We can now give proper meaning to previous descriptions of initial values $(\tau,\xi)$ as natural, physically meaningful, etc.~for the IVPs of \autoref{prob:ivp}, given our focus on volatility $\sqrt{\vp'}$.

\begin{definition}[Natural initial values]\label{def:natural_point}
    For $f\in\uF$, the initial value $(\tau,\xi)\in\RR^2$ of the IVP $x'=f(t,x)$, $x(\tau)=\xi$ will be called natural if the corresponding TVP admits a strictly increasing solution $\vp\in\uC^1((T,\tau],\RR)$ for some $T\in[-\infty,\tau)$, so that $\sqrt{\vp'(t)}\ge0$ over $(T,\tau]$.
\end{definition}

Given the strictly increasing TVP solution in \autoref{fig:nonuniqueness2}, it is clear that the point $(0,0)$ is natural, unlike in \autoref{fig:nonuniqueness}, despite having $f(0,0)=0$ in both cases. Natural initial values of course define a subset of those covered by \autoref{thm:include_zero}. To see this, take a strictly increasing TVP solution $\vp\in\uC^1((T,\tau],\RR)$, then having $f(t,\vp(t))\ge0$ for $t\in(T,\tau]$ ensures that $f(\tau,x)>0$ for $x\in(\vp(T),\xi)$. That is, $f(t,x)>0$ on the vertical line $\cX_{c,\ep}(\tau,\xi)$ with $c=0$ and $\ep=\xi-\vp(T)$, in fact any $c\ge0$ and $\ep\le \xi-\vp(T)$. So either the conditions of \autoref{thm:include_zero} or \autoref{def:natural_point} ensures the IVP is well-posed in every sense of \autoref{chap:wellposed}. 

Notice that if $f(\tau,\xi)=0$, then $\phi(\xi)=\tau$, where $\phi\in\uC(\RR,\bRR)$ is the path from \autoref{lem:zeros1}, with $f(\phi(x),x)=0$ whenever $\phi(x)\in\RR$. So, before moving on to the main simplifying purpose of this section, we clarify what the existence of the line $\cX_{c,\ep}(\tau,\xi)$ from \autoref{thm:include_zero} means for $\phi$. Keep in mind the paths $\phi$ illustrated in \autoref{fig:nonuniqueness} and \autoref{fig:nonuniqueness2} for $x\in(-1,0)$.

Assuming $f(t,x)>0$ on $\cX_{c,\ep}(\tau,\xi)$, then with every $f(\cdot,x)$ being strictly increasing, the path where $f(t,x)=0$ resides to the left of $\cX_{c,\ep}(\tau,\xi)$. Specifically, we find $\phi(x)-\phi(\xi)<c(\xi-x)$ for $x\in(\xi-\ep,\xi)$, and so the existence of $\cX_{c,\ep}(\tau,\xi)$ provides this one-sided Lipschitz condition on $\phi$ \emph{at the point} $(\tau,\xi)$. Conversely, if $\phi(x)-\phi(\xi)<L(\xi-x)$ for $x$ in some $(\xi-\ep,\xi)$, with $L\in\RR$ and $\phi(\xi)=\tau$, then we must have $f(t,x)>0$ on every $\cX_{c,\ep}(\tau,\xi)$ with $c\ge L$. So the existence of $\cX_{c,\ep}(\tau,\xi)$ and $\phi$ having this one-sided Lipschitz condition are equivalent when $f(\tau,\xi)=0$. Although this one-sided condition is related to that in \autoref{lem:strict_upper}, which \emph{precludes} finding points where $\vp'(t)=0$, these conditions apply in opposing $x$ directions.

\vspace{3mm}\textbf{Simplified problems.} Although \autoref{thm:include_zero} is an informative theoretical result, the repetition of its condition depending on such a line $\cX_{c,\ep}(\tau,\xi)$ from \autoref{eq:line_def}, or indeed the weaker condition in \autoref{def:natural_point}, is superfluous, given we can proceed more pragmatically.

Towards this, we are firstly going to henceforth fix the initial value $(\tau,\xi):=(0,0)$. It has thus far been helpful to have the freedom to vary $(\tau,\xi)\in\RR^2$, but there is now no need to, given that having $\vp$ solve $x'=f(t,x)$, $x(\tau)=\xi$ is equivalent to having $\vp_{\tau,\xi}$ solve the shifted version $x'=f_{\tau,\xi}(t,x)$, $x(0)=0$, where $\vp_{\tau,\xi}(t):=\vp(t+\tau)-\xi$ and $f_{\tau,\xi}(t,x):=f(t+\tau,x+\xi)$.

Secondly, we will no longer consider functions $f\in\uF$ defined from all of $\RR^2$, but rather the forthcoming functions $g\in\uG$ just from $\RR^2_+$. This is natural given we are just interested in strictly increasing IVP solutions. Needless to say, we will always be able to map conclusions drawn for IVPs $x'=g(t,x)$, $x(0)=0$ back to cases of \autoref{prob:ivp} with the same unique maximal solution. As an example, consider $x'=f(t,x)$, $x(0)=0$ where $f\in\uF$ is defined by
\begin{equation}\label{eq:func_extension}
    f(t,x):=
    \begin{cases}
        g(t,0\vee x) & \text{ if } t\ge0,\\
        2g(0,0\vee x) - g(-t,0\vee x) & \text{ if } t<0.
    \end{cases}
\end{equation}

Finally, as mentioned now many times, it is going to be very helpful to ensure that maximal solutions $\vp\in\uC^1_0([0,T_*),[0,X_*))$ of such IVPs do not just verify $T_*\vee X_*=\infty$ as in \autoref{thm:bijectivity_new}, but rather define spatially unbounded global solutions, i.e.~$T_*= X_*=\infty$, so $\vp\in\uC^1_0(\RR_+,\RR_+)$. To ensure this, conditions like those of \autoref{cor:exist_summary} will be imposed. As introduced in \autoref{ch:intro}, we then arrive at the following subset $\uG\subset\uC(\RR_+^2,\RR)$, related to $\uF$.

\textbf{\autoref{def:mod_driving_func}} (Set $\uG$ of functions)\textbf{.}
    Let the subset $\uG\subset\uC(\RR_+^2,\RR)$ contain the functions $g$ which are such that: 1.~$g(0,0)\ge0$; 2.~$g(\cdot,x)$ is strictly increasing for each $x\in\RR_+$, and;
    \begin{equation}
        \text{3.~}\inf_{x\in\RR_+}g(t,x)<0\ \ \forall t\in\RR_+; \quad\text{4.~} \sup_{t\in\RR_+}g(t,x)>0\ \ \forall x\in\RR_+.
    \end{equation}

The related problem was already provided in \autoref{ch:intro}, but is repeated here for convenience.

\textbf{\autoref{prob:ivp2}} (IVPs of \autoref{chap:solutions})\textbf{.}
    For $g\in \uG$, find a global solution $\vp\in\uC^1_0(\RR_+,\RR_+)$ of the IVP $x'=g(t,x)$, $x(0)=0$. That is, $\vp$ verifying $\vp'(t)=g(t,\vp(t))$ for $t\in\RR_+$ and $\vp(0)=0$.

Notice that, by defining functions in $\uG$ only from $\RR_+^2$, it follows from nothing but definitions that such a global solution $\vp\in\uC^1_0(\RR_+,\RR)$ \emph{cannot} be negative, and so all the careful considerations leading to \autoref{thm:include_zero}, which serves to preclude this possibility, is pragmatically avoided. It is an understatement to say that this analysis can simply be forgotten, however. All that remains now is to consolidate results applicable to solutions of this \autoref{prob:ivp2}.

This next consolidatory result makes several specific statements applicable to \autoref{prob:ivp2}, although, more broadly, the point is that all results in \autoref{chap:wellposed} applicable to bijective maximal solutions $\vp\in\uC^1([\tau,T_*),[\xi,X_*))$ of \autoref{prob:ivp} apply also to solutions of \autoref{prob:ivp2}, after simply fixing the initial values $\tau=\xi=0$ and the interval end points $T_*=X_*=\infty$. 

For brevity, now set $\uC^1_0:=\uC^1_0(\RR_+,\RR_+)$ and $\uD:=\uD(\RR_+,\RR_+)$. Recall the norm $\Vert\cdot\Vert_{\RR_+}$ on $\uC^1_0$ from \autoref{def:uni_metric}, which characterises uniform convergence over compacts, and similarly define the following norms on sets $\Pi(\RR_+)$ and $\uG$ of partitions and functions respectively
\begin{equation}
    \Vert \pi \Vert_{\RR_+} := \sum_{n\in\NN} 2^{-n}(1\wedge \Vert \pi \Vert_{[0,n]}),\quad \Vert g \Vert_{\RR_+^2}:= \sum_{n\in\NN} 2^{-n}(1\wedge \Vert g \Vert_{[0,n]^2}).
\end{equation}

\begin{theorem}[Well-posedness for \autoref{prob:ivp2}]\label{thm:wellposed}
    Assume $g\in\uG$ from \autoref{def:mod_driving_func}. Then, regarding any such IVP $x'=g(t,x)$, $x(0)=0$ in \autoref{prob:ivp2}, the following results hold\emph{:}
    
    \emph{1 (Global existence and uniqueness).} This IVP has a unique maximal solution $\vp$, which is a strictly increasing and unbounded path in $\uC^1_0$, so is always the unique \emph{global} solution\emph{;}

    \emph{2 (Upper bound).} This unique global solution $\vp$ is bounded above by the strictly increasing and unbounded c\`adl\`ag path $\bvp\in\uD$, well-defined over $\RR_+$ by $\bvp(t):=\inf\{x>0:g(t,x)<0\}$\emph{;}
    
    \emph{3 (Continuous dependence).} The solution map of \autoref{prob:ivp2}, from $\uG$ to $\uC^1_0$, is continuous w.r.t.~uniform convergence over compacts. That is, for $\{g_n\}_{n\in\NN_0}$ with solutions $\{\vp_n\}_{n\in\NN_0}$,
    \begin{equation}\label{eq:new_cont_statement}
        \Vert g_0-g_n\Vert_{\RR_+^2}\xrightarrow{n\to\infty}0 \implies \Vert \vp_0-\vp_n\Vert_{\RR_+}\xrightarrow{n\to\infty}0;
    \end{equation}
    
    \emph{4 (Simulation convergence).} Any sequence $\{\vp_n\}_{n\in\NN_0}$ of forward Euler polygons from \autoref{alg:forward_euler}, using instead $\vp_\pi(t) := (\vp_\pi(t_k) + (t-t_k) g(t_k,\vp_\pi(t_k)))_+$ in \autoref{alg:equation}, where $\cdot_+:=0\vee \cdot$, converges uniformly over compacts to $\vp$ as $n\to\infty$, provided $\Vert\pi_n\Vert_{\RR_+}\xrightarrow{n\to\infty}0$.
\end{theorem}
\begin{proof}
    Comparing the related sets $\uG$ from \autoref{def:mod_driving_func} with $\uF$ from \autoref{def:driving_func}, it is clear that, given any $g\in\uG$, a function $f\in\uF$ can always be constructed which coincides with $g$ on $\RR_+^2$ and is extended into $\RR^2\setminus\RR^2_+$ in a way which meets the conditions of \autoref{thm:include_zero}. Indeed, \autoref{eq:func_extension} provides one example of this. This ensures that the IVPs $x'=f(t,x)$, $x(0)=0$ and $x'=g(t,x)$, $x(0)=0$ have the same unique bijective maximal solution $\vp\in\uC^1_0([0,T_*),[0,X_*))$. Now points 1--4 can be confirmed by using results from \autoref{chap:wellposed}.
    
    1. By \autoref{thm:bijectivity_new}, any maximal solution $\vp$ of the IVP $x'=g(t,x)$, $x(0)=0$ defines a strictly increasing bijection in some set $\uC^1([0,T_*),[0,X_*))$ with $T_*\vee X_*=\infty$. This is unique, using \autoref{thm:global_uniqueness}. We find that $T_*=X_*=\infty$ by \autoref{cor:exist_summary}, the conditions of which are met by the assumptions of $\sup_{t\in\RR_+}g(t,x)>0$ and $\inf_{x\in\RR_+}g(t,x)<0$ in \autoref{def:mod_driving_func}.
    
    2. The function $\bvp$ defines a strictly increasing path in $\uD(\RR_+,\bRR_+)$ by \autoref{lem:spatial_bounds}, but the assumption $\inf_{x\in\RR_+}g(t,x)<0$ in \autoref{def:mod_driving_func} in fact ensures $\bvp(t)<\infty$, so actually $\bvp\in\uD$. The bound $\vp(t)\le\bvp(t)$ then holds by \autoref{lem:spatial_bounds}, and since $\vp$ is unbounded, so too is $\bvp$.
    
    3. This follows from \autoref{thm:continuity_solution_map} when setting $\tau=\xi=0$ and $T_0=X_0=\infty$ there. For clarity, from the assumption $\Vert g_0-g_n\Vert_{\RR_+^2}\xrightarrow{n\to\infty}0$ we have $\Vert g_0 - g_n \Vert_{[0,T]\times[0,X]}$ for any $T,X>0$, thus $\Vert \vp_0 - \vp_n \Vert_{[0,T]}$ for any $T>0$ by \autoref{thm:continuity_solution_map}, and therefore $\Vert\vp_0-\vp_n\Vert_{\RR_+}\xrightarrow{n\to\infty}0$.
    
    4. This follows from \autoref{thm:euler_converge}. The use of $\vp_\pi(t) := (\cdot)_+$ just ensures that polygons $\vp_n$ never escape the domain $\RR^2_+$ where $g$ is defined. One could instead use $\vp_\pi(t) := \vp_\pi(t_k) + (t-t_k) g(t_k,\vp_\pi(t_k))_+$, which also ensures polygons are non-decreasing, like the limit $\vp$.
\end{proof}

With \autoref{prob:ivp2} and these well-posedness results in \autoref{thm:wellposed}, we are now in the robust setting of the remainder of this chapter. This setting extends naturally to \autoref{chap:framework} also, once an appropriate probability space has been defined supporting \emph{random} functions in $\uG$.

% \clearpage
\subsection{The problem's solution map}\label{sec:solution_space}

This section studies the solution map of \autoref{prob:ivp2}, i.e.~the map which takes each function $g\in\uG$ to the global solution $\vp\in\uC^1_0:=\uC^1_0(\RR_+,\RR_+)$ of the IVP $x'=g(t,x)$, $x(0)=0$. From points 1.~and 3.~of \autoref{thm:wellposed}, we already know this map to be well-defined and continuous from $\uG$ to the problem's solution set, w.r.t.~uniform convergence over compacts (equivalently, w.r.t~the norms in \autoref{eq:new_cont_statement}). The first focus is to establish what this solution set actually is. Clearly understanding this set is important in practice, because it precisely describes the paths $\vp$ which, in theory, we can model using \autoref{prob:ivp2}, and therefore the corresponding volatility $\sqrt{\vp'}$. We already know from point 1.~of \autoref{thm:wellposed} that this solution set is contained in the subset $\Phi\subset\uC^1_0$ introduced in \autoref{ch:intro}, but repeated here for convenience. Of course, ideally, the solution set would be the entirety of this set $\Phi$.

\textbf{\autoref{def:solutionset}} (Set $\Phi$ of paths)\textbf{.}
    Let the set $\Phi$ contain the bijective paths in $\uC^1_0(\RR_+,\RR_+)$.

Paths in $\Phi$ of course satisfy $\vp(0)=0$ and are strictly increasing. As usual, let $\vp^{-1}\in\uC_0$ denote the inverse of any path $\vp\in\Phi$, which satisfies $\vp^{-1}(\vp(t))=t$ and $\vp(\vp^{-1}(x))=x$ for all $(t,x)\in\RR_+^2$. This inverse is clearly well-defined, and is similarly strictly increasing and bijective. As discussed in \autoref{ch:intro}, this set $\Phi$ precisely captures the possible future trajectories of a price process's cumulative variance which we are interested in modelling. This next result clarifies that \autoref{prob:ivp2} is, uncoincidentally, well suited to this task.

\begin{theorem}[The solution set]\label{thm:solutionset}
    The global solution set of \autoref{prob:ivp2} is $\Phi$. In particular, fixing any $\vt\in\uC_0(\RR_+,\RR)$, then each $\vp\in\Phi$ solves the IVP $x' = g(t,x)$, $x(0)=0$ when
    \begin{equation}\label{eq:g_rep}
        g(t,x):=\vp'(\vp^{-1}(x)) + \vt(t) - \vt(\vp^{-1}(x)).
    \end{equation}
    This IVP provides an example of \autoref{prob:ivp2}, i.e.~$g\in\uG$ from \autoref{def:mod_driving_func}, if $\vt$ is strictly increasing with $\sup_{t\in\RR_+}\vt(t)-\vp'(t)=\infty$. In this case, $\vp$ is this IVP's \emph{unique} global solution.
\end{theorem}
\begin{proof}
    Firstly, notice that any solution $\vp$ of \autoref{prob:ivp2} is in $\Phi$, using \autoref{thm:wellposed}. The solution set of \autoref{prob:ivp2} will be \emph{precisely} $\Phi$ if any $\vp\in\Phi$ can be constructed as claimed. 
    
    Fixing $\vp\in\Phi$, we always have $\vp(0)=0$, so for $\vp$ to be the global solution of $x' = g(t,x)$, $x(0)=0$ we require $\vp'(t)=g(t,\vp(t))$ over $\RR_+$. Substitution from \autoref{eq:g_rep} provides
    \begin{equation}
        g(t,\vp(t)) =\vp'(\vp^{-1}(\vp(t))) + \vt(t) - \vt(\vp^{-1}(\vp(t))) = \vp'(t)
    \end{equation}
    for each $t\in\RR_+$, where $\vp^{-1}(\vp(t))=t$ is used twice. Notice that this holds for any $\vt\in\uC_0(\RR_+,\RR)$, and at this point we may \emph{not} have $g\in\uG$, and $\vp$ may \emph{not} be this IVP's only solution. If we \emph{do} have $g\in\uG$, then \autoref{thm:wellposed} ensures $\vp$ is the unique global solution. So it just remains to show $g\in\uG$ when $\vt$ is strictly increasing with $\sup_{t\in\RR_+}\vt(t)-\vp'(t)=\infty$.
    
    To establish $g\in\uG$, we just have to check the three requirements from \autoref{def:mod_driving_func}. Firstly, $g(0,0)=\vp'(0) + \vt(0)-\vt(0) =\vp'(0)\ge0$, which uses $\vp^{-1}(0)=0$ and that $\vp'(0)\ge0$ for every $\vp\in\Phi$. Next, each $g(\cdot,x)$ is clearly strictly increasing because $\vt$ is. The assumption $\sup_{t\in\RR_+}\vt(t)-\vp'(t)=\infty$ provides $\sup_{t\in\RR_+}\vt(t)=\infty$, given $\vp'(t)\ge0$, and therefore also the requirement $\sup_{t\in\RR_+}g(t,x)>0$ for each $x\in\RR_+$. Finally we require $\inf_{x\in\RR_+}g(t,x)<0$ for each $t\in\RR_+$. Given that $\sup_{t\in\RR_+}\vt(t)=\infty$, this demands $\sup_{x\in\RR_+}\vt(\vp^{-1}(x)) - \vp'(\vp^{-1}(x))=\infty$. But given $\sup_{x\in\RR_+}\vp^{-1}(x)=\infty$, this is equivalent to the condition $\sup_{t\in\RR_+}\vt(t) - \vp'(t)=\infty$. This is exactly what is assumed, so we find $g$ in $\uG$, and the proof is thus complete.
\end{proof}

This next result follows trivially from \autoref{thm:solutionset}, just using the definitions of injective and surjective maps, and the fact that in \autoref{thm:solutionset} actually an \emph{infinitude} of IVPs from \autoref{prob:ivp2} are provided which generate any chosen path $\vp\in\Phi$ as the unique global solution. For example, we can always define an infinite set of paths $\vt$, with the required properties in \autoref{thm:solutionset}, like any defined by $\vt(t):=at + \sup_{s\in[0,t]}\vp'(s)-\vp'(0)$ with $a>0$.

\begin{corollary}[Solution map surjectivity]\label{cor:surjective_map}
    The solution map of \autoref{prob:ivp2}, taking each $g\in\uG$ to the solution $\vp\in\Phi$ of the IVP $x'=g(t,x)$, $x_0=0$, is non-injective and surjective.
\end{corollary}

Now we clarify that if subsets of $\uG$ which have related temporal structures are considered, specifically assuming $g\in\uG$ admits the separable representation $g(t,x) = \vt(t)-w(x)$ for some fixed $\vt$, like in \autoref{eq:g_rep}, then this solution map in \autoref{cor:surjective_map} becomes bijective. Importantly, this bijectivity is obtained without compromising the solution set very much. Notice that each subset $\uG_\vt\subset\uG$ defined in \autoref{thm:solutionmap} is related to the subset $\uF_\vt\subset\uF$ from \autoref{ex:main_example}, containing the Heston example in \autoref{eq:hest_ode} when $\vt(t):=\kappa\theta t$.

\begin{theorem}[Solution map bijectivity]\label{thm:solutionmap}
    Fix any strictly increasing $\vt\in\uC_0(\RR_+,\RR)$ with $\lim_{t\to\infty}\vt(t)=\infty$. Let $\Phi_\vt\subset\Phi$ contain the paths $\vp$ which verify $\sup_{t\in\RR_+}\vt(t)-\vp'(t)=\infty$, and let $\uG_\theta\subset\uG$ contain functions $g$ with representation $g(t,x):=\vt(t)-w(x)$ for some $w\in\uC(\RR_+,\RR)$ with $w(0)\le0$ and $\sup_{x\in\RR_+}w(x)=\infty$. Then the map which takes each $g\in\uG_\theta$ to the solution $\vp\in\Phi_\vt$ of the case $x' = g(t,x)$, $x(0)=0$ of \autoref{prob:ivp2} is bijective.
\end{theorem}
\begin{proof}
    We first show that $\uG_\theta$ is indeed a subset of $\uG$. For this fix any such $\vt$ and $w$, and define $g(t,x):=\vt(t)-w(x)$. Checking the properties in \autoref{def:mod_driving_func}, we have $g(0,0) = \vt(0) - w(0)=-w(0)\ge0$ as required. Each $g(\cdot,x)$ is clearly strictly because $\vt$ is, and also we have $\sup_{t\in\RR_+}g(t,x) =\infty >0$ for each $x\in\RR_+$ because $\sup_{t\in\RR_+}\vt(t) =\infty$. Finally we find $\inf_{x\in\RR_+}g(t,x)=-\infty<0$ for each $t\in\RR_+$ because $\sup_{x\in\RR_+}w(x)=\infty$. So $\uG_\vt\subset\uG$.

    Having $g\in\uG$ ensures the IVP $x'=g(t,x)$, $x(0)=0$ has a unique global solution $\vp\in\Phi$ by \autoref{thm:wellposed}, and now we show that this is also in the subset $\Phi_\theta\subset\Phi$. For this we require $\sup_{t\in\RR_+}\vt(t)-\vp'(t)=\infty$. Notice that any solution must verify $\vp'(t)=\vt(t)-w(\vp(t))$, so that $\vt(t)-\vp'(t) = w(\vp(t))$. From this we obtain the requirement as follows, using the fact that $\vp$ defines a bijection from and to $\RR_+$, and using the assumption $\sup_{x\in\RR_+}w(x)=\infty$, 
    \begin{equation}
        \sup_{t\in\RR_+}\vt(t)-\vp'(t) = \sup_{t\in\RR_+}w(\vp(t)) = 
        \sup_{x\in\RR_+}w(x) = \infty.
    \end{equation}
    
    Now we show that there is a unique $g\in\uG_\vt$ which generates any $\vp\in\Phi_\vt$ as a solution. Fixing $\vp\in\Phi_\theta$, to be a solution we again require $\vp'(t)=\vt(t)-w(\vp(t))$ for some such $w$. But there is clearly only one such $w$, defined by $w(x):=\vt(\vp^{-1}(x)) - \vp'(\vp^{-1}(x))$ for each $x\in\RR_+$. This is in $\uC(\RR_+,\RR)$ with $w(0)=-\vp'(0)\le0$ and satisfies the requirement
    \begin{equation}
        \sup_{x\in\RR_+}w(x) = \sup_{x\in\RR_+}\vt(\vp^{-1}(x)) - \vp'(\vp^{-1}(x)) = \sup_{t\in\RR_+}\vt(t)-\vp'(t)=\infty,
    \end{equation}
    using the assumption $\sup_{t\in\RR_+}\vt(t)-\vp'(t)=\infty$. This establishes the bijectivity claim.
\end{proof}

For convenience, let the subsets $\Theta,\uW\subset\uC(\RR_+,\RR)$ contain paths with the properties of $\vt$ and $w$ in \autoref{thm:solutionmap} respectively, like we did in \autoref{ex:zeros_example}. Now \autoref{thm:solutionmap} is very powerful, for it tells us that after fixing \emph{any} $\vt\in\Theta$, a very wide subset $\Phi_\theta$ of the solution set $\Phi$ of \autoref{prob:ivp2} can be generated bijectively, simply by varying the path $w\in\uW$, which governs the spatial behaviour of the function $g(t,x):=\vt(t)-w(x)$. As discussed in \autoref{ch:intro}, every one of these subsets $\Phi_\vt\subset\Phi$ contains (but is not limited to) the paths $\vp\in\Phi$ which verify $\liminf_{t\to\infty}\vp'(t)<\infty$, so is wide enough for volatility modelling given \emph{any} $\vt\in\Theta$. The following is a straightforward consequence of \autoref{thm:solutionmap}, but should not be taken for granted, because if we allow $w$ to denote a sample path of Brownian motion over $\RR_+$, then we cannot make such pathwise statements for e.g.~It\^o SDE solution maps.

\begin{corollary}\label{cor:bijec_2}
    Let $\uW$ contain the paths $w$ in \autoref{thm:solutionmap}. Then fixing any $\vt$ there, the map which takes $w\in\uW$ to the solution $\vp\in\Phi_\vt$ of $x' = \vt(t)-w(x)$, $x(0)=0$ is bijective.
\end{corollary}

Of course, these solution map surjections and bijections are also continuous w.r.t.~uniform convergence over compacts, in the sense of \autoref{eq:new_cont_statement} to be precise. We conclude this section with some consequences of these results for the derivative $\vp'$ of a solution $\vp\in\Phi$ of \autoref{prob:ivp2}. This is important given $\sqrt{\vp'}$ will constitute a realisation of a price's volatility, so the possible derivative paths $\vp'$ tell us how wide our volatility modelling framework is.

\begin{definition}[Set $\Phi'$ of paths]\label{def:solutionderivset}
    Let the subset $\Phi'\subset\uC(\RR_+,\RR_+)$ contain paths $\vp'$ such that for every $(a,b)\subset\RR_+$ there exists $t\in(a,b)$ where $\vp'(t)>0$, and $\lim_{t\to\infty}\int_0^t\vp'(s)\dd s = \infty$.
\end{definition}

Observe that if $\vp\in\Phi$, so by definition $\vp$ is a bijective path in $\uC^1_0(\RR_+,\RR_+)$, then clearly $\vp'\in\uC(\RR_+,\RR_+)$, meaning $\vp'(t)\in\RR_+$ for every $t\in\RR_+$. But actually, we find $\vp\in\Phi$ if and only if $\vp'\in\Phi'$, and so the set of solutions' derivatives $\vp'$ for \autoref{prob:ivp2} is precisely $\Phi'$. 

To see this, first assume $\vp\in\Phi$, so $\vp'\in\uC(\RR_+,\RR_+)$. If $\vp'(t)=0$ over any $(a,b)\subset\RR_+$, then clearly $\vp(b)=\vp(a)$ despite $b>a$, violating the strictly increasing nature of $\vp$. Also, $\lim_{t\to\infty}\int_0^t\vp'(s)\dd s = \lim_{t\to\infty}\vp(t)= \infty$, so indeed $\vp'\in\Phi'$. Conversely, if $\vp'\in\Phi'$, then in any $(a,b)$ we find some $\vp'(t)>0$. The continuity of $\vp'$ then ensures an open subinterval of $(a,b)$ where $\vp'(t)>0$, so clearly $\vp(b)-\vp(a)= \int_a^b\vp'(s)\dd s>0$, clarifying $\vp$ is strictly increasing. Similarly, $\lim_{t\to\infty}\vp(t)=\infty$, so $\vp$ defines a bijection in $\uC_0^1(\RR_+,\RR_+)$, and $\vp'\in\Phi'\!\iff\!\vp\in\Phi$.

Despite this characterisation in \autoref{def:solutionderivset} of the set $\Phi'$ of derivatives from \autoref{prob:ivp2}, it is still not easy to appreciate the full diversity of this subset of $\uC(\RR_+,\RR_+)$, and so nor the volatility paths $\sqrt{\vp'}$ which can (theoretically) be modelled using \autoref{prob:ivp2}. For example, it is surprising that despite finding $\vp'(t)>0$ in any $(a,b)\subset\RR_+$, the Lebesgue measure of the set of points in $(a,b)$ where $\vp'(t)=0$ can be arbitrarily close to $b-a$. This is demonstrated in Section 6.5 of \cite{Royden_2010}, where the authors \emph{specify} $\vp'(t)=0$ on so-called `fat Cantor' subsets of $(a,b)$, and show that the integral $\vp$ remains strictly increasing. 

This tells us that, for practical purposes, we can essentially model \emph{any} continuous non-negative volatility path $\sqrt{\vp'}$ using \autoref{prob:ivp2}. Given that this is achieved without compromising the well-posedness properties of \autoref{thm:wellposed} or solution map properties obtained in this section, we have clearly arrived at a modelling framework very well-suited to volatility modelling, in fact the modelling of any paths in $\Phi$ or $\Phi'$, whatever the application.

\subsection{The uniform exit-time space} \label{sec:exit_topology}

We have thus far taken for granted that we want to model \emph{continuous} volatility paths, like those deriving from the set $\Phi'$ in \autoref{def:solutionderivset}, of \emph{continuous} prices. Clearly more work is required to reconcile our volatility modelling framework, summarised by \autoref{prob:ivp2}, with \emph{discontinuous} price paths, like those of the NIG process introduced in the \hyperlink{prologue}{Prologue}. This reconciliation will be achieved through limit theorems, which can strengthen and generalise the Heston and NIG relationship in \autoref{thm:mech_pro}. In this section we define the most important `exit-time' metric space for these limit theorems, and look at some of its properties. 

Before getting into details, it should be reassuring to keep in mind that the product of this section is a metric space $(\bPhi,d_{\bPhi})$ on a subset $\bPhi\subset \uD(\RR_+,\RR_+)$ of c\`adl\`ag paths, which is isometric to the metric space $(\uN,d)$, where $\uN$ simply contains all non-decreasing and unbounded paths in $\uC_0(\RR_+,\RR_+)$, and $d$ is the metric from \autoref{def:uni_metric} which characterises uniform convergence over compacts. So, despite this exit-time space $(\bPhi,d_{\bPhi})$ seeming unusual at first, everything is justified by the fact that it is considerably simpler to understand and work with compared to the alternatives which can be defined on $\bPhi$ through the metrics of \cite{Skorokhod_1956}. This will be clear to anyone who has worked directly with these alternative metrics, so also the sets of parametric representations on which they depend.

To draw some comparisons, this exit-time space $(\bPhi,d_{\bPhi})$ is not only separable and complete (which is straightforward to check, once this isometry is established), but \autoref{thm:m1_relationship} shows it to be finer than Skorokhod's $\uM_1$ space. So, like on $\uM_1$, convergence on $\bPhi$ w.r.t.~$d_{\bPhi}$ is additionally stronger than pointwise a.e.~and all $\uL_p$ convergences. This space seems the best we can hope for, because our primary interest is the convergence of \emph{differentiable} solutions of \autoref{prob:ivp2} to discontinuous paths in $\uD(\RR_+,\RR_+)$. This is not possible on Skorokhod's most popular $\uJ_1$ space, on which only discontinuous sequences can find such limits. The text \cite{Whitt_2002} has been a fantastic resource for these matters, especially Chapter 11.

\vspace{3mm}\textbf{The inverse metric.} To prepare for the exit-time metric $d_{\bPhi}$, consider an unconventional `inverse metric' $d^{-1}:\Phi\times\Phi\to[0,1]$ defined on the solution set of \autoref{prob:ivp2} according to 
\begin{equation}\label{eq:inv_metric}
     d^{-1}(\vp_1,\vp_2) := \Vert \vp_2^{-1} - \vp^{-1}_1\Vert_{\RR_+}.
\end{equation}
Here, $\Vert\cdot\Vert_{\RR_+}$ is the norm from \autoref{def:uni_metric}, which characterises uniform convergence over compacts. This metric $d^{-1}$ clearly differs from the related metric $d$ through its application to inverses. It is not difficult to see that $(\Phi,d^{-1})$ is a bona fide (separable but incomplete) metric space, because, defining the set $\Phi^{-1}:=\{\vp^{-1}:\vp\in\Phi\}$ of inverses, then the inverse map clearly defines an isometry to $(\Phi^{-1},d)$, given $d(\vp_1^{-1},\vp_2^{-1})=d^{-1}(\vp_1,\vp_2)$. This isometry is additionally an involution, i.e.~is its own inverse, given that also $d^{-1}(\vp_1^{-1},\vp_2^{-1}) = d(\vp_1,\vp_2)$.

Note that convergence on $(\Phi,d)$ provides convergence on $(\Phi,d^{-1})$, proof of which is straightforward via moduli of continuity. So in point 3.~of \autoref{thm:wellposed} we could actually write
\begin{equation}\label{eq:inverse_cont}
    \Vert g_0-g_n\Vert_{\RR_+^2}\xrightarrow{n\to\infty}0 \implies \Vert \vp_0^{-1}-\vp_n^{-1}\Vert_{\RR_+}=:d^{-1}(\vp_n,\vp_0)\xrightarrow{n\to\infty}0.
\end{equation}

We should think of the inverse metric $d^{-1}$ as measuring distances in \emph{time} between paths in $\Phi$, rather than the usual distances in space. Restricted to the set $\Phi$, the inverse and exit-time metrics will coincide, but the exit-time metric $d_{\bPhi}$ is defined, unlike $d^{-1}$, on a superset $\bPhi\supset\Phi$ containing the discontinuous paths which we are interested in modelling, like those of the IG L\'evy process arising in \autoref{eq:ig_intro}. The corresponding space $(\bPhi,d_{\bPhi})$ is a natural generalisation of $(\Phi,d^{-1})$, with an exit-time map providing the new isometry into $(\uN,d)$.

\vspace{3mm}\textbf{The exit-time functional.} We take more care with defining the exit-time functional $E$ now, compared with the notational use in \autoref{chap:wellposed}, given that this functional is going to constitute the isometry just discussed between $(\bPhi,d_{\bPhi})$ and $(\uN,d)$. Section 13.6 of \cite{Whitt_2002} can be consulted for extensive details on exit-time paths $E(\bvp)\in\uD(\RR_+,\RR_+)$ when $\bvp\in\uD(\RR_+,\RR)$ verifies $\bvp(0)\ge0$ and $\sup_{t\in\RR_+}\bvp(t)=\infty$. Note that a `right inverse' notation $\bvp^{-1}$ is used there. For consistency elsewhere, e.g.~\autoref{eq:exit_def}, we do not impose $\bvp(0)\ge0$.

\begin{definition}[The exit-time functional]\label{def:exit_functional}
    Let the subset $\uD_*\subset \uD(\RR_+,\RR)$ contain only the positively unbounded paths, i.e.~those $\bvp\in\uD(\RR_+,\RR)$ verifying $\sup_{t\in\RR_+}\bvp(t)=\infty$. Then let the functional $E:\uD_*\to\uD_*$ map each path $\bvp$ to the exit-time $E(\bvp)$ defined over $\RR_+$ by 
    \begin{equation}\label{eq:sep_cases}
        E(\bvp)(x) := \inf\{t>0 : \bvp(t)>x\}.
    \end{equation}
\end{definition}

% We have already discussed, prior to \autoref{lem:zeros2}, why the such paths $E(\bvp)$ are found in $\uD_*$. Indeed, using the reasoning from \cite{Whitt_2002}, we also showed there that if $\bvp\in\uC(\RR_+,\RR)$ with $\bvp(0)\le 0$, then $E(\bvp)$ is strictly increasing, and \autoref{lem:whitt} below covers the converse. With this functional properly defined, we recall the subset $\bPhi\subset\uD_*$ introduced in \autoref{ch:intro}.
We have already discussed, before \autoref{lem:zeros2}, why $E(\bvp)$ is found in the subset $\uD_*\subset\uD$ whenever $\bvp$ is in $\uD_*$. This clarifies that the exit-time functional $E:\uD_*\to\uD_*$ is well-defined. Indeed, using the reasoning from \cite{Whitt_2002}, we also showed there that if $\bvp\in\uC(\RR_+,\RR)$ with $\bvp(0)\le 0$, then $E(\bvp)$ is strictly increasing, and \autoref{lem:whitt} below covers the converse. With this functional properly defined, we recall the subset $\bPhi\subset\uD_*$ introduced in \autoref{ch:intro}.

\textbf{\autoref{def:setE}} (Set $\bPhi$ of paths)\textbf{.}
    Let the superset $\bPhi\supset\Phi$ contain the strictly increasing c\`adl\`ag paths $\bvp$ in $\uD(\RR_+,\RR_+)$ which are also unbounded, i.e.~which verify $\lim_{t\to\infty}\bvp(t)=\infty$.

It is clear that indeed $\bPhi\supset\Phi$, because the set $\Phi$ from \autoref{def:solutionset} contains precisely the paths $\vp\in\bPhi$ which are additionally differentiable with $\vp(0)=0$. It is not coincidental that this set $\bPhi$ contains all the paths $\bvp$ which can arise as upper bounds in \autoref{thm:wellposed}, such as that from \autoref{eq:path_bound} in the Heston case, illustrated in \autoref{fig:function_zeros}. Now the following subset $\uN\subset\uC_0(\RR_+,\RR)$ similarly defines a superset of the set $\Phi^{-1}:=\{\vp^{-1}:\vp\in\Phi\}$ of inverse paths.

\begin{definition}[Set $\uN$ of paths]\label{def:setN}
    $\!$Let the superset $\uN\supset\Phi^{-1}$ contain only the non-decreasing and positively unbounded elements of $\uC_0(\RR_+,\RR_+)$, i.e.~those $\vp$ with $\lim_{t\to\infty}\vp(t)=\infty$.
\end{definition}

Again, the inclusion $\Phi^{-1}\subset\uN$ is clear. The main point for defining these supersets $\bPhi\supset\Phi$ and $\uN\supset\Phi^{-1}$ is the next result, which clarifies that the exit-time map $E$ between $\bPhi$ and $\uN$ generalises the inverse map between $\Phi$ and $\Phi^{-1}$. This just consolidates known properties of the exit-time map, as discussed following \autoref{def:exit_functional}, specifically provided in Lemmas 13.6.2 and 13.6.5 of \cite{Whitt_2002}, related to \cite{Whitt_1971} and \cite{Puhalskii_1997}.

\begin{lemma}[Exit-time bijectivity]\label{lem:whitt}
    From $\bPhi$ to $\uN$ and from $\uN$ to $\bPhi$, the exit-time functional $E$ defines a bijective involution. In particular, $(E\circ E)(\bvp)=\bvp$ for each $\bvp$ in either $\bPhi$ or $\uN$.
\end{lemma}

Now that some neat properties of the exit-time functional are understood, we can define the exit-time metric which depends upon it and generalises the inverse metric in \autoref{eq:inv_metric}.

\vspace{3mm}\textbf{The exit-time metric.} Supplementing the mapping properties between sets just covered, the exit-time functional $E$ is going to define an involutive isometry between the exit-time space $(\bPhi,d_{\bPhi})$ and the seemingly simpler (but isometric) space $(\uN,d)$. This again generalises the inverse map, which defines an isometry between the subspaces $(\Phi,d^{-1})$ and $(\Phi^{-1},d)$.

\begin{definition}[Exit-time metric]\label{def:exit_metric}
    For $\bvp_1,\bvp_2\in\bPhi$, define the exit-time metric $d_{\bPhi}$ by
    \begin{equation}\label{eq:exit_metric}
        d_{\bPhi}(\bvp_1,\bvp_2) := \Vert E(\bvp_2) - E(\bvp_1) \Vert_{\RR_+}.
    \end{equation}
\end{definition}

Of course the metric $d_{\bPhi}$ is also well-defined on $\uN$, and, given \autoref{lem:whitt}, it is straightforward to see that $E$ defines the claimed involutive isometry between $(\bPhi,d_{\bPhi})$ and $(\uN,d)$, because
\begin{multline}\label{eq:exit_isometry}
    d(E(\bvp_1),E(\bvp_2)) := \Vert E(\bvp_2) - E(\bvp_1) \Vert_{\RR_+} =: d_{\bPhi}(\bvp_1,\bvp_2),\\
    d_{\bPhi}(E(\vp_1),E(\vp_2)) = \Vert \vp_2 - \vp_1 \Vert_{\RR_+} =: d(\vp_1,\vp_2),
\end{multline}
where we have used $(E\circ E)(\vp)=\vp$. Given that the norm $\Vert\cdot\Vert_{\RR_+}=d$ defined in \autoref{def:uni_metric} characterises \emph{spatial} uniform convergence over compact subsets of \emph{time}, we see from \autoref{def:exit_metric} that the exit-time metric $d_{\bPhi}$ instead characterises \emph{temporal} uniform convergence over compact subsets of \emph{space}. This is why we call $(\bPhi,d_{\bPhi})$ the uniform exit-time metric space. We could also define pseudometrics on $\bPhi$ over compacts, but give preference instead to seminorms. For example, notice that from \autoref{eq:exit_metric} and \autoref{def:uni_metric} we have
\begin{equation}
    d_{\bPhi}(\bvp_1,\bvp_2) = \sum_{n\in\NN} 2^{-n}(1\wedge \Vert E(\bvp_2) - E(\bvp_1) \Vert_{[0,n]})\le \sum_{n\in\NN} 2^{-n} = 1,
\end{equation}
and therefore, using the monotonicity of $\Vert \cdot \Vert_{[0,n]}$ in $n$ and $\sum_{n=N+1}^\infty 2^{-n} = 2^{-N}$, we obtain 
\begin{equation}\label{eq:metric_bound}
    d_{\bPhi}(\bvp_1,\bvp_2) \le n\Vert E(\bvp_2) - E(\bvp_1) \Vert_{[0,n]} + 2^{-n},
\end{equation}
which was noted following \autoref{def:uni_metric} and will be used shortly. Given $(\uN,d)$ is both separable and complete, so too is $(\bPhi,d_{\bPhi})$, unlike $(\bPhi,d_{\uS})$ where $d_\uS$ is any of the $\uJ_{1,2}$, $\uM_{1,2}$ metrics from \cite{Skorokhod_1956}. Thus $d_{\bPhi}$ induces a Polish topology on $\bPhi$, which enables Prokhorov's popular approach to probabilistic limit theorems, as outlined in \cite{Jacod_2003}. Finally, $\Phi$ is dense in $(\bPhi,d_{\bPhi})$, i.e.~the completion of $\Phi\subset\bPhi$ w.r.t~$d_{\bPhi}$ is the entirety of $\bPhi$. This follows from $\bPhi$ itself being complete, and the fact that in the next section we will explicitly construct any path in $\bPhi$ from a practicable sequence of solutions $\vp_n\in\Phi$ of \autoref{prob:ivp2}.

\vspace{3mm}\textbf{Skorokhod's M$\mathbf{{}_1}$ space.} This part serves to relate our exit-time space $(\bPhi,d_{\bPhi})$ to Skorokhod's $\uM_1$ space restricted to $\bPhi$. Compared with $d_{\bPhi}$, the $\uM_1$ metric is relatively complicated to define, so thankfully we will not explicitly rely on the relationship between the two. This relationship is nevertheless informative, and provides access to some consequences, like \autoref{cor:ae_converge} here. To help define $\uM_1$, we use Section 3.3 of \cite{Whitt_2002}. For an intuitive introduction to all Skorokhod $\uJ_{1,2}$, $\uM_{1,2}$ metrics, Section 11.5.2 should be consulted instead.

For $\bvp\in\uD(\RR_+,\RR)$, define the completed graph of $\bvp$ over $[0,T]\subset\RR_+$ to be the set of points 
\begin{equation}\label{eq:graph_def}
    \Gamma_T(\bvp) := \{(t,x)\in[0,T]\times\RR: x\in[\bvp(t_-)\wedge\bvp(t),\bvp(t_-)\vee\bvp(t)]\}
\end{equation}
where as usual $\bvp(t_-):=\lim_{s\uparrow t}\bvp(s)$, and also $\bvp(0_-):=0$. Notice that for $\bvp\in\bPhi$ we could simply use $x\in[\bvp(t_-),\bvp(t)]$ in \autoref{eq:graph_def}, given $\bvp$ is strictly increasing. The effect of defining $\bvp(0_-):=0$ is that the line between the points $(0,0)$ and $(0,\bvp(0))$ is included in $\Gamma_T(\bvp)$. This idea was introduced in \cite{Puhalskii_1997}, to relax the impractical requirement $\bvp_n(0)\xrightarrow{n\to\infty}\bvp_0(0)$ from the setting of \cite{Skorokhod_1956}. To alleviate similar problems caused by discontinuities at the endpoint $T$, define the graph $\Gamma^*_T(\bvp):=\Gamma_T(\bvp)\cup\{T\}\times[\bvp(T),\infty)$, so that the line between points $(T,\bvp(T))$ and $(T,\infty)$ is similarly included.

Now call $(\tau,\sigma)$ a parametric representation of $\Gamma^*_T(\bvp)$ if $\tau\in\uC(\RR_+,[0,T])$ is non-decreasing, $\sigma\in\uC(\RR_+,\RR)$ and $(\tau,\sigma):\RR_+\to\Gamma^*_T(\bvp)$ is bijective. Let $\Pi^*_T(\bvp)$ be the set of such parametric representations. For $\bvp_{1,2}\in\uD(\RR_+,\RR)$, the $\uM_1$ pseudometric over $[0,T]$ is then defined by
\begin{equation}\label{eq:m1_metric}
    d_{\uM_1,T}(\bvp_1,\bvp_2) := \inf\{\Vert \tau_2 - \tau_1 \Vert_{\RR_+}\vee\Vert \sigma_2 - \sigma_1 \Vert_{\RR_+} : (\tau_i,\sigma_i)\in\Pi^*_T(\bvp_i)\},
\end{equation}
and then as usual the $\uM_1$ metric on $\uD(\RR_+,\RR)$ by $d_{\uM_1}(\bvp_1,\bvp_2) :=\sum_{n\in\NN}2^{-n} d_{\uM_1,n}(\bvp_1,\bvp_2)$. 

Now we can establish that the metric space $(\bPhi,d_{\bPhi})$ is finer than $(\bPhi,d_{\uM_1})$. Letting $\id$ denote the identity path in $\bPhi$, then the key observation is that for any $\bvp\in\bPhi$, the path $(E(\bvp),\id)$ parameterises a \emph{completed} graph, given that both $E(\bvp)$ and $\id$ are in $\uN\subset\uC_0(\RR_+,\RR_+)$ by \autoref{lem:whitt}. This contrasts $(\id,\bvp)$, which is \emph{incomplete} whenever $\bvp$ has discontinuities. To be precise, the proof of \autoref{thm:m1_relationship} relies upon the fact that for any $\bvp\in\bPhi$, a specific element of $\Pi^*_T(\bvp)$ is obtained when $\tau^*\in\uC(\RR_+,[0,T])$ and $\sigma^*\in\uC(\RR_+,\RR)$ take the form
\begin{equation}\label{eq:para_rep}
    \tau^* := E(\bvp)\wedge T,\quad \sigma^* :=\id.
\end{equation}
This simple parametric representation is only possible because we have included the line $\{T\}\times[\bvp(T),\infty)$ in the graph $\Pi^*_T(\bvp)$. Otherwise, treating the endpoint $T$ is more complicated without any practical gain, given the goal is to define a metric over the entirety of $\RR_+$.

\begin{theorem}[Relationship with $\uM_1$]\label{thm:m1_relationship}
    The identity map is continuous from the exit-time metric space $(\bPhi,d_{\bPhi})$ to Skorokhod's $(\bPhi,d_{\uM_1})$. Equivalently, for a sequence $\{\bvp_n\}_{n\in\NN_0}\subset\bPhi$,
    \begin{equation}
        d_{\bPhi}(\bvp_n,\bvp_0)\xrightarrow{n\to\infty}0 \implies d_{\uM_1}(\bvp_n,\bvp_0)\xrightarrow{n\to\infty}0.
    \end{equation}
\end{theorem}
\begin{proof}
    We first show $d_{\uM_1,T}(\vp_n,\vp_0)\xrightarrow{n\to\infty}0$ for $T>0$, then the claim essentially follows by definition of $d_{\uM_1}$. Let $(\tau^*_n,\sigma^*_n)\in\Pi^*_T(\bvp_n)$ be defined as in \autoref{eq:para_rep} for $n\in\NN_0$, then
    \begin{align}\label{eq:m1_limit}
        d_{\uM_1,T}(\vp_n,\vp_0) := \inf\{\Vert \tau_2 - \tau_1 & \Vert_{\RR_+}\vee\Vert \sigma_2 - \sigma_1 \Vert_{\RR_+} : (\tau_1,\sigma_1)\in\Pi^*_T(\vp_n),\ (\tau_2,\sigma_2)\in\Pi^*_T(\vp_0)\} \nonumber\\
        & \le\Vert \tau^*_0 - \tau^*_n \Vert_{\RR_+}\vee\Vert \sigma^*_0 - \sigma^*_n \Vert_{\RR_+} \nonumber\\
        &:= \Vert E(\vp_0)\wedge T - E(\vp_n)\wedge T \Vert_{\RR_+} \vee \Vert \id - \id \Vert_{\RR_+} \nonumber\\
        &\le \Vert E(\vp_0) - E(\vp_n) \Vert_{\RR_+} =: d_{\bPhi}(\vp_n,\vp_0).
    \end{align}
    So the assumption $d_{\bPhi}(\vp_n,\vp_0)\xrightarrow{n\to\infty}0$ with the inequalities $d_{\uM_1,T}(\vp_n,\vp_0) \le d_{\bPhi}(\vp_n,\vp_0)$ not only provides $d_{\uM_1,T}(\vp_n,\vp_0)\xrightarrow{n\to\infty}0$ for all $T$, but also the claim of $d_{\uM_1}(\vp_n,\vp_0)\xrightarrow{n\to\infty}0$, given that e.g.~from \autoref{eq:metric_bound} we have the bounds $d_{\uM_1}\le T d_{\uM_1,T} + 2^{-T}$ for all $T\in\NN$.
\end{proof}

It certainly seems plausible that the spaces $(\bPhi,d_{\bPhi})$ and $(\bPhi,d_{\uM_1})$ are actually topologically equivalent, which requires also the converse statement in \autoref{thm:m1_relationship}. We don't pursue this, however, given there is no direct dependence on the $\uM_1$ metric henceforth, and it is considerably more difficult to work with than the exit-time metric. It is certainly worth noting that, while $(\bPhi,d_{\bPhi})$ is both separable and complete, $(\bPhi,d_{\uM_1})$ is separable but not complete. For example, setting $\vp_n(t):=n^{-1}t$ for $n\in\NN$ and $\vp_0(t):=0$, then we have $\vp_n\in\bPhi$ but $\vp_0\not\in\bPhi$ given that $\vp_0$ is not strictly increasing, yet $d_{\uM_1}(\vp_n,\vp_0)\le d(\vp_n,\vp_0)\xrightarrow{n\to\infty}0$.

Given that the discontinuities of paths in $\uD(\RR_+,\RR)$ are at most countable, see e.g.~Section 13 of \cite{Billingsley_1999}, then it becomes straightforward to show that convergence in $(\bPhi,d_{\bPhi})$ is stronger than a.e.~pointwise convergence, and therefore also convergence in any $\uL_p$ space. Indeed, convergence in $(\bPhi,d_{\uM_1})$ is stronger than in these senses too, as specifically clarified in Section 11.5.2~of \cite{Whitt_2002}, so we can just consider the following a consequence of \autoref{thm:m1_relationship}. Let Leb denote the Lebesgue measure and note $\vp_n^p(t)=|\vp_n(t)|^p$ for $\vp\in\bPhi$.

\begin{corollary}[A.e.~pointwise and $\uL_p$ convergence]\label{cor:ae_converge}
    Suppose the convergence $\vp_n\xrightarrow{n\to\infty}\vp_0$ takes place on the exit-time space $(\bPhi,d_{\bPhi})$, as in \autoref{thm:m1_relationship}. Then, for any $T,p\in\RR_+$,
    \begin{equation}
        \mathrm{Leb}\left[t\in[0,T]:\vp_n(t)\xrightarrow{n\to\infty}\vp_0(t)\right]=T \text{ and } \int_{[0,T]}\vp^p_n(t)\dd t \xrightarrow{n\to\infty}\int_{[0,T]}\vp^p_0(t)\dd t.
    \end{equation}
\end{corollary}

Now we can move on to the main limit theorems of this chapter, applicable to solutions of \autoref{prob:ivp2} and taking place on the exit-time space $(\bPhi,d_{\bPhi})$, designed especially for them.

\subsection{Uniform exit-time solution limits}\label{sec:bounds_limits}

This section shows how sequences of IVPs, each providing an example of \autoref{prob:ivp2} distinguished only by a single parameter, can be set up so that solutions converge to \emph{any} chosen limit on the exit-time metric space $(\bPhi,d_{\bPhi})$. From a practical perspective, the consequences of this are far-reaching:~although we work in a framework where the cumulative variance $\vp\in\Phi$ of price paths are differentiable (so the corresponding volatility $\sqrt{\vp'}$ exists), we can construct \emph{any} discontinuous trajectory $\bvp\in\bPhi$, e.g.~that of a L\'evy subordinator, as a limit.

\begin{figure}[H]
    \centering
    \includegraphics[width=0.48\linewidth]{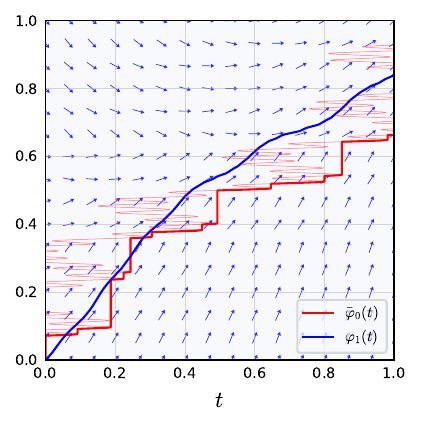}
    \includegraphics[width=0.48\linewidth]{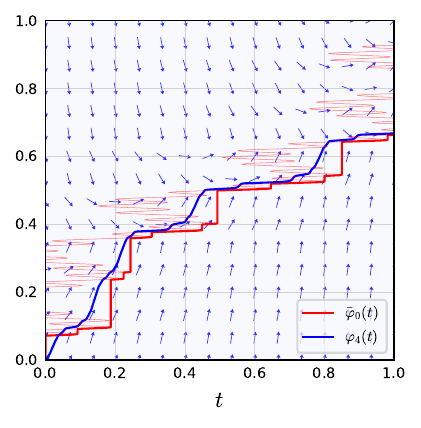}
    
    \includegraphics[width=0.48\linewidth]{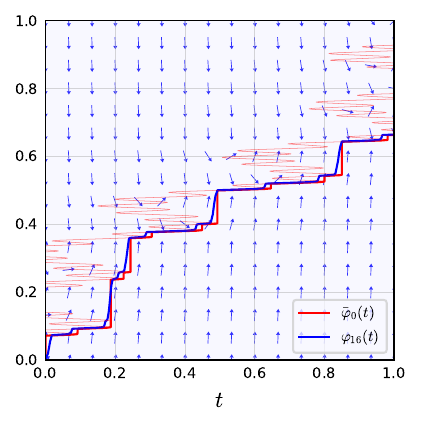}
    \includegraphics[width=0.48\linewidth]{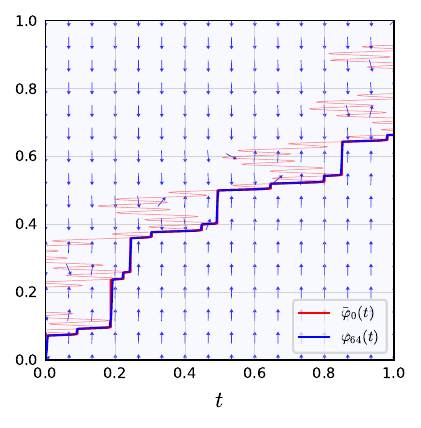}
    \caption{The convergence $\vp_n\xrightarrow{n\to\infty}\bvp_0$ on $(\bPhi,d_{\bPhi})$ of integrated CIR paths $\vp_n$ to an IG L\'evy path $\bvp_0$, as in \autoref{ex:CIR_to_IG}, is illustrated for $n=1,4,16$ and 64.}
    \label{fig:exit_converge}
\end{figure}

As an example which is both theoretically surprising and practically valuable, we conclude this section with \autoref{ex:CIR_to_IG}, which specifically establishes the convergence of integrated CIR paths to those of the IG L\'evy process. This is illustrated graphically in \autoref{fig:exit_converge}, and provides an intuitive yet deep foundation for understanding the Heston and NIG relationship discussed in the \hyperlink{prologue}{Prologue}, allowing us to precisely characterise notions of functional convergence between these two specific popular models in the next section. The combination of \autoref{thm:exit_limit} and \autoref{cor:construct_limit} here, however, generalises this specific connection widely.  

First, in \autoref{cor:exit_lower}, we clarify a straightforward property of the exit-times of the paths $\bvp\in\bPhi$ from \autoref{thm:wellposed}, which are upper bounds to solutions of \autoref{prob:ivp2}. Following this, \autoref{lem:bounds_converge} establishes a certain convergence of such bounds on $(\bPhi,d_{\bPhi})$. Then we will be ready to combine these to prove \autoref{thm:exit_limit}, which is the main result of this section.

It is both harmonious and clarifying to now supplement the exit-time functional $E$ from \autoref{def:exit_functional} with the supremum functional $M:\uD(\RR_+,\RR)\to\uD(\RR_+,\RR)$. We define this by
\begin{equation}\label{eq:maximal_func}
    M(\bvp)(t) := \sup\{0\vee \bvp(s):s\in[0,t]\}.
\end{equation}
This is denoted by $S$ in \cite{Whitt_1971}, which we avoid because this denotes our price processes, and by $\vp^{\uparrow}$ in \cite{Whitt_2002}. Properties of this functional are specifically analysed in \cite{Whitt_1980} and Section 13.4 of \cite{Whitt_2002}, and connections with the exit-time functional in Section 13.6 of \cite{Whitt_2002}. The elegant `dual' relationships of $E\circ E=M$ and $E\circ M=E$ should be noted, see again \cite{Whitt_1971}. Our use of `$0\vee\bvp(s)$' in \autoref{eq:maximal_func} preserve these relationships when $M$ is defined from all of $\uD(\RR_+,\RR)$, not requiring $\bvp(0)\ge0$. We clearly have the identity equivalence $M=\id$ on any non-decreasing subset of $\uD(\RR_+,\RR_+)$, like $\bPhi$ or $\uN$, so from such subsets we simply find $E\circ E=\id$, as already noted in \autoref{lem:whitt}. 

Finally, notice that if the paths $\phi\in\uC(\RR_+,\RR)$ and $\bvp\in\uD(\RR_+,\RR_+)$ are defined as in \autoref{sec:function_zeros} but from $g\in\uG$, as in \autoref{eq:zero_paths} with $\sup\varnothing:=0$, then the paths $M(\phi)$ and $\bvp$ are (uncoincidentally) found in the sets $\uN$ and $\bPhi$ from \autoref{def:setN} and \autoref{def:setE}.

\begin{corollary}[Exit-time lower bound]\label{cor:exit_lower}
    Fix $g\in\uG$, let $\vp$ be the unique global solution of the IVP $x'=g(t,x)$, $x(0)=0$, and let $\phi\in\uC(\RR_+,\RR)$ and $\bvp\in\bPhi$ be defined as usual by
    \begin{equation}\label{eq:zero_paths}
        \phi(x) := \sup\{t\in\RR_+:g(t,x)<0\},\quad \bvp(t):=\inf\{x>0:g(t,x)<0\}.
    \end{equation}
    Then $E(\phi)=\bvp$, $E(\bvp)=M(\phi)$ and $M(\phi)(x) = E(\bvp)(x) \le E(\vp)(x)=\vp^{-1}(x)$ for $x\in\RR_+$.
\end{corollary}
\begin{proof}
    Recall from \autoref{lem:zeros1} the path $\phi$ is in $\uC(\RR,\bRR)$ when deriving from a function $f\in\uF$. The property $\sup_{t\in\RR_+}g(t,x)>0$ of functions in $\uG$ ensures we now find $\phi\in\uC(\RR_+,\RR)$. Similarly, $\inf_{x\in\RR_+}g(t,x)<0$ ensures $\bvp\in\uD(\RR_+,\RR_+)$, and $\bvp=E(\phi)$ by \autoref{lem:zeros2}. We also get $E(\bvp)=(E\circ E)(\phi)=M(\phi)$ using the general functional relationship $E\circ E= M$. The equivalence $E(\vp)=\vp^{-1}$ is obvious, so it just remains to show that the lower temporal bound $E(\bvp)(x) \le E(\vp)(x)$ holds over $\RR_+$. This follows by applying the functional $E$ to the upper spatial bound $\vp(t)\le\bvp(t)$ from \autoref{thm:wellposed}, which simply inverts the ordering.
\end{proof}

This next result, applicable to \emph{bounds}, will do half the work towards \autoref{thm:exit_limit}. It utilises $\phi$ from \autoref{eq:zero_paths}, which is of course related to that in \autoref{lem:zeros1}, \emph{characterising} the zeros of a function $f\in\uF$ according to $f(\phi(x),x)=0$ when $\phi(x)\in\RR$. With $\phi$ defined as in \autoref{eq:zero_paths}, we instead have $g(\phi(x),x)=0$ when $\phi(x)>0$, and $g(0,x)\ge0$ otherwise.

\begin{lemma}[Convergence of bounds]\label{lem:bounds_converge}
    Assume $\{g_n\}_{n\in\NN_0}\subset\uG$ and let $\{\bvp_n\}_{n\in\NN_0}\subset\bPhi$ be the usual IVP solution bounds defined, as in \autoref{cor:exit_lower}, from each $g_n$. Then we have
    \begin{equation}
        \Vert g_0 - g_n \Vert_{\RR^2_+}\xrightarrow{n\to\infty}0 \implies d_{\bPhi}(\bvp_n,\bvp_0) := \Vert E(\bvp_0) - E(\bvp_n) \Vert_{\RR_+}\xrightarrow{n\to\infty}0.
    \end{equation}
\end{lemma}
\begin{proof}
    We focus on establishing $\Vert \phi_0 - \phi_n\Vert_{\RR_+} \xrightarrow{n\to\infty}0$, where $\{\phi_n\}_{n\in\NN_0}\subset\uC(\RR_+,\RR)$ are the paths defined, as in \autoref{eq:zero_paths}, from each $g_n$. The claim will then follow by applying the continuous map $M$ and appealing to the relationships $M(\phi_n) = E(\bvp_n)$ from \autoref{cor:exit_lower}. 
    
    Given these paths $\phi_n$ are related to each $g_n$ via $g_n(\phi_n(x),x)=0$ when $\phi_n(x)>0$, to ask if
    \begin{equation}\label{eq:zero_converge}
        \Vert g_0 - g_n \Vert_{\RR_+^2}\xrightarrow{n\to\infty}0 \implies \Vert \phi_0 - \phi_n \Vert_{\RR_+}\xrightarrow{n\to\infty}0
    \end{equation}
    is to essentially ask if the zeros of $g_n$, considered as a graph over the $x$ axis, converge uniformly over compacts to those of $g_0$. This need not be the case for general $\{g_n\}_{n\in\NN_0}\subset\uC(\RR_+^2,\RR)$, e.g.~take $g_0:=0$ and $g_n:=n^{-1}$. In our setting, with every $g_n(\cdot,x)$ being strictly increasing also with $\sup_{t\in\RR_+}g_n(t,x)>0$, \autoref{eq:zero_converge} can be intuited and may be clear.
    
    Towards this, first suppose we are in the more relevant setting of $\phi_0(x)>0$ over some bounded $\cX\subset\RR_+$. We will show that for any $\ep>0$, we have $\Vert\phi_0-\phi_n\Vert_{\cX}<\ep$ as $n\to\infty$. Given that $g_0(\phi_0(x),x)=0$ for all $x\in\cX$, and every $g_0(\cdot,x)$ is strictly increasing, note that
    \begin{equation}
        g_0(\phi_0(x)-\ep,x)< 0 < g_0(\phi_0(x)+\ep,x)
    \end{equation}
    over $\cX$, where we can w.l.o.g.~assume $\phi_0(x)-\ep\ge 0$. From this we get the related ordering 
    \begin{equation}\label{eq:bounds_zeros}
        g_n(\phi_0(x)-\ep,x)< 0 < g_n(\phi_0(x)+\ep,x)
    \end{equation}
    over $\cX$ whenever $n$ is greater than some $N\in\NN$, using the assumption $\Vert g_0 - g_n \Vert_{\RR_+^2}\xrightarrow{n\to\infty}0$. But given $\phi_n$ is the unique path defined over $\cX$ where $g_n(\phi_n(x),x)=0$, \autoref{eq:bounds_zeros} tells us that, over $\cX$, $\phi_n$ falls between $\phi_0\pm \ep$ whenever $n>N$. So $\Vert\phi_0-\phi_n\Vert_{\cX}<\ep$ whenever $n>N$, and therefore $\Vert\phi_0-\phi_n\Vert_{\cX}\xrightarrow{n\to\infty}0$ holds for any such bounded $\cX$ where $\phi_0(x)>0$. 
    
    In the alternative setting where $\phi_0(x)=0$ over bounded $\cX\subset\RR_+$, so $g_0(0,x)\ge0$, we find
    \begin{equation}
        0\le g_0(0,x) = g_0(\phi_0(x),x) < g_n(\phi_0(x)+\ep,x) = g_n(\ep,x)
    \end{equation}
    in place of \autoref{eq:bounds_zeros}, so $\Vert\phi_0-\phi_n\Vert_{\cX} = \Vert\phi_n\Vert_{\cX}<\ep$ whenever $n>N$. By splitting any compact $[0,X]\subset\RR_+$ into points where $\phi_0(x)>0$ or not, we get $\Vert\phi_0-\phi_n\Vert_{[0,X]}\xrightarrow{n\to\infty}0$. Therefore, $\Vert E(\bvp_0)-E(\vp_n)\Vert_{[0,X]} = \Vert M(\phi_0)-M(\phi_n)\Vert_{[0,X]}\xrightarrow{n\to\infty}0$ by the mapping properties of $M$. Now, essentially by definition, we have the claim of $d_{\bPhi}(\bvp_n,\bvp_0)\xrightarrow{n\to\infty}0$, for which we could use $d_{\bPhi}(\bvp_n,\bvp_0) \le N\Vert E(\bvp_0) - E(\bvp_n) \Vert_{[0,N]} + 2^{-N}$ from \autoref{eq:metric_bound}.
\end{proof}

Now we are ready to prove the main result of this section, which will constitute the main tool towards establishing probabilistic limits on the topology induced by $(\bPhi,d_{\bPhi})$, like the convergence of the integrated CIR process to the IG L\'evy subordinator. On first reading, it should be helpful to set $g_n:=g_0$ for every $n\in\NN$, as the generalisation to a sequence $\{g_n\}_{n\in\NN_0}\subset\uG$ is not so difficult, with most of the work for this part done by \autoref{lem:bounds_converge}.

The fact that in \autoref{thm:exit_limit} we set each $\vp_n$ as the IVP solution deriving from $n g_n$, rather than $g_n$, cannot be overlooked. Indeed, this is practically the entire point:~the exploding values of $n g_n$, thus $\vp'_n$, as $n\to\infty$, generate discontinuous limits on $(\bPhi,d_{\bPhi})$. If we use instead $g_n$, then we get the less practically valuable continuity result in \autoref{eq:inverse_cont}, on the subspace $(\Phi,d_{\bPhi})=(\Phi,d^{-1})$. Notice that the zeros of each $n g_n$ coincide with those of $g_n$;~this is what ensures non-degenerate limits $\bvp_0$ of the solutions $\vp_n$ always exist, unlike for $\vp'_n$. For complete clarity, such limits $\bvp_0$ in \autoref{thm:exit_limit} are defined as usual from $g_0$ by
\begin{equation}
    \bvp_0(t):=\inf\{x>0:g_0(t,x)<0\}.
\end{equation}
\begin{theorem}[Uniform exit-time limits]\label{thm:exit_limit}
    $\!$Suppose $\{g_n\}_{n\in\NN_0}\!\subset\!\uG$, let $\{\vp_n\}_{n\in\NN}\!\subset\!\Phi$ solve each IVP $x'\!=\!ng_n(t,x)$, $x(0)\!=\!0$ respectively, and let $\bvp_0\!\in\!\bPhi$ derive as usual from $g_0$.$\!$ Then,
    \begin{equation}\label{eq:converge_metric}
        \Vert g_0 - g_n \Vert_{\RR_+^2}\xrightarrow{n\to\infty}0 \implies d_{\bPhi}(\vp_n,\bvp_0) = \Vert E(\bvp_0) - \vp^{-1}_n \Vert_{\RR_+} \xrightarrow{n\to\infty} 0.
    \end{equation}
\end{theorem}
\begin{proof}
    The claimed convergence $d_{\bPhi}(\vp_n,\bvp_0) \xrightarrow{n\to\infty} 0$ will follow from the bound $d_{\bPhi}(\vp_n,\bvp_0) \le N\Vert E(\bvp_0) - E(\vp_n) \Vert_{[0,N]} + 2^{-N}$ in \autoref{eq:metric_bound} if we can establish the uniform convergence~
    \begin{equation}
        \Vert E(\bvp_0) - E(\vp_n) \Vert_{[0,X]} = \Vert E(\bvp_0) - \vp^{-1}_n \Vert_{[0,X]} \xrightarrow{n\to\infty} 0
    \end{equation}
    for any $X\in\RR_+$. Notice $E(\vp_n)=\vp^{-1}_n$ since, restricted to $\Phi$, $E$ coincides with the inverse map. Now for any $\ep>0$, we will show $\Vert E(\bvp_0) - \vp_n^{-1} \Vert_{[0,X]}< \ep$ for sufficiently large $n$. For the most part, a sufficiently tight upper bound to each $\vp^{-1}_n$ is established through differential inequalities, then \autoref{cor:exit_lower} and \autoref{lem:bounds_converge} will be invoked to provide also a lower bound. Toward this, define the line $\lambda\in\uC([0,X],[\frac12\ep,\ep])$ between points $(\frac12\ep,0)$ and $(\ep,X)$,
    \begin{equation}
        \lambda(x) := \frac12\ep\left(1 + \frac{x}{X}\right),
    \end{equation}
    and notice $\Vert\lambda\Vert_{[0,X]}=\lambda(X)=\ep$. Define also the shifted function $\mu := E(\bvp_0) + \lambda$, noting $\Vert\mu - E(\bvp_0)\Vert_{[0,X]} =\ep$. Recall from \autoref{lem:whitt} that $E(\bvp_0)$ defines a non-decreasing path in $\uC_0(\RR_+,\RR)$, so $\mu\in\uC([0,X],[\frac12\ep,T])$ defines a strictly increasing bijective path with inverse $\mu^{-1}$, where $T:=E(\bvp_0)(X)+\ep$. Since $\lambda$ has gradient $\frac{\ep}{2X}>0$, $\mu$ has the one-sided Lipschitz property
    $\mu(x)-\mu(u)\ge\frac{\ep}{2X}(x-u)$ for $x\ge u$ in $[0,X]$. So $\mu^{-1}$ has the reciprocated version
    \begin{equation}\label{eq:lipschitz_bound}
        \mu^{-1}(t) - \mu^{-1}(s) \le L(t-s),\quad L:= 2X\ep^{-1} <\infty
    \end{equation}
    for $t\ge s$ in $[\frac12\ep,T]$. Note that $\mu^{-1}(\frac12\ep)=0$ follows from $\mu(0)=\frac12\ep$, so set $\mu^{-1}(t):=0$ over $[0,\frac12\ep)$ to define a path in $\uC([0,T],[0,X])$ which retains the property of \autoref{eq:lipschitz_bound}. 
    
    Having $\mu^{-1}(t):=0$ over $[0,\frac12\ep)$ clearly makes $\mu^{-1}$ a strict lower bound to any (strictly increasing) IVP solution $\vp_n$ over $(0,\frac12\ep)$, and the Lipschitz property in \autoref{eq:lipschitz_bound} will allow us to extend this relationship also over $[\frac12\ep,T]$ whenever the differential inequality
    \begin{equation}\label{eq:ineq_bound}
        n g_n(t,\mu^{-1}(t)) > L
    \end{equation}
    is verified over this interval. For clarity, this is because a touching point $\vp_n(t)=\mu^{-1}(t)$ here leads to the usual $L<L$ contradiction: $L < n g_n(t,\mu^{-1}(t)) = n g_n(t,\vp(t)) = \vp'(t) \le L$.
    
    Now we show \autoref{eq:ineq_bound} is indeed verified for sufficiently large $n$, which can already be intuited given these $g_n$ find the limit $g_0$ as $n\to\infty$. Towards this, define the line of points
    \begin{equation}
        \cX := \{(t,\mu^{-1}(t)):t\in[\tfrac12\ep,T]\} = \{(\mu(x),x):x\in[0,X]\}
    \end{equation}
    and set $m_0:=\min_\cX g_0$. Given $\mu(x)>E(\bvp_0)(x)=M(\phi_0)(x)$ over $[0,X]$ and $g_0(\phi_0(x),x)\ge0$, then the strictly increasing nature of $g_0(\cdot,x)$ ensures $m_0>0$. Now the assumption $\Vert g_0 - g_n \Vert_{\RR^2_+}\xrightarrow{n\to\infty}0$ of course provides $\Vert g_0 - g_n \Vert_\cX \xrightarrow{n\to\infty}0$ and therefore $\min_\cX g_n=:m_n \xrightarrow{n\to\infty} m_0$, so we can proceed w.l.o.g.~assuming that $m_n > \frac12 m_0$ for every $n\in\NN$. But now we see the inequality of \autoref{eq:ineq_bound} is verified whenever $n>2L/m_0$, because then
    \begin{equation}
         n g_n(t,\mu^{-1}(t)) \ge n \min_\cX g_n =: nm_n > \frac12 nm_0 > L.
    \end{equation}
    So again we can proceed w.l.o.g.~assuming this holds for every $n\in\NN$, enforcing the bound $\mu^{-1}(t)<\vp_n(t)$ over $[0,T]$ for every $n$, and equivalently $\mu(x)>\vp_n^{-1}(x)$ over $[0,X]$. Recall we have ensured this bound is sufficiently close to $E(\bvp_0)$, precisely meaning $\Vert\mu - E(\bvp_0)\Vert_{[0,X]} =\ep$.
    
    A corresponding lower bound to each $\vp_n^{-1}$ was established in \autoref{cor:exit_lower}, namely each $E(\bvp_n)$. Notice there is no difference between defining each $\bvp_n$ through $ng_n$ or $g_n$, provided $n>0$. Importantly, \autoref{lem:bounds_converge} establishes the convergence $\Vert E(\bvp_0) - E(\bvp_n) \Vert_{[0,X]}\xrightarrow{n\to\infty}0$ of bounds, so, supplementing the ordering $E(\bvp_n)(x)\le \vp^{-1}_n(x)<\mu(x)$ over $[0,X]$, we can also set $n$ sufficiently high to ensure $\Vert E(\bvp_0) - E(\bvp_n) \Vert_{[0,X]}<\ep$. Doing this, we then obtain
    \begin{equation}
        \Vert E(\bvp_0) - \vp^{-1}_n \Vert_{[0,X]} \le \Vert E(\bvp_0) - \mu \Vert_{[0,X]} \vee \Vert E(\bvp_0) - E(\bvp_n) \Vert_{[0,X]} \le \ep.
    \end{equation}
    That is, given this ordering, the distance from $\vp^{-1}_n$ to $E(\bvp_0)$ cannot be more than the maximum distance from $E(\bvp_0)$ to each of $E(\bvp_n)$ and $\mu$. Having shown that, for any $X,\ep>0$, $\Vert E(\bvp_0) - \vp^{-1}_n \Vert_{[0,X]} < \ep$ holds for sufficiently large $n$, we therefore have $\Vert E(\bvp_0) - \vp^{-1}_n \Vert_{[0,X]}\xrightarrow{n\to\infty}0$ by definition, and the desired conclusion of $d_{\bPhi}(\vp_n,\bvp_0)\xrightarrow{n\to\infty}0$ also.
\end{proof}

Having established this limit theorem on the exit-time space $(\bPhi,d_{\bPhi})$, we immediately obtain the convergence $\vp_n\xrightarrow{n\to\infty}\bvp_0$ on Skorokhod's $\uM_1$ space by \autoref{thm:m1_relationship}, as well as convergence a.e.~pointwise and on all $\uL_p$ spaces, in the sense of \autoref{cor:ae_converge} to be precise.

It is now tempting to cover an example of this result in action, and if the reader prefers they may skip ahead to \autoref{ex:CIR_to_IG} and \autoref{fig:exit_converge} for this. But we keep the theoretical momentum up here and cover the last result of this section. This compliments the previous by demonstrating how we can explicitly \emph{construct} any chosen limit in the exit-time space $(\bPhi,d_{\bPhi})$, and relates also to the construction from \autoref{thm:solutionset} of any IVP solution in $\Phi$. 

Given that $(\bPhi,d_{\bPhi})$, being isometric to $(\uN,d)$, is a complete metric space, the following construction also conveniently clarifies that the completion of the solution set $\Phi$ of \autoref{prob:ivp2} is the entirety of $\bPhi$, as mentioned following \autoref{def:exit_metric}. That is, $\Phi$ is dense in $(\bPhi,d_{\bPhi})$.

Recall finally the subsets $\Theta,\uW\subset\uC(\RR_+,\RR)$ from \autoref{thm:solutionmap} and \autoref{cor:bijec_2}. For convenience: $\Theta$ denotes the set of strictly increasing paths $\vt\in\uC_0(\RR_+,\RR)$ with $\lim_{t\to\infty}\vt(t)=\infty$, and $\uW$ the set of paths $w\in\uC(\RR_+,\RR)$ with $w(0)\le0$ and $\sup_{x\in\RR_+}w(x)=\infty$. As shown in \autoref{thm:solutionmap}, the IVP $x'=g(t,x)$, $x(0)=0$ with $g(t,x):=\vt(t)-w(x)$ then provides an example of \autoref{prob:ivp2} with $g\in\uG_\vt\subset\uG$, and so has a unique global solution $\vp\in\Phi_\vt\subset\Phi$.

\begin{theorem}[Construction of exit-time limits]\label{cor:construct_limit}
    Converse to \autoref{thm:exit_limit}, fix any $\bvp\in\bPhi$, $\vt\in\Theta$ and $w\in\uW$ satisfying $M(w)=\vt\circ E(\bvp)$, e.g.~take $w:=\vt\circ E(\bvp)$. Define $g_0\in\uG_\vt$ by $g_0(t,x):=\vt(t)-w(x)$, and let $\{g_n\}_{n\in\NN}$ verify $\Vert g_0-g_n\Vert_{\RR_+^2}\xrightarrow{n\to\infty}0$, e.g.~set $g_n:=g_0$. Then $d_{\bPhi}(\vp_n,\bvp) \xrightarrow{n\to\infty} 0$, where each $\vp_n$ solves the IVP $x'=ng_n(t,x)$, $x(0)=0$. Thus every $\bvp\in\bPhi$ can be constructed as a limit of such solutions $\vp_n$ of \autoref{prob:ivp2}, on $(\bPhi,d_{\bPhi})$.
\end{theorem}
\begin{proof}
    The conditions of \autoref{thm:exit_limit} are met, so we obtain $d_{\bPhi}(\vp_n,\bvp_0) \xrightarrow{n\to\infty} 0$, where
    \begin{equation}\label{eq:limit_rep}
        \bvp_0(t):=\inf\{x>0:g_0(t,x)<0\} = \inf\{x>0:w(x)>\vt(t)\}.
    \end{equation}
    So we just need to show that any such $w\in\uW$ with $M(w)=\vt\circ E(\bvp)$ ensures the equivalence $\bvp = \bvp_0$. Towards this, using the relationship $E\circ M=E$ we can also write \autoref{eq:limit_rep} as
    \begin{multline}\label{eq:limit_rep2}
        \bvp_0(t) = \inf\{x>0:M(w)(x)>\vt(t)\} \\ = \inf\{x>0:(\vt^{-1}\circ M(w))(x)>t\}=:E (\vt^{-1}\circ M(w))(t).
    \end{multline}
    So in general we have the representation $\bvp_0 = E(\vt^{-1}\circ M(w))$, supplementing the simpler expression $\bvp_0=E(w)\circ\vt$ from \autoref{eq:limit_rep}. Now applying the assumption of $M(w)=\vt\circ E(\bvp)$, we see indeed $\bvp_0=(E\circ E)(\bvp)=M(\bvp)=\bvp$, which uses the general relationship $E\circ E=M$ and the fact that $\bvp$ is strictly increasing. Given $\bvp_0=\bvp$, the proof is complete.
\end{proof}

This result demonstrates a tremendous amount of freedom in generating a chosen limit $\bvp$ on $(\bPhi,d_{\bPhi})$, given that we can choose any $\vt\in\Theta$, any $w\in\uW$ with $M(w)=\vt\circ E(\bvp)$, and any such sequence $\{g_n\}_{n\in\NN}$. If we fix a temporal structure via $\vt$, then unlike the bijectivity result of \autoref{thm:solutionmap}, there are now many $w\in\uW$ generating the same limit $\bvp$, each converging with different rates. Geometrically, we see that the given example $w:=\vt\circ E(\bvp)$ is the unique \emph{non-decreasing} path which generates the limit $\bvp$, but will do so at the slowest possible rate.

Given it is habitual to imagine paths in $\uD(\RR_+,\RR)$, thus $\bPhi$, as being \emph{strictly} discontinuous, e.g.~forgetting $\uC^\infty \subset \uC^1 \subset \uAC \subset\uD$, it is finally worth pointing out that \autoref{cor:construct_limit} is not just a result enabling us to construct any discontinuous cumulative variance path $\bvp\in\bPhi$ as a limit of solutions $\vp\in\Phi$ of \autoref{prob:ivp2}, but also provides the means to interpolate between continuous paths, like those of Black-Scholes and a richer rough volatility model.

\begin{figure}[H]
    \centering
    \includegraphics[width=0.48\linewidth]{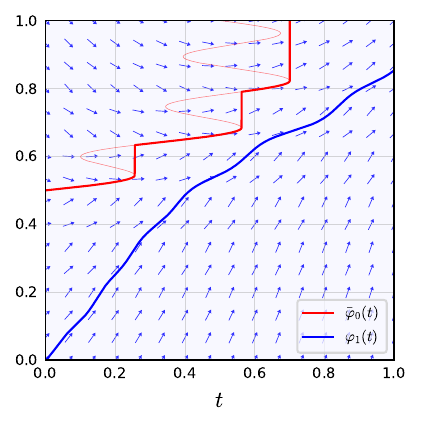}
    \includegraphics[width=0.48\linewidth]{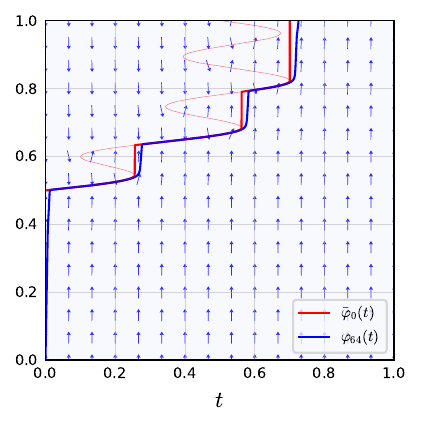}
    \caption{The convergence in \autoref{fig:exit_converge} is repeated using $g_n(t,x) := \sigma w(x) + \kappa\left(\theta t - x\right) + v$ in \autoref{ex:CIR_to_IG} instead and truncating $w$ sooner, like in \autoref{fig:main_examples2}.}
    \label{fig:exit_converge2}
\end{figure}

\vspace{4mm} In \autoref{ex:CIR_to_IG}, we now show how \autoref{thm:exit_limit} establishes the convergence of IVP solutions to paths of the IG L\'evy subordinator, first discussed following \autoref{eq:path_bound}. The illustrations in \autoref{fig:exit_converge} and \autoref{fig:exit_converge2} show that this example relates to those from \autoref{chap:wellposed}, and evidently the limit obtained is like the c\`adl\`ag solution bound $\bvp$ in \autoref{fig:global_exist_examples}. From the first part of \autoref{ch:intro} leading up to \autoref{eq:ig_intro}, it will be clear why we consider \autoref{ex:CIR_to_IG} to demonstrate the pathwise convergence of the integrated CIR process. Put simply, this provides the deepest imaginable understanding of \autoref{thm:mech_pro} from \cite{Mechkov_2015}. The \hyperlink{epilogue}{Epilogue}, which the reader now has the tools to consider, generalises this example to obtain other L\'evy process limits, and works probabilistically rather than pathwise.

\begin{example}[Pathwise integrated CIR to IG]\label{ex:CIR_to_IG}
    Assume $\sigma,\kappa,\theta,v>0$ and $w\in\uC_0(\RR_+,\RR)$, and define $\{g_n\}_{n\in\NN}\subset\uC(\RR_+^2,\RR)$ by $g_n(t,x) := \sigma w(x) + \kappa\left(\theta t - x\right) + v/n$ respectively. Note $g_1$ coincides with the Heston function in \autoref{eq:hest_ode}. Then $\Vert g_0-g_n\Vert_{\RR_+^2}\xrightarrow{n\to\infty}0$, where
    \begin{equation}
        g_0(t,x) := \sigma w(x) + \kappa\left(\theta t - x\right).
    \end{equation}
    Using \autoref{def:mod_driving_func}, it is straightforward to confirm $\{g_n\}_{n\in\NN_0}\subset\uG$ provided $w$ verifies the condition $\sup_{x\in\RR_+}\kappa x - \sigma w(x)=\infty$. Equivalently, provided the function $\bvp_0$ specified by
    \begin{equation}\label{eq:exit_limit}
        \bvp_0(t) := \inf\{x>0:g_0(t,x)<0\} = \inf\{x>0: \kappa x - \sigma w(x) > \kappa\theta t\}
    \end{equation}
    is a well-defined path in $\uD(\RR_+,\RR)$. Note these conditions are met a.s.~when $w=W^1(\omega)$ is a sample path of Brownian motion. The conditions to apply \autoref{thm:exit_limit} are thus met. This tells us that $\vp_n\xrightarrow{n\to\infty} \bvp_0$ on the exit-time space $(\bPhi,d_{\bPhi})$, where each $\vp_n$ solves the IVP $x'=ng_n(t,x)$, $x(0)=0$. That is, $\vp_n$ is the unique path in $\Phi$ which verifies $\vp_n(0)=0$ and
    \begin{equation}\label{eq:ode_verify}
        \vp'_n(t) = n\sigma w(\vp_n(t)) - n\kappa(\theta t - \vp_n(t)) + v.
    \end{equation}
\end{example}

Since each $\vp_n$ can be considered a sample path of an integrated CIR process $\int_0^\cdot V_s(\omega)\dd s$ as in \autoref{eq:tce}, and the limit $\bvp_0$ a sample path of the IG Le\'vy process as in \autoref{eq:ig_intro}, we have thus established the pathwise convergence of such processes on $(\bPhi,d_{\bPhi})$, and provided a pathwise origin for \autoref{thm:mech_pro} applicable to the related Heston and NIG processes.

This convergence is illustrated in \autoref{fig:exit_converge}, with $w$ a truncated Weierstrass path from \autoref{eq:truncated_weierstrass}. For the graphs labelled $\bvp_0(t)$ we actually use those of the interval-valued paths $\bvp_*(t):=[\bvp_0(t_-),\bvp_0(t)]$ like in \autoref{fig:function_zeros}, which coincide with the graphs of $E(\bvp_0)(x)=M(\phi_0)(x)$ by \autoref{cor:exit_lower} and help visualise $d_{\bPhi}(\vp_n,\bvp_0) \xrightarrow{n\to\infty} 0$, i.e.~uniform convergence in \emph{time} over compacts in \emph{space}. In \autoref{fig:exit_converge2} we assume $g_n(t,x) := \sigma w(x) + \kappa\left(\theta t - x\right) + v = g_0(t,x)$ instead. Contrasting \autoref{eq:exit_limit} and \cite{Mechkov_2015}, a discontinuity at $t=0$ is then ensured for any chosen $w\in\uC_0(\RR_+,\RR)$ provided $v>0$, and this discontinuity has size
\begin{equation}
    \bvp_0(0) = \inf\{x>0: \kappa x - \sigma w(x) > v\}.
\end{equation}

\subsection{Excursionary limits}\label{sec:excursionary_limits}

The previous section was concerned with limits of sequences of solutions $\vp$ of the IVPs in \autoref{prob:ivp2}. As outlined in \autoref{ch:intro}, in \autoref{chap:framework} we define a framework in which price processes admit the representation $S=\exp(W^\rho_X-\frac12 X)$, where the cumulative variance process $X=\{X_t\}_{t\in\RR_+}$ here solves \autoref{prob:ivp2} on a pathwise basis. Notice that $S$ just derives from $X$ through a simple (albeit random) composition with geometric Brownian motion $\Lambda_x:=\exp(W^\rho_x-\frac12 x)$, i.e.~$S=\Lambda\circ X$. So to understand how paths of $S$ behave under the exit-time limits of \autoref{sec:bounds_limits}, we focus here primarily on the behaviour of compositions $w\circ\vp$ for some $w\in\uC(\RR_+,\RR)$, possibly related to $\vp$, under the limits of \autoref{sec:bounds_limits}. Such compositions $w\circ\vp$ do not \emph{have} to represent price paths, but could be a functional of such, like a derivative payoff. We find that instantaneous \emph{excursions} can develop in such composite limits, and these conclusions will allow us the extend the pathwise CIR and IG relationship in \autoref{ex:CIR_to_IG} to answer the Heston and NIG related questions raised in the \hyperlink{prologue}{Prologue}. 

Towards understanding these composite limits, the following is reassuring and should be kept in mind. This just uses the fact that the discontinuities of a limit $\bvp_0$ on $(\bPhi,d_{\bPhi})$ are at most countable, as discussed prior to \autoref{cor:ae_converge}. Thus $\bvp_0$ is a.e.~continuous, and this can be extended to this corollary essentially by definition of $w$ being assumed continuous.
\begin{corollary}[A.e.~composite convergence]\label{cor:ae_composite}
    Suppose $\vp_n\xrightarrow{n\to\infty}\bvp_0$ on $(\bPhi,d_{\bPhi})$ and also $w\in\uC(\RR_+,\RR)$. Then for any $T\in\RR_+$ the following pointwise convergence a.e.~takes place 
    \begin{equation}
        \Leb\left[t\in[0,T]: (w\circ\vp_n)(t)\xrightarrow{n\to\infty}(w\circ\bvp_0)(t)\right] =T.
    \end{equation}
\end{corollary}
This clearly leads also to integral convergences like the $\uL_p$ statements of \autoref{cor:ae_converge}. Although reassuring, the limitations of this result must be understood, because it turns out related convergences of path-dependent functionals can be violated, for example we find
\begin{equation}\label{eq:barrier_violation}
    \lim_{n\to\infty}\sup_{t\in[0,T)}(w\circ\vp_n)(t) \ge \sup_{t\in[0,T)}(w\circ\bvp_0)(t)
\end{equation}
with a strict inequality in cases of practical interest. This is an important example in practice, as it relates to the payoff of a common barrier option, albeit continuously-monitored. The analysis of this section will allow us to understand such limits, and in this example we obtain the following surprising yet fairly elegant convergence on $(\RR,|\cdot|)$ for any $T>0$
\begin{equation}\label{eq:def_barrier}
    \sup_{t\in[0,T)}(w\circ\vp_n)(t) \xrightarrow{n\to\infty} \sup_{t\in[0,T)}\sup\ (w\bullet\bvp_0)(t),\quad (w\bullet\bvp_0)(t) := \{w(x):x\in[\bvp_0(t_-),\bvp_0(t)]\}.
\end{equation}
The path $w\bullet\bvp_0$ is set-valued, hence $\sup\sup$ above, in fact compact \emph{interval}-valued. Indeed we have the following equivalent representation, noting $\bvp_*$ was utilised earlier in \autoref{fig:function_zeros},
\begin{equation}
    (w\bullet\bvp_0)(t) = \left[\min_{x\in\bvp_*(t)}w(x),\max_{x\in\bvp_*(t)}w(x)\right],\quad \bvp_*(t):=[\bvp_0(t_-),\bvp_0(t)].
\end{equation}
Such paths $w\bullet\bvp_0$ were discussed briefly in \autoref{ch:intro}, and now the emergence of an interval-valued process from the Heston model, as discussed in the \hyperlink{prologue}{Prologue}, should seem plausible.

So in general the convergence of compositions $w\circ\vp_n\xrightarrow{n\to\infty}w\circ\bvp_0$ is violated on all reasonable metric spaces on $\uD(\RR_+,\RR)$, like those of \cite{Skorokhod_1956}, despite having $\vp_n\xrightarrow{n\to\infty}\bvp_0$ on $(\bPhi,d_{\bPhi})$ as in \autoref{sec:bounds_limits}. Hence the introduction of interval-valued paths, which we name \emph{excursionary}. This naming is inspired by Chapter 15 of \cite{Whitt_2002}, the aim of which is to understand processes with similar paths with excursions, arising in queuing applications.

\vspace{3mm}\textbf{The excursionary space.} We now properly define our set $\uE$ of excursionary paths, then the relevant metric $d_\uE$ on it, which should be thought of as characterising the convergence of graphs over compacts in $\RR_+\times\RR$ w.r.t.~Hausdorff distances. On the resulting excursionary space $(\uE,d_\uE)$, we will obtain the functional convergence $w\circ\vp_n \xrightarrow{n\to\infty} w\bullet\bvp_0$ which \emph{cannot} be obtained on $\uD(\RR_+,\RR)$, generalising the specific limit given in \autoref{eq:def_barrier} on $(\RR,|\cdot|)$.

\begin{definition}[Set $\uE$ of paths]\label{def:excursionary_paths}
    Let the set $\uE$ contain real compact interval-valued paths $\ve$ over $\RR_+$, i.e.~which for each $t\in\RR_+$ returns a compact interval $\ve(t)=:[\ve_-(t),\ve_+(t)]\subset\RR$.
\end{definition}

For such a path $\ve$, having $\ve_-(t)=\ve_+(t)$ is acceptable, in which case $\ve(t)$ returns a singleton. Alternatively, supposing $\ve\in\uD:=\uD(\RR_+,\RR)$, then the path which returns a singleton $\{\ve(t)\}$ for each $t\in\RR_+$ is in $\uE$, which will still be labelled $\ve$ when convenient. In this sense, $\uD\subset\uE$. 

We are especially interested in paths $\ve_\bullet\in\uE$ which for some $\ve_\circ\in\uD$ verifies $\ve_\bullet(t)=\{\ve_\circ(t)\}$ whenever $\ve_\circ(t_-)=\ve_\circ(t)$ and $[\ve_\circ(t_-)\wedge \ve_\circ(t),\ve_\circ(t_-)\vee \ve_\circ(t)]\subseteq \ve_\bullet(t) = [\ve_-(t),\ve_+(t)]$ otherwise. Recall that such inclusions were discussed following \autoref{eq:intro_limit_def}, which defines an excursionary generalisation $S^\bullet_t$ of the exponentiated NIG process arising from the Heston model. Unlike $\ve\in\uE$ in general, these excursionary paths $\ve_\bullet$ related to some $\ve_\circ\in\uD$ fall into the setting of Section 15.4 in \cite{Whitt_2002}, and Theorem 15.4.1 there provides conditions for such paths to define a separable space when equipped with our Hausdorff metric $d_\uE$.

Towards defining this metric $d_\uE$, for $\ve\in\uE$ let its graph $\Gamma_T(\ve)$ over $[0,T]\subset\RR_+$ be defined
\begin{equation}\label{eq:graphs_deriv}
    \Gamma_T(\ve):= \{(t,x)\in[0,T]\times\RR:x\in \ve(t)\},
\end{equation}
and then define the extended graph $\Gamma^*_T(\ve):=\Gamma_T(\ve)\cup\{T\}\times\RR$ to alleviate issues at the arbitrary endpoint $T$, like we did to help define $d_{\uM_1}$ in \autoref{thm:m1_relationship}. This explains why the endpoint was removed manually in the example of \autoref{eq:def_barrier}. Now $d_\uE$ is defined thus.

\begin{definition}[Excursionary metric]\label{def:excursionary_metric}
    For $\ve_1,\ve_2\in\uE$ and $T\in\RR_+$, let the excursionary pseudometric $d_{\uE,T}$ return the Hausdorff distance $d_\uH$ between graphs $\Gamma_{1,2}:=\Gamma^*_T(\ve_{1,2})$, i.e.~
    \begin{multline}\label{eq:hausdorff_dist}
        d_{\uE,T}(\ve_1,\ve_2) := d_{\uH}(\Gamma_1,\Gamma_2) :=\\ \max\left\{\sup_{(t,x)\in\Gamma_1}\inf_{(s,u)\in\Gamma_2} |(t,x) - (s,u)|,\sup_{(t,x)\in\Gamma_2}\inf_{(s,u)\in\Gamma_1} |(t,x) - (s,u)|\right\}.
    \end{multline}
    Then define the excursionary metric $d_\uE$ on $\uE$ by $d_\uE(\ve_1,\ve_2):=\sum_{n\in \NN}2^{-n}(1\wedge d_{\uE,n}(\ve_1,\ve_2))$.
\end{definition}

Note that, like each $d_{\uE,T}$, $d_\uE$ actually defines another pseudometric on $\uE$, which is often the case with Hausdorff distances. To see this, simply consider $\ve_{1,2}\in\uE$ with $\ve_1(t):=[-1,1]$ over $\RR_+$ but $\ve_2(t):=[-1,1]$ only at the rationals $\QQ_+$, with $\ve_2(t):=\{0\}$ otherwise. Then clearly $d_\uE(\ve_1,\ve_2)=0$ but $\ve_1\neq \ve_2$. The pseudometric space $(\uE,d_\uE)$ can be upgraded to a bona fide metric space by the usual consideration of \emph{equivalence classes} of paths in $\uE$. Doing so explicitly is not actually necessary, however, given that $d_\uE$ certainly induces a Borel $\sigma$-algebra $\cE$ as usual on $\uE$, making the (pseudometrisable) topological space $(\uE,\cE)$ also a measurable space, thus suitable for our probabilistic volatility-related applications.

\vspace{2mm} Before characterising excursionary limits of composite paths $w\circ\vp$, where $\vp\in\Phi$ solves \autoref{prob:ivp2}, we first study limits of time derivatives $\vp'$. Such limits are not only helpful in their own right, given they capture the behaviour of volatility paths $\sqrt{\vp'}$, but can often be \emph{related} to composite limits. This can be seen in the Heston example in \autoref{eq:ode_verify}, or more generally when $g(t,x):=\vt(t)-w(x)$ as in \autoref{thm:solutionmap}, because then we simply find
\begin{equation}\label{eq:deriv_composite}
    (w\circ\vp)(t) = \vt(t) - \vp'(t).
\end{equation}
As already noted, in this section we will for convenience use $\vp'$ and $w\circ\vp$ to also denote the \emph{singleton}-valued paths in $\uE$, the latter coinciding with $w\bullet\vp$ defined in \autoref{eq:def_barrier}. The theory developed here will always be applied in the setting of \autoref{thm:exit_limit}, so although not strictly required the reader may adopt the assumptions there throughout: $\{g_n\}_{n\in\NN_0}\subset\uG$ is such that $\Vert g_0 - g_n \Vert_{\RR_+^2}\xrightarrow{n\to\infty}0$, $\{\vp_n\}_{n\in\NN}\subset\Phi$ respectively solve $x'=ng_n(t,x)$, $x(0)=0$ from \autoref{prob:ivp2}, and most importantly $d_{\bPhi}(\vp_n,\bvp_0)\xrightarrow{n\to\infty}0$, where $\bvp_0\in\bPhi$ is as in \autoref{thm:wellposed}.

\vspace{3mm}\textbf{Derivatives' limits.} \autoref{cor:construct_limit} constructs \emph{any} path $\bvp_0\in\bPhi\subset\uD(\RR_+,\RR)$ as a limit of solutions $\vp_n$ to \autoref{prob:ivp2}, on $(\bPhi,d_{\bPhi})$. Supposing $\bvp_0$ has a discontinuity in $(0,T)$, then we must find $\Vert \vp'_n \Vert_{[0,T]}\xrightarrow{n\to\infty}\infty$, and it is clear we should not attempt to find a limit of $\vp'_n$ as $n\to\infty$ on $\uC$, $\uD$ or even $\uE$, so neither for volatility $\sqrt{\vp'_n}$. The best we achieve here are limits of the scaled paths $n^{-1}\vp'_n$ on $(\uE,d_{\uE})$, and the approach towards this via explicit parametric representations is applied also to compositions $w\circ\vp_n$, often related like in \autoref{eq:deriv_composite}.

For the next result, keep in mind that if $\vp\in\Phi$ is a global solution of \autoref{prob:ivp2}, then the derivative $\vp'$ admits the trivial parametric representation $(\id,\vp')$, and this is in the same equivalence class as $(\vp^{-1},\vp'\circ\vp^{-1})$. This just amounts to the following equivalence in $\RR_+^2$
\begin{equation}\label{eq:deriv_equiv}
    \{(t,\vp'(t)):t\in\RR_+\} = \{(\vp^{-1}(x),\vp'(\vp^{-1}(x))):x\in\RR_+\}.
\end{equation}
Given that limits for temporal components like $\vp^{-1}$ are understood through \autoref{thm:exit_limit}, the focus in \autoref{thm:para_deriv} is the spatial component $\vp'\circ\vp^{-1}$. For clarity, in \autoref{eq:para_limit} define the path $g_0(E(\bvp_0),\id):x\mapsto g_0(E(\bvp_0)(x),x)$ in $\uC(\RR_+,\RR)$, and let the norm $\Vert\cdot\Vert_{\RR_+}$ be defined as usual using \autoref{def:uni_metric}, which characterises uniform convergence over compacts.

\begin{lemma}[Parametric derivative limits]\label{thm:para_deriv}
    Adopt the assumptions of \autoref{thm:exit_limit}, so $\vp_n\xrightarrow{n\to\infty}\bvp_0$ on $(\bPhi,d_{\bPhi})$. Then the following convergence of derivatives $\{\vp'_n\}_{n\in\NN}$ takes place~
    \begin{equation}\label{eq:para_limit}
        \left\Vert g_0(E(\bvp_0),\id) - n^{-1}\vp'_n\circ\vp_n^{-1}\right\Vert_{\RR_+} \xrightarrow{n\to\infty}  0.
    \end{equation}
\end{lemma}
\begin{proof}
    Given $\vp_n$ is the unique global solution of the IVP $x'=ng_n(t,x)$, $x(0)=0$, then $\vp'_n(t)=ng_n(t,\vp_n(t))$ is verified for each $t\in\RR_+$. Substituting $t=\vp^{-1}_n(x)$ into this, we see
    \begin{equation}
        n^{-1}\vp_n'(\vp_n^{-1}(x)) = g_n(\vp^{-1}_n(x),x)
    \end{equation}
    for each $x\in\RR_+$. The proof will thus be complete, just by definitions, if the convergence 
    \begin{equation}\label{eq:converge_composite}
        \delta_n := \left\Vert g_0(E(\bvp_0),\id) - g_n(\vp^{-1}_n,\id)\right\Vert_{[0,X]} \xrightarrow{n\to\infty}  0
    \end{equation}
    takes place for all $X\in\RR_+$. We already have both $\Vert E(\bvp_0)-\vp^{-1}_n \Vert_{[0,X]}\xrightarrow{n\to\infty} 0$ and $\Vert g_0 - g_n \Vert_{[0,T]\times[0,X]}\xrightarrow{n\to\infty} 0$ for all $T,X\in\RR_+$ from \autoref{thm:exit_limit}, and these indeed combine to give \autoref{eq:converge_composite}. A modulus of continuity for $g_0\in\uC(\RR^2,\RR)$ will be utilised to show this.
    
    Toward this, fix $X\in\RR_+$ and $T>E(\bvp_0)(X)$. Then $\Vert E(\bvp_0)-\vp^{-1}_n \Vert_{[0,X]}<T-E(\bvp_0)(X)$ for all sufficiently large $n$, so assume w.l.o.g. $\Vert\vp_n^{-1}\Vert_{[0,X]} = \vp_n^{-1}(X)<T$ for all $n$. This means the paths of all $(\vp_n^{-1},\id)$ over $[0,X]$, which appear as arguments in \autoref{eq:converge_composite}, are bounded into the rectangle $\cX:=[0,T]\times[0,X]$. Notice that the triangle inequality gives 
    \begin{equation}\label{eq:triangle_deriv}
        \delta_n \le \Vert g_0(E(\bvp_0),\id) - g_n(E(\bvp_0),\id)\Vert_{[0,X]} + \Vert g_n(E(\bvp_0),\id) - g_n(\vp^{-1}_n,\id)\Vert_{[0,X]}.
    \end{equation}
    Treating the first component here is simple, as it is clearly bounded by $\Vert g_0 - g_n\Vert_\cX$. For the second, let $w_0:\RR_+\to\RR_+$ be a modulus of continuity of $g_0$ over $\cX$, so we have the bound
    \begin{equation}\label{eq:mod_cont2}
        |g_0(t,x) - g_0(s,u)| \le w_0(|(t,x) - (s,u)|)
    \end{equation}
    for all $(t,x),(s,u)\in\cX$, and $w_0(\ep)\xrightarrow{\ep\downarrow0}0$. Using the triangle inequality twice, the relationship of \autoref{eq:mod_cont2} can be extended to each $g_n$ and $w_n$, provided we define $w_n:=w_0 + 2\Vert g_0-g_n\Vert_\cX$. Note that $w_n$ is \emph{not} a modulus of continuity for $g_n$ over $\cX$, because $w_n(\ep)\xrightarrow{\ep\to 0}2\Vert g_0-g_n\Vert_\cX\neq0$. Nevertheless, we can now bound the last term in \autoref{eq:triangle_deriv}, using
    \begin{equation}
        \Vert g_n(E(\bvp_0),\id) - g_n(\vp^{-1}_n,\id)\Vert_{[0,X]} \le w_0(\Vert E(\bvp_0)-\vp^{-1}_n \Vert_{[0,X]}) + 2\Vert g_0 - g_n\Vert_\cX.
    \end{equation}
    So in full, the claimed convergence in \autoref{eq:converge_composite} takes place because for every $X\in\RR_+$
    \begin{equation}
        \left\Vert g_0(E(\bvp_0),\id) - n^{-1}\vp'_n\circ\vp_n^{-1}\right\Vert_{[0,X]} = \delta_n \le w_0(\Vert E(\bvp_0)-\vp^{-1}_n \Vert_{[0,X]}) + 3\Vert g_0 - g_n\Vert_\cX \xrightarrow{n\to\infty}  0.
    \end{equation}
    Given $X$ is arbitrary, this extends to the claim in \autoref{eq:para_limit} w.r.t.~the norm $\Vert\cdot\Vert_{\RR+}$.
\end{proof}

Letting $\{\ve_n\}_{n\in\NN}\subset\uE$ be defined by the singletons $\ve_n(t):=\{n^{-1}\vp'(t)\}$, thus capturing the behaviour of volatility $\sqrt{\vp'}$, in \autoref{cor:hausdorff_deriv} we now combine \autoref{thm:exit_limit} and \autoref{thm:para_deriv} to obtain a surprising limit $\ve_n\xrightarrow{n\to\infty}\ve_0$ on the excursionary space $(\uE,d_\uE)$. Our approach to this, which will be repeated for the composite paths $w\circ\vp_n$ in \autoref{cor:graph_hausdorff}, is to define helpful parametric representations $(\tau_n,\sigma_n)$ of each $\ve_n$, and to establish the product uniform convergence (over compacts) of these to a limit $(\tau_0,\sigma_0)$, finally interpreting this limit as a path $\ve_0$ in $\uE$. It is intuitive that this uniform convergence of parametric representations indeed provides convergence on $\uE$ w.r.t.~Hausdorff distances, and this can be confirmed easily enough by noting that, within the definition of $d_\uE$ in \autoref{eq:hausdorff_dist} we have the bound 
\begin{multline}
    \sup_{(t,x)\in\Gamma_n}\inf_{(s,u)\in\Gamma_0} |(t,x) - (s,u)| = \sup_{s\in[0,1)}\inf_{u\in[0,1)} |(\tau_n(s),\sigma_n(s)) - (\tau_0(u),\sigma_0(u))|\\
    \le \sup_{s\in[0,1)} |(\tau_n(s),\sigma_n(s)) - (\tau_0(s),\sigma_0(s))| =: \Vert(\tau_0,\sigma_0) - (\tau_n,\sigma_n)\Vert_{[0,1)}
\end{multline}
where we have assumed w.l.o.g.~that the domain of all parametric representations $(\tau_n,\sigma_n)$ have been conveniently transformed into $[0,1)$. This specifically demonstrates the bound $d_{\uE,T}(\ve_n,\ve_0)\le \Vert(\tau_0,\sigma_0) - (\tau_n,\sigma_n)\Vert_{[0,1)}$, which may be extended to what we will use, namely~
\begin{equation}\label{eq:para_limit_to_excursion}
    \Vert(\tau_0,\sigma_0) - (\tau_n,\sigma_n)\Vert_{\RR_+}\xrightarrow{n\to\infty}0 \implies d_\uE(\ve_n,\ve_0)\xrightarrow{n\to\infty}0.
\end{equation}

\begin{theorem}[Excursionary derivative limits]\label{cor:hausdorff_deriv}
    Adopt the assumptions of \autoref{thm:exit_limit}, so $d_{\bPhi}(\vp_n,\bvp_0)\xrightarrow{n\to\infty}0$, and define $\{\ve_n\}_{n\in\NN}\subset\uE$ by $\ve_n(t):=\{n^{-1}\vp'_n(t)\}$ respectively. Then $d_\uE(\ve_n,\ve_0)\xrightarrow{n\to\infty}0$, where $\ve_0\in\uE$ returns the singleton $\{0\}$ a.e.~and is precisely defined by
    \begin{equation}\label{eq:def_deriv_excursion}
        \ve_0(t) := \left[0, \ve_0^+(t)\right],\quad \ve_0^+(t):=\max_{x\in\bvp_*(t)} g_0(t,x),\quad \bvp_*(t):=[\bvp_0(t_-),\bvp_0(t)].
    \end{equation}
\end{theorem}
\begin{proof}
    Combining \autoref{thm:exit_limit} and \autoref{thm:para_deriv}, we obtain the product convergence
    \begin{equation}\label{eq:uni_prod_limit}
        \Vert(E(\bvp_0),g_0(E(\bvp_0),\id)) - (\vp_n^{-1},n^{-1}\vp'_n\circ\vp_n^{-1})\Vert_{\RR_+}\xrightarrow{n\to\infty}0.
    \end{equation}
    Now set $(\tau_n,\sigma_n):=(\vp_n^{-1},n^{-1}\vp'_n\circ\vp_n^{-1})$, which, given the equivalence of graphs in \autoref{eq:deriv_equiv}, define parametric representations of each $\ve_n$ respectively. So we will obtain the claimed convergence $d_\uE(\ve_n,\ve_0)\xrightarrow{n\to\infty}0$ using \autoref{eq:para_limit_to_excursion} and \autoref{eq:uni_prod_limit} \emph{if} the path $\ve_0$ defined in \autoref{eq:def_deriv_excursion} is similarly parameterised by $(\tau_0,\sigma_0):=(E(\bvp_0),g_0(E(\bvp_0),\id))$.
    
    Now $E(\bvp_0)$ is in $\uN$ from \autoref{def:setN}, i.e.~defines a non-decreasing and spatially unbounded path in $\uC_0(\RR_+,\RR)$. So the graph $\Gamma_0$ of this parametric representation $(\tau_0,\sigma_0)$ has the form
    \begin{equation}\label{eq:graph_rep}
        \Gamma_0 = \left\{(t,g_0(t,x))\in\RR_+\times\RR: x\in\bvp_*(t)\right\}.
    \end{equation}
    Using the continuity of $g_0$ and the compactness of $\bvp_*(t):=[\bvp_0(t_-),\bvp_0(t)]$, we then obtain
    \begin{equation}
        \Gamma_0 = \left\{(t,x)\in\RR_+\times\RR: x\in\left[\min_{u\in\bvp_*(t)} g_0(t,u),\max_{u\in\bvp_*(t)} g_0(t,u)\right]\right\}.
    \end{equation}
    Now if we can show that $\ve^-_0(t):=\min_{x\in\bvp_*(t)} g_0(t,x)=0$ for every $t$, then we finally obtain 
    \begin{equation}
        \Gamma_0 = \left\{(t,x)\in\RR_+\times\RR: x\in\left[0,\ve_+(t)\right]\right\}.
    \end{equation}
    This will provide $\Gamma_0=\{(t,x):x\in\ve_0(t)\}$, clarifying that indeed $(\tau_0,\sigma_0)$ parameterises $\ve_0$, and therefore $d_\uE(\ve_n,\ve_0)\xrightarrow{n\to\infty}0$. To confirm $\ve^-_0(t)=0$, consider \autoref{fig:exit_converge2}. At discontinuities of $\bvp_0(t)$, we see that $g(t,x)\ge0$ is ensured for all $x\in\bvp_*(t)$. Given that $g(t,\bvp_0(t_-))=g(t,\bvp_0(t)) = 0$ also follows from the continuity of $g$, then indeed $\ve^-_0(t)=0$. 
    
    It just remains to confirm that $\ve_0(t)=\{0\}$ a.e., and this follows from $\bvp_0(t_-)=\bvp_0(t)$ a.e., given discontinuities are countable, and therefore $\ve^+_0(t)=0$ a.e., completing the proof.
\end{proof}

\begin{figure}[H]
    \centering
    \includegraphics[width=0.48\linewidth]{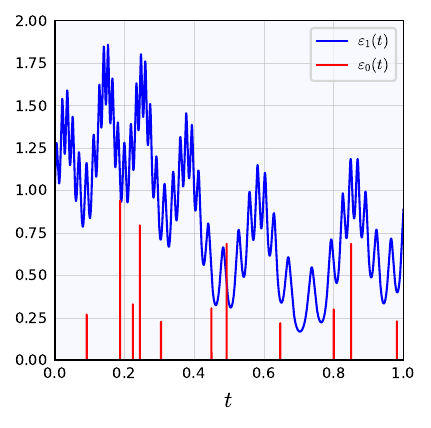}
    \includegraphics[width=0.48\linewidth]{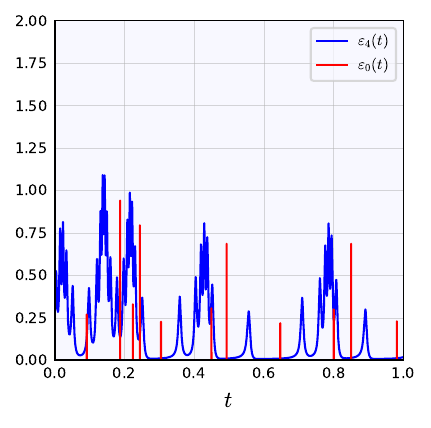}
    \includegraphics[width=0.48\linewidth]{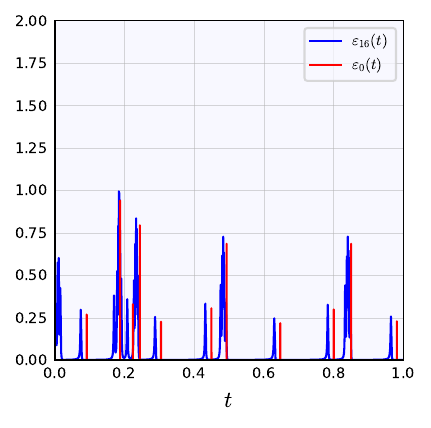}
    \includegraphics[width=0.48\linewidth]{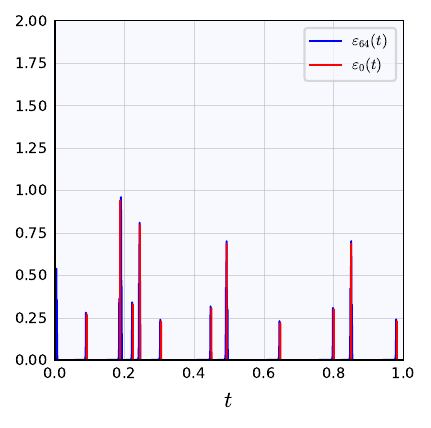}
    \caption{The convergence $\ve_n\xrightarrow{n\to\infty}\ve_0$ on $(\uE,d_\uE)$ is shown, where $\ve_n(t):=\{n^{-1}\vp'_n(t)\}$ are scaled CIR paths from \autoref{ex:CIR_to_IG}, and $\ve_0$ is defined in \autoref{eq:ex_limit_def}.}
    \label{fig:hausdorff_deriv}
\end{figure}

\vspace{2mm}We now demonstrate \autoref{cor:hausdorff_deriv} using again \autoref{ex:CIR_to_IG}. Notice that whenever a limit of $n^{-1}\vp'_n(t)$ is characterised in the setting of \autoref{thm:exit_limit}, we are equivalently characterising that of $g_n(t,\vp_n(t))=n^{-1}\vp'_n(t)$. Given \autoref{eq:ode_verify}, in \autoref{ex:CIR_to_IG} this takes the form
\begin{equation}
    \sigma (w\circ\vp_n)(t) - \kappa(\theta t - \vp_n(t)) + v/n = n^{-1}\vp'_n(t),
\end{equation}
and these paths coincide with those of a scaled CIR process $n^{-1}V(\omega):=n^{-1}\vp_n$, recalling we identified $\int_0^\cdot V_s(\omega)\dd s := \vp_n$ in \autoref{ex:CIR_to_IG}. The limit $\ve_0\in\uE$ in \autoref{cor:hausdorff_deriv} reads
\begin{equation}\label{eq:ex_limit_def}
    \ve_0(t) := \left[0, \max_{x\in\bvp_*(t)} \sigma w(x) + \kappa\left(\theta t - x\right)\right],\quad \bvp_0(t):=\inf\{x>0: \kappa x - \sigma w(x) > \kappa\theta t\},
\end{equation}
so we can interpret the instantaneous excursions of $\ve_0$ in \emph{space} as manifesting from certain excursions of the path $w$ in \emph{time} over intervals in space. The convergence of $\ve_n(t):=\{n^{-1}\vp'_n(t)\}$ to $\ve_0$ on $(\uE,d_\uE)$, as implied by \autoref{cor:hausdorff_deriv}, is demonstrated in \autoref{fig:hausdorff_deriv}. Comparing this with \autoref{fig:exit_converge}, notice excursions in $\ve_n$ indeed develop only at jumps of $\bvp_0$.

\vspace{3mm}\textbf{Composite limits.} Applicable to composite paths $w\circ\vp_n$, like those of our price processes $S_t:=\exp(W^\rho_{X_t}-\frac12 X_t)$ defined later, the structure of this part mirrors the last, clarifying our approach towards limiting results which derive from \autoref{thm:exit_limit}. Specifically, \autoref{thm:para_uniform} here provides the convergence of some helpful parametric representations (uniformly over compacts), and \autoref{cor:graph_hausdorff} reduces this to a result on the excursionary space $(\uE,d_\uE)$. Recall that in \autoref{cor:ae_composite} we have already demonstrated the pointwise convergence $(w\circ\vp_n)(t)\xrightarrow{n\to\infty}(w\circ\bvp_0)(t)$ a.e., so here we generalise this to (surprising) functional statements. Although we will work with a fixed path $w\in\uC:=\uC(\RR_+,\RR)$, this may be generalised to a sequence $\{w_n\}_{n\in\NN_0}$ verifying $\Vert w_0 - w_n\Vert_{\RR_+}\xrightarrow{n\to\infty} 0$ without difficulty. 

For a global solution $\vp\in\Phi$ of \autoref{prob:ivp2} with inverse $\vp^{-1}$, notice that $(\id,w\circ\vp)$ and $(\vp^{-1},w)$ are in the same equivalence class of parametric representations, precisely meaning
\begin{equation}\label{eq:para_equiv}
    \{(t,w(\vp(t))):t\in\RR_+\} = \{(\vp^{-1}(x),w(x)):x\in\RR_+\}.
\end{equation}
This equivalence is analogous to that in \autoref{eq:deriv_equiv}, applicable instead to derivatives $\vp'$. The next result uses this to obtain a trivial extension of \autoref{thm:exit_limit} which nevertheless encodes what we need to appreciate the behaviour of composite paths $w\circ\vp_n$ as $n\to\infty$.

\begin{corollary}[Parametric composite limits]\label{thm:para_uniform}
    Adopt the assumptions of \autoref{thm:exit_limit}, so that $d_{\bPhi}(\vp_n,\bvp_0)\xrightarrow{n\to\infty}0$, and fix $w\in\uC(\RR_+,\RR)$. Then the sequence $\{(\vp^{-1}_n,w)\}_{n\in\NN}$ verifies
    \begin{equation}
         \left\Vert (E(\bvp_0),w) - (\vp^{-1}_n,w)\right\Vert_{\RR_+} \xrightarrow{n\to\infty}  0.
    \end{equation}
\end{corollary}
\begin{proof}
    Given the spatial components here clearly verify $\Vert w-w\Vert_{\RR_+}=0$, the claim follows just from the conclusion $d_{\bPhi}(\vp_n,\bvp_0)=\Vert E(\bvp_0)-\vp_n^{-1}\Vert_{\RR_+}\xrightarrow{n\to\infty}0$ within \autoref{thm:exit_limit}.
\end{proof}

Letting $\{\ve_n\}_{n\in\NN}\subset\uE$ be defined by the singletons $\ve_n(t):=\{(w\circ\vp_n)(t)\}$, in \autoref{cor:graph_hausdorff} we now establish a limit $\ve_n\xrightarrow{n\to\infty}\ve_0$ on the excursionary space $(\uE,d_\uE)$, thus describing the limiting behaviour of $w\circ\vp_n$ and therefore price paths also. As discussed in \autoref{ch:intro}, the limit found is of course the filled composition $\ve_0:=w\bullet\bvp_0$ defined by $(w\bullet\bvp)(t) := \{w(x):x\in\bvp_*(t)\}$ where as usual $\bvp_*(t):=[\bvp_0(t_-),\bvp_0(t)]$. Note $\ve_0(t)$ returns the singleton $\{(w\circ\bvp_0)(t)\}$ a.e., and to draw comparisons with \autoref{eq:def_deriv_excursion} we have the equivalent representation
\begin{equation}\label{eq:def_comp_excursion}
    \ve_0(t) := \left[\ve_0^-(t), \ve_0^+(t)\right],\quad \ve_0^-(t):=\min_{x\in\bvp_*(t)} w(x),\quad \ve_0^+(t):=\max_{x\in\bvp_*(t)} w(x).
\end{equation}

\begin{theorem}[Excursionary composite limits]\label{cor:graph_hausdorff}
    Adopt the assumptions of \autoref{thm:exit_limit}, so that $d_{\bPhi}(\vp_n,\bvp_0)\xrightarrow{n\to\infty}0$, and fix $w\in\uC(\RR_+,\RR)$. Define the set $\{\ve_n\}_{n\in\NN}\subset\uE$ respectively by the singletons $\ve_n(t):=\{(w\circ\vp_n)(t)\}$, and $\ve_0:=w\bullet\bvp_0\in\uE$. Then $d_\uE(\ve_n,\ve_0)\xrightarrow{n\to\infty}0$.
\end{theorem}
\begin{proof}
    From \autoref{thm:para_uniform}, we have the product convergence $\Vert(\tau_0,\sigma_0) - (\tau_n,\sigma_n) \Vert_{\RR_+} \xrightarrow{n\to\infty}  0$, where $(\tau_0,\sigma_0):=(E(\bvp_0),w)$ and $(\tau_n,\sigma_n):=(\vp^{-1}_n,w)$. So we can use the approach from \autoref{eq:para_limit_to_excursion} to obtain the claim of $d_\uE(\ve_n,\ve_0)\xrightarrow{n\to\infty}0$ \emph{if} every $(\tau_n,\sigma_n)$ parameterises $\ve_n$.
    
    When $n\neq0$, $(\tau_n,\sigma_n)$ clearly parameterises $\ve_n$, using the equivalence in \autoref{eq:para_equiv}. To see that $(\tau_0,\sigma_0)$ also parameterises $\ve_0$, we manipulate the graph $\Gamma_0$ of $(\tau_0,\sigma_0)$ to obtain
    \begin{multline}
        \Gamma_0 = \left\{(E(\bvp_0)(x),w(x)): x\in\RR_+\right\}\\
        = \left\{(t,w(x))\in\RR_+\times\RR: x\in[\bvp_0(t_-),\bvp_0(t)]\right\}\\
        = \left\{(t,x)\in\RR_+\times\RR: x\in(w\bullet\bvp_0)(t)\right\}.
    \end{multline}
    Given $\ve_0:=w\bullet\bvp_0$, then we may equivalently write $\Gamma_0=\{(t,x):x\in\ve_0(t)\}$ to see that $(\tau_0,\sigma_0)$ indeed parameterises $\ve_0$. So we obtain $d_\uE(\ve_n,\ve_0)\xrightarrow{n\to\infty}0$, completing the proof.
\end{proof}

\begin{figure}[H]
    \centering
    \includegraphics[width=0.48\linewidth]{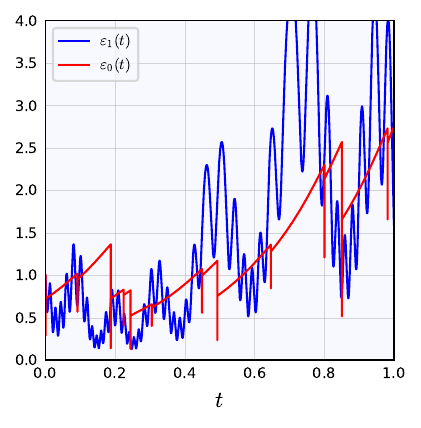}
    \includegraphics[width=0.48\linewidth]{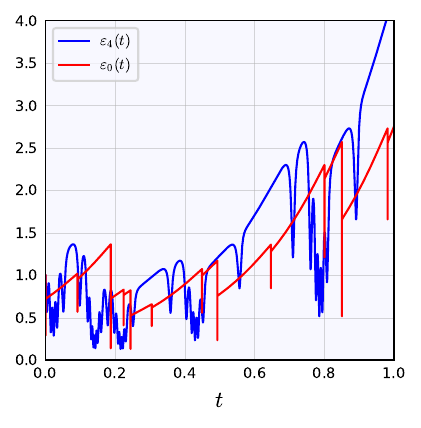}
    \includegraphics[width=0.48\linewidth]{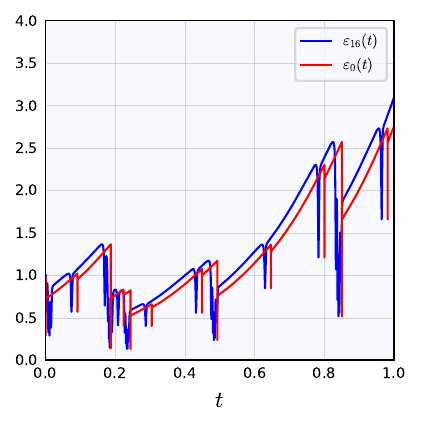}
    \includegraphics[width=0.48\linewidth]{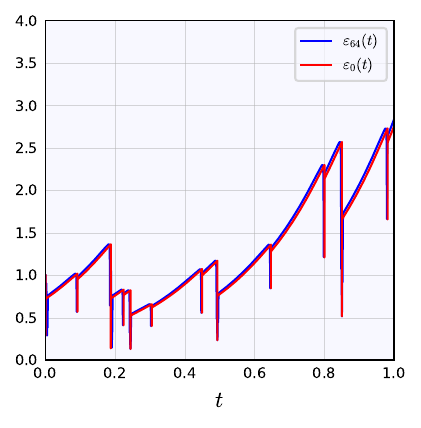}
    \caption{The convergence $\ve_n\xrightarrow{n\to\infty}\ve_0$ on $(\uE,d_\uE)$ is shown, where $\ve_n$ are Heston price paths and $\ve_0$ is an interval-valued generalisation of an exponentiated NIG path, both defined in \autoref{ex:heston_to_NIG}.}
    \label{fig:hausdorff_composite}
\end{figure}

To conclude this chapter, an example of \autoref{cor:graph_hausdorff} is provided which extends the pathwise CIR and IG limiting relationship from \autoref{ex:CIR_to_IG} to the Heston and NIG models, as discussed in the \hyperlink{prologue}{Prologue}. This provides a deep foundation for strengthening and generalising \autoref{thm:mech_pro} from \cite{Mechkov_2015}, as we will do in \autoref{sec:price_limits}. The resulting convergence $\ve_n\xrightarrow{n\to\infty}\ve_0$ of price paths is shown in \autoref{fig:hausdorff_composite}, and is consistent with \autoref{fig:hausdorff_deriv}.

\begin{example}[Pathwise Heston to NIG]\label{ex:heston_to_NIG}
    Fix paths $w_{0,1}\in\uC_0:=\uC_0(\RR_+,\RR)$ and for some $\rho\in[-1,1]$ define $w_\rho:=\rho w_1 + \sqrt{1-\rho^2}w_0$. These can be interpreted as sample paths of the Brownian motion which defines the Heston model in \autoref{eq:tce2}, e.g.~$w_{\rho}:=W^{\rho}(\omega)$. Now let $\vp_n\in\Phi$ solve the IVPs $x'=ng_n(t,x)$, $x(0)=0$ in \autoref{ex:CIR_to_IG} with $w:=w_1$, so that
    \begin{equation}
        \vp'_n(t) = n\sigma (w_1\circ\vp_n)(t) - n\kappa(\theta t - \vp_n(t)) + v.
    \end{equation}
    Let each singleton-valued path $\ve_n\in\uE$ be defined from $w_\rho$ and the solution $\vp_n$ according to
    \begin{equation}
        \ve_n(t) := \left\{\exp\left((w_\rho\circ\vp_n)(t) - \frac12\vp_n(t)\right)\right\}.
    \end{equation}
    Using again \autoref{eq:tce2}, notice that $\ve_n$ and $\vp_n$ may then be considered as sample paths of the Heston price process and its cumulative variance respectively. As in \autoref{ex:CIR_to_IG}, \autoref{thm:exit_limit} can be applied to obtain $\vp_n\xrightarrow{n\to\infty}\bvp_0$ on $(\bPhi,d_{\bPhi})$, where $\bvp_0$ is an IG L\'evy path. Defining the geometric Brownian path $w(x):=\exp(w_\rho(x)-\frac12 x)$, we can apply \autoref{cor:graph_hausdorff} to additionally obtain the convergence $\ve_n\xrightarrow{n\to\infty}\ve_0$ on $(\uE,d_{\uE})$, where $\ve_0:=w\bullet\bvp_0$. In full, 
    \begin{equation}\label{eq:pricelimit_def}
        \ve_0(t) := \left\{\exp\left(w_\rho(x)-\frac12 x\right): x\in[\bvp_0(t_-),\bvp_0(t)]\right\}.
    \end{equation}
\end{example}
    
    Notice that $\ve_0$ is a path of the limit $S^\bullet$ in \autoref{eq:intro_limit_def}, which we claim is an interval-valued generalisation of the exponentiated NIG process $S^\circ$ there. This will be covered in \autoref{lem:NIG_reduc}, but notice that when $\bvp_0(t_-)=\bvp_0(t)$, which is a.e., then $\ve_0(t)$ just contains
    \begin{equation}
        \exp\left((w_\rho\circ\bvp_0)(t)-\frac12 \bvp_0(t)\right) = \exp\left(\sqrt{1-\rho^2}(w_0\circ\bvp_0)(t) + \frac{2\rho-\sigma}{2\sigma}\bvp_0(t) -\frac{\rho\theta}{\sigma} t\right)
    \end{equation}
    where the final expression uses the relationship $\kappa \bvp_0(t) - \sigma (w_1\circ\bvp_0)(t)=\kappa\theta t$ to eliminate $w_1\circ\bvp_0$, which follows from the definition of $\bvp_0$. Now clearly this expression coincides with a path of the exponentiated NIG process $S^\circ$. We have thus demonstrated the convergence of Heston price paths $\ve_n$ to paths $\ve_0$ of an interval-valued exponentiated NIG generalisation.
    
\vspace{4mm}For consistency, in \autoref{fig:hausdorff_composite} we fix the path $w_1$ in \autoref{ex:heston_to_NIG} to be the Weierstrass path driving \autoref{fig:hausdorff_deriv}, and set $\rho=-1$ (not unreasonable for equity prices) so there is actually no dependence on the additional path $w_0$. An interesting effect of setting $\rho=-1$ is that only \emph{downwards} excursions develop in the Heston price paths $\ve_n$, which are upwards if instead $\rho=1$. These downwards excursions are clearly evident in the generalised NIG limit $\ve_0$ in \autoref{fig:hausdorff_composite}. Through the relevant ODE, e.g.~\autoref{eq:deriv_composite}, these downwards price excursions in \autoref{fig:hausdorff_composite} can be related directly to the \emph{upwards} excursions of volatility from \autoref{fig:hausdorff_deriv}.

%% file: chapters/4_framework.tex
\clearpage
\section{A pathwise volatility modelling framework}\label{chap:framework}

We are now ready to employ the new ODE theory from the previous two chapters to build a \emph{probabilistic} volatility modelling framework. We use `pathwise' (which may be considered an antonym of `probabilistic') to describe this framework as a reminder that all models within it, which will exist on some probability space $(\Omega,\cF,\PP)$, are well-defined on an explicit subset $\Omega_*\subseteq\Omega$ of outcomes (or `paths') with full $\PP$-measure. Proof of this will just follow from our probability-free well-posedness theory applied to each outcome $\omega\in\Omega_*$. This situation is more helpful than a model just being a.s.~well-defined, since this need not \emph{explicitly} provide $\Omega_*$. We do not consider there to be a standardised meaning of `pathwise' in general; see e.g.~\cite{Vovk_2016} for a short background on its varied use in the context of It\^o-type integrals.

Equipped with $\Omega_*$ where $\PP[\Omega_*]=1$, our models thus remain a.s.~well-defined under any other measure $\PP_*$ verifying $\PP_*[\Omega_*]=1$, so by definition under any $\PP_*\!\ll\!\PP$. Practically, this enables us to e.g.~\emph{replace} Brownian motion in the Heston model's representation from \autoref{eq:tce} with a vast range of other (irregular) stochastic processes, \emph{with no additional well-posedness analysis required}. This is discussed following \autoref{thm:solutionmap_rand} and made precise within \autoref{def:gen_heston_frame}, where alternative volatility drivers $Z$ are defined. We make use of these when defining the RLH model in \autoref{sec:RLH_model}, where one of the Gaussian processes from \autoref{cor:gauss_heston_mgf} is used. Recall that this ability to replace Brownian motion is one of the motivations behind the pathwise theory of \cite{Friz_2010} and \cite{Friz_2014}.

We can only consider e.g.~the Heston SDEs from \autoref{eq:heston_intro} in an similar `pathwise' sense if we invoke pathwise It\^o calculus, originating from \cite{Follmer_1981}. This is an active line of research; see \cite{Davis_2018}, \cite{Lochowski_2018} and \cite{Cont_2019} for developments, all similarly motivated by problems in finance. Our pathwise framework built from ODEs was not designed to compete with these rough path and pathwise It\^o alternatives (rather, it emerged when treating the problems outlined in the \hyperlink{prologue}{Prologue}). But a significant benefit of it compared with these is its relative simplicity, given it does not depend on the non-Riemannian integrals at the core of these alternatives.

The programme of this chapter is now outlined before our probabilistic set-up is specified.

\vspace{3mm}\textbf{From framework to model.} Building on the ODE theory from the previous two chapters, the first half of this one serves as a three-stage funnel. A very general framework for modelling price processes on a probability space $(\Omega,\cF,\PP)$ is first defined in \autoref{sec:base_frame}. As discussed in \autoref{ch:intro}, this depends on the solutions of \autoref{prob:ivp2} for each $\omega\in\Omega$. So as per \autoref{cor:surjective_map} it is built from a non-injective and surjective solution map which by \autoref{thm:wellposed} is continuous w.r.t.~uniform convergence over compacts. This general framework is then reduced to two distinct sub-frameworks in \autoref{sec:heston_frame} and \autoref{sec:martingale}, which respectively contain generalised Heston and martingale models. Finally the funnel's specific product, the RLH model, is defined in \autoref{sec:RLH_model}, which resides in the intersection of these two sub-frameworks. This was illustrated in \autoref{fig:venn}, repeated here for convenience.

\vspace{2mm}
\begin{figure}[H]
    \centering
    \includegraphics[width=0.60\linewidth]{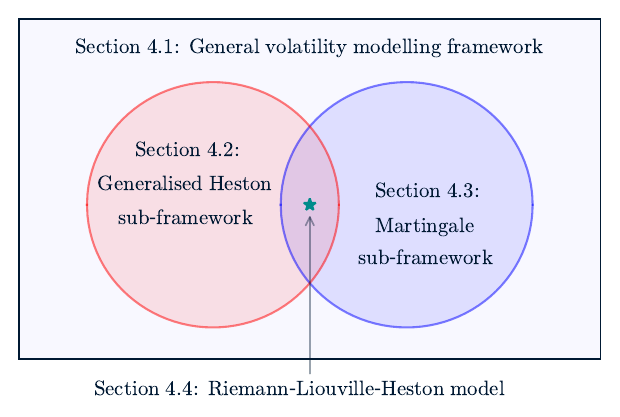}
    \label{fig:venn2}
\end{figure}

\vspace{-6mm}

{\centering\autoref{fig:venn}: Venn diagram showing the frameworks and model defined in this chapter.\\}

\vspace{4mm}

Much care has been taken in the selection of these funnelling stages for presentation here, because we are trying to attain several goals without compromising any. Of course, we want to end up with a volatility model which exhibits some properties of the leading counterparts from other more established frameworks. But on the other hand, it is arguably more important that the route to this model is identifiable in some respect, given that the ODE-based foundations on which the framework here rests are unconventional in finance. Subject to these requirements, we must additionally show how the limiting results of \autoref{chap:solutions} can be applied to precisely characterise the Heston-NIG relationship discussed in the \hyperlink{prologue}{Prologue}. By doing this, this new framework's ability to teach us practically valuable things about others is unquestionable, not least because the Heston and NIG models are two of the most popular in finance, respectively deriving from differing (continuous and pure jump) frameworks.

This all considered, we let the Heston model take a somewhat central role throughout this chapter. Specifically, the sub-framework in \autoref{sec:heston_frame} produces a price process which is equal in distribution to the Heston model's when a process $Z$ which drives volatility is Brownian motion, but allows for this process to be replaced by essentially any other random element of $\uC_0(\RR_+,\RR)$. (A reader comfortable with stochastic processes existing only up to a non-zero but random explosion time, see e.g.~Definition 2.1 in \cite{Ikeda_1989}, can omit `essentially' here.) This ability to simply replace Brownian motion cannot be taken for granted, and as just discussed is reminiscent of rough path theory advertisements.

The RLH model, the focus of \autoref{sec:RLH_model}, then constitutes the special case within this generalised Heston sub-framework where the volatility-driving Brownian motion is replaced by its Riemann-Liouville fractional derivative of some order in $(0,\frac12)$. As a result, the price process from this model coincides (in distribution) with that from the Heston model when (and only when) this derivative order is zero. Because the Riemann-Liouville fractional derivative map defines a continuous isomorphism between H\"older spaces, see e.g.~\cite{Samko_1993}, it is straightforward to reconcile this RLH model with the growing evidence that volatility typically exhibits H\"older regularities much lower than that of Brownian motion.

\vspace{3mm}\textbf{Specific application choices.} The second half of this chapter focuses on applications, the last of which are the limits already mentioned, covered in \autoref{sec:price_limits}. These limits are treated in the specific case of the RLH model for maximum clarity, given the probability-free theory for any other model is provided in \autoref{chap:solutions}. As should be clear after \autoref{ex:heston_to_NIG} and  \autoref{fig:hausdorff_composite}, these limits are not just mathematical curiosities, but will provide precise answers to the questions in the \hyperlink{prologue}{Prologue}, regarding the popular Heston and NIG models.

Before this we show in \autoref{sec:vol_surfaces} how derivative prices can be simulated under the RLH model. Theoretically, this depends on both the martingale theory from \autoref{sec:martingale} and the probability-free simulation convergence from \autoref{thm:wellposed}. A background to the relevance of martingales for derivative pricing is also provided in \autoref{sec:martingale}, following the very pragmatic approaches of \cite{Cont_2003} and \cite{Guyon_2013}. It is specifically \autoref{cor:RLH_martingality} which establishes the RLH price process to be a martingale, by bringing together several other results, existing and new. Towards this, \autoref{thm:mgf_existence} should be noted, clarifying how martingale prices can be related to the thickness of a volatility-driving process's marginal tails, and in the wider generalised Heston sub-framework from \autoref{sec:heston_frame}.

There are two reasons for prioritising a simulation-based approach at this stage of the framework's development. Firstly, it provides a standalone framework-wide solution for pricing derivatives (or other applications like hedging or forecasting), rather than depending on model-specific probabilistic analysis, which we leave for the future. Secondly, recent research has shown that \emph{alongside} simulation, neural networks offer an alternative approach to the problems classically treated by probabilistic analysis. See e.g.~\cite{Buehler_2019} for hedging and \cite{Horvath_2021} for model calibration.

To aid our simulation convergence, the variance reduction methods recommended in \cite{McCrickerd_2018} are utilised, for which we have no statistical biases to report. By reconciling simulated results with analytically-available classical Heston counterparts, we gain confidence that simulations are implemented correctly and have converged sufficiently.
Concise \texttt{python} code is also provided in the \hyperlink{appendix}{Appendix} to help others implement our models.

\vspace{3mm}\textbf{Probabilistic set-up.} Given the probability-free foundations in the previous two chapters, much of this chapter could also be presented without reference to a probability measure. However, most practical applications, like ours depending on the martingales in \autoref{sec:martingale}, or the weak convergence results in both \autoref{sec:vol_surfaces} and \autoref{sec:price_limits}, are inseparable from probability. It is thus clearer to start introducing the probabilistic necessities immediately. 

To this end, we will always work generally on a probability space $(\Omega,\cF,\PP)$ supporting all random elements referred to, and let $\omega$ denote an arbitrary element of $\Omega$. Often it will be possible to construct these random elements on a \emph{fixed} probability space, although for the sake of brevity we will not repeatedly do so. As an example, the RLH model from \autoref{sec:RLH_model} can be constructed on the canonical probability space supporting just a two-dimensional (2d) standard Brownian motion $W=(W^0,W^1)$ over $\RR_+$. Accordingly, we could fix $\Omega:=\uC_0(\RR_+,\RR)\times\uC_0(\RR_+,\RR)$, let $\cF:=\cB(\Omega)$ be the specific Borel $\sigma$-algebra that characterises uniform convergence over compacts, let $\PP:=\WW$ be the Wiener measure on $(\Omega,\cF)$ and let $W$ be the canonical process on $(\Omega,\cF,\PP)$, defined simply by $W(\omega):=\omega$ for each $\omega\in\Omega$. This clarifies that each outcome $\omega$ need not just be connected indirectly with a path of $W$, but it may \emph{actually be} a path of $W$. Both of the Heston and NIG processes from \autoref{eq:heston_prologue2} and \autoref{eq:nig_prologue} can be likewise constructed on this fixed space $(\Omega,\cF,\PP)$, because like all models in our framework these are built from a pathwise unique map.

We assume $(\Omega,\cF,\PP)$ supports such a 2d Brownian motion $W$, which will often be indexed by the variable $x\in\RR_+$, e.g.~$W\!=\!\{W_x\}_{x\in\RR_+}$. To do otherwise can be confusing when $W$ governs the \emph{spatial} behaviour of the random fields $Y=\{Y_{t,x}\}_{(t,x)\in\RR^2_+}$ introduced shortly, and is thereafter composed with a random IVP solution $X$, like $W^\rho_{X}$ in the Heston representation in \autoref{eq:tce2}. For consistency we will then use $\{\cF_x\}_{x\in\RR_+}$ to denote the natural filtration of $W$, and $\{\cG_t\}_{t\in\RR_+}$ for a \emph{different} filtration w.r.t.~which our price processes are martingales.

\subsection{A general price process framework}\label{sec:base_frame}

Loosely, we now want to define random counterparts $X$ of the IVP solutions $\vp$ to \autoref{prob:ivp2}. Recall from \autoref{thm:solutionset} that the solution set $\Phi$ of this problem is precisely the bijective paths in $\uC^1_0(\RR_+,\RR_+)$. A price process $S$ will be obtained from these paths via composition with geometric Brownian motion, specifically $S:=\exp(W^\rho_X - \frac12 X)$, so we call $X$ the cumulative variance of $S$, $X'$ instantaneous variance and $\sqrt{X'}$ volatility. As usual, $W^\rho$ is the 1d Brownian motion on $(\Omega,\cF,\PP)$ defined by $W^\rho:=\sqrt{1-\rho^2}W^0 + \rho W^1$ for some correlation $\rho\in[-1,1]$. There is no need to constrain how $X$ and $W$ are related, via $Y$, yet. 

It is worth elaborating on this last point. We do not impose such constraints at this stage because the well-posedness of our framework does not require it, unlike others. For example, in order to even exist, the It\^o integral $\int_0^t\sqrt{V_s}\dd W_s^\rho$ from the Heston model in \autoref{eq:heston_intro2} requires that $V$ is adapted to the natural filtration of $W^\rho$. We manage to defer introducing corresponding constraints until \autoref{def:spatial_adapted}, only when we consider martingale prices.

A benefit of this deferral is that if we are not working under the constraints of martingales, e.g.~if our application is volatility forecasting rather than derivative pricing, then we do not have to check the condition in \autoref{def:spatial_adapted}. What we pay for this freedom is that in full generality, where $Y$ and $W$ are merely random elements on the same space $(\Omega,\cF,\PP)$, the correlation $\rho$ and process $W^\rho$ are theoretically redundant. We choose to continue using these to define our price processes, however, because in our applications we will use them consistently with their introduction in \autoref{eq:heston_intro}. Namely, we will use $\rho$ to control the correlation between a price $S$ and its volatility $\sqrt{X'}$, often referred to as a leverage effect in equity markets. This effect may be detected in the at-the-money implied volatility skews in \autoref{fig:heston_comparison_skewed}, defined in \autoref{eq:skew_curv}, and also in the paths of $S$ and $V:=X'$ in \autoref{fig:simulation_output}.

\vspace*{2mm}
A question now arising from this loose description is: in what sense should $X$ (and $S$) be considered a bona fide stochastic process, e.g.~into which function topology does $X$ actually define a measurable map from $(\Omega,\cF,\PP)$? Recall from \autoref{thm:wellposed} that the solution map of \autoref{prob:ivp2} is continuous between $\uG\subset\uC(\RR_+^2,\RR)$ and $\uC(\RR_+,\RR)$ w.r.t.~the norms $\Vert\cdot\Vert_{\RR_+^d}$ which characterise uniform convergence over compacts. So this solution map is clearly measurable between the induced $\sigma$-algebras (topologies). So provided the random counterparts of the functions $g\in\uG$ in \autoref{prob:ivp2} are measurable from $(\Omega,\cF,\PP)$, then $X$ (and $S$) will also be. This random counterpart of $g\in\uG$ is called a random field, introduced in \autoref{def:rand_field}.

We only invoke \autoref{thm:wellposed} above because we can, and the measurability of $X$ and $S$ can be established directly. For this a sequence of (measurable) forward Euler polygon processes with vanishing mesh can be utilised, the convergence of which is ensured by \autoref{thm:euler_converge}. This approach mirrors that in Section 2.1.2 of \cite{Han_2017}, where Picard-Lindel\"of sequences are used because the counterparts to our functions $g\in\uG$ are spatially Lipschitz.

\vspace{3mm}\textbf{Random fields and IVPs.} In this part the random counterpart to \autoref{prob:ivp2} is stated, for which we introduce continuous random fields. In our setting, these will be random elements of $\uC(\RR_+^2,\RR)$, but the meaning from other domains will be clear. We utilise the notation $Y=\{Y_{t,x}\}_{(t,x)\in\RR_+^2}$ from \cite{Nielsen_2018} to denote these, despite the application there being to `ambit stochastics'. Now recall the norm $\Vert\cdot\Vert_{\RR_+^2}$ on $\uC(\RR_+^2,\RR)$ used in \autoref{thm:wellposed}, which induces the topology of uniform convergence over compacts.

\begin{definition}[Continuous random field]\label{def:rand_field}
    Let a continuous random field $Y=\{Y_{t,x}\}_{(t,x)\in\RR_+^2}$ be any random element of $\uC(\RR_+^2,\RR)$. That is, any measurable map from $(\Omega,\cF,\PP)$ to the set $\uC(\RR_+^2,\RR)$ equipped with the Borel $\sigma$-algebra induced by the norm $\Vert\cdot\Vert_{\RR_+^2}$ on $\uC(\RR_+^2,\RR)$.
\end{definition}

Using continuous random fields (hereafter just random field), random ODEs and IVPs, and their solutions, can be defined as a natural extension of their non-random counterparts. We define solutions over all of $\RR_+$ because we are most interested in IVPs like \autoref{prob:ivp2} where maximal solutions are global. Reducing this to compact subsets of $\RR_+$ is straightforward.

\begin{definition}[Random IVP]\label{def:rand_ivp}
    For a random field $Y=\{Y_{t,x}\}_{(t,x)\in\RR_+^2}$ on $(\Omega,\cF,\PP)$, call a stochastic process $X=\{X_t\}_{t\in\RR_+}$ a solution of the random ODE `$x'=Y_{t,x}$' if $X$ a.s.~verifies $X'_t = Y_{t,X_t}$ over $\RR_+$. Call $X$ a solution of the random IVP `$x'=Y_{t,x}$, $x_0=0$' if also $X_0=0$.
\end{definition}

As noted in \autoref{ch:intro}, our definition is consistent with the `SP' (sample path) formulation of random ODEs in \cite{Strand_1970}, which is based on the author's PhD thesis \cite{strand_1968}. This should be contrasted with the definition given e.g.~in \cite{Han_2017}, which is consistent with those from \cite{Soong_1973} and \cite{Sussman_1978}, all extending the definition from \cite{Srinivasan_1971}. Specifically, fixing a stochastic process $Z=\{Z_t\}_{t\in\RR_+}$ and function $h\in\uC(\RR^2,\RR)$, then \cite{Han_2017} would ask that a random ODE solution $X=\{X_t\}_{t\in\RR_+}$ verifies an expression like $x'=h(Z_t,x)$ over $\RR_+$, i.e.~$X'_t=h(Z_t,X_t)$.

We discussed briefly in \autoref{ch:intro} why this is too restrictive for volatility modelling, because even in the Heston case of \autoref{eq:tce2} we instead have $x'=h(t,Z_x)$, so $X'_t=h(t,Z_{X_t})$. Classical random ODE theory avoids such cases for good reason, because most desirable (random) functions of type $h(t,Z_\cdot)$ violate the Lipschitz condition which is relied upon for well-posedness properties. In the Heston case, $h(t,Z_\cdot)$ inherits the regularity of Brownian motion so is only H\"older continuous of orders in $(0,\frac12)$. This kind of reasoning motivates similarly pessimistic remarks in \cite{Soong_1973} regarding spatially Lipschitz random ODEs.

Following \autoref{def:rand_ivp}, the class of problems considered in this chapter may be obvious, but worth stating clearly before clarifying their well-posedness. Recall the subset $\uG\subset\uC(\RR_+^2,\RR)$ of functions from \autoref{def:mod_driving_func}, to which most results in \autoref{chap:wellposed} and \autoref{chap:solutions} apply. 

\begin{problem}\label{prob:random_ivp}
    Fix a random field $Y=\{Y_{t,x}\}_{(t,x)\in\RR^2_+}$ which is a.s.~in the set $\uG$. Then find a stochastic process $X=\{X_t\}_{t\in\RR_+}$ which solves the random IVP $x'=Y_{t,x}$, $x_0=0$ over $\RR_+$.
\end{problem}

For each outcome $\omega\in\Omega$, the (non-random) IVP $x'=Y_{t,x}(\omega), x(0)=0$ then a.s.~provides an example $x'=g_\omega(t,x), x(0)=0$ of \autoref{prob:ivp2}, given a.s.~$g_\omega(t,x)=Y_{t,x}(\omega)\in\uG$. From a probabilistic perspective, our use of `Fix\dots Then\dots' in \autoref{prob:random_ivp} should be noted. This is because we are in the privileged situation where we \emph{can} first fix $Y$ and \emph{will} be able to find a solution $X$, never having to seek the couple $(X,Y)$ simultaneously. To borrow terminology from SDEs, we are seeking only \emph{unique strong} solutions. If found for arbitrary $Y$ in $\uG$, this is `the best possible situation', as described in Figure 1.1 from \cite{Cherny_2005}.

In a related vein, whenever we refer to stochastic processes like $X$, e.g.~a unique solution of \autoref{prob:random_ivp}, we are as usual referring to an equivalence class of indistinguishable stochastic processes. Only \autoref{thm:rand_well_posed} acknowledges this explicitly, by constructing one solution $X$ to \autoref{prob:random_ivp} and clarifying that any other $X^*$ is indistinguishable, i.e.~a.s.~verifies $X^*=X$.

\vspace{3mm}\textbf{Well-posedness.} By considering the random IVPs in \autoref{prob:random_ivp} driven only by random fields $Y$ a.s.~in $\uG$, we can draw upon the probability-free analysis in the previous two chapters. The next result specifically clarifies the consequences of \autoref{thm:wellposed} for \autoref{prob:random_ivp}. We will henceforth omit the repetition of `a.s.' when it is clear to do to so, e.g.~writing $Y\in\uG$. Such an assumption \emph{implies} that the set $\{\omega\in\Omega:Y(\omega)\in\uG\}$ is measurable, i.e.~in $\cF$. In turn any countable intersection of full-measure sets is measurable by the properties of $\sigma$-algebras, and retains full-measure by \autoref{eq:full_measure}. This is precisely why our probability-free theory can be applied on a pathwise basis to obtain a.s.~results, as the next proof demonstrates.
\begin{theorem}[Well-posedness for \autoref{prob:random_ivp}]\label{thm:rand_well_posed}
    All of the probability-free statements in \autoref{thm:wellposed}, applicable to a solution of \autoref{prob:ivp2}, apply on an a.s.~basis to a solution of \autoref{prob:random_ivp}, i.e.~to a solution of a random IVP $x'=Y_{t,x}$, $x_0=0$ with $Y\in\uG$. Specifically\emph{:}
    
    \emph{1 (Global existence and uniqueness).} There exists a unique solution $X=\{X_t\}_{t\in\RR_+}$ of any such random IVP. This solution has paths in the set $\Phi\subset\uC^1_0(\RR_+,\RR_+)$ from \autoref{def:solutionset}\emph{;}

    \emph{2 (Upper bound).} This solution $X$ is dominated by the process $\bX=\{\bX_t\}_{t\in\RR_+}$ defined by $\bX_t=\inf\{x>0:Y_{t,x}<0\}$, which has paths in the set $\bPhi\subset\uD(\RR_+,\RR_+)$ from \autoref{def:setE}\emph{;}
    
    \emph{3 (Continuous dependence).}$\!$ The solution map of \autoref{prob:random_ivp} is continuous from $\uG$ to $\Phi$ w.r.t.~uniform convergence over compacts. That is, if $\{Y^n\}_{n\in\NN_0}$ generate solutions $\{X^n\}_{n\in\NN_0}$,~
    \begin{equation}\label{eq:cont_dep_sim}
        \Vert Y^0-Y^n\Vert_{\RR_+^2}\xrightarrow[\mathrm{a.s.}]{n\to\infty}0\quad \implies\quad \Vert X^0- X^n\Vert_{\RR_+}\xrightarrow[\mathrm{a.s.}]{n\to\infty}0.
    \end{equation}
\end{theorem}
\begin{proof}
    Let the subset $\Omega_*\subset\Omega$ of outcomes be defined by $\Omega_*:=\{\omega\in\Omega:Y(\omega)\in\uG\}$. Given that $Y\in\uG$ by assumption, we know this set $\Omega_*$ has full measure, i.e.~$\PP[\Omega_*]=1$. For each $\omega\in\Omega_*$, the (non-random) IVP $x'=g_\omega(t,x):=Y_{t,x}(\omega)$, $x(0)=0$ constitutes an example of \autoref{prob:ivp2}, so adheres to the well-posedness results of \autoref{thm:wellposed}. In particular, for each $\omega\in\Omega_*$, this IVP has a unique solution $\vp=\vp_\omega$ which is bounded above by the path $\bvp=\bvp_\omega$. Checking $(\vp,\bvp)\in\Phi\times\bPhi$ is straightforward given \autoref{thm:wellposed} and these sets' definitions.
    
    Now by simply defining $X(\omega):=\vp$ and $\bX(\omega):=\bvp$ for each $\omega\in\Omega_*$, processes $X$ and $\bX$ are constructed with the claimed properties in points 1.~and 2. here. There are technically other processes $X^*$ which solve \autoref{prob:random_ivp} for this choice of field $Y$, but assuming these \emph{are not} indistinguishable from $X$ provides a subset of $\Omega_*$ with positive measure where the uniqueness statement in \autoref{thm:wellposed} is violated. So indistinguishability $X^*=X$ is ensured.
    
    The continuous dependence statement in \autoref{eq:cont_dep_sim} follows in a similar way by applying \autoref{thm:wellposed} on a subset $\Omega_*\subset\Omega$ of outcomes with full measure. Specifically, we can define
    \begin{equation}
        \Omega_n:=\{\omega\in\Omega: Y^n(\omega)\in\uG\},\quad \Omega_*:=\cap_n\Omega_n \cap \{\omega\in\Omega: \Vert Y^0(\omega) - Y^n(\omega)\Vert_{\RR_+^2}\xrightarrow{n\to\infty}0\}
    \end{equation}
    then obtain $\Vert X^0(\omega) - X^n(\omega)\Vert_{\RR_+}\xrightarrow{n\to\infty}0$ for each $\omega\in\Omega_*$ by applying \autoref{thm:wellposed}. Provided $Y^n\in\uG$ and $\Vert Y^0-Y^n\Vert_{\RR_+^2}\xrightarrow[\mathrm{a.s.}]{n\to\infty}0$, then since $\Omega_*$ is a countable intersection of full-measure sets we have $\PP[\Omega_*]=1$ by \autoref{eq:full_measure}. We have thus shown $\Vert X^0- X^n\Vert_{\RR_+}\xrightarrow[\mathrm{a.s.}]{n\to\infty}0$.
\end{proof}

In addition to the continuity statement given in \autoref{eq:cont_dep_sim}, we can alternatively use \autoref{thm:wellposed} to get a statement \emph{not} related to the same outcomes of different random elements, but instead different outcomes of fixed ones. E.g.~for outcomes $\{\omega_n\}_{n\in\NN_0}\subset\Omega_*$,
\begin{equation}\label{eq:rand_cont_dep}
    \Vert Y(\omega_0) - Y(\omega_n) \Vert_{\RR^2_+}\xrightarrow{n\to\infty}0 \implies \Vert X(\omega_0) - X(\omega_n) \Vert_{\RR_+}\xrightarrow{n\to\infty}0.
\end{equation}
This pathwise statement differs from the probability-free one in \autoref{eq:new_cont_statement} only through its applicability to a full-measure set $\Omega_*$, and having the ability to make such statements on \emph{explicit} full-measure sets is why we describe our framework as `pathwise' itself. The statement in \autoref{eq:cont_dep_sim} suggests more practical value than \autoref{eq:rand_cont_dep}, though. E.g.~suppose we would like to simulate a random IVP solution $X^0$ but cannot simulate $Y^0$. Then we may utilise approximating fields $\{Y^n\}_{n\in\NN}$ and at least generate a converging sequence $\{X^n\}_{n\in\NN}$.

\vspace{3mm}\textbf{The solution space.} We now clarify two more properties of the solution map of \autoref{prob:random_ivp}, like point 3.~in \autoref{thm:rand_well_posed} but instead deriving from \autoref{chap:solutions}. Proofs are not provided for these results because they follow from \autoref{thm:solutionset} and \autoref{thm:solutionmap} respectively, just like \autoref{thm:rand_well_posed} does from \autoref{thm:wellposed}. That is, by defining the appropriate full-measure set $\Omega_*$, then applying \autoref{thm:solutionset} and \autoref{thm:solutionmap} for each outcome $\omega\in\Omega_*$.

Extending the use of $\vp^{-1}$ in \autoref{thm:solutionset}, we now let the process $X^{-1}=\{X^{-1}_x\}_{x\in\RR_+}$ denote the unique inverse of any $X\in\Phi$, like solutions of \autoref{prob:random_ivp}. This inverse is well-defined, has bijective paths in $\uC_0(\RR_+,\RR_+)$ like $X$, and verifies $X^{-1}_{X_t}=t$ and $X_{X^{-1}_x}=x$ for $(t,x)\in\RR_+^2$.

\begin{corollary}[The solution set]\label{thm:solutionset_rand}
    The solution set of \autoref{prob:random_ivp} is precisely all stochastic processes $X=\{X_t\}_{t\in\RR_+}$ with paths in $\Phi$. In particular, fixing any process $\vt=\{\vt_t\}_{t\in\RR_+}$ with paths in $\uC_0(\RR_+,\RR)$, then each $X\in\Phi$ solves the random IVP $x' = Y_{t,x}$, $x_0=0$ when
    \begin{equation}\label{eq:g_rep_rand}
        Y_{t,x}:=X'_{X^{-1}_x} + \vt_t - \vt_{X^{-1}_x}.
    \end{equation}
    This random IVP provides an example of \autoref{prob:random_ivp}, i.e.~$Y\in\uG$, when $\vt$ is strictly increasing with $\sup_{t\in\RR_+}\vt_t-X'_t=\infty$. In this case, $X$ is this random IVP's unique solution.
\end{corollary}

Notice that, for each fixed $x\in\RR_+$, the temporal structure of the random field in \autoref{eq:g_rep_rand} is governed entirely by the process $\vt$. This next result follows from \autoref{thm:solutionmap} and tells us that the solution set of \autoref{prob:random_ivp} is not compromised very much if we reduce this process $\vt$ to a \emph{fixed} function. Specifically, the solution set $\Phi$ reduces to all stochastic process with paths in the subset $\Phi_\vt\subset\Phi$ defined in \autoref{thm:solutionmap}. As discussed thereafter, \emph{any} such subset $\Phi_\vt$ contains all paths $\vp\in\Phi$ with the additional property of $\liminf_{t\to\infty}\vp'(t)<\infty$.

\begin{corollary}[Solution map bijectivity]\label{thm:solutionmap_rand}
    Fix any strictly increasing function $\vt\in\uC_0(\RR_+,\RR)$ with $\sup_{t\in\RR_+}\vt(t)=\infty$. Let $\Phi_\vt\subset\Phi$ contain the paths $\vp$ which verify $\sup_{t\in\RR_+}\vt(t)-\vp'(t)=\infty$, and let $\uG_\vt\subset\uG$ contain functions $g$ with representation $g(t,x):=\vt(t)-w(x)$ for some $w\in\uC(\RR_+,\RR)$ with $w(0)\le0$ and $\sup_{x\in\RR_+}w(x)=\infty$. Then the map taking each random field $Y\in\uG_\vt$ to the solution $X\in\Phi_\vt$ of the case $x' = Y_{t,x}$, $x_0=0$ of \autoref{prob:random_ivp} is bijective.
\end{corollary}

Like in the proof of \autoref{thm:solutionmap}, the unique field $Y\in\uG_\vt$ which generates the chosen process $X\in\Phi_\vt$ as the solution of \autoref{prob:random_ivp} is now given in terms of a \emph{process} $Z=\{Z_x\}_{x\in\RR_+}$ by 
\begin{equation}\label{eq:separable_field}
    Y_{t,x}:=\vt(t) - Z_x,\quad Z_x := \vt(X^{-1}_x) - X'_{X^{-1}_x}.
\end{equation}
This process has paths in $\uC(\RR_+,\RR)$ and satisfies $Z_0\le0$ and $\sup_{x\in\RR_+}Z_x=\infty$. The solution map bijectivity in \autoref{thm:solutionmap_rand} of course supplements this map being continuous from $\uG_\vt$ to $\Phi_\vt$ w.r.t.~uniform convergence over compacts, like in point 3.~of \autoref{thm:rand_well_posed}. It is no coincidence that such fields $Y\in\uG_\vt$ from \autoref{eq:separable_field} are closely related to the functions in $\uF_\vt$ first introduced in \autoref{ex:main_example} and containing the Heston case defined in \autoref{eq:cir_ivp_well_posed}. 

We will advocate the use of such fields $Y_{t,x}:=\vt(t)-Z_x$ for volatility modelling more generally, where $\vt\in\uC_0(\RR_+,\RR)$ is strictly increasing, $\sup_{t\in\RR_+}\vt(t)=\infty$, $Z\in\uC(\RR_+,\RR)$, $Z_0\le0$ and $\sup_{x\in\RR_+}Z_x=\infty$. This is because \autoref{thm:solutionmap_rand} says that even if we do fix the temporal structure of a random field $Y\in\uG$ via a function $\vt$, the solution set of \autoref{prob:random_ivp} \emph{only} reduces to the processes in $\Phi_\vt$. All these processes satisfy $\sup_{t\in\RR_+}\vt(t)-X'_t=\infty$, which is ensured by $\sup_{t\in\RR_+}\vt(t)=\infty$ when $\liminf_{t\to\infty}X'_t<\infty$. Given we will shortly define the process $\sqrt{X'}$ to be a price process's volatility, this condition $\liminf_{t\to\infty}X'_t<\infty$ is not only weak but actually desirable, given $\liminf_{t\to\infty}X'_t=\infty$ a.s.~is clearly unrealistic.

Now recall the set $\Phi'$ from \autoref{def:solutionderivset} and the following discussion. This set characterises the instantaneous variance processes $X'$ we can theoretically model with \autoref{prob:random_ivp}. So if we use fields of type $Y_{t,x}=\vt(t)-Z_x$ then in full we can model any $X'$, thus volatility $\sqrt{X'}$, which satisfies $\liminf_{t\to\infty}X'_t<\infty$ and $\lim_{t\to\infty}\int_0^t X'_s\dd s=\infty$, and is not zero over intervals.

Now that we understand why \autoref{prob:random_ivp} is so promising for volatility modelling, we are finally ready to properly define the modelling frameworks which have this problem at their heart. We have clearly not yet consolidated all probability-free results which apply to \autoref{prob:random_ivp} on a pathwise basis. The remainder, like the simulation convergence in \autoref{thm:wellposed} and the exit-time limits in \autoref{thm:exit_limit}, will be introduced instead when they are needed.

\vspace{3mm}\textbf{A price process framework.} At the beginning of this section we described our general framework for modelling price processes $S=\{S_t\}_{t\in\RR_+}$, via the expression $S=\exp(W^\rho_X-\frac12 X)$. This framework is properly defined here in \autoref{def:price_frame}, depending on \autoref{prob:random_ivp} which has a unique solution $X=\{X_t\}_{t\in\RR_+}$ by \autoref{thm:rand_well_posed}. Following this definition we can finally call $\sqrt{X'}$ volatility in this framework, and then the framework is better described as a `general volatility modelling framework', like it is labelled in the Venn diagram of \autoref{fig:venn}.

\begin{definition}[Price process framework]\label{def:price_frame}
    Let the space $(\Omega, \cF,\PP)$ support a 2d Brownian motion $W=(W^0,W^1)$ over $\RR_+$ and random field $Y\in\uG$. Let $X=\{X_t\}_{t\in\RR_+}$ be the unique solution of the random IVP $x'=Y_{t,x}$, $x_0=0$, then define the price process $S=\{S_t\}_{t\in\RR_+}$ simply by $S_t:=\exp(W^\rho_{X_t}-\frac12 X_t)$, where $W^\rho:=\sqrt{1-\rho^2}W^0 + \rho W^1$ for some fixed $\rho\in[-1,1]$.
\end{definition}

Like the Heston model's representation in \autoref{eq:rand_ode}, specific models for $S$ and $X$ in this general framework will be summarised by the equations which they uniquely verify, namely
\begin{equation}
    X'_t = Y_{t,X_t},\quad S_t:=\exp(W^\rho_{X_t}-\tfrac12 X_t).
\end{equation}

We have already discussed at the beginning of this section why these processes $X$ and $S$ are indeed bona fide stochastic processes; because they both define measurable maps from $(\Omega, \cF,\PP)$ to the set $\uC(\RR_+,\RR)$ equipped with the Borel $\sigma$-algebra which characterises uniform convergence over compacts. In fact these maps are \emph{continuous} in this sense if $(\Omega, \cF,\PP)$ is defined appropriately. E.g.~let $(\Omega, \cF,\PP)$ be the canonical product space supporting Brownian motion $W=(W^0,W^1)$ and random field $Y$, so $\Omega:=\uC(\RR_+,\RR)^2\times\uC(\RR_+^2,\RR)$, then continuity of $S$ is confirmed by extending the assumption in \autoref{eq:rand_cont_dep} to the product convergence
\begin{multline}\label{eq:joint_continuity}
    (\Vert W^0(\omega_0) - W^0(\omega_n)\Vert_{\RR_+} , \Vert W^1(\omega_0) - W^1(\omega_n)\Vert_{\RR_+}, \Vert Y(\omega_0) - Y(\omega_n)\Vert_{\RR^2_+}) \xrightarrow{n\to\infty} (0,0,0).
\end{multline}
From this we obtain $\Vert S(\omega_0) - S(\omega_n) \Vert_{\RR_+}\xrightarrow{n\to\infty}0$ provided $\{Y(\omega_n)\}_{n\in\NN_0}\!\subset\!\uG$, which is a.s.~in the framework of \autoref{def:price_frame} given the assumption $Y\in\uG$. Note that no constraints on the relationship between $W$ and $Y$ have been imposed yet. Contrasting this, recall the Heston case $Y_{t,x}:=\sigma W^1_x + \kappa(\theta t - x) + v$ from \autoref{eq:rand_ode}, where $Y$ is constructed linearly from $W^1$ and the third component in \autoref{eq:joint_continuity} is thus redundant, being implied by the second. Like in this Heston case, the next two sections impose constraints on $W$ and $Y$ in order to define sub-frameworks in which the price process $S$ inherits desirable properties.

Now that a price process framework is fully specified in \autoref{def:price_frame}, only now can we precisely say what we mean by the framework-dependent stochastic process called volatility.
\begin{definition}[Volatility]\label{def:volatility}
    Let $S=\{S_t\}_{t\in\RR_+}$ be a price process constructed in the framework of \autoref{def:price_frame}. Then let the volatility $\sigma=\{\sigma_t\}_{t\in\RR_+}$ of $S$ be defined by $\sigma := \sqrt{X'}$.
\end{definition}

While this definition may seem at odds with the more recognisable relationship of $\sigma^2_t = \frac{\dd}{\dd t}[\log S]_t$, we will prove consistency between the two in the martingale setting of \autoref{sec:martingale}, i.e.~there we show $[\log S] = X$ in \autoref{them:price_martingale}, so also $\frac{\dd}{\dd t}[\log S]_t = X'_t$. It makes sense to treat this consistency alongside the martingality of $S$, given that the existence and properties of quadratic variations (in the conventional probabilistic sense) are intricately related to martingales. We will henceforth prioritise the use of $\sqrt{X'}$ to denote volatility, to avoid clashes of notation with Heston's volatility of volatility parameter first introduced in \autoref{eq:heston_prologue}.

\subsection{A generalised Heston sub-framework}\label{sec:heston_frame}

This section moves down the funnel described at the beginning of this chapter, reducing the general volatility modelling framework from \autoref{def:price_frame} to one of the sub-frameworks in \autoref{fig:venn}. The models in this sub-framework are generalisations of the popular stochastic volatility model from \cite{Heston_1993}, which was introduced informally in the \hyperlink{prologue}{Prologue}. Specifically, \autoref{thm:heston_ode} demonstrates how to recover this model's price process distribution.

Besides clarifying consequences of \autoref{thm:rand_well_posed} and \autoref{thm:solutionset_rand} from the previous section, the main contributions of this section are the conditions in \autoref{thm:mgf_existence} and \autoref{cor:gauss_heston_mgf} which ensure the existence of moment generating functions (MGFs) $\EE[e^{pX_t}]$ within this sub-framework, where $X$ is a solution of \autoref{prob:random_ivp}. These results illustrate how valuable the dominating process $\bX$ from \autoref{thm:rand_well_posed} is, and are some of the first intrinsically probabilistic contributions of this thesis, given everything thus far may be reduced to the probability-free results from \autoref{chap:wellposed} and \autoref{chap:solutions} on a pathwise basis. Although informative in their own right, these MGF existence results are critically important towards establishing the martingality of corresponding price processes $S = \exp(W^\rho_X - \frac12 X)$ in \autoref{sec:martingale}, given the line we take there via Novikov's condition for martingales, provided in \autoref{thm:novikov}. We now recall the volatility model from \cite{Heston_1993}, preparing for generalisations.

\vspace{3mm}\textbf{The classical Heston model.} As usual, let our space $(\Omega,\cF,\PP)$ support a fixed standard 2d Brownian motion $W=(W^0,W^1)$ over $\RR_+$. Constructed from $W$, we can then define the classical Heston model as follows. For completeness, the celebrated pathwise uniqueness result of \cite{Yamada_1971} can be invoked to show that the CIR SDE in \autoref{eq:heston_cir} has a unique strong solution, so the model specified here is indeed well-defined. 

\begin{definition}[Classical Heston model]\label{def:heston_stoch_vol}
    For fixed parameters $\sigma,\kappa,\theta,v > 0$, let the process $V=\{V_t\}_{t\in\RR_+}$ be the unique solution of the CIR SDE depending on $W^1$, i.e.~verifying
    \begin{equation}\label{eq:heston_cir}
        \dd V_t = \sigma\sqrt{V_t}\dd W^1_t + \kappa(\theta - V_t)\dd t,\quad V_0=v.
    \end{equation}
    Then, for fixed $\rho\in[-1,1]$, let the Heston price process $S=\{S_t\}_{t\in\RR_+}$ be defined by
    \begin{equation}\label{eq:heston_stock}
        S_t := \exp\left(\int_0^t\sqrt{V_s}\dd W^\rho_s - \frac12 \int_0^t V_s\dd s\right),\quad W^\rho := \sqrt{1-\rho^2} W^0 + \rho W^1.
    \end{equation}
\end{definition}

This relatively simple model has been analysed to a tremendous degree since its formulation, yet it continues to inform some of the most cutting edge volatility modelling developments, like the rough Heston model of \cite{El_Euch_2019} and its quadratic variant from \cite{Gatheral_2020}. This considered, it is surprising that the relationship between this model and random ODEs has not been taken seriously before now. The obvious reason for this is that existing ODE theory does not immediately provide well-posedness for the resulting random ODE, but \autoref{chap:wellposed} has now dealt with this obstacle.

\vspace{3mm}\textbf{A generalised Heston framework.} A modelling framework is now defined which constitutes a sub-framework of that from \autoref{def:price_frame}, and features in \autoref{fig:venn}. The relationship with the Heston model is briefly deferred until \autoref{thm:heston_ode}, although by comparing \autoref{def:gen_hest_field} below with \autoref{eq:heston_cir} above, this can be intuited when $Z:=W^0$ and $\vt(t):=\theta t$.

\begin{definition}[Generalised Heston framework]\label{def:gen_heston_frame}
    Let $\vt$ be a
    bijective path in $\uC_0(\RR_+,\RR_+)$, and $Z=\{Z_x\}_{x\in\RR_+}$ any process in $\uC_0(\RR_+,\RR)$ verifying the condition $\sup_{x\in\RR_+}\kappa x - \sigma Z_x=\infty$ for parameters $\sigma,\kappa > 0$. Let the random field $Y=\{Y_{t,x}\}_{(t,x)\in\RR_+^2}$ in $\uG$ be then defined by
    \begin{equation}\label{def:gen_hest_field}
        Y_{t,x} := \sigma Z_x + \kappa(\vt(t) - x) + v,
    \end{equation}
    for $v\ge0$, let $X=\{X_t\}_{t\in\RR_+}$ be the unique solution of the random IVP $x'=Y_{t,x}$, $x_0=0$, and let the price process $S=\{S_t\}_{t\in\RR_+}$ be defined by $S_t:=\exp(W^\rho_{X_t}-\frac12 X_t)$ for fixed $\rho\in[-1,1]$.
\end{definition}

When helpful, specific models in this framework will be summarised using the equations
\begin{equation}\label{eq:model_summary}
    X'_t = \sigma Z_{X_t} + \kappa(\vt(t) - X_t) + v,\quad S_t:=\exp(W^\rho_{X_t}-\tfrac12 X_t)
\end{equation}
but to help draw comparisons with the CIR SDE in \autoref{eq:heston_cir}, notice we could write
\begin{equation}\label{eq:model_summary2}
    V_t = \sigma Z_{\int_0^t V_s\dd s} + \kappa\left(\vt(t) - \int_0^t V_s\dd s\right) + v\ \implies\ \dd V_t = \sigma \dd Z_{\int_0^t V_s\dd s} + \kappa(\vt'(t) - V_t)\dd t
\end{equation}
where $V:=X'$, and the right-hand equation assumes the equivalence $\vt(t)=\int_0^t\vt'(s)\dd s$, which is technically not required in \autoref{def:gen_heston_frame}, i.e.~$\vt$ need not be absolutely continuous. Contrasting the Heston case where $Z:=W^0$ and $\vt(t):=\theta t$, we still do not need to impose a link between $(W^0, W^1)$ and $Y$ via $Z$, as discussed in the more general setting of \autoref{sec:base_frame}.

It is of course worth clarifying the implicit claim in \autoref{def:gen_heston_frame} that any such field $Y$ in \autoref{def:gen_hest_field} is indeed found in $\uG$. It is certainly clear from \autoref{def:gen_hest_field} that $Y$ defines a random element of $\uC(\RR^2_+,\RR)$, but using the definition of $\uG$ from \autoref{def:mod_driving_func}, we require~
\begin{equation}
    1.~Y_{0,0}\ge 0\quad 2.~Y_{\cdot,x} \text{ strictly increasing }\quad 3.~\inf_{x\in\RR_+}Y_{t,x}<0\quad 4.~\sup_{t\in\RR_+}Y_{t,x}>0.
\end{equation}
Now 1.~$Y_{0,0} =v >0$ follows from $Z_0=\vt(0)=0$, 2.~$Y_{\cdot,x}$ is strictly increasing for each fixed $x\in\RR_+$ because $\vt$ is strictly increasing, 3.~$\inf_{x\in\RR_+}Y_{t,x}=-\infty<0$ for each $t\in\RR_+$ because of the growth assumption $\sup_{x\in\RR_+}\kappa x - \sigma Z_x=\infty$ and finally 4.~$\sup_{t\in\RR_+}Y_{t,x}=\infty>0$ for each $x\in\RR_+$ because the bijectivity of $\vt$ gives $\sup_{t\in\RR_+}\vt(t)=\infty$. Notice that these checks are just like those performed in \autoref{thm:solutionmap}, because of course the settings here and there are uncoincidentally similar. We have thus shown $Y\in\uG$, and also that the generalised Heston framework from \autoref{def:gen_heston_frame} indeed defines a sub-framework of that from \autoref{def:price_frame}.

More specifically $Y\in\uG$ tells us that the random IVP in \autoref{def:gen_heston_frame} is an example of \autoref{prob:random_ivp}, so the cumulative variance process $X$ has all the properties from \autoref{chap:solutions} consolidated in \autoref{thm:rand_well_posed}. We will shortly return to some of these properties, but now clarify how the classical Heston process is recovered in this generalised Heston framework.

\vspace{3mm}\textbf{The Heston relationship.} We now clarify how the generalised Heston framework of \autoref{def:gen_heston_frame} and the classical Heston model in \autoref{def:heston_stoch_vol} are related. Except for the parameters $\sigma,\kappa,v,\rho$, notice that a specific model in our framework is defined through choices of a path $\vt$ and process $Z$. So the primary task here is to explicitly make such choices which produces a price process with distribution equal to that of the classical Heston model's.

The main tool towards achieving this is the following result originating from \cite{Dambis_1965} and \cite{Dubins_1965}, but stated here like Theorem 5.1.6 in \cite{Revuz_1999}. Similar statements can be found in \cite{Karatzas_1998} and \cite{Ikeda_1989}.

\begin{theorem}[Dambis, Dubins-Schwarz]\label{thm:DDS}
    Let $M$ be a continuous local martingale on $(\Omega,\cF,\{\cF_t\}_{t\in\RR_+},\PP)$ with $M_0=0$ and $[M]_\infty=\infty$, and define the process $T$ by $T_t:=\inf\{s>0:[M]_s>t\}$. Then $B_t:=M_{T_t}$ defines an $\cF_{T_t}$-Brownian motion which verifies $B_{[M]_t}=M_t$.
\end{theorem}

Notice how in this result $t\in\RR_+$ is allowed to index \emph{both} the filtration $\{\cF_t\}_{t\in\RR_+}$ and the process $T$ with which this filtration is composed, in $\cF_{T_t}$. Although mathematically palatable, this can lead to poor intuition for the relationship between the processes $M$ and $T$, and their physical relevance. Except when discussing such existing results, this is why we index our Brownian motion $W$ with the variable $x\in\RR_+$ instead; both to avoid a repetition of indices and to highlight the physical interpretation of this as a \emph{spatial} variable, like in \autoref{def:gen_hest_field}.

Now the continuous local martingales to which we would like to apply \autoref{thm:DDS} are the components $M^i_t:=\int_0^t \sqrt{V_s}\dd W^i_s$ for $i=0,1$ in \autoref{def:heston_stoch_vol}, so $[M^i]_t=\int_0^t V_s\dd [W^i]s=\int_0^t V_s\dd s$. As stated in \autoref{thm:DDS}, this requires the a.s.~limit $[M^i]_\infty=\lim_{t\to\infty}\int_0^t V_s\dd s=\infty$. Although it is straightforward to verify this \emph{once we have expressed} the Heston model in the framework of \autoref{def:price_frame} (using the unboundedness of $X$ in \autoref{thm:wellposed}), such an argument would be circular. This circularity may be avoided by localisation to a compact time horizon, then extending this to infinity. Alternatively, the ergodicity of the CIR process may be used, covered generally in \cite{Papoulis_2002} or specifically in \cite{Jin_2019}. Proof via moment generating functions is also possible, as \autoref{lem:int_cir_unbound} below outlines. Except for minor notational differences, the expressions given in \autoref{eq:int_cir_char_func} agree with those e.g.~obtained in \cite{Dufresne_2001} and \cite{Carr_2003}.

\begin{lemma}[Integrated CIR unboundedness]\label{lem:int_cir_unbound}
    Let the CIR process $V=\{V_t\}_{t\in\RR_+}$ verify the SDE in \autoref{eq:heston_cir}. Then the convergence $\int_0^t V_s\dd s\asc\infty$ takes place as $t\to\infty$.
\end{lemma}
\begin{proof}
    Define the sequence of random variables $\{X_n\}_{n\in\NN}$ by $X_n:=\int_0^n V_s\dd s$. We will first establish $n^{-1}{X_n}\cd \theta$ as $n\to\infty$, then extend this to the claim. Towards this, the moment generating function of each variable $n^{-1}{X_n}$ is given by $\EE[e^{pn^{-1}X_n}] = e^{\vp_0^n + \vp_1^n v}$, wherein
    \begin{equation}\label{eq:int_cir_char_func}
        \vp_0^n:=\frac{\kappa^2\theta n}{\sigma^2}- \frac{2\kappa\theta}{\sigma^2}\log\left(\cosh\left(\frac{\lambda n}{2}\right) + \frac{\kappa}{\lambda}\sinh\left(\frac{\lambda n}{2}\right)\right),\quad \vp_1^n:= \frac{2pn^{-1}}{\kappa + \lambda\coth\left(\frac{\lambda n}{2}\right)},
    \end{equation}
    and $\lambda := \sqrt{\kappa^2-2\sigma^2pn^{-1}}$. We could restrict $p\in\RR$ to ensure $\lambda>0$, but this is ensured naturally as $n\to\infty$. Now it is straightforward to check $\vp_1^n\xrightarrow{n\to\infty}0$, but less straightforward to see $\vp_0^n\xrightarrow{n\to\infty}p\theta$. For this, we first write the following linear expansion in $n$ as $n\to\infty$~
    \begin{equation}\label{eq:expansion}
        \log\left(\cosh\left(\frac{\lambda n}{2}\right) + \frac{\kappa}{\lambda}\sinh\left(\frac{\lambda n}{2}\right)\right) = \frac{\kappa n}{2} - \frac{\sigma^2 p}{2\kappa} + \ve(n),
    \end{equation}
    then by utilising further expansions $\frac{\lambda n}{2} = \frac{\kappa n}{2} - \frac{\sigma^2 p}{2\kappa} + O(n^{-1})$, $1+\frac{\kappa}{\lambda} = 2 + O(n^{-1})$ and $1-\frac{\kappa}{\lambda} = O(n^{-1})$ the requirement of $\ve(n)\xrightarrow{n\to\infty}\log(1)=0$ becomes clear from the representation
    \begin{equation}\label{eq:error_estimate}
        \ve(n)=\log\left(\frac{(1+\frac{\kappa}{\lambda})e^{\frac{\lambda n}{2}} + (1 -\frac{\kappa}{\lambda})e^{-\frac{\lambda n}{2}}}{2e^{\frac{\kappa n}{2} - \frac{\sigma^2 p}{2\kappa}}}\right).
    \end{equation}
    The claim of $\vp_0^n\xrightarrow{n\to\infty}p\theta$ then follows from the cancellation of $\frac{\kappa^2\theta n}{\sigma^2}$ in \autoref{eq:int_cir_char_func}, i.e.~
    \begin{equation}
        \vp_0^n = \frac{\kappa^2\theta n}{\sigma^2}- \frac{2\kappa\theta}{\sigma^2}\left(\frac{\kappa n}{2} - \frac{\sigma^2 p}{2\kappa}+ \ve(n)\right) \xrightarrow{n\to\infty} p\theta.
    \end{equation}
    So we find $\EE[e^{pn^{-1}X_n}]\xrightarrow{n\to\infty}e^{p\theta}$. With $e^{p\theta}$ being the moment generating function of the constant $\theta$, we get $n^{-1}{X_n}\cd \theta$ as $n\to\infty$ by L\'evy's continuity theorem. This provides $n^{-1}{X_n}\cp \theta$, and also $n_k^{-1}{X_{n_k}}\asc \theta$ as $k\to\infty$ for a subsequence $\{n_k\}_{k\in\NN}$. Given that the sequence $\{X_{n_k}\}_{k\in\NN}$ is non-decreasing, this provides $X_{n_k}\asc \infty$, because $\lim_{k\to\infty}X_{n_k}< \infty$ yields the contradiction $n_k^{-1}{X_{n_k}}\asc 0<\theta$. So we have shown $\int_0^{n_k} V_s\dd s\asc \infty$ as $k\to\infty$, and this extends to $\int_0^t V_s\dd s\asc \infty$ as $t\to\infty$ given that $\int_0^tV_s\dd s$ is also non-decreasing.
\end{proof}

Numerical tests support an intuitive estimate $\ve(n)=O(n^{-1})$ in \autoref{eq:expansion}, but utilising the exact expression in \autoref{eq:error_estimate} clearly suffices to establish the priority $\ve(n)\xrightarrow{n\to\infty}0$. At this point it is worth considering the proof of \autoref{lem:int_cir_unbound} and especially its relative complexity compared with the counterpart in our framework. Even if not applying \autoref{thm:wellposed} directly, this counterpart goes as follows: the classical Heston random field $Y_{t,x}:=\sigma W^1_x+\kappa(\theta t - x)+v$ satisfies $\inf_{x\in\RR_+}Y_{t,x}<0$ and $\sup_{t\in\RR_+}Y_{t,x}>0$. Applying \autoref{cor:exist_summary} on a pathwise basis, the random IVP solution $X$ has bijective paths in $\uC_0^1(\RR_+,\RR_+)$.

Contrary to \autoref{lem:int_cir_unbound}, it is straightforward to show the process $X_t=\int_0^t V_s\dd s$ is \emph{strictly} increasing, which means that $T$ in \autoref{thm:DDS} coincides with the inverse $X^{-1}$, justifying its use in \autoref{lem:heston_timechange}. Following the discussion after \autoref{def:solutionderivset}, $X$ is strictly increasing if $V$ cannot be zero over intervals. But assuming such an interval $(a,b)$ leads to a violation of the SDE in \autoref{eq:heston_cir}, because this then just reads $0=\kappa\theta(t-a)>0$ for any $t\in(a,b)$.

Following this next result, which uses \autoref{lem:int_cir_unbound} to apply \autoref{thm:DDS} to the classical Heston model, we will be ready to recover this model from within the generalised Heston framework in \autoref{def:gen_heston_frame}. We have not yet defined time-changes properly, covered in \autoref{sec:martingale}, so the description in \autoref{lem:heston_timechange} can be considered non-mathematical for now.

\begin{lemma}[Classical Heston time-change]\label{lem:heston_timechange}
    Let $W$ and $V$ be as in the classical Heston model from  \autoref{def:heston_stoch_vol}, and define also $X_t:=\int_0^t V_s\dd s$. Then $B=\{B_x\}_{x\in\RR_+}$ defined by
    \begin{equation}
        B_x := \int_0^{X^{-1}_x}\sqrt{V_s}\dd W_s
    \end{equation}
    is another 2d Brownian motion on $(\Omega, \cF,\PP)$, and this verifies $B_{X_t} = \int_0^t\sqrt{V_s}\dd W_s$ over $\RR_+$.
\end{lemma}
\begin{proof}
    Let $\{\cF^i_t\}_{t\in\RR_+}$ be the natural filtration of each component $W^i$ for $i=0,1$ and define the local martingales $M^i_t:=\int_0^t \sqrt{V_s}\dd W^i_s$ on $(\Omega,\cF,\{\cF_t\}_{t\in\RR_+},\PP)$. These clearly verify $M^i_0=0$, and $[M^i]_t=\int_0^t V_s\dd s =: X_t$. From \autoref{lem:int_cir_unbound}, we also have $\lim_{t\to\infty}\int_0^tV_s\dd s=[M^i]_\infty=\infty$, so \autoref{thm:DDS} can be applied as stated for each of $i=0,1$. This provides that $B_x:=M_{X^{-1}_x}$ defines a Brownian motion which verifies $B_{X_t}=M_t$, and this is precisely the claim here.
\end{proof}

This next result brings precise meaning to the manipulations at the beginning of \autoref{ch:intro}.

\begin{theorem}[Classical Heston recovery]\label{thm:heston_ode}
    Let the price process $S=\{S_t\}_{t\in\RR_+}$ derive from the generalised Heston framework of \autoref{def:gen_heston_frame}, in the specific case where we choose
    \begin{equation}\label{eq:heston_field}
        \vt(t) := \theta t\quad\text{and}\quad Z:=W^1.
    \end{equation}
    Then the distribution of $S$ coincides with that of the classical Heston process in \autoref{def:heston_stoch_vol}, with the same parameters, $\sigma,\kappa,\theta,v>0$ and $\rho\in[-1,1]$. In fact, if this generalised Heston process is constructed \emph{not} from Brownian motion $W$, but instead $B$ from \autoref{lem:heston_timechange} (additionally using $Z:=B^1$), then it is \emph{indistinguishable} from the classical Heston process.
\end{theorem}
\begin{proof}
    Let $V$ and $S$ be as in the classical Heston model, so $V$ verifies the integral equation~
    \begin{equation}\label{eq:cir_intergal_eqn}
        V_t = \sigma\int_0^t\sqrt{V_s}\dd W^1_s + \kappa\int_0^t(\theta - V_s)\dd s + v.
    \end{equation}
    Prioritising the Brownian motion $B$ from \autoref{lem:heston_timechange} and defining $B^\rho := \sqrt{1-\rho^2} B^0 + \rho B^1$ like $W^\rho$, we can equivalently write \autoref{eq:cir_intergal_eqn} and the price process in \autoref{eq:heston_stock} as 
    \begin{equation}
        V_t = \sigma B^1_{\int_0^t V_s\dd s} + \kappa \left(\theta t - \int_0^t V_s\dd s\right) + v\quad\text{and}\quad  S_t = \exp\left(B^\rho_{\int_0^t V_s\dd s} - \frac12 \int_0^t V_s\dd s\right).
    \end{equation}
    Now prioritising the process $X_t:=\int_0^t V_s\dd s$, this reduces to a specific case of \autoref{eq:model_summary}:
    \begin{equation}\label{eq:time_change_rep}
        X'_t = \sigma B^1_{X_t} + \kappa \left(\theta t - X_t\right) + v,\quad  S_t := \exp\left(B^\rho_{X_t} - \tfrac12 X_t\right).
    \end{equation}
    So $S$ is nothing else than the specific model within the framework of \autoref{def:gen_heston_frame}, constructed from $B$ rather than $W$ and with $\vt(t) := \theta t$ and $Z:=B^1$. So we have first arrived at the \emph{indistinguishability} claim. The \emph{distributional} claim follows by replacing $B$ with $W$ in \autoref{eq:time_change_rep}. In the terminology of SDEs, every model in the framework of \autoref{def:gen_heston_frame} has a unique strong solution, so the distribution of $S$ is invariant to such replacements.
    
    For completeness we must verify that the classical Heston choices $\vt(t):=\theta t$ and $Z:=W^1$ verify the requirements in \autoref{def:gen_heston_frame}, namely that $\vt$ is a bijective path in $\uC_0(\RR_+,\RR_+)$ and that $Z$ is in $\uC_0(\RR_+,\RR)$ and verifies $\sup_{x\in\RR_+}\kappa x-\sigma Z_x=\infty$. This final growth condition, i.e.~$\sup_{x\in\RR_+}\kappa x-\sigma W^1_x=\infty$, is the only non-trivial requirement, but this follows e.g.~from the fact that Brownian motion is a.s.~recurrent at zero, as covered in \cite{Sato_1999}, meaning that for every $N>0$ there a.s.~exists $x>N\kappa^{-1}$ where $W^0_x=0$, and therefore $\kappa x-\sigma W^0_x>N$.
\end{proof}

It is clear from the above proof that this result does not only recover the Heston price process $S$, but also its cumulative variance $X_t:=\int_0^t V_s\dd s$, and in fact the processes $(X,S)$ jointly.

\vspace{3mm}\textbf{The solution map.} When working in the generalised Heston sub-framework from \autoref{def:gen_heston_frame}, the specification of a random field $Y$ is reduced to that of parameters $\sigma,\kappa,v,\rho$, a path $\vt$ and volatility-driving process $Z$. It is worth covering consequences of this on solution map results like \autoref{thm:solutionmap_rand} which are applicable in the wider framework of \autoref{def:price_frame}.

\begin{corollary}\label{cor:gen_hest_solution}
    Fix parameters $\sigma,\kappa,v$ and path $\vt$ in the generalised Heston framework of \autoref{def:gen_heston_frame}. Let $\Phi_{v,\vt}\subset\Phi$ contain paths $\vp$ with $\vp'(0)=v$ and $\sup_{t\in\RR_+}\vt(t)-\vp'(t)=\infty$. Then the map taking each process $Z$ to the random IVP solution $X\in\Phi_{v,\theta}$ is bijective and continuous w.r.t.~uniform convergence over compacts. Specifically, $X$ is generated when
    \begin{equation}\label{eq:noise_choice}
        Z_x := \sigma^{-1}\left( X'_{X^{-1}_x}-\kappa(\vt(X^{-1}_x) - x) - v \right).
    \end{equation}
\end{corollary}

This result follows from \autoref{thm:solutionmap_rand}, except for the continuity statement which follows from point 3.~in \autoref{thm:rand_well_posed}. Note that nothing changes if we allow the parameter $v$ to be a random variable in $\RR_+$, and this widens the solution set from processes in $\Phi_{v,\vt}$ to those in $\Phi_{\vt}$ from \autoref{thm:solutionmap_rand}. In addition to $Z$ given by \autoref{eq:noise_choice}, we then also require the random selection $v:=X'_0$ to generate a chosen process $X\in\Phi_\vt$ as the random IVP solution. 

As covered in the discussion following \autoref{thm:solutionmap_rand}, recall that the set of processes satisfying the condition $\sup_{t\in\RR_+}\vt(t)-X'(t)=\infty$ in \autoref{cor:gen_hest_solution} is wider than those in $\Phi$ which verify the more natural condition $\liminf_{t\to\infty}X'_t<\infty$. So \autoref{cor:gen_hest_solution} tells us that even in the generalised Heston sub-framework of \autoref{def:gen_heston_frame} we can still, through the selection of $Z$, theoretically model any price process $S$ accepting the representation $S_t=\exp(W^\rho_{X_t}-\frac12 X_t)$, where $X$ is \emph{any} bijective process in $\uC^1_0(\RR_+,\RR_+)$ with $X'_0=v$ and $\liminf_{t\to\infty}X'_t<\infty$.

This is precisely why we advocated the use of additively separable fields of type $Y_{t,x}=\vt(t)-Z_x$ following \autoref{thm:solutionmap_rand}, and indeed why we introduced \autoref{ex:main_example}. If we take the generalised Heston random field from \autoref{def:gen_hest_field}, then it is easy to see the connection~
\begin{equation}\label{eq:add_sep}
    Y_{t,x} = \tilde\vt(t) - \tilde Z_x,\quad\text{where}\quad \tilde\vt(t):=\kappa\vt(t),\quad \tilde Z_x:=\kappa x - \sigma Z_x - v.
\end{equation}
So the generalised Heston framework is actually just a framework of additively separable fields presented in a recognisable manner to those familiar with the classical Heston model, with the precise connection given by \autoref{thm:heston_ode}. The less recognisable representation in \autoref{eq:add_sep} can be helpful for mathematical manipulations, as shown in \autoref{thm:mgf_existence}.

\vspace{3mm}\textbf{General MGF existence.} Except for results like \autoref{thm:heston_ode} which relate to existing and intrinsically probabilistic theory, everything in this thesis thus far can be reduced to the probability-free results of \autoref{chap:solutions}. Contrasting this, the main contribution of this section regards the existence of (intrinsically probabilistic) MGFs $M_X(p,t):=\EE[e^{pX_t}]$. Here, $X$ is a random IVP solution restricted to the generalised Heston framework from \autoref{def:gen_heston_frame}, but following the discussion after \autoref{cor:gen_hest_solution}, this is not much of a restriction at all.

Most tangibly, this MGF existence will help to establish the martingality of price processes $S=\exp(W^\rho_X-\frac12 X)$ in \autoref{sec:martingale}, which can be intuited given the expectation of the component $\exp(\frac12 X_t)$ here coincides with $M_X(\frac12,t)$. But more generally, use of the process $\bX$ in \autoref{thm:mgf_existence} and \autoref{cor:gauss_heston_mgf} demonstrates the power of \emph{always} having this process $\bX$ from \autoref{thm:rand_well_posed}, which dominates $X$. This is especially helpful because $\bX_t:=\inf\{x>0:Y_{t,x}<0\}$ derives \emph{directly} from the random IVP's underlying random field, enabling us to draw probabilistic conclusions on the random IVP solution $X$ \emph{without analysing random IVPs}. This makes our framework more accessible to probabilists less familiar with ODEs.

Before the next result it is worth clarifying that by succeeding in establishing the existence of $\EE[e^{pX_t}]<\EE[e^{p\bX_t}]<\infty$ for some $p>0$, we immediately obtain $\bX_t<\infty$. As shown after \autoref{def:gen_heston_frame}, the point of the growth assumption $\sup_{x\in\RR_+}\kappa x - \sigma Z_x=\infty$ is to ensure the property $\inf_{x\in\RR_+}Y_{t,x}<0$ of fields in $\uG$, which is equivalent to $\bX_t<\infty$. So if we obtain $\EE[e^{p\bX_t}]<\infty$ for $t\in\RR_+$ and $p>0$, we do not have to check $\sup_{x\in\RR_+}\kappa x - \sigma Z_x=\infty$ as well.

\begin{theorem}[General MGF existence]\label{thm:mgf_existence}
    Let the random field and IVP solution $Y$ and $X$ be as in the generalised Heston framework from \autoref{def:gen_heston_frame}, so that we can write $Y_{t,x}=\tilde{\vt}(t)-\tilde{Z}_x$ where $\tilde{\vt}:=\kappa\vt$ and $\tilde{Z}_x:=\kappa x - \sigma Z_x - v$. Fix $p,T>0$, then provided that the left tails of $\tilde{Z}$ (thus right tails of $Z$) are thin enough to be dominated in the sense of
    \begin{equation}\label{eq:integrable_assumption}
        \exists\ a,b,c>0\quad s.t.\quad \log\PP[\tilde{Z}_x < \tilde{\vt}(T)] < a - (p+b) x\quad \forall\ x\in[c,\infty),
    \end{equation}
    then the MGF $M_{\bX}(p,t):=\EE[e^{p\bXX_t}]$ exists for $t\in[0,T]$. Likewise for $M_{X}(p,t):=\EE[e^{pX_t}]$.
\end{theorem}
\begin{proof}
    \autoref{thm:rand_well_posed} establishes that $\bX$ dominates $X$ in the sense of $\bX_t\ge|X_t|=X_t$, so the conclusion regarding $X$ follows immediately from that regarding $\bX$. If $Y$ derives from \autoref{def:gen_heston_frame}, then $\bX$ is given elegantly by the exit-time of $\tZ$ from $(-\infty,\vt(t)]$. Specifically,~
    \begin{equation}\label{eq:bound_rep}
        \bX_t := \inf\{x>0:Y_{t,x}<0\} = \inf\{x>0:\tZ_x > \tvt(t)\} =: E_{\tvt(t)}(\tZ).
    \end{equation}
    Given that $\bXX$ has non-negative strictly increasing paths, the conclusion holds for $t\in[0,T]$ provided that it holds for the final time $T$. So let $\mu:=\PP \bX_T^{-1}$ denote the distribution of $\bX_T$, satisfying $\mu(\bRR_+)=1$, where as usual $\bRR_+=\RR_+\cup\{\infty\}$. At this stage, the singleton $\{\infty\}$ being an atom of $\mu$, i.e.~$\mu(\{\infty\})>0$, should not be ruled out. We are required to establish
    \begin{equation}\label{eq:integrable}
        \EE[e^{p\bX_T}] := \int_{\bRR_+}e^{px}\mu(\dd x) <\infty,
    \end{equation}
    and once this is achieved then clearly we will have $\mu(\{\infty\})=0$, so that $\RR_+$ supports $\mu$. Using the expansion $e^{px} = 1 + p\int_0^x e^{pu}\dd u$ in \autoref{eq:integrable} then Tonelli's theorem, we get
    \begin{multline}\label{eq:integrable2}
        \int_{\bRR_+}e^{px}\mu(\dd x) = \underbrace{\mu(\bRR_+)}_{=1} + p\int_{\bRR_+}\int_{[0,x]} e^{pu}\dd u\mu(\dd x)\\ = 1 + p\int_{\bRR_+} e^{pu}\int_{[u,\infty]} \mu(\dd x)\dd u = 1 + p\int_{\bRR_+} e^{pu}\mu([u,\infty])\dd u,
    \end{multline}
    where at this stage these expressions could feasibly read `$\infty=\infty$'. Now for $c\ge0$, define the integrals $I_c:=\int_{[c,\infty]} e^{pu}\mu([u,\infty])\dd u$. Then \autoref{eq:integrable2} shows \autoref{eq:integrable} will be verified if $I_0<\infty$. But actually \autoref{eq:integrable} will be verified if $I_c <\infty$ for any $c$, because
    \begin{equation}
        I_0 - I_c = \int_0^c e^{pu}\mu([u,\infty])\dd u \le e^{pc}\int_0^c \mu([u,\infty])\dd u \le ce^{pc} < \infty.
    \end{equation}
    To establish $I_c<\infty$ and complete the proof, define $M_x(\tZ):=\max_{u\in[0,x]}\tZ_u$ and note that
    \begin{equation}\label{eq:integrable3}
        \mu([x,\infty]) = \PP[ E_{\tilde\vt(T)}(\tZ) \ge x] = \PP[M_x(\tZ) \le \tilde\vt(T)] \le \PP[\tZ_x \le \tilde\vt(T)].
    \end{equation}
    The central equality here follows from the general equivalence $\inf\{u\!>\!0:f(u)\!>\!t\}\ge x\iff\sup_{u\in[0,x]}f(u)\le t$ for continuous $f$ with $f(0)\le0$, as e.g.~utilised in \cite{Meerschaert_2004}, and the final inequality follows just from $M_x(Z)\ge Z_x$. Now \autoref{eq:integrable3} relates $\mu([x,\infty])$ appearing in $I_c$ with our assumption on $\PP[\tZ_x \le \tilde\vt(T)]$ in \autoref{eq:integrable_assumption}, providing $\mu([x,\infty]) \le e^{a-(p+b)x}$ for $x\ge c$. Substituting this into $I_c$, we find $I_c$ thus $\EE[e^{p\bX_T}]$ exists if
    \begin{equation}\label{eq:integral_existence}
        \int_c^\infty e^{a - bx}\dd x = b^{-1}e^{a-bc} < \infty.
    \end{equation}
    Since this is clearly the case for positive constants $a,b$ and $c$, then we have demonstrated the MGF existence $\EE[e^{p X_t}] \le \EE[e^{p\bX_t}]<\infty$ for all $t\in[0,T]$, and the proof is thus complete.
\end{proof}

The integral obtained in \autoref{eq:integral_existence} being so clearly finite demonstrates that our assumption on the growth of $Z$ in \autoref{eq:integrable_assumption} is by no means optimal. Indeed, the priority is to provide a preparatory result for \autoref{cor:gauss_heston_mgf}, which regards a class of Gaussian processes $Z$ that are already known to be helpful in volatility modelling. This class is then reduced to a specific example in the RLH model defined in \autoref{sec:RLH_model}. Should one need to improve on \autoref{thm:mgf_existence} then the following equivalence from our proof provides a good starting point
\begin{equation}
    \EE[e^{p\bX_t}] = 1 + p \int_{\bRR_+}e^{px} \PP[M_x(\tZ) \le \tilde\vt(t)] \dd x.
\end{equation}

Since \autoref{thm:mgf_existence} depends only on the process $\bX$, its proof actually applies \emph{to all} random fields $Y\in\uG$ which generate the same process $\bX$ in \autoref{eq:bound_rep}, even though these do not generate the same random IVP solution $X$. E.g.~let $\lambda\in\uC_0(\RR,\RR)$ be a strictly increasing and bijective process, then \autoref{thm:mgf_existence} applies to all fields $Y^\lambda_{t,x}:=\lambda(\tilde\vt(t)-\tZ_x)$, because
\begin{equation}
    \bX^\lambda_t := \inf\{x>0:Y^\lambda_{t,x}<0\} = \inf\{x>0:Y_{t,x}<0\} =: \bX_t.
\end{equation}

\vspace{3mm}\textbf{Gaussian MGF existence.} Models in the generalised Heston framework of \autoref{def:gen_heston_frame} are identified by the equations that the price $S$ and cumulative variance $X$ uniquely verify,
\begin{equation}\label{eq:model_summary3}
    X'_t = \sigma Z_{X_t} + \kappa(\vt(t) - X_t) + v,\quad S_t=\exp(W^\rho_{X_t}-\tfrac12 X_t).
\end{equation}
We now show how \autoref{thm:mgf_existence} can be applied to give the existence of $M_X(p,t):=\EE[e^{p X_t}]$ assuming that the volatility-driving process $Z$ is Gaussian, with a variance growth constraint.

What makes \autoref{cor:gauss_heston_mgf} particularly surprising is that, besides this constrain on $Z$, no \emph{additional} restrictions are placed on the parameters $\sigma, \kappa,\vt,\rho$, yet the conclusion holds for all $(p,t)\in\RR\times\RR_+$. This is essentially achieved by assuming the variance of $Z_x$ is dominated by that of Brownian motion as $x\to\infty$. Crucially, the variance of $Z_x$ is still free to grow at an arbitrary rate over a fixed compact, and this provides the \emph{global} freedom required to reconcile observations. This is validated in \autoref{sec:vol_surfaces}, but if the point is not clear, note that \emph{all} historic, and most future, volatility observations can be reproduced even by a \emph{bounded} volatility process. This global freedom supplements our existing \emph{local} freedom, given $Z$ has no local constraints other than its continuity, e.g.~need not be H\"older regular of any order.

Of course, this next result will apply to the specific RLH model defined shortly in \autoref{sec:RLH_model}. Taking guidance from recent rough volatility modelling developments, this model supposes $Z$ is a (H\"older continuous) fractional Gaussian process verifying $\EE[Z^2_x]=x^{\gamma}$ for some $\gamma\in(0,1)$.

\begin{theorem}[Gaussian MGF existence]\label{cor:gauss_heston_mgf}
    Let $X=\{X_t\}_{t\in\RR_+}$ be as in the generalised Heston framework of \autoref{def:gen_heston_frame}. Provided the process $Z=\{Z_x\}_{x\in\RR_+}$ is centred Gaussian and verifies $\EE[Z_x^2] < \alpha + \beta x^{\gamma}$ for some $\alpha,\beta \ge 0$, $\gamma\in(0,1)$, then $M_X(p,t):=\EE[e^{p X_t}]$ exists globally, i.e.~for all $(p,t)\in\RR\times\RR_+$, regardless of how the parameters $\sigma,\kappa,\vt,v,\rho$ are chosen. 
\end{theorem}
\begin{proof}
    Given $X_t$ is non-negative, $M_X(p,t)$ is clearly in $[0,1]$ when $p\le0$, so we can now assume $p>0$. In order to apply \autoref{thm:heston_ode} for the \emph{global} result here, the condition in \autoref{eq:integrable_assumption} must hold for any $p,T>0$. So fixing any $p,T>0$, we seek $a,b,c>0$ with
    \begin{equation}\label{eq:log_bound0}
        \log\PP[\tZ_x < \tvt(T)] := \log\PP[\kappa x - \sigma Z_x - v < \kappa\vt(T)] < a-(p+b) x
    \end{equation}
    for $x\in[c,\infty)$. Now define the constant $\vt_T:=\vt(T)+\kappa^{-1}v>0$, so \autoref{eq:log_bound0} becomes
    \begin{equation}\label{eq:log_bound}
        \log\PP[\sigma Z_x > \kappa(x-\vt_T)] < a-(p+b) x.
    \end{equation}
    It helps to seek only $c>\vt_T$, which means also $x>\vt_T$. Fixing $x$, then having $\kappa(x-\vt_T)>0$ in \autoref{eq:log_bound} makes this a condition directly on the \emph{positive} tail of $Z_x$. Given $Z_x$ is a centred Gaussian random variable with variance less than $\alpha + \beta x^{\gamma}$, \autoref{eq:log_bound} holds if
    \begin{equation}\label{eq:log_bound1}
        \log\PP\left[\phi > \frac{\kappa(x - \vt_T)}{\sigma\sqrt{\alpha + \beta x^{\gamma}}}\right] < a-(p+b) x,
    \end{equation}
    where $\phi$ is a standard Gaussian number. By invoking the popular Gaussian bound $\PP[\phi>u]<e^{-\frac12 u^2}$ for $u\ge0$, we obtain this requirement in \autoref{eq:log_bound1} if for some such $a,b>0$
    \begin{equation}\label{eq:ineq_require}
        -\frac12\frac{\kappa^2(x - \vt_T)^2}{\sigma^2(\alpha + \beta x^{\gamma})} < a - (p+b) x
    \end{equation}
    for all $x$ greater than some $c\ge\vt_T$. Taking expansions in \autoref{eq:ineq_require} as $x\to\infty$, we see
    \begin{equation}\label{eq:ineq_require1}
        O(x^{2-\gamma}) = \frac12\frac{\kappa^2(x - \vt_T)^2}{\sigma^2(\alpha + \beta x^{\gamma})}\quad \text{and}\quad (p+b) x - a = O(x).
    \end{equation}
    Given that $2-\gamma>1$ follows from the assumption $\gamma\in(0,1)$, the existence of such $a,b,c$ finally becomes plausible, \emph{regardless} of $\kappa,\sigma,\vt,\alpha,\beta,T$ or $p$. Indeed, in the Gaussian setting here we can actually first fix any $a,b>0$, then basic manipulations of \autoref{eq:ineq_require} demonstrate that this is satisfied for all $x\in[c,\infty)$, as required, provided we select $c>1\vee\vt_T\vee d$, where 
    \begin{equation}
        d := \left( 2\vt_T + 2\frac{\sigma^2}{\kappa^2}(p+b)(\alpha + \beta) \right)^{\frac{1}{1-\gamma}}<\infty.
    \end{equation}
    Having found such values $a,b,c>0$, \autoref{thm:heston_ode} provides the existence of both $M_\bX(p,t)$ and $M_X(p,t)$ for $(p,t)\in\RR\times[0,T]$. This extends to all  $(p,t)\in\RR\times\RR_+$ given $T$ is arbitrary.
\end{proof}
% For completeness, the `basic manipulations' referred to in the above proof are as follows
% \begin{align}
%     & &\frac12\frac{\kappa^2(x - \vt_T)^2}{\sigma^2(\alpha + \beta x^{\gamma})} > (p+b) x - a\\
%     &\impliedby & \frac12\frac{\kappa^2}{\sigma^2}(x - \vt_T)^2 > (\alpha + \beta x^{\gamma})((p+b) x - a)\\
%     &\impliedby & \frac12\frac{\kappa^2}{\sigma^2}(x - 2\vt_T) > (\alpha + \beta x^{\gamma})(p+b)\\
%     &\impliedby & \frac12\frac{\kappa^2}{\sigma^2}(x^2 - 2\vt_T) > (\alpha + \beta x^{\gamma})(p+b)
% \end{align}
% Now assuming $x>1$, multiply $\alpha$ and $\vt_T$ by $x^\gamma$ and rearrange for the result.

Note that the Gaussian bound used to obtain \autoref{eq:ineq_require} may be found in \cite{Feller_1968}, along with the tighter one $\PP[\phi>x]<\frac{1}{x\sqrt{2\pi}}e^{-\frac12 x^2}$ as $x\to\infty$. This tighter bound may help to deal with the boundary case $\gamma=1$ in \autoref{cor:gauss_heston_mgf} if ever required, although does not lend itself to straightforward manipulations after composition with $\log$ in \autoref{eq:log_bound1}.

% {\color{red}See notebook from 2020-09-22 for computational check on inequalities in \autoref{eq:ineq_require}.}

That concludes our theory for these generalised Heston models from \autoref{def:gen_heston_frame}, which define a sub-framework of the general one from \autoref{def:price_frame}. In \autoref{sec:RLH_model}, this theory will be applied to the specific RLH model in this sub-framework. But first, we look at the martingale sub-framework also shown in \autoref{fig:venn}, and in which the RLH model also resides.

\subsection{A martingale sub-framework}\label{sec:martingale}

In this section a sub-framework of that in \autoref{def:price_frame} is defined which accommodates \emph{only} price processes $S=\{S_t\}_{t\in\RR_+}$ which are martingales w.r.t.~some filtration $\{\cG_t\}_{t\in\RR_+}$ of $(\Omega,\cF,\PP)$. Shown in \autoref{fig:venn}, this martingale framework can be characterised by fields $Y\in\uG$ which exhibits two additional properties. These properties respectively ensure $S$ verifies the adaptedness and integrability conditions which, like in \autoref{def:martingale}, any martingale must. 

Like the MGFs just covered, there is an atypical value to these properties of a field $Y$, which is that they can be checked \emph{immediately} following its specification, i.e.~do not require probabilistic analysis of the \emph{solution} $X$ of the associated random IVP $x'=Y_{t,x}$, $x_0=0$ driven by $Y$. So although martingales are inseparable from probability and cannot be established on a pathwise basis, we are still able to maintain our probabilistically uncomplicated approach.

The general importance of martingales in finance, and so the value of this martingale framework, is related to the practice of arbitrage-free derivative pricing, which is explained now. As the goal here is to present a succinct exposition of practical value rather than a technical mathematical one, we draw primarily upon the concise reasoning in \cite{Cont_2003}.

\vspace{3mm}\textbf{Derivative pricing means measures.} Let the time $t=0$ denote the present, and consider the possible future paths of a real-world stock price $S_t>0$ (e.g.~any published price) as a continuous stochastic process $S=\{S_t\}_{t\in\RR_+}$ on a probability space $(\Omega,\cF,\PP)$. Let a filtration $\{\cG_t\}_{t\in\RR_+}$ contain information relating to $S$ over each interval $[0,t]$, like $\cF$ does over $\RR_+$, and assume any available price history $\{S_t\}_{t\in[-T,0)}$ is fixed and in the present information $\cG_0$.

For our purposes, a financial derivative on $S$ is a contract between two parties to exchange a cash amount (payoff) at a finite future time $T>0$ (maturity), which depends on the behaviour of $S$ over $[0,T]$. So for now let a derivative be a bounded map from $\{S_t\}_{t\in[0,T]}$ to a payoff $\#\in\RR$ (measurable with respect to Borel $\sigma$-algebras). E.g.~consider $\#:=\mathbbm{1}_{S_T\ge S_0}$ or $\max\{K - \int_0^T S_t\dd t, 0\}$ for $K>0$ (strike). \autoref{sec:vol_surfaces} will focus on the case $\max\{K - S_T, 0\}$.

Starting with a fixed sum of cash at time $0$, assume that all market participants' future investment activity amounts to being able to buy or sell any finite amount of this stock $S$, or enter into such derivative contracts with other parties for agreed prices, both at any time. Additionally assume that any cash left over after such activities remains constant over time. The relevant question is then: how should a party go about assigning prices to derivatives?

Considering this question only at time 0 (of course the argument generalises), it is answered by another map $\Pi$ (pricing rule) from derivative payoffs $\#$ to prices $\Pi(\#)\in\RR$. It is \emph{convenient} (not necessary) to specify $\Pi$ via expectations of payoffs under a probability measure $\QQ$ on $(\Omega,\cF)$, with $\QQ$ being recovered from $\Pi$ via indicator payoffs $\mathbbm{1}_{A}$ for $A\in\cF$,
\begin{equation}
    \Pi(\#) := \EE^\QQ[\#]\quad\implies\quad \QQ[A] = \Pi(\mathbbm{1}_A).
\end{equation}
Note that our boundedness assumption on $\#$ ensures the existence of $\EE^\QQ[\#]$, but this can be ensured (if desirable) via the selection of $\QQ$ otherwise. Now there are two important points to be stressed. Firstly, this \emph{convenience} of specifying pricing rules $\Pi$ via measures $\QQ$ is not merely such. Under very natural constraints on the map $\Pi$, like positivity and linearity:
\begin{equation}\label{eq:pos_and_lin}
    \#\ge 0\implies \Pi(\#)\ge0\quad\text{and}\quad \Pi(\sum_{i=1}^n \#_i) = \sum_{i=1}^n\Pi(\#_i),
\end{equation}
the specification of $\Pi$ or $\QQ$ are mathematically \emph{equivalent}, provided we utilise the relationship $\Pi(\#)=\EE^\QQ[\#]$. This should not be a complete surprise, given that probability measures exhibit properties very similar to those in \autoref{eq:pos_and_lin}, but for subsets of a $\sigma$-algebra.

Secondly, this equivalence between specifying pricing rules and measures should not be interpreted as more than a mathematical fact. E.g.~there is, at this stage at least, no direct relationship between the real-world measure $\PP$, and any of the possible pricing measures $\QQ$.

\vspace{3mm}\textbf{Arbitrage-free means martingales.} Recall that the map $\Pi(\#)=\EE^{\QQ}[\#]$ only assigns derivative prices at time 0, and note that this can be equivalently written $\Pi_0(\#)=\EE^{\QQ}[\#|\cG_0]$ if $\QQ$ agrees with $\PP$ on the information $\cG_0$ begin fixed. Now this pricing relationship between $\Pi$ and $\QQ$ is extended \emph{consistently} over times $t\in[0,T]$ when utilising $\Pi_t(\#)=\EE^\QQ[\#|\cG_t]$, and each price $\Pi_t(\#)$ then, like $S_t$, defines a real-world stochastic process up to its maturity.

Now we want to additionally ensure that prices $\Pi_t(\#)=\EE^\QQ[\#|\cG_t]$, assigned by selecting a pricing measure $\QQ$, do not accommodate the apparent generation of risk-free wealth under the real-world measure $\PP$. Prices set in accordance with this principle are called arbitrage-free. We omit a strict mathematical definition of arbitrage in favour of a sufficient example.

Working from any time $t\in[0,T]$, consider a derivative with payoff $\#:=S_T$ at maturity $T$. To ensure the derivative price $\Pi_t(\#)=\EE^\QQ[\#|\cG_t]$ actually exists, we must relax the earlier boundedness assumption on $\#$ to an integrability condition on measures: $\EE^\QQ[S_T|\cG_t]<\infty$. 

At time $T$, this derivative's price $\EE^{\QQ}[S_T|\cG_T]=S_T$ coincides with the stock's, regardless of the measure $\QQ$ selected. So if we can sell this derivative at time $t$ using a measure which verifies $\EE^\QQ[S_T|\cG_t]>S_t$, a profit of $\EE^\QQ[S_T|\cG_t] - S_t>0$ is ensured by simultaneously buying the stock at price $S_t$. This simple strategy demonstrates arbitrage, and can only be prohibited, for all parties and times, if $\QQ$ is selected such that $\EE^\QQ[S_T|\cG_t]=S_t$ for all $t,T\in\RR_+$ with $t\le T$.

Any measure $\QQ$ verifying this property $\EE^\QQ[S_T|\cG_t]=S_t$ can be called risk-neutral, because it suggests there is no expected benefit or cost associated with the risk of buying the stock $S_t$. But this property is more importantly the main feature of martingales, and more generally, we call $\QQ$ a martingale measure if $S$ defines a martingale on $(\Omega,\cF,\{\cG_t\}_{t\in\RR_+},\QQ)$. The rigorous definition of a martingale is deferred until \autoref{def:martingale} to maintain the practical focus.

Recalling that arbitrage is a real-world notion, i.e. relating to the measure $\PP$, it surprisingly turns out that \emph{all} arbitrage, not just the simple example above, is prohibited if derivative prices are set using a martingale measure $\QQ$ and map $\Pi_t[\#]=\EE^\QQ[\#|\cG_t]$, \emph{provided} that $\QQ$ is additionally equivalent to $\PP$, meaning that for any event $A\in\cF$, $\PP[A]=0\implies \QQ[A]=0$. This equivalence generalises our earlier assumption that $\QQ$ agrees with $\PP$ on $\cG_0$ begin fixed.

The complete relationship between arbitrage and martingales runs deeper than this, and is an astounding achievement of mathematical finance, often referred to as the fundamental theorem of asset pricing. This result additionally establishes that, should we wish to prohibit arbitrage, we actually have no choice but to do so (explicitly or implicitly) via such an equivalent martingale measure $\QQ$. \cite{Cont_2003} can be consulted for more details.

\vspace{3mm}\textbf{Pricing in practice.} In the above reasoning, we have deliberately played down the role of the real-world measure $\PP$ in derivative pricing, as compared with conventional expositions. Recall that we assumed $S$ is a stochastic process over \emph{continuous} time $t\in\RR_+$, even though it defines a model for a \emph{discretely} published price. Consequentially, the empirical verification of real-world properties like $\PP[A]=0$, as opposed to $\PP[A]=10^{-9}$, are theoretically impossible. We continue in accordance with this remark from Emile Borel, also in \cite{Cont_2003}.

\vspace{-6mm}
\begin{adjustwidth}{18pt}{18pt}
\begin{remark}\label{rem:borel}
    \emph{It might be possible to prove certain theorems [about probability], but they might not be of any interest since, in practice, it would be impossible to verify whether the assumptions are fulfilled.}
\end{remark}
\end{adjustwidth}

\vspace{2mm}

In practice, derivative pricers often focus on developing and utilising successful martingale models directly, i.e.~characterising martingale measures, while neglecting some real world implications. One of the rare successful counterexamples to this was provided recently by rough volatility. In this case, researchers developed martingale models, e.g.~that of \cite{Bayer_2015}, to specifically accommodate their real-world belief that volatility can exhibit H\"older regularities much lower than that of Brownian motion, see e.g.~\cite{Gatheral_2020}. 

However motivated, the manner in which we generally assess the practical performance of such new martingale models is through their ability to reconcile ever-larger sets of existing real-world derivative price quotes. Recall that a derivative price $\Pi_t(\#)=\EE^\QQ[\#|\cG_t]$ defines a stochastic process under $\PP$, like $S_t$. But under $\QQ$, by applying the tower property of conditional expectations, this price $\Pi_t(\#)$ is seen to share the martingale property with $S_t$,
\begin{equation}
    \EE^\QQ[\Pi_T(\#)|\cG_t] := \EE^\QQ[\EE^\QQ[\#|\cG_T]|\cG_t] = \EE^\QQ[\#|\cG_t] =: \Pi_t(\#).
\end{equation}
In this way, we can think of derivative prices like stock prices. And, just like being able to trade a stock at two different prices would constitute the most simple of arbitrages, so setting derivative prices which are inconsistent with reliable existing quotes would too. This justifies using the reconciliation of sets of existing prices as a model performance measure. Indeed this measure can be circular, but to employ it in this way (unintentionally) demonstrates a lack of ability to choose sensible derivative sets, which can be more of an art than science.

Before moving on, it is worth pointing out that many common definitions relating to the real-world measure $\PP$ have been omitted here, such as admissible and self-financing trading strategies, and the notions of buyers' and sellers' prices. To understand how these concepts relate to martingale measures via real-world super-replication and market completeness, the practical yet mathematically elegant text \cite{Guyon_2013} is recommended.

\vspace{3mm}\textbf{Novikov's martingale condition.} We now work towards a martingale sub-framework of that from \autoref{def:price_frame}. We will leave all connections with a real-world probability measure, as described above, for future work, so reintroduce our filtered space $(\Omega,\cF,\{\cG_t\}_{t\in\RR_+},\PP)$ from the beginning of this section, understanding that $\PP$ will characterise an abstract model, not the real world. It will soon become clear why we use $\cG_t$ to denote our general filtration.

First, we define a \emph{continuous} martingale on this space properly. Towards this, recall that a continuous stochastic process $M=\{M_t\}_{t\in\RR_+}$ on $(\Omega,\cF,\{\cG_t\}_{t\in\RR_+},\PP)$ is called adapted if $M_t$ is $\cG_t$-measurable for every $t\in\RR_+$. If we let the index $t$ denote time, then this essentially says we do not require information from the future, i.e.~in some set $\cG_T\setminus\cG_t$ with $T>t$, to construct $M_t$. Likewise, because of the inclusivity $s\le t\implies\cG_s\subseteq\cG_t$ of filtrations, if we \emph{can} construct $M_t$ from $\cG_t$, then we can additionally construct the entire history $\{M_s:s\in[0,t)\}$.

\begin{definition}[Continuous martingale]\label{def:martingale}
    With respect to a filtered probability space $(\Omega,\cF,\{\cG_t\}_{t\in\RR_+},\PP)$, a continuous martingale is a continuous process $M=\{M_t\}_{t\in\RR_+}$ which is adapted and verifies both $\EE[|M_t|]<\infty$ and $\EE[M_t|\cG_s]=M_s$ for every $s,t\in\RR_+$ with $s\le t$.
\end{definition}

This definition will be summarised by writing that $M$ is a $\cG_t$-martingale. If $M$ additionally has paths in $\uC(\RR_+,\RR_+)$, i.e.~paths which are non-negative and bounded over compacts, then the integrability condition $\EE[|M_t|]<\infty$ is redundant when $M_t$ does not depend on $\cG_0$, i.e.~verifies $\EE[M_t] = \EE[M_t|\cG_0]$. Then we always find $\EE[|M_t|]=\EE[M_t] = \EE[M_t|\cG_0]=M_0<\infty$.

Regarding martingale \emph{price} processes, we will always be in this setting just described. To see this, recall that our price processes $S=\{S_t\}_{t\in\RR_+}$ in the general framework of \autoref{def:price_frame} take the exponentiated form $S=\exp(W^\rho_X-\frac12 X)$, with $X\in\Phi\subset\uC^1_0(\RR_+,\RR_+)$ from \autoref{def:solutionset}. So paths of $S$ are strictly positive and finite, with $S_0=1$ given $W^\rho_0=X_0=0$. Our main tool towards establishing the martingality of such a price process is the following, accredited to \cite{Novikov_1972} although presented here like in \cite{Ikeda_1989}.

\begin{theorem}[Novikov martingale condition]\label{thm:novikov}
    Let $L=\{L_t\}_{t\in\RR_+}$ be a continuous local martingale on $(\Omega, \cF,\{\cG_t\}_{t\in\RR_+},\PP)$ with $L_0=0$, and define the process $M = \{M_t\}_{t\in\RR_+}$ by $M_t := \exp(L_t - \frac12[L]_t)$. Then provided $\EE[e^{\frac12[L]_t}]<\infty$ for every $t\in\RR_+$, $M$ is a $\cG_t$-martingale.
\end{theorem}

Note that our statement of Novikov's condition technically omits an \emph{implicit} local square-integrability assumption in \cite{Ikeda_1989}. This assumption is superfluous, i.e. is satisfied by any such process $L$ here, as clarified in Chapter 5 of \cite{Rogers_1994}.

Of course we have not actually defined the \emph{local} martingales $L=\{L_t\}_{t\in\RR_+}$ and related quadratic variations $[L]=\{[L]_t\}_{t\in\RR_+}$ on which \autoref{thm:novikov} depends. But this is because the application of existing `time-change' results, as covered shortly, will enable us to apply \autoref{thm:novikov} in our framework without direct dependence on these complicated objects. Specifically, comparing the representation $\exp(L_t - \frac12[L]_t)$ in \autoref{thm:novikov} with the price process $S_t=\exp(W^\rho_{X_t}-\frac12 X_t)$ in \autoref{def:price_frame}, the requirements to apply Novikov's condition is clear: the random IVP solution $X$ must be such that $W^\rho_X=\{W^\rho_{X_t}\}_{t\in\RR_+}$ defines a $\cG_t$-local martingale for such a filtration, with also $[W^\rho_X]=X$ and $\EE[e^{\frac12 X_t}]<\infty$ for $t\in\RR_+$.

For those familiar with time-changes, it is important to recognise that properties like $[W^\rho_X]=X$ are by no means verified for any such random IVP solution $X$ depending on arbitrary random field $Y\in\uG$ on $(\Omega,\cF,\PP)$. This is equivalent to saying that the solutions of \autoref{prob:random_ivp} are not \emph{merely} time-changes in disguise, clarified with an example following \autoref{thm:time_changed_bm}. So if we want to apply \autoref{thm:novikov} with time-change theory, then we must select only random fields $Y$ with additional properties compared with those in \autoref{def:price_frame}. Some such properties of a random field will essentially reveal themselves, once we understand related properties applicable to a general process like $X$, not necessarily a random IVP solution.

\vspace{3mm}\textbf{Time-changed Brownian motions.} For this part we use Section 1 of Chapter 5 in \cite{Revuz_1999}. Very similar sections can be found in other popular texts, like \cite{Ikeda_1989}, \cite{Karatzas_1998} and \cite{Rogers_1994}, but by using \cite{Revuz_1999} we can deal with the time-change-related issues above most succinctly.

This said, the notation used for indexing in all these texts can be confusing in our setting. This confusion can be foreseen intuitively by noting that our goal is to conclude that $S$ is a $\cG_t$-martingale, i.e.~we want to draw a \emph{conclusion} regarding a process and filtration indexed by `time' $t\in\RR_+$. However, after a \emph{change} of time (thus index if we want to avoid duplicating its use), this simple goal will not be achieved if we \emph{start} on a space $(\Omega,\cF,\{\cF_t\}_{t\in\RR_+},\PP)$ indexed by $t\in\RR_+$, like we usually do. Of course there are plenty of settings where duplicating the use of an arbitrary index is fine, but this is not our setting, because the indices which would be duplicated correspond directly with those indexing our random fields $Y=\{Y_{t,x}\}_{(t,x)\in\RR_+^2}$. 

We have already discussed the natural resolution of this minor issue when describing our probabilistic setting at the beginning of this chapter. We just need to start on the space $(\Omega,\cF,\PP)$ supporting our 2d Brownian motion $W$, and index this process with the \emph{spatial} variable $x\in\RR_+$, i.e.~$W=\{W_x\}_{x\in\RR_+}$. Then $\{\cF_x\}_{x\in\RR_+}$ denotes the natural filtration of $W$.

We now properly define a time-change. Contrasting Definition 1.2 in \cite{Revuz_1999}, we consider here only \emph{continuous} time-changes, which simplifies presentation. For our applications we actually require only \emph{strictly} increasing and \emph{differentiable} time-changes, like our random IVP solutions $X\in\Phi$. Recall first that a random variable $\tau\in\RR_+$ on $(\Omega,\cF,\{\cF_x\}_{x\in\RR_+},\PP)$ is called an $\cF_x$-stopping time if the event $\{\tau\le x\}$ is in $\cF_x$. We continue to use stopping \emph{time} for such variables, even though in our setting stopping \emph{level} would be more appropriate.

\begin{definition}[Continuous time-change]\label{def:time_change}
    A continuous time-change on the filtered probability space $(\Omega,\cF,\{\cF_x\}_{x\in\RR_+},\PP)$ is a stochastic process $X=\{X_t\}_{t\in\RR_+}$ which has increasing paths in $\uC_0(\RR_+,\RR_+)$ and is such that each random variable $X_t$ defines an $\cF_x$-stopping time.
\end{definition}

Now we are able to state part of Proposition 1.5 in \cite{Revuz_1999} succinctly, as follows.

\begin{theorem}[Time-changed Brownian motion]\label{thm:time_changed_bm}
    Let $X=\{X_t\}_{t\in\RR_+}$ be a time-change on $(\Omega,\cF,\{\cF_x\}_{x\in\RR_+},\PP)$. Then $W_X=\{W_{X_t}\}_{t\in\RR_+}$ is an $\cF_{X_t}$-local martingale with $[W_X]=X$.
\end{theorem}

So now a route towards establishing our price processes $S:=\exp(W^\rho_X-\frac12 X)$ to be $\cG_t:=\cF_{X_t}$-martingales is revealed: we need the random IVP solution $X=\{X_t\}_{t\in\RR_+}$ to be a time-change as per \autoref{def:time_change}. Then we will be able to combine \autoref{thm:novikov} and \autoref{thm:time_changed_bm} without needing to consider properties of local martingales or quadratic variations directly.

Following \autoref{def:time_change} and \autoref{thm:rand_well_posed}, it is clear that any such random IVP solution $X$ in $S:=\exp(W^\rho_X-\frac12 X)$ is a time-change provided that each $X_t$ is an $\cF_x$-stopping time. For clarity, this requires that for each $x\in\RR_+$, we find the event $\{X_t \le x\}$ in $\cF_x$, where $\{\cF_x\}_{x\in\RR_+}$ is the natural filtration of $W$. It is important to see that, given in \autoref{prob:ivp2} we place \emph{literally} no constraints on the relationship between a driving random field $Y$ and the Brownian motion $W$ generating $\cF_x$, this stopping time property is by no means exhibited by $X$ naturally: our random IVP solutions are not merely time-changes in disguise; time-change theory is just convenient for us to establish our martingale framework. To confirm this, for any $c>0$ consider the field $Y\in\uG$ in the generalised Heston framework defined by
\begin{equation}\label{eq:counterexample_time_change}
    Y_{t,x} := \sigma Z_x + \kappa(\vt(t) - x) + v,\quad Z_x := W^1_{|c - x|} - W^1_c.
\end{equation}
Then whenever $x\in[0,c)$, $\{X_t\le x\}$ is in $\cF_{c-x}\setminus\cF_x$ and not $\cF_x$, so $X$ is \emph{not} a time-change.

So in pursuit of a martingale sub-framework of \autoref{def:price_frame}, the task is now to characterise a subset of fields $Y\in\uG$ which ensure that each $X_t$ defines an $\cF_x$-stopping time, thus time-change. Conditions which ensure this are by no means difficult to obtain, given the simple relationship $X'_t=Y_{t,X_t}$, and some are formalised in the adaptedness \autoref{def:spatial_adapted} shortly.

Now it is practically informative to note that, in a setting where $X$ \emph{does} define a time-change, then using $\log S = W^\rho_X-\frac12 X$ with \autoref{thm:time_changed_bm} we see $[\log S] = [W^\rho_X] = X$. So then our general notion of volatility $\sigma:=\sqrt{X'}$ from \autoref{def:volatility} reconciles with the conventional relationship $\sigma^2_t=\frac{\dd}{\dd t}[\log S]_t$. This is the case \emph{whether or not} \autoref{thm:novikov} can be applied, e.g.~we could find $\EE[e^{\frac12 X_t}]=\infty$ and then $S := \exp(W^\rho_X-\frac12 X)$ might not be a martingale.

\vspace{3mm}\textbf{The martingale framework.} This part defines the martingale sub-framework from \autoref{fig:venn} in \autoref{def:martingale_frame}, and culminates with \autoref{them:price_martingale}, which actually proves that this indeed generates martingale price processes $S=\{S_t\}_{t\in\RR_+}$. This result is stated with an integrability assumption $M_{\bX}(\frac12,t):=\EE[\exp(\frac12 \bX_t)]<\infty$, and we have already shown how this can be verified for generalised Heston models, through \autoref{thm:mgf_existence} and \autoref{cor:gauss_heston_mgf}.

This next definition just formalises the idea that, given the Brownian motion $W=\{W_x\}_{x\in\RR_+}$ over the subinterval $[0,x]\subset\RR_+$, we want to be able to construct the random field $Y=\{Y_{t,x}\}_{(t,x)\in\RR_+^2}$ over the subdomain $\RR_+\times[0,x]\subset\RR_+^2$. Recall $\{\cF_x\}_{x\in\RR_+}$ denotes the natural filtration of $W$ and let $\cR$ be the Borel $\sigma$-algebra of $\RR$ induced e.g.~by the Euclidean distance.

\begin{definition}[Spatially adapted field]\label{def:spatial_adapted}
    On $(\Omega,\cF,\{\cF_x\}_{x\in\RR_+},\PP)$, call a random field $Y=\{Y_{t,x}\}_{(t,x)\in\RR_+^2}$ spatially adapted if $Y_{t,x}\!:\!(\Omega,\cF_x)\to(\RR,\cR)$ is measurable for each $(t,x)\in\RR^2_+$.
\end{definition}

Note that it is the ordering property $u\le x\implies\cF_u\subseteq\cF_x$ of filtrations which ensures that if $Y_{t,x}:(\Omega,\cF_x)\to(\RR,\cR)$ is measurable, then so too is $Y_{t,u}:(\Omega,\cF_x)\to(\RR,\cR)$ for each $u\in[0,x)$. This is to say, if a random field $Y$ is spatially adapted as defined here, then we can indeed construct $Y$ over the entirety of $\RR_+\times[0,x]$ provided we are given $W$ over $[0,x]$.

It may be clear from this ability to construct $Y$ over $\RR_+\times[0,x]$, i.e.~\emph{for all times}, when given $W$ over $[0,x]$, that we will assume $Y_{\cdot,x} - Y_{0,x}$ defines a deterministic function for each fixed $x\in\RR_+$. Of course this is the case in the generalised Heston framework from \autoref{def:gen_heston_frame}. Given the general goal to ensure that each $X_t$ defines an $\cF_x$-stopping time, it is plausible that \autoref{def:spatial_adapted} could be generalised considerably by utilising stopping times directly. This could enable us to retain the martingality of price processes when constructing the underlying random field from \emph{another} random IVP solution, but this amounts to considering higher dimensional random IVPs, and of course it makes sense to explore the 1d case first.

This next result confirms the value of spatially adapted fields, showing that these ensure the stopping time property $\{X_t \le x\}\in\cF_x$ for $(t,x)\in\RR_+^2$. Towards this, it can help to first observe that the event $\{X_t \le x\}:=\{\omega\in\Omega:X_t(\omega) \le x\}$ coincides with $\{X^{-1}_x\ge t\}$, given that paths of the random IVP solution $X$ define bijections from and to $\RR_+$ by \autoref{thm:rand_well_posed}.

\begin{lemma}[Time-change solutions]\label{lem:time_change_soln}
    Let $Y\in\uG$ be a spatially adapted field on the space $(\Omega,\cF,\{\cF_x\}_{x\in\RR_+},\PP)$. Then the solution of \autoref{prob:random_ivp}, $x'=Y_{t,x}$, $x_0=0$, is a time-change.
\end{lemma}
\begin{proof}
    Using \autoref{def:time_change}, it is clear from \autoref{thm:rand_well_posed} that a solution $X\in\Phi$ has all the properties necessary to be a time-change except that in general each $X_t$ does not have to be an $\cF_x$-stopping time. This was clarified with the counterexample in \autoref{eq:counterexample_time_change}.
    
    This stopping time condition requires that for each $x\in\RR_+$, we find $\{X_t \le x\}\in\cF_x$. Given that the field $Y$ is spatially adapted as in \autoref{def:spatial_adapted}, each restriction $\{Y_{t,u}\}_{(t,u)\in\RR_+\times[0,x]}$ is $\cF_x$-measurable, i.e.~from $\cF_x$ we can construct the field $Y$ over $\RR_+\times[0,x]$. So, given the IVP relationship $X'_t=Y_{t,X_t}$, $X_0=0$ between $X$ and $Y$, from $\cF_x$ we can clearly also construct the process $X$ up to the same level $x\in\RR_+$, which is reached at the random time $X^{-1}_x$.
    
    Given $\cF_x$, this ability to construct the strictly increasing process $X$ up to the level $x\in\RR_+$ clarifies that for any $t\in\RR_+$, the event $\{X_t \le x\}$ is known from the information in $\cF_x$: we just measure the random time $X^{-1}_x$, and then use $\{X_t \le x\} = \{X^{-1}_x\ge t\}$. This shows that $\{X_t \le x\}\in\cF_x$, so by definition each $X_t$ is an $\cF_x$-stopping time, completing the proof.
\end{proof}

We are finally ready to define the martingale framework shown in \autoref{fig:venn}. This is a sub-framework of the general one from \autoref{def:price_frame}, with $Y$ assumed spatially adapted and $\EE[\exp(\frac12 \bX_t)]<\infty$ where $\bX_t\!:=\!\inf\{x>0\!:\!Y_{t,x}<0\}$, but is fully defined here for more clarity.

\begin{definition}[Martingale price framework]\label{def:martingale_frame}
    Let $(\Omega, \cF,\PP)$ support a Brownian motion $W=(W^0,W^1)$ over $\RR_+$, let $\{\cF_x\}_{x\in\RR_+}$ be the natural filtration of $W$ and $Y$ be a spatially adapted random field in $\uG$, satisfying $M_\bX(\frac12,t):=\EE[\exp(\frac12 \bX_t)]<\infty$ over $\RR_+$. Let $X=\{X_t\}_{t\in\RR_+}$ be the solution of the random IVP $x'=Y_{t,x}$, $x_0=0$, then define the price process $S=\{S_t\}_{t\in\RR_+}$ by $S:=\exp(W^\rho_{X}-\frac12 X)$, where $W^\rho=\sqrt{1-\rho^2}W^0 + \rho W^1$ for some $\rho\in[-1,1]$.
\end{definition}

The next concluding result has essentially been established over the course of this section, but is still consolidated here. Although it will be clear that the integrability assumption in \autoref{def:martingale_frame} serves only to ensure $M_X(\frac12,t)<\infty$, the former is prioritised for good reason: it is very practically valuable, given this is a condition which can be checked directly from $Y$, \emph{not} requiring analysis of random IVPs. This was demonstrated in \autoref{thm:mgf_existence}.

\begin{theorem}[Martingale price process]\label{them:price_martingale}
    Any price process $S=\exp(W^\rho_X-\frac12 X)$ deriving from the framework in \autoref{def:martingale_frame} is a $\cG_t:=\cF_{X_t}$-martingale, and verifies $[\log S] = X$.
\end{theorem}
\begin{proof}
    Given that $M_X(\frac12,t)\le M_\bX(\frac12,t)$, and $M_\bX(\frac12,t)<\infty$ is ensured by assumption, then $\EE[e^{\frac12 X_t}]<\infty$ and the Novikov condition from \autoref{thm:novikov} can be invoked to conclude that $S$ is a martingale \emph{if} $W^\rho_X$ is an $\cF_{X_t}$-local martingale which verifies $[W^\rho_X]=X$. These properties are given precisely by \autoref{thm:time_changed_bm} \emph{if} $X$ is a time-change on $(\Omega,\cF,\{\cF_x\}_{x\in\RR_+},\PP)$, and then $[\log S] = [W^\rho_X] =X$ follows also, as discussed after \autoref{thm:time_changed_bm}. The sole purpose of \autoref{lem:time_change_soln} was to establish that $X$ indeed defines a time-change as required, provided $Y$ is spatially adapted as in \autoref{def:spatial_adapted}, so applying this lemma completes the proof.
\end{proof}

\subsection{The Riemann-Liouville-Heston model}\label{sec:RLH_model}

The main purpose of this section is to define and clarify properties of a specific model in the \emph{intersection} of the two price process sub-frameworks just covered, as shown in \autoref{fig:venn}. That is, a generalised Heston \emph{and} martingale model, as per \autoref{def:gen_heston_frame} and \autoref{def:martingale_frame}. 

Although the generalised Heston sub-framework provides much freedom through the selection of a volatility-driving processes $Z\!=\!\{Z_x\}_{x\in\RR_+}$ (recall \autoref{cor:gen_hest_solution} and the following discussion), the second purpose here is to demonstrate how, via the selection of $Z$, we can accommodate rough volatility research with ease and mathematical harmony. Specifically, how we can accommodate H\"older continuous volatility models for any fixed order in $(0,\frac12)$.

To foresee this harmony, first recall that models in the generalised Heston framework can be summarised by the equations uniquely verified by a price $S$ and its volatility $\sqrt{X'}$, namely
\begin{equation}\label{eq:gen_mod_summary}
    X'_t = \sigma Z_{X_t} + \kappa(\vt(t) - X_t) + v,\quad S_t=\exp(W^\rho_{X_t}-\tfrac12 X_t),
\end{equation}
where as usual $W^\rho=\sqrt{1-\rho^2}W^0+ \rho W^1$, and $(W^0,W^1)$ is a standard 2d Brownian motion. Recall also from \autoref{thm:heston_ode} that the distribution of $S$ here coincides with the classical Heston model's when selecting $\vt(t) = \theta t$ and $Z=W^1$. Then, the new model defined here embodies the idea to simply \emph{replace} the classical Brownian motion $Z=W^1$ selection with its Riemann-Liouville fractional derivative $Z=D^\alpha(W^1)=:W^{\alpha}$ of some order $\alpha$ in $(0,\frac12)$.

This is a new generalisation of the Heston model, and will be labelled the Riemann-Liouville-Heston (RLH) model for obvious reasons. This model can be summarised by setting $Z=W^{\alpha}$ in \autoref{eq:gen_mod_summary}, and the classical Heston model is then simply recovered in the omitted boundary case where $\alpha=0$. Assuming $\vt$ in \autoref{eq:gen_mod_summary} to be e.g.~Lipschitz, then the variance process $X'$ inherits the H\"older continuity of $Z:=W^\alpha$, i.e.~$\frac12-\alpha-\ep$ for any $\ep>0$. 

The fact that this classical replacement is acceptable in our ODE-based framework, with no additional well-posedness work required, cannot be overlooked. Indeed, this demonstrates our framework's stability deriving from the results of \autoref{chap:wellposed}, contrasting e.g.~the It\^o-based framework of the classical Heston model, in which this harmonious replacement idea has essentially no meaning without material additional work. See e.g.~\cite{Keller_2018} and \cite{Larsson_2019} for research applicable to an alternative `rough Heston' generalisation, still not known to have a unique strong solution.

The eager reader can skip ahead to \autoref{fig:simulation_output}, which demonstrates sample paths from our RLH model, but this section is primarily devoted to defining this model rigorously, starting with Riemann-Liouville-type fractional derivatives. Consequences of results from \autoref{chap:wellposed} and \autoref{chap:solutions} will then be clarified, before confirming that the RLH price process defines a martingale, so generates arbitrage-free derivative prices, as covered in the previous section.

\vspace{3mm}\textbf{Fractional derivatives.} Riemann-Liouville (RL) fractional derivatives of orders in $(0,1)$ are now introduced, and a continuous mapping property between H\"older spaces is emphasised. This property will help later, firstly with understanding related mapping properties of the RLH model, and then in establishing the convergence of its simulation via \autoref{thm:euler_converge}. 

For any $\lambda\in(0,1)$, let $\uH_\lambda\subset\uC_0(\RR_+,\RR)$ denote the set of functions $w$ starting from 0 and which over any compact subinterval $\II=[0,I]\subset\RR_+$ verify the H\"older condition of order $\lambda$:
\begin{equation}
    \Vert w \Vert^\lambda_\II := \sup_{x\in\II}|w(x)| + \sup_{\substack{x,u\in\II \\ x\neq u}}\frac{|w(x) - w(u)|}{|x - u|^\lambda}<\infty.
\end{equation}
Recall that the space $(\uH_\lambda,\Vert\cdot\Vert^\lambda_\II)$ (containing \emph{restrictions} of each $w\in\uH_\lambda$ to $\II$) is a \emph{non-separable} Banach space. It proves very convenient for us that Brownian motion can be constructed on a separable subspace $(\uH^0_\lambda,\Vert\cdot\Vert^\lambda_\II)$, introduced later like in \cite{Hamadouche_2000}.

\begin{definition}[Riemann-Liouville fractional derivative]\label{def:RL_deriv}
    For any path $w\in\uH_\lambda$ and order $\alpha\in(0,\lambda)$, the $\alpha$-fractional derivative of $w$ is the path $D^\alpha(w)\in\uH_{\lambda-\alpha}$ defined over $\RR_+$ by
    \begin{equation}\label{eq:RL_deriv}
        D^\alpha(w)(x) := \frac{1}{\Gamma(1-\alpha)}\frac{\dd}{\dd x}\int_0^x \frac{w(u)}{(x-u)^{\alpha}}\dd u.
    \end{equation}
\end{definition}

The operator $D^\alpha$ defined here proves \emph{well}-defined by the classical results of \cite{Hardy_1932}, consolidated neatly in Theorem 8 of \cite{Hamadouche_2000}. The implication that $D^\alpha(w)(0)=0$ holds, following $w(0)=0$, should not be overlooked. For reference, $D^\alpha$ coincides with a \emph{left-handed} RL fractional derivative from Definition 2.2 in \cite{Samko_1993}, denoted there by $\cD^\alpha_{0+}$. This popular text is however not recommended for our purposes.

% It is not until Lemma 13.1 in \cite{Samko_1993} that $\cD^\alpha_{0+}$ is shown to indeed map any such $w\in\uH_0^\lambda$ to an element of $\uH_0^{\lambda-\alpha}$, with $\cD^\alpha_{0+}(w)(0)=0$ being ensured when $w(0)=0$. This depends on the equivalence of $\cD^\alpha_{0+}$ with the so-called \emph{Marchaud} derivative $\uD^\alpha_{0+}$ on such Holder sets $\uH^\lambda_0$. See the remarks following Lemma 13.1.

The following continuity result is also due to \cite{Hardy_1932}, although a slick proof is also provided as Proposition 2 in \cite{Hamadouche_2000}. This proof also clarifies that $D^\alpha:\uH_\lambda\to\uH_{\lambda-\alpha}$ is bijective, so also defines an isomorphism with respect to H\"older norms.

\begin{theorem}[H\"older continuity of fractional derivatives]\label{thm:fractional_continuity}
    For $\lambda\in(0,1)$, $\{w_n\}_{n\in\NN_0}\subset\uH_\lambda$, $\alpha\in(0,\lambda)$ and $\II=[0,I]\subset\RR_+$, the operator $D^\alpha$ is H\"older continuous in the sense that
    \begin{equation}
        \Vert w_0 - w_n \Vert^\lambda_\II\xrightarrow{n\to\infty}0 \implies \Vert D^\alpha(w_0) - D^\alpha(w_n) \Vert^{\lambda-\alpha}_\II\xrightarrow{n\to\infty}0.
    \end{equation}
\end{theorem}

Towards reconciling rough volatility observations; that volatility exhibits H\"older regularities much lower than that of Brownian motion, we are simply going to drive our volatility process in the generalised Heston framework of \autoref{def:gen_heston_frame} by an RL fractional derivative process. 

\begin{definition}[Riemann-Liouville process]\label{def:RL_process}
    From the Brownian motion $W^1=\{W^1_x\}_{x\in\RR_+}$ on $(\Omega,\cF,\PP)$, define the process $W^\alpha=\{W^\alpha_x\}_{x\in\RR_+}$ by $W^\alpha=D^\alpha(W^1)$, where $\alpha\in(0,\frac12)$. I.e.,
    \begin{equation}\label{eq:RL_process}
        W^\alpha_x := \frac{1}{\Gamma(1-\alpha)}\frac{\dd}{\dd x}\int_0^x \frac{W^1_u}{(x-u)^{\alpha}}\dd u.
    \end{equation}
\end{definition}

Because paths of $W^1$ are a.s.~in $\uH_\lambda$ for $\lambda\in(0,\frac12)$, then we find $W^\alpha$ in $\uH_\lambda$ for $\lambda\in(0,\frac12-\alpha)$. So we can reduce the H\"older regularity of $W^\alpha$, thus $X'$, by simply raising the derivative order $\alpha$ as required. Omitting constants, this process $W^\alpha$ is actually indistinguishable from the It\^o integral $\int_0^x(x-u)^{-\alpha}\dd W^1_u$, introduced by \cite{Levy_1953} and related to the fractional Brownian motion of \cite{Mandlebrot_1968}. Specifically for rough volatility modelling, such indistinguishable relationships were generalised in \cite{Horvath_2017}. 

So that the connection with fractional derivatives' properties is clearest, the representation of $W^\alpha$ from \autoref{def:RL_process} will be prioritised. This also continues emphasising our lack of any direct dependence on stochastic calculus. The full covariance structure of $W^\alpha$ can be found in \cite{Jacquier_2018}, but the following summarises what we need.

\begin{lemma}[Riemann-Liouville process properties]\label{lem:RL_properties}
    For $\alpha\in(0,\frac12)$, the Riemann-Liouville process $W^\alpha$ is Gaussian with paths in $\uH_\lambda$ for every $\lambda\in(0,\frac12-\alpha)$, and for all $x\in\RR_+$ verifies~
    \begin{equation}
        \EE[W^\alpha_x]=0 \quad\text{and}\quad \EE[(W^\alpha_x)^2] = \frac{x^{1-2\alpha}}{\Gamma(1-\alpha)^2(1-2\alpha)}.
    \end{equation}
\end{lemma}

At this point it is worth noting that the process $W^\alpha$ has sublinear variance growth in the sense that $\EE[(W^\alpha_x)^2]<a+b x^c$ for some fixed $a,b\ge0$, $c\in(0,1)$ and all $x\in\RR_+$. For example, recalling that $\alpha\in(0,\frac12)$ and using \autoref{lem:RL_properties}, take any $a>0$, $b>\Gamma(1-\alpha)^{-2}(1-2\alpha)^{-1}$ and $c=1-2\alpha$. Notice that this enables the application of \autoref{cor:gauss_heston_mgf}, for several purposes.

\vspace{3mm}\textbf{The RLH model.} The RLH price process model is that within the generalised Heston framework of \autoref{def:gen_heston_frame}, where we make the fractional derivative selection $Z=W^\alpha:=D^\alpha(W^1)$. It is thus well-defined in full as follows, on any probability space $(\Omega,\cF,\PP)$ supporting the usual 2d Brownian motion $W=(W^0,W^1)$ over $\RR_+$. No confusion should arise from our symbolic use of $W^\alpha$ and $W^\rho$ to denote different processes, e.g.~$\rho=\alpha\centernot\implies W^\rho=W^\alpha$.

\begin{definition}[Riemann-Liouville-Heston model]\label{def:RLH_model}
    Let $\vt$ be any bijection in $\uC_0(\RR_+,\RR_+)$, and let $W^\alpha=\{W^\alpha_x\}_{x\in\RR_+}$ be the fractional derivative $W^\alpha=D^\alpha(W^1)$ of order $\alpha\in(0,\frac12)$. For some fixed parameters $\sigma,\kappa,v > 0$, define the random field $Y=\{Y_{t,x}\}_{(t,x)\in\RR_+^2}$ in $\uG$ by
    \begin{equation}\label{def:gen_hest_field3}
        Y_{t,x} := \sigma W^\alpha_x + \kappa(\vt(t) - x) + v,
    \end{equation}
    then let $X=\{X_t\}_{t\in\RR_+}$ be the solution of the random IVP $x'=Y_{t,x}$, $x_0=0$, and let the price process $S=\{S_t\}_{t\in\RR_+}$ be defined by $S:=\exp(W^\rho_{X}-\frac12 X)$ for some fixed $\rho\in[-1,1]$.
\end{definition}

The RLH model can thus be summarised by the equations which $X$ and $S$ uniquely verify:
\begin{equation}\label{eq:RLH_model_summary}
    X'_t = \sigma W^\alpha_{X_t} + \kappa(\vt(t) - X_t) + v,\quad S_t=\exp(W^\rho_{X_t}-\tfrac12 X_t),\quad W^\rho = \sqrt{1-\rho^2}W^0 + \rho W^1,
\end{equation}
and, using \autoref{thm:heston_ode}, the distribution of the price process $S$ coincides with that of the classical Heston model when $\vt(t) =  \theta t$ and when $\alpha=0$, by noting $D^0(W^1)=W^1$. Like with \autoref{eq:model_summary2}, our volatility process $\sqrt{V}=\sqrt{X'}$ in this model equivalently verifies
\begin{equation}\label{eq:RLH_model_summary2}
    V_t = \sigma W^\alpha_{\int_0^t V_s\dd s} + \kappa\left(\vt(t) - \int_0^t V_s\dd s\right) + v\ \implies\ \dd V_t = \sigma \dd W^\alpha_{\int_0^t V_s\dd s} + \kappa(\vt'(t) - V_t)\dd t,
\end{equation}
where the second equation assumes absolute continuity of $\vt$, i.e.~$\vt(t)=\int_0^t\vt'(s)\dd s$. Note that $\vt$ being strictly increasing ensures its a.e.~differentiability, so $V=X'$ a.e.~inherits the $(\frac12-\alpha-\ep)$-H\"older continuity of $W^\alpha$. If $\vt$ is additionally $(\frac12-\alpha-\ep)$-H\"older continuous (e.g.~Lipschitz), then so is $X'$ (everywhere, not just a.e.). So finally volatility $\sqrt{X'}$ inherits $(\frac12-\alpha-\ep)$-H\"older continuity on intervals where $X'_t>0$, and is $(\frac14-\frac12\alpha - \ep)$-H\"older otherwise.

\vspace{3mm}\textbf{Well-posedness.} We need to confirm the implication $Y\in\uG$ in \autoref{def:RLH_model}. This is achieved if the RLH model is in the generalised Heston sub-framework from \autoref{def:gen_heston_frame}. It certainly looks so, but note we have omitted the requirement $\sup_{x\in\RR_+}\kappa x - \sigma Z_x=\infty$ there. As discussed before \autoref{thm:mgf_existence}, this condition is equivalent to the existence of the process $\bX=\{\bX_t\}_{t\in\RR_+}$ from \autoref{thm:rand_well_posed} over $\RR_+$, given the bijective nature of $\vt$ and that
\begin{equation}
    \bX_t := \inf\{x>0:Y_{t,x}<0\} = \inf\{x>0:\kappa x - \sigma Z_x > \kappa\vt(t) + v\} < \infty.
\end{equation}
So we need to confirm that $\bX_t<\infty$ for all $t\in\RR_+$ when $Z:=W^\alpha$. For complete clarity, this means $\PP[X_t<\infty]=1\ \forall t\in\RR_+$, rather than $\PP[X_t<\infty\ \forall t\in\RR_+]=1$, although in our setting these conditions are equivalent anyway because $X$ is a.s.~strictly increasing. As mentioned before \autoref{thm:mgf_existence}, we will obtain $\bX_t<\infty$ for all $t\in\RR_+$ if we have the stronger MGF existence $\EE[e^{p\bX_t}]<\infty$ for all $t\in\RR_+$ and some $p>0$. This is confirmed by this next result.

\begin{corollary}[RLH MGF existence]\label{cor:rlh_mgf_exist}
    Let the random IVP solution $X=\{X_t\}_{t\in\RR_+}$ and its upper bound $\bX=\{\bX_t\}_{t\in\RR_+}$ be those from the RLH model in \autoref{def:RLH_model}. Then the MGFs $M_X(p,t):=\EE[e^{pX_t}]$ and $M_\bX(p,t):=\EE[e^{p\bX_t}]$ exist globally, that is for all $(p,t)\in\RR\times\RR_+$.
\end{corollary}
\begin{proof}
    The claim will be established if we can apply \autoref{cor:gauss_heston_mgf}. For this, we require that the process $Z=W^\alpha$ under the RLH model is centred Gaussian and verifies $\EE[(W^\alpha_x)^2]<a + bx^c$ for some $a,b\ge0$, $c\in(0,1)$ and all $x\in\RR_+$. The variance of $W^\alpha$ given in \autoref{lem:RL_properties} shows that this is indeed the case for any $a>0$, $b>\Gamma(1-\alpha)^{-2}(1-2\alpha)^{-1}$ and $c=1-2\alpha$.
\end{proof}

Given that $\bX_t<\infty$ follows for all $t\in\RR_+$, then equivalently $\sup_{x\in\RR_+}\kappa x - \sigma W^\alpha_x=\infty$, and so $Y\in\uG$ and the RLH model is indeed one of our generalised Heston models from \autoref{def:gen_heston_frame}. For completeness, this means the well-posedness of the RLH model follows directly from \autoref{thm:rand_well_posed}, which can be summarised by saying that the defining equations in \autoref{eq:RLH_model_summary} have a pathwise unique solution. More specifically, the RLH paths $X(\omega)$ and $S(\omega)$ exist uniquely over $\RR_+$ for every $\omega\in\Omega_*:=\{\omega\in\Omega: \sup_{x\in\RR_+}\kappa x -\sigma W^\alpha_x(\omega)=\infty\}$.

\vspace{3mm}\textbf{Solution map continuity.} We now briefly consolidate continuity statements like those in \autoref{thm:rand_well_posed} and \autoref{eq:joint_continuity}, for the RLH model. Such statements ultimately derive from the results of \autoref{chap:wellposed}, specifically \autoref{thm:continuity_solution_map}. First note that implicit in the set $\Omega_*$ just defined is the assumption that each $W^\alpha(\omega)$ actually exists. For convenience we can reduce this set to contain only outcomes for which $W^1(\omega)\in \uH_\lambda\ \forall\lambda\in(0,\tfrac12)$. This has full measure given $W^1$ is Brownian motion, ensures that each $W^\alpha(\omega)$ exists by \autoref{def:RL_deriv}, and also clarifies that the forthcoming H\"older norms $\Vert\cdot\Vert^\lambda_{\RR_+}:=\sum_{n\in\NN} 2^{-n}(1\wedge \Vert\cdot\Vert^\lambda_{[0,n]})$ exist.

\begin{theorem}[RLH solution map continuity]\label{cor:RLH_continuity}
    Let $(X,S)$ be the processes defined in the RLH model, constructed from $(W^0,W^1)$.
    Then for outcomes $\{\omega_n\}_{n\in\NN_0}\subset\Omega_*$ and $\lambda\in(\alpha,\frac12)$,
    \begin{multline}\label{eq:product_converge}
        (\Vert W^0(\omega_0) - W^0(\omega_n)\Vert_{\RR_+},\
        \Vert W^1(\omega_0) - W^1(\omega_n)\Vert_{\RR_+}^\lambda)
        \xrightarrow{n\to\infty}(0,0)\\
        \implies
        (\Vert X(\omega_0) - X(\omega_n) \Vert_{\RR_+},\ \Vert S(\omega_0) - S(\omega_n) \Vert_{\RR_+})
        \xrightarrow{n\to\infty}(0,0).
    \end{multline}
\end{theorem}
\begin{proof}
    Firstly note that since $\alpha\in(0,\frac12)$, we always have  $\lambda\in(0,\frac12)$, and therefore $W^1(\omega_n)\in\uH_\lambda$ for every $n\in\NN_0$. This clarifies that the norms $\Vert\cdot\Vert^\lambda_{\RR_+}$ here exist. Now from the assumption $\Vert\cdot\Vert^\lambda_{\RR_+}\xrightarrow{n\to\infty}0$ here, \autoref{thm:fractional_continuity} provides $\Vert W^\alpha(\omega_0) - W^\alpha(\omega_n)\Vert^{\lambda-\alpha}_{\RR_+}\xrightarrow{n\to\infty}0$, again noting that $\lambda-\alpha\in(0,\frac12)$ is ensured. So the limiting assumption in \autoref{eq:product_converge} is stronger than~
    \begin{equation}\label{eq:assumption_rlh_cont}
        (\Vert W^0(\omega_0) - W^0(\omega_n)\Vert_{\RR_+},\
        \Vert W^1(\omega_0) - W^1(\omega_n)\Vert_{\RR_+}, W^\alpha(\omega_0) - W^\alpha(\omega_n)\Vert_{\RR_+})
        \xrightarrow{n\to\infty}(0,0,0),
    \end{equation}
    i.e.~stronger than product \emph{uniform} convergence over compacts. Using \autoref{def:RLH_model}, we then obtain the RLH random field convergence $\Vert Y(\omega_n)-Y(\omega_0)\Vert_{\RR_+^2}\xrightarrow{n\to\infty}0$. Since the assumptions of both \autoref{eq:rand_cont_dep} and \autoref{eq:joint_continuity} are now confirmed, we obtain the consequences of these. These coincide precisely with the claim here, so complete the proof.
\end{proof}

\vspace{3mm}\textbf{Martingality.} Letting $\{\cF_x\}_{x\in\RR_+}$ denote the natural filtration of $W=(W^0,W^1)$ as usual, we now confirm that the RLH price process $S=\{S_t\}_{t\in\RR_+}$ from \autoref{def:RLH_model} is a martingale on the filtered space $(\Omega,\cF,\{\cG_t\}_{t\in\RR_+},\PP)$, where $\cG_t:=\cF_{X_t}$. Like in \autoref{them:price_martingale}, we also obtain the relationship $X=[\log S]$ which volatility processes $\sqrt{X'}$ conventionally satisfy.

To achieve this, the more general martingality result of \autoref{them:price_martingale} will be applied. This depends on an MGF $\EE[e^{p\bX_t}]$ existence condition and the spatially adapted condition of $Y$, from \autoref{def:spatial_adapted}. To help with the latter, the following is provided first, which applies to all models in the generalised Heston sub-framework, so the RLH model specifically. 

\begin{lemma}[Generalised Heston adaptedness]\label{lem:gen_hest_adapted}
    Let the random field $Y=\{Y_{t,x}\}_{(t,x)\in\RR_+^2}$ take the generalised Heston form in \autoref{def:gen_heston_frame}, i.e.~$Y_{t,x} = \sigma Z_x + \kappa(\vt(t) - x)+v$ for some $Z=\{Z_x\}_{x\in\RR_+}$ and path $\vt$.
    If $Z$ is adapted to $\{\cF_x\}_{x\in\RR_+}$, then $Y$ is spatially adapted.
\end{lemma}
\begin{proof}
    Selecting a process $Z$ which is $\cF_x$-adapted means $Z_x$ is $\cF_x$-measurable for every $x\in\RR_+$. That is, $Z_x:(\Omega,\cF_x)\to(\RR,\cR)$ defines a measurable map, where $\cR$ is the Borel $\sigma$-algebra of $\RR$ e.g.~induced by the Euclidean distance. In the generalised Heston case, where $Y$ takes the form $Y_{t,x} = \sigma Z_x + \kappa(\vt(t) - x)+v$, this assumption extends to $Y_{t,x}:(\Omega,\cF_x)\to(\RR,\cR)$ being measurable for every $(t,x)\in\RR^2_+$, given that $\vt$ is a fixed continuous function. So now just using \autoref{def:spatial_adapted}, $Y$ is spatially adapted as claimed, and the proof is complete.
\end{proof}

Now \autoref{them:price_martingale}, \autoref{cor:rlh_mgf_exist} and \autoref{lem:gen_hest_adapted} come together to provide the following.

\begin{corollary}[Martingality of RLH model]\label{cor:RLH_martingality}
    The RLH price process $S=\{S_t\}_{t\in\RR_+}$ from \autoref{def:RLH_model} is a martingale on the filtered space $(\Omega,\cF,\{\cG_t\}_{t\in\RR_+},\PP)$, where $\cG_t:=\cF_{X_t}$.
\end{corollary}
\begin{proof}
    \autoref{them:price_martingale} will provide the claim, after the assumptions there are confirmed as being applicable here. For this, firstly the evaluated MGF $\EE[e^{\frac12\bX_t}]$ must exist over $\RR_+$, and we have already confirmed this in \autoref{cor:rlh_mgf_exist}. Secondly and finally, the RLH field $Y$ must be spatially adapted. For this we can apply \autoref{lem:gen_hest_adapted}, applicable to all generalised Heston models, provided the RLH fractional derivative selection $Z:=W^\alpha:=D^\alpha(W^1)$ is adapted to $\{\cF_x\}_{x\in\RR_+}$. It is clear from the integral representation in \autoref{def:RL_process}, namely~
    \begin{equation}
        W^\alpha_x := \frac{1}{\Gamma(1-\alpha)}\frac{\dd}{\dd x}\int_0^x\frac{W^1_u}{(x-u)^\alpha}\dd u,
    \end{equation}
    that $W^\alpha$ is not just $\cF_x$-adapted, but adapted to the natural filtration of just the component $W^1$. So we can apply \autoref{them:price_martingale} to complete the proof, and also confirm $X=[\log S]$.
\end{proof}

\subsection{Derivative pricing by simulation}\label{sec:vol_surfaces}

Now our attention turns to approximating the \emph{theoretical} RLH price process $S=\{S_t\}_{t\in\RR_+}$ from \autoref{def:RLH_model} with a computationally practicable process that can be simulated using the forward Euler scheme from \autoref{alg:forward_euler}. Although, for the sake of specificity, we focus on the RLH model here, the approach taken clarifies how the flexible convergence result of \autoref{thm:euler_converge} can be applied to other models in the general framework of \autoref{def:price_frame}.

Given that \autoref{them:price_martingale} establishes the martingality of $S$ on a space $(\Omega, \cF,\{\cG_t\}_{t\in\RR_+},\PP)$, the primary application in mind is the evaluation of (arbitrage-free) derivative prices. Following \autoref{sec:martingale}, we will thus be concerned with approximating expectations $\EE[\#|\cG_0]=\EE[\#]$ for a real, bounded and continuous derivative payoff $\#=\#(S)$, by Monte-Carlo simulation. \cite{Glasserman_2004} and \cite{Asmussen_2007} provide backgrounds to this objective.

Towards this, a sequence $\{\hS^n\}_{n\in\NN}$ of RLH \emph{polygon} processes will be defined which can be simulated and indeed verify $\EE[\#(\hS^n)]\xrightarrow{n\to\infty}\EE[\#(S)]$ for any such payoff $\#$. This is, by definition, equivalent to establishing the weak convergence of random elements $\hS^n\cw S$, or weak convergence of induced probability measures; see e.g. Section 1 of \cite{Billingsley_1999}.

Next we treat the fact that for any such approximating process $\hS^n$, the expectation $\EE[\#(\hS^n)]$ can itself only be approximated, by an estimator $N^{-1}\sum_{i=1}^N \#(\hS^n_i)$ depending on finite realisations $\{\hS_i^n\}_{i=1}^N$. This joint approximation is an under-emphasised issue in Monte-Carlo theory, but by neglecting one of these approximations, both theoreticians and practitioners rarely state the notion in which \emph{actual} computer simulations converge. In \autoref{thm:conv_prices}, we provide a tractable and intuitive joint convergence statement, justifying existing practises.

\vspace{3mm}\textbf{Preparatory results.} Considering the forward Euler convergence result \autoref{thm:euler_converge}, at the heart of the approach here will be a practicable sequence $\{\hY^n\}_{n\in\NN}$ of random fields converging uniformly over compacts to the RLH random field $Y_{t,x}:=\sigma W^\alpha_x+\kappa(\vt(t)-x)+v$.

Although having a weak convergence result, i.e.~$\hY^n\cw Y$, would suffice, in order to apply the pathwise simulation convergence result in \autoref{thm:wellposed} most clearly, we will move to a purely abstract probability space on which a.s.~convergence results can be established. Like in \autoref{thm:skorokhod_rep}, random elements on this space will usually be indicated by the use of $\tilde{{\color{white}X}}$. This approach via a.s.~convergence is one of those suggested in \cite{Billingsley_1999}.

We now provide some preparatory results which will enable this. The first is Skorokhod's powerful representation theorem from \cite{Skorokhod_1956}, stated here as in \cite{Billingsley_1999}. To interpret this properly, recall that implicit in convergence statements $X_n\cw X_0$ on a normed vector space $(\cX,\Vert\cdot\Vert_\cX)$ is the measurability of maps $X_n:(\Omega,\cF,\PP)\to(\cX,\cB(\cX))$, where $\cB(\cX)$ is the Borel $\sigma$-algebra of $\cX$ induced by $\Vert\cdot\Vert_\cX$, and a \emph{support} of $X_n$ is any set $A\in\cB(\cX)$ such that $\mu_{X_n}[A]=1$, where $\mu_{X_n}:=\PP X_n^{-1}$ is the distribution of $X_n$. Finally, for such a set to be \emph{separable} means it has a countable subset which is dense in $(\cX,\Vert\cdot\Vert_\cX)$.

\begin{theorem}[Skorokhod's representation theorem]\label{thm:skorokhod_rep}
    Suppose $X_n\cw X_0$ on $(\cX,\Vert\cdot\Vert_\cX)$, and $X_0$ has a separable support. Then there exists random elements $\tX_n$ on a common probability space, such that $X_n\ed\tX_n$ for every $n\in\NN_0$, yet $\tX_n\xrightarrow[\mathrm{a.s.}]{n\to\infty} \tX_0$ on $(\cX,\Vert\cdot\Vert_\cX)$.
\end{theorem}

This result will be combined with Lamperti's invariance principle for Brownian motion, from \cite{Lamperti_1962}. We state this as in \cite{Hamadouche_2000}, emphasising the subsets $\uH^0_\lambda\subset\uH_\lambda$ containing paths $w\in\uH_\lambda$ with the additional continuity property $\omega_\lambda(w,\delta)\xrightarrow{\delta\to0}0$, where
\begin{equation}
    \omega_\lambda(w,\delta) := \sup_{\substack{x,u\in\II \\ 0<|x-u|\le\delta}}\frac{|w(x) - w(u)|}{|x-u|^\lambda},
\end{equation}
and $\II\subset\RR_+$ is any compact interval. Related results, e.g. the characterising limit theorem of \cite{Rackauskas_2004}, apply to these subsets. The point is that $(\uH^0_\lambda,\Vert\cdot\Vert^\lambda_{\RR_+})$ is a \emph{separable} Banach space, as shown in \cite{Ciesielski_1960}, so \autoref{thm:skorokhod_rep} can be applied without modification. For clarity, separability with respect to $\Vert\cdot\Vert^\lambda_{\RR_+}:=\sum_{n=1}^\infty2^{-n}(1\wedge\Vert\cdot\Vert^\lambda_{[0,n]})$ follows from the stability of separability on infinite-dimensional product spaces, see e.g.~\cite{Billingsley_1999}. Here we let $W=\{W_x\}_{x\in\RR_+}$ denote a standard 1d Brownian motion.

\begin{theorem}[Lamperti's invariance principle]\label{thm:lamperti}
    Let $\{\zeta_k\}_{k\in\NN}$ be a sequence of i.i.d. random variables with $\EE[\zeta_k]=0$, $\EE[\zeta^2_k]=\sigma^2$ and $\EE[|\zeta_k|^\gamma]<\infty$ for some $\gamma>2$. Define the sequence $\{\hW^n\}_{n\in\NN}$ of piecewise linear processes $\hW^n=\{\hW^n_x\}_{x\in\RR_+}$ respectively using 
    \begin{equation}\label{eq:lamperti_polygon}
        \hW^n_x = \frac{1}{\sigma\sqrt{n}}\left[\sum_{k=1}^{\lfloor nx \rfloor}\zeta_k + (nx - \lfloor nx \rfloor)\zeta_{\lfloor nx \rfloor+1}\right].
    \end{equation}
    Then the weak convergence $\hW^n\cw W$ takes place on $(\uH^0_\lambda,\Vert\cdot\Vert^\lambda_{\RR_+})$, for all $\lambda\in(0,\frac12-\frac{1}{\gamma})$.
\end{theorem}

Since in practice such polygons from \autoref{eq:lamperti_polygon}  will be considered  for fixed $n\in\NN$ and $\sigma=1$, it is helpful to note that $\hW^n$ is nothing more than the linear interpolation between the values $\hW^n_{x_k}:=\sqrt{\upsilon_n}\sum_{j=1}^k\zeta_j$, where the points $x_k:=k\upsilon_n$ have a step size $\upsilon_n:=n^{-1}$. Clearly we may also invert this simple relationship, to make use of  $\sqrt{\upsilon_n}\zeta_k = \hW^n_{x_k} - \hW^n_{x_{k-1}}$.

Finally we provide the following lemmas, which simplify fractional derivatives $D^\alpha(w)$ for polygon paths $w\in\uAC_0(\RR_+,\RR)$, like those of $\hW^n$ from \autoref{thm:lamperti}. The evaluation points $x^*_k\in(x_k,x_{k+1})$ derived here coincide with those from \cite{Bennedsen_2017}, contrasting those of \cite{Horvath_2017}, both concerned with approximating the related integral $\int_0^x(x-u)^{-\alpha}\dd W_u$. The simple connection here between the points $x^*_k$ and polygons is novel, and will be leveraged alongside the H\"older continuity of $D^\alpha$ from \autoref{thm:fractional_continuity}.

\begin{lemma}[Polygon fractional derivatives]\label{lem:RL_polygons}
    Let the path $w\in\uAC_0(\RR_+,\RR)$ be linear between the points $(x_k,w(x_k))$, for $x_k:=k\upsilon$, $k\in\NN_0$ and some $\upsilon>0$. Then for any $\alpha\in(0,1)$, the derivative $D^\alpha(w)$ admits the following representation at the points $\{x_k\}_{k\in\NN}$
    \begin{equation}\label{eq:rl_polygon_rep}
        D^\alpha(w)(x_k) = \frac{1}{\Gamma(1-\alpha)}\sum_{j=1}^k (x^*_{k-j})^{-\alpha} (w(x_j)-w(x_{j-1})),\quad (x^*_k)^{-\alpha} := \frac{x_{k+1}^{1-\alpha} - x_k^{1-\alpha}}{(1-\alpha)\upsilon}.
    \end{equation}
\end{lemma}
\begin{proof}
    Since $w$ is in $\uAC(\RR_+,\RR)$ with $w(0)=0$, Lemma 2.2 of \cite{Samko_1993} provides
    \begin{equation}
        \Gamma(1-\alpha)D^\alpha(w)(x):=\frac{\dd}{\dd x}\int_0^x w(u)(x-u)^{-\alpha}\dd u = \int_{[0,x]}w'(u)(x-u)^{-\alpha}\dd u.
    \end{equation}
    Using the a.e. equivalence $w'(u) = \sum_{j=1}^k \mathbbm{1}_{u\in[x_{j-1},x_j)}\frac{w(x_j) - w(x_{j-1})}{x_j - x_{j-1}}$ over $[0,x_k]$, we then have
    \begin{multline}\label{eq:rl_polygon}
        \Gamma(1-\alpha)D^\alpha(w)(x_k) = \int_0^{x_k}\sum_{j=1}^k \mathbbm{1}_{u\in[x_{j-1},x_j)}\frac{w(x_j) - w(x_{j-1})}{x_j - x_{j-1}}(x-u)^{-\alpha}\dd u \\
        = \sum_{j=1}^k \int_{x_{j-1}}^{x_j}(x-u)^{-\alpha}\dd u \frac{w(x_j) - w(x_{j-1})}{x_j - x_{j-1}}.
    \end{multline}
    Now evaluating the integrals $\int_{x_{j-1}}^{x_j}(x-u)^{-\alpha}\dd u$ provides the representation in \autoref{eq:rl_polygon_rep}, noting that $x_k - x_j = x_{k-j}$ and $x_j - x_{j-1}=\upsilon$ follows from the equipartition $x_k:=k\upsilon$.
\end{proof}

Finally the following is helpful in practice, as it allows us to make use of computationally convenient polygons \emph{between} the fractional derivative points $(x_k,D^\alpha(w)(x_k))$ of \autoref{lem:RL_polygons}. 

\begin{lemma}[Convergence of polygons]\label{eq:polygon_converge}
    Let $\{w_n\}_{n\in\NN}\subset\uAC_0(\RR_+,\RR)$ be linear between the points $(x_{n,k},w_n(x_{n,k}))$, for $x_{n,k}:=kn^{-1}$, $k\in\NN_0$ with $\Vert w_n-w_0\Vert^\lambda_{\RR_+}\xrightarrow{n\to\infty}0$ for some $w_0\in\uH_\lambda$, $\lambda\in(0,1)$. For $\alpha\in(0,\lambda)$, let $\{w_{\alpha,n}\}_{n\in\NN}\subset\uAC_0(\RR_+,\RR)$ be linear between the points $(x_{n,k},w_{\alpha,n}(x_{n,k}))$, where $w_{\alpha,n}(x_{n,k}):=D^\alpha(w_n)(x_{n,k})$. Then $\Vert D^\alpha(w_0)- w_{\alpha,n}\Vert_{\RR_+}\xrightarrow{n\to\infty}0$.
\end{lemma}
\begin{proof}
    By \autoref{thm:fractional_continuity} $\Vert D^\alpha(w_0)- D^\alpha(w_n)\Vert^{\lambda-\alpha}_{\RR_+}\xrightarrow{n\to\infty}0$ holds, so the uniform convergence $\Vert D^\alpha(w_0)- D^\alpha(w_n)\Vert_{\II}\xrightarrow{n\to\infty}0$ for any $\II=[0,I]\subset\RR_+$ also. The triangle inequality gives 
    \begin{equation}
        \Vert D^\alpha(w_0)- w_{\alpha,n}\Vert_{\II}\le\Vert D^\alpha(w_0)- D^\alpha(w_n)\Vert_{\II} + \Vert D^\alpha(w_n)- w_{\alpha,n}\Vert_{\II}.
    \end{equation}
    Suppose $\Vert D^\alpha(w_0)- D^\alpha(w_n)\Vert_\II=\ep>0$. Then since $w_{\alpha,n}(x):=D^\alpha(w_n)(x)$ for $x=kn^{-1}$, and $w_{\alpha,n}$ is linear between these points of distance $n^{-1}$, we have $\Vert D^\alpha(w_n)- w_{\alpha,n}\Vert_{\II}\le\ep+\omega_0(n^{-1})$, where $\omega_0$ is the modulus of continuity of $D^\alpha(w_0)$ over $\II$. So for any such interval $\II$, we have
    \begin{equation}
        \Vert D^\alpha(w_0)- w_{\alpha,n}\Vert_{\II}\le 2\Vert D^\alpha(w_0)- D^\alpha(w_n)\Vert_{\II} + \omega_0(n^{-1}) \xrightarrow{n\to\infty}0
    \end{equation}
    and the claim then follows just by definition of the norm $\Vert\cdot\Vert_{\RR_+}:=\sum_{n=1}^\infty2^{-n}(1\wedge\Vert\cdot\Vert_{[0,n]})$.
\end{proof}

For clarity we finally reduce the forward Euler convergence results from \autoref{thm:euler_converge} and \autoref{thm:wellposed} to a probabilistic corollary which can be applied directly in the setting here. By analogy with \autoref{alg:forward_euler}, define the forward Euler process $X=\{X_t\}_{t\in\RR_+}$ for the random IVP $x'=Y_{t,x}$ $x_0=0$, with step size $\Delta>0$, to be the linearly interpolating process between $X_0=0$ and the variables $X_{t_{k+1}} = X_{t_k} + Y_{t_k,X_{t_k}}\Delta$, where $t_k:=k\Delta$ and $k\in\NN_0$.

\begin{corollary}[Forward Euler convergence]\label{cor:euler_converge}
    Let $\{Y^n\}_{n\in\NN_0}$ be random fields in $\uG$, let $\{X^n\}_{n\in\NN}$ be the forward Euler processes for the random IVPs $x'=Y^n_{t,x}$, $x_0=0$ using step sizes $n^{-1}\Delta$ for some $\Delta>0$, and let $X^0$ solve the random IVP $x'=Y^0_{t,x}$, $x_0=0$. Then,
    \begin{equation}
        \Vert Y^0 - Y^n \Vert_{\RR_+^2} \xrightarrow[\mathrm{a.s.}]{n\to\infty} 0 \implies \Vert X^0 - X^n\Vert_{\RR_+} \xrightarrow[\mathrm{a.s.}]{n\to\infty} 0.
    \end{equation}
\end{corollary}

Notice the double approximation taking place in \autoref{cor:euler_converge}: the field $Y^0$ is being approximated by a convenient sequence $Y^n$, and from these approximations, we build approximating forward Euler processes. This coincides with the assumptions of  \autoref{thm:euler_converge}, only here we have reduced the general partitions $\pi_n$ there to those with fixed step sizes $n^{-1}\Delta$.

\vspace{3mm}\textbf{Price process simulation.} Now recall the five processes $\bS:=(W^0,W^1,W^\alpha,X,S)$, all over $\RR_+$, in the RLH model from \autoref{def:RLH_model}, which are related through the equations
\begin{equation}
    X'_t = \sigma W^\alpha_{X_t} + \kappa(\vt(t) - X_t)+v,\quad S_t=\exp(W^\rho_{X_t}-\tfrac12 X_t),\quad W^\rho = \sqrt{1-\rho^2}W^0 + \rho W^1.
\end{equation}
An approximating process $\hat\bS:=(\hW^0,\hW^1,\hW^\alpha,\hX,\hS)$ will now be defined which, unlike $\bS$, can be simulated (over compacts) exactly on a computer. At the core of this will be the forward Euler scheme from \autoref{alg:forward_euler}, for approximating solutions of random IVPs $x'=Y_{t,x}$, $x_0=0$. The RLH random field $Y_{t,x}:=\sigma W^\alpha_x+\kappa(\vt(t)-x)+v$ will essentially be approximated on a discrete equipartitioned grid, with practicable interpolations between.

\begin{definition}[RLH polygon]\label{def:rlh_polygon}
    Fix admissible RLH parameters $\sigma, \kappa,v>0$, $\alpha\in(0,\frac12)$, $\rho\in[-1,1]$ and path $\vt\in\uC_0(\RR_+,\RR_+)$ as in \autoref{def:RLH_model}. Fix temporal and spatial step sizes $\tau,\upsilon>0$ and for $k\in\NN_0$ define $t_k:=k\tau$, $x_k:=k\upsilon$. For $i=0,1$, let $\{\zeta^i_n\}_{n\in\NN}$ be sequences of i.i.d.~standard Gaussian random variables. Now the following five steps deal with approximating the RLH processes $(W^0,W^1,W^\alpha,X,S)$ respectively, with polygons. 
    
    \textbf{Step 1.} Define the process $\hW^0$ by linear interpolation between the point $\hW^0_0:=0$ and the variables $\hW^0_{x_k}:=\sqrt{\upsilon}\sum_{j=1}^k\zeta^0_j$. That is, over each interval $(x_k,x_{k+1})$ of length $\upsilon$, define
    \begin{equation}
        \hW^0_{x} := \hW^0_{x_k} + \upsilon^{-1}(\hW^0_{x_{k+1}} - \hW^0_{x_k})(x - x_k).
    \end{equation}
    
    \textbf{Step 2.} Define $\hW^1$ similarly, only constructed from $\{\zeta^1_k\}_{k\in\NN}$ rather than $\{\zeta^0_k\}_{k\in\NN}$.
    
    \textbf{Step 3.} Define the process $\hW^\alpha$ by linear interpolation between $\hW^\alpha_0:=0$ and the variables
    \begin{equation}\label{eq:RL_polygon}
        \hW^\alpha_{x_k} := \frac{\sqrt{\upsilon}}{\Gamma(1-\alpha)}\sum_{j=1}^k(x^*_{k-j})^{-\alpha}\zeta^1_j,\quad (x^*_k)^{-\alpha} := \frac{x_{k+1}^{1-\alpha} - x_k^{1-\alpha}}{(1-\alpha)\upsilon}.
    \end{equation}
    
    \textbf{Step 4.} Define the random field $\hY$ by $\hY_{t,x}:=\sigma \hW^\alpha_x + \kappa(\vt(t) - x)+v$ and $\hX$ to be the forward Euler polygon process for the random IVP $x'=\hY_{t,x}$, $x_0=0$ with step size $\tau$. That is, define $\hX_0:=0$ then $\hX$ by linear interpolation between the variables $\hX_{t_{k+1}}:= \hX_{t_k} + \hY_{t_k,\hX_{t_k}}\tau$.
    
    \textbf{Step 5.} Define the exp-polygon $\hS$ by $\hS_t:=\exp(\hW^\rho_{\hX_t} - \frac12\hX_t)$, where $\hW^\rho:=\sqrt{1-\rho}\hW^0 + \rho\hW^1$.
\end{definition}

Now call the process $\hat\bS:=(\hW^0,\hW^1,\hW^\alpha,\hX,\hS)$ an RLH polygon process with step sizes $\tau,\upsilon$. 

Our primary concern now is with the \emph{theoretical} convergence of a sequence of RLH polygon processes, but so it is clear that we have not lost touch with practicalities, succinct \texttt{python} code is provided in the \hyperlink{appendix}{Appendix}, which illustrates how these RLH polygons from \autoref{def:rlh_polygon} may be simulated. A sample path of the process $\hS$ is also shown in \autoref{fig:simulation_output}.

Here and in \autoref{thm:rlh_convergence} denote $\uC:=\uC(\RR_+,\RR)$. The main result of this section is one of weak convergence on the \emph{product} topology of uniform convergence over compacts, supporting paths of the process $\bS\in\uC^5$ and its polygonal approximation $\hat\bS$. For specificity, equip such finite product sets $\uC^d$ with the product norm $\Vert w\Vert_{\RR_+}:=\sum_{i=1}^d \Vert w_i \Vert_{\RR_+}$, where $w=(w_i)_{i=1}^d\in\uC^d$. Recall, e.g.~\cite{Billingsley_1999}, that the separability and completeness of such product spaces $(\uC^d,\Vert\cdot\Vert_{\RR_+})$ is inherited from the underlying spaces $(\uC,\Vert\cdot\Vert_{\RR_+})$, and separability ensures the Borel $\sigma$-algebra $\cB(\uC^d)$ of this product is precisely the product of Borel $\sigma$-algebras $\cB(\uC)^d$.

\begin{theorem}[RLH polygon convergence]\label{thm:rlh_convergence}
    Let $\bS:=(W^0,W^1,W^\alpha,X,S)$ be the RLH price processes and $\{\bS^n:=(W^{0,n},W^{1,n},W^{\alpha,n},X^n,S^n)\}_{n\in\NN}$ be a sequence of RLH polygon processes generated with temporal and spatial step sizes $\tau_n:=n^{-1}\Delta$ and $\upsilon_n:=n^{-1}$ for $\Delta>0$. Then the weak convergence $\bS^n\cw\bS$ takes place on the product space $(\uC^5,\Vert\cdot\Vert_{\RR_+})$.
\end{theorem}
\begin{proof}
    The main idea is to move to a probability space supporting processes $\tilde\bS^n\ed\bS^n$ and $\tilde\bS\ed\bS$ and to establish the convergence $\tilde\bS^n\xrightarrow{\mathrm{a.s.}}\tilde\bS$ (as $n\to\infty)$ on $(\uC^5,\Vert\cdot\Vert_{\RR_+})$. While not necessarily required, this enables a clear application of \autoref{thm:euler_converge} via \autoref{cor:euler_converge}.

    \textbf{Step 1.} As clarified following \autoref{thm:lamperti}, the processes $\{W^{1,n}\}_{n\in\NN}$ coincide with those in \autoref{eq:lamperti_polygon}, when setting $\sigma=1$ and $\zeta_k=\zeta^1_k$. Since each $\zeta^1_k$ is Gaussian with $\EE[|\zeta^1_k|^\gamma]<\infty$ for all $\gamma>2$, then \autoref{thm:lamperti} provides $W^{1,n}\cw W^1$ on $(\uH^0_\lambda,\Vert\cdot\Vert^\lambda_{\RR_+})$ for all $\lambda\in(0,\frac12)$.
    
    Since each $(\uH^0_\lambda,\Vert\cdot\Vert^\lambda_{\RR_+})$ is separable, apply \autoref{thm:skorokhod_rep} to move to another space supporting $\tW^{1,n}\ed W^{1,n}$ and $\tW^1\ed W^1$, with $\tW^{1,n}\asc \tW^1$ on $(\uH^0_\lambda,\Vert\cdot\Vert^\lambda_{\RR_+})$. Let this space support another Brownian motion $\tW^0$ independent from $\tW^1$, and define the sequence $\{\tW^{0,n}\}_{n\in\NN}$ by linear interpolation between the points of $\tW^0$ separated by step sizes $\upsilon_n=n^{-1}$ respectively. So now $\tW^{0,n}\ed W^{0,n}$ and $\tW^0\ed W^0$ but continuity of $\tW^0$ gives $\tW^{0,n}\asc \tW^0$ on $(\uC,\Vert\cdot\Vert_{\RR_+})$.
    
    \textbf{Step 2.} Let $\tW^{\alpha,n}$ be defined, like $W^{\alpha,n}$, by linear interpolation between the variables
    \begin{equation}
        \tW^{\alpha,n}_{x_k} := \frac{\sqrt{\upsilon_n}}{\Gamma(1-\alpha)}\sum_{j=1}^k (x^*_{n,k-j})^{-\alpha}\tilde\zeta^{1,n}_k,\ \ \ \tilde\zeta^{1,n}_k:= \frac{\tW^{1,n}_{x_j} - \tW^{1,n}_{x_{j-1}}}{\sqrt{\upsilon_n}},\ \ \ (x^*_{n,k})^{-\alpha} := \frac{x_{k+1}^{1-\alpha} - x_k^{1-\alpha}}{(1-\alpha)\upsilon_n}.
    \end{equation}
    By design of the points $x^*_{n,k}$ from \autoref{lem:RL_polygons}, $\tW^{\alpha,n}$ coincides with $D^\alpha(\tW^{1,n})$ at the points $x_{n,k}=k\upsilon_n$, and \autoref{eq:polygon_converge} gives $D^\alpha(\tW^{1,n})\neq \tW^{\alpha,n} \asc \tW^\alpha:=D^\alpha(\tW^1)$ on $(\uC,\Vert\cdot\Vert_{\RR_+})$.
    
    \textbf{Step 3.} Let $\tX^n$ be defined, like $X^n$, to be forward Euler polygons of the random IVPs $x'=\tY^n_{t,x}$, $x_0=0$ with step size $\tau_n$, where $\tY^n_{t,x}:=\sigma \tW^{\alpha,n}_x + \kappa(\vt(t) - x)+v$. Let $\tX$ solve the random IVP $x'=\tY_{t,x}$, $x_0=0$ where $\tY_{t,x}:=\sigma \tW^{\alpha}_x + \kappa(\vt(t) - x)+v$. Given $\tW^{\alpha,n} \asc \tW^\alpha$ on $(\uC,\Vert\cdot\Vert_{\RR_+})$, then $\Vert \tY - \tY^n \Vert_{\RR_+^2} \xrightarrow{\mathrm{a.s.}} 0$ and \autoref{cor:euler_converge} provides $\tX^n \asc \tX$ on $(\uC,\Vert\cdot\Vert_{\RR_+})$.
    
    \textbf{Step 4.} Define $\tS^n$ and $\tS$ respectively by $\tS^n:=\exp(\tW^{\rho,n}_{\tX^n} - \frac12\tX^n)$ and $\tS:=\exp(\tW^\rho_{\tX} - \frac12\tX)$, then $\tS^n \asc \tS$ on $(\uC,\Vert\cdot\Vert_{\RR_+})$ follows from having $\tW^{\rho,n} \asc \tW^\rho$ and $\tX^n \asc \tX$ here also. 
    
    \textbf{Step 5.} We have established a sequence $\tilde\bS^n:=(\tW^{0,n},\tW^{1,n},\tW^{\alpha,n},\tX^n,\tS^n)\ed\bS^n$ of RLH polygons and the RLH process $\tilde\bS:=(\tW^0,\tW^1,\tW^\alpha,\tX,\tS)\ed \bS$ such that $\tilde\bS^n\asc\tilde\bS$ takes place on the product space $(\uC^5,\Vert\cdot\Vert_{\RR_+})$. So the claim of $\bS^n\cw\bS$ on $(\uC^5,\Vert\cdot\Vert_{\RR_+})$ follows.
\end{proof}

Recall that, since coordinate-wise projections are continuous, the weak convergence $\bS^n\cw\bS$ on  $(\uC^5,\Vert\cdot\Vert_{\RR_+})$ immediately provides weak coordinate-wise convergence, i.e.~$W^{0,n}\cw W^0$, \dots, $S^n\cw S$ each on $(\uC,\Vert\cdot\Vert_{\RR_+})$, although the converse is generally not true. 

So in particular, for any continuous and bounded derivative payoff $\#:(\uC,\Vert\cdot\Vert_{\RR_+})\to(\RR,|\cdot|)$ we now have the convergence of derivative prices $\EE[\#(S^n)]\xrightarrow{n\to\infty}\EE[\#(S)]$. But notice that this remains a \emph{theoretical} result, since in practice we must approximate these approximating expectations $\EE[\#(S^n)]$, using a i.i.d.~sample $\{S^n_i\}_{i=1}^N$ and estimator $N^{-1}\sum_{i=1}^N \#(S^n_i)$. Because of this double approximation, manifesting theoretically as a double limit $n,N\to\infty$, we cannot \emph{directly} apply the laws of large numbers as $N\to\infty$ to establish the limit $\EE[\#(S)]$.

\vspace{3mm}\textbf{Derivative pricing.} The final mathematical goal of this section is to extend the theoretical weak convergence $S^n\cw S$ result of \autoref{thm:rlh_convergence} to a computationally realisable one based on finite simulation samples $\{S_i^n\}_{i=1}^N$ for some $n,N\in\NN$. This is achieved quite simply by combining weak convergence with laws of large numbers, but doing so is often neglected, with most authors focusing \emph{either} on drawing theoretical weak convergence statements like \autoref{thm:rlh_convergence} or on applying Monte-Carlo theory as if exact simulation of $\{S_i\}_{i=1}^N$ is possible. \cite{Horvath_2017} and \cite{McCrickerd_2018} provide recent examples of this.

It should be clear that this next result actually applies to arbitrary random elements $\{S^n\}_{n\in\NN_0}$ of a set $\cX$ provided $S^n\cw S^0$ on a space $(\cX,\Vert\cdot\Vert_\cX)$ and $\#:(\cX,\Vert\cdot\Vert_\cX)\to(\RR,|\cdot|)$. A statement based on the \emph{strong} law of large numbers is prioritised here, see e.g.~\cite{Dekking_2005}, but the corresponding weak statement is given after.

\begin{theorem}[Convergence of derivative prices]\label{thm:conv_prices}
    Suppose $S^n\cw S$ on $(\uC,\Vert\cdot\Vert_{\RR_+})$ as in \autoref{thm:rlh_convergence} and let $\{S_i^n\}_{i=1}^N$ denote i.i.d.~replications of $S^n$. Then for any bounded and continuous $\#\!:\!(\uC,\Vert\!\cdot\!\Vert_{\RR_+})\!\to\!(\RR,|\cdot|)$ and tolerance $\ep>0$, there exists $n_*\!=\!n_*(\#,\ep)$ such that
    \begin{equation}\label{eq:sim_convergence}
        \lim_{N\to\infty}\left|\EE[\#(S)] - \frac{1}{N}\sum_{i=1}^N \#(S^n_i) \right|<\ep\ \text{ a.s.~for any } n>n_*.
    \end{equation}
\end{theorem}
\begin{proof}
    By definition of $S^n\cw S$ on $(\uC,\Vert\cdot\Vert_{\RR_+})$ we have $\EE[\#(S^n)]\xrightarrow{n\to\infty}\EE[\#(S)]$, so there exists $n_* = n_*(\#,\ep)$ such that $|\EE[\#(S)] - \EE[\#(S^n)] |<\ep$ for all $n>n_*$. For any $n\in\NN$, the strong law of large numbers provides the a.s.~convergence $N^{-1}\sum_{i=1}^N \#(S^n_i)\xrightarrow{N\to\infty}\EE[\#(S^n)]$ where the existence of $\EE[\#(S^n)]$ is ensured given $\#$ is bounded. Continuity of the function $f(x)=|\EE[\#(S)] - x|$ then provides the a.s.~claim in \autoref{eq:sim_convergence} for any $n>n_*$:
    \begin{multline}
        \lim_{N\to\infty}\left|\EE[\#(S)] - \frac{1}{N}\sum_{i=1}^N \#(S^n_i) \right| =\\ \left|\EE[\#(S)] - \lim_{N\to\infty}\frac{1}{N}\sum_{i=1}^N \#(S^n_i) \right|=|\EE[\#(S)] - \EE[\#(S^n)] |<\ep.
    \end{multline}
\end{proof}

The statement analogous to \autoref{eq:sim_convergence} but deriving instead from the weak law is 
\begin{equation}\label{eq:sim_convergence2}
    \PP\left[\left|\EE[\#(\bS)] - \frac{1}{N}\sum_{i=1}^N \#(\bS^n_i) \right|>\ep\right] \xrightarrow{N\to\infty} 0 \text{ for any } n>n_*.
\end{equation}
Practically, this reads: for any fixed tolerance $\ep$, it is possible to set our simulation quality high enough, via $n$, and thereafter diminish the probability of realising a derivative price error greater than $\ep$ to zero, via $N$. This notion of convergence is stronger than the iterated limit
$\lim_{n\to\infty}\lim_{N\to\infty}\PP\left[\cdot\right] = 0$,
which does not guarantee that any of the $N\to\infty$ limits in \autoref{eq:sim_convergence2} are actually zero. This convergence is however weaker than the joint convergence in probability as $n,N\to\infty$, for which it would be sufficient to establish some uniformity in the separate limits' convergence, so that the Moore-Osgood theorem applies.

In practice we are often more concerned with setting such tolerances $\ep$ \emph{not} directly on prices $\EE[\#(S)]$, but on convenient functions $\Psi:\RR\to\RR$ thereof. When such functions are continuous, \autoref{thm:conv_prices} provides the following corollary. Proof of this is essentially immediate via the modulus of continuity $\omega_\Psi$ of $\Psi$, which necessarily satisfies $\omega_\Psi(\ep)\xrightarrow{\ep\downarrow 0}0$.

\begin{corollary}\label{cor:vol_converge} 
    In the setting of \autoref{thm:conv_prices}, let $\Psi:\RR\to\RR$ be continuous when restricted to an open ball containing $\EE[\#(S)]$. Then there exists $n_*=n_*(\#,\Psi,\ep)$ such that
        \begin{equation}\label{eq:sim_convergence3}
            \lim_{N\to\infty}\left|\Psi\left(\EE[\#(S)]\right) - \Psi\left(\frac{1}{N}\sum_{i=1}^N \#(S^n_i)\right) \right|<\ep\ \text{ a.s.~for any } n>n_*.
        \end{equation}
\end{corollary}

In the next part we will focus on the simple case of the put option $\#(S):=\max\{K-S_T,0\}$ for a range of fixed strikes and maturities $K,T>0$. As is common practice, we will map the estimated values of the put option price $\EE[\#(S)]$ onto Black-Scholes implied volatilities $\mathrm{IV}$, like we did in \autoref{fig:vols_converge}. Regarding \autoref{cor:vol_converge}, we thus set $\Psi=\mathrm{IV}:=\mathrm{BS}^{-1}$, where
\begin{equation}\label{eq:bs_formula}
    \mathrm{BS}(\sigma) := K \mathrm{N}(-d_-) - \mathrm{N}(-d_+),\quad d_\pm = d_\pm(\sigma) := -\frac{\log(K)}{\sigma\sqrt{T}} \pm \frac{\sigma\sqrt{T}}{2},
\end{equation}
and $\mathrm{N}$ is the standard Gaussian CDF. The text \cite{Gatheral_2006} provides more details on this implied volatility map IV, and confirms it to be continuous as required by \autoref{cor:vol_converge}.

\begin{figure}[H]
    \centering
    \includegraphics[height=0.32\linewidth]{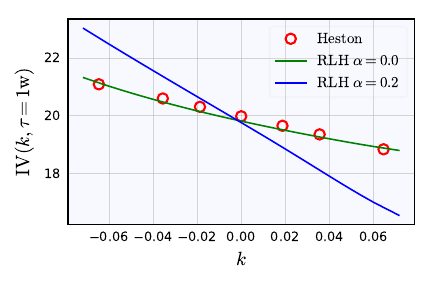}
    \includegraphics[height=0.32\linewidth]{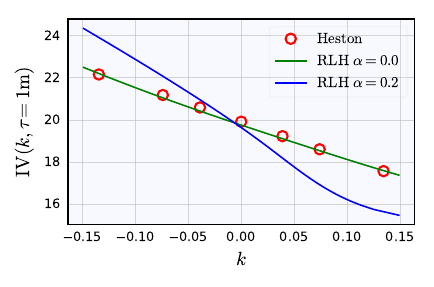}
    \includegraphics[height=0.32\linewidth]{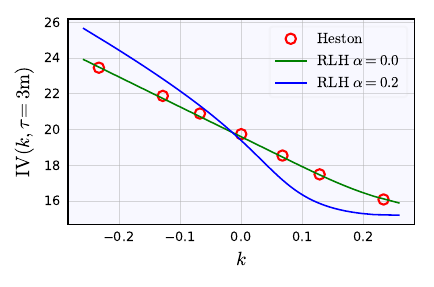}
    \includegraphics[height=0.32\linewidth]{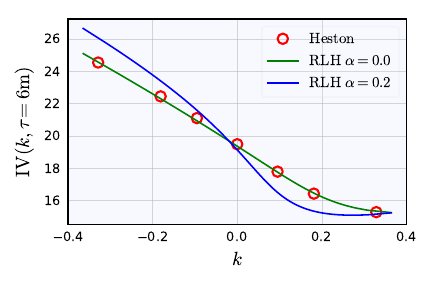}
    \hspace*{-2mm}\includegraphics[height=0.32\linewidth]{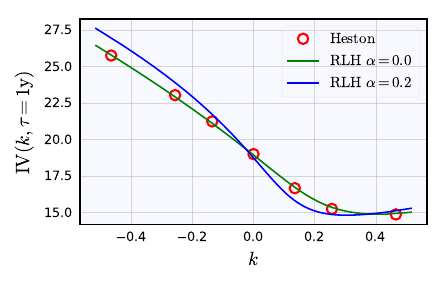}
    \hspace*{-2mm}\includegraphics[height=0.32\linewidth]{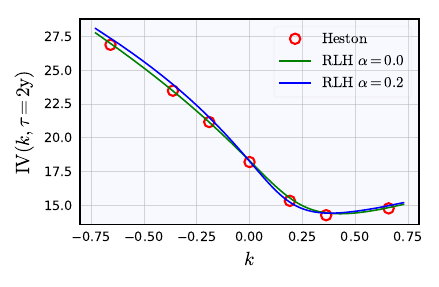}
    \caption{Implied volatilities $\mathrm{IV}(k,\tau)$ from the classical Heston and RLH models, defined in \autoref{eq:heston_plots} and \autoref{eq:RLH_plots} respectively, are shown. Parameters are set to $\sigma = \kappa = 0.2$, $\theta=v=0.04$, $\rho=-0.7$, with the fractional derivative $\alpha$ as shown.}
    \label{fig:heston_comparison_skewed}
\end{figure}

\begin{figure}[H]
    \centering
    \includegraphics[height=0.32\linewidth]{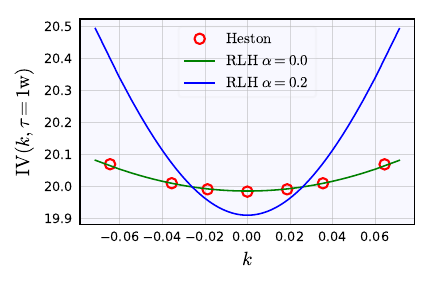}
    \includegraphics[height=0.32\linewidth]{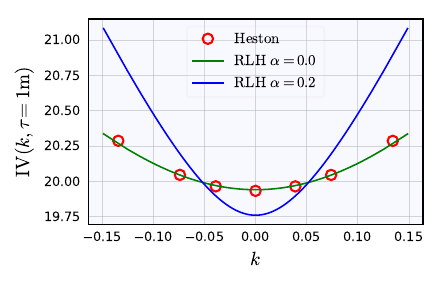}
    \includegraphics[height=0.32\linewidth]{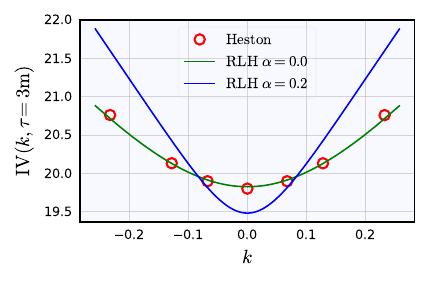}
    \includegraphics[height=0.32\linewidth]{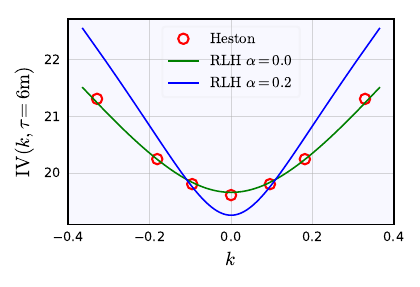}
    \includegraphics[height=0.32\linewidth]{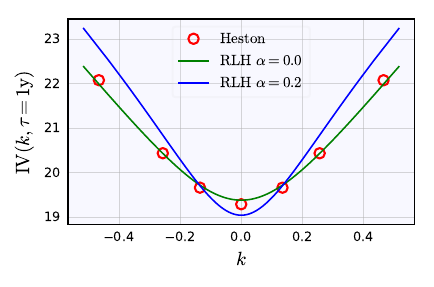}
    \includegraphics[height=0.32\linewidth]{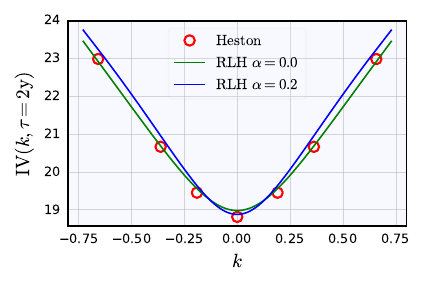}
    \caption{A reproduction of \autoref{fig:heston_comparison_skewed} is shown, setting instead $\rho=0$.}
    \label{fig:heston_comparison_symmetric}
\end{figure}

\begin{figure}[H]
    \centering
    \includegraphics[width=0.75\linewidth]{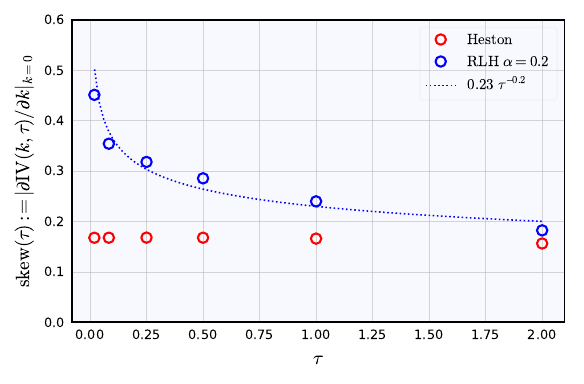}
    \caption{At-the-money implied volatility skews approximated by central finite difference from the data in \autoref{fig:heston_comparison_skewed}.}
    \label{fig:skews}
\end{figure}

\begin{figure}[H]
    \centering
    \includegraphics[width=0.74\linewidth]{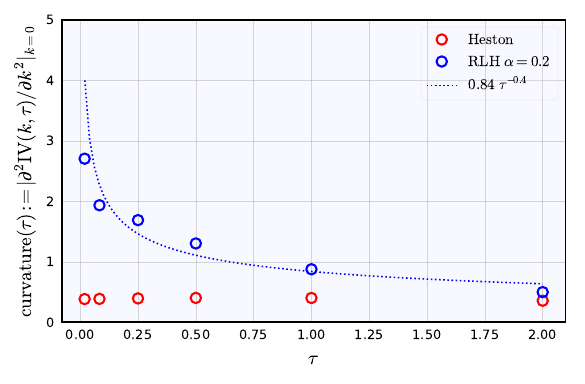}
    \caption{At-the-money implied volatility curvatures approximated by central finite difference from the data in \autoref{fig:heston_comparison_symmetric}.}
    \label{fig:curvatures}
\end{figure}

\vspace{3mm}\textbf{RLH implied volatilities.} We now simulate RLH implied volatilities, using the scheme from \autoref{def:rlh_polygon}. The priority is to confirm that these coincide with those of the classical Heston model when the RLH fractional derivative $\alpha\in(0,0.5)$ is on the zero boundary (recall this is a consequence of \autoref{thm:heston_ode}), and to then show the effect of increasing $\alpha$ to 0.2. These comparisons between $\alpha = 0$ and $\alpha=0.2$ are given in both \autoref{fig:heston_comparison_skewed} and \autoref{fig:heston_comparison_symmetric}, under different correlation regimes. In \autoref{fig:skews} and \autoref{fig:curvatures}, we then take a closer look at at-the-money (ATM) skews and curvatures, showing how the RLH model appears to generate explosive power-laws for these important quantities, like the leading rough volatility models.

\begin{figure}[H]
    \centering
    \includegraphics[width=0.48\linewidth]{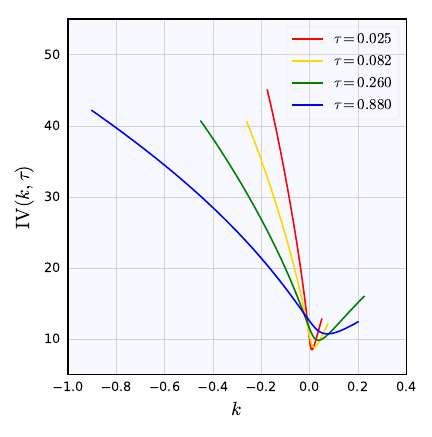}
    \includegraphics[width=0.48\linewidth]{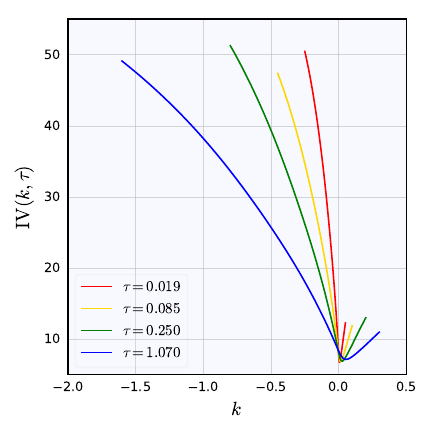}
    \caption{Implied volatilities from the modified RLH model in \autoref{eq:rough_heston_RLH}, which compare to those of rough Heston. Parameters are $H=0.1216,\nu=0.2910,\rho=-0.6714$ in the left panel and $H=0.0474,\nu=0.4061,\rho=-0.6710$ in the right.}
    \label{fig:rough_heston_comparison}
\end{figure}

We go on to speculate that the RLH model is similar to the celebrated rough Heston model, first defined in \cite{El_Euch_2019}. This speculation is justified by \autoref{fig:rough_heston_comparison}, which displays similar implied volatilities to those from the rough Heston models in \cite{El_Euch_2019b}. To aid this comparison, we first write down a reduced version of the RLH model in \autoref{eq:rough_heston_RLH} which prioritises the three rough Heston parameters, $H$, $\nu$ and $\rho$. For now this similarity remains empirical, however. This is because in order to draw these comparisons we must set the RLH fractional derivative close to its upper bound of 0.5, and more numerical evidence is required until we can be sure that our relatively simple forward Euler-based simulation scheme from \autoref{def:rlh_polygon} is still sufficiently converged.

Recall the classical Heston model from \autoref{def:heston_stoch_vol}, in which the price process $S$ verifies
\begin{equation}\label{eq:heston_plots}
    \dd V_t = \sigma\sqrt{V_t}\dd W^1_t + \kappa(\theta - V_t)\dd t,\quad V_0=v,\quad S_t := \exp\left(\int_0^t\sqrt{V_s}\dd W^\rho_s - \frac12 \int_0^t V_s\dd s\right),
\end{equation}
and recall the related RLH model from \autoref{def:RLH_model}, in which the price process $S$ verifies~
\begin{equation}\label{eq:RLH_plots}
    X'_t = \sigma\theta^\alpha W^\alpha_{X_t} + \kappa(\vt(t) - X_t)+v,\quad S_t :=\exp(W^\rho_{X_t}-\tfrac12 X_t),
\end{equation}
and where in both cases $W^\rho := \sqrt{1-\rho^2}W^0 + \rho W^1$. Notice the inclusion of the coefficient $\theta^\alpha$ in \autoref{eq:RLH_plots}. This helps to draw the comparison as $\alpha$ changes, and can be justified theoretically by the self-similarity of the fractional derivative process $W^\alpha$, see e.g.~\cite{Jacquier_2018}. Theory aside, the last ($\tau=2$) panels of \autoref{fig:heston_comparison_skewed} and \autoref{fig:heston_comparison_symmetric} show obvious similarities, and so a clearer comparison is possible for other panels, i.e.~for earlier maturities.

For simplicity we set $v=\theta$, so that the expectations $\EE[\int_0^t V_s \dd s]=-2\EE[\log S_t] = \theta t$ are linear in time under the Heston model. The RLH curve $\vt$ is then sought numerically so that the analogous relationship $\EE[X_t]=-2\EE[\log S_t] = \theta t$ holds, and so that all implied volatilities have the value $\sqrt{\theta}$ on average. (`Average' can be made precise; see e.g.~Figure 9 in \cite{McCrickerd_2018} and the related discussion.) By succeeding in finding such a $\vt$, using \autoref{eq:RLH_plots} and Tonelli's theorem for $\EE[X'_t]=\theta$ we obtain the representation
\begin{equation}\label{eq:theta_curve}
    \vt(t) = \theta t - \frac{\sigma \theta^\alpha}{\kappa}\EE[W^\alpha_{X_t}].
\end{equation}
We observe no changes in output when utilising \autoref{eq:theta_curve} to obtain $\vt$ on the fly \emph{during} a simulation, which removes $\vt$ as an input to the model whenever we are instead given a target `forward variance' curve $\xi(t)=\EE[X'_t]$, such as $\xi(t)=\theta$ here. We note that $\EE[W^\alpha_{X_t}]\neq 0$ for $t>0$ and $\alpha\neq0$ (optional stopping theory only applies when $\alpha=0$, given $W^\alpha$ is not a local martingale otherwise). But empirically we observe that $-\EE[W^\alpha_{X_t}]$ is strictly increasing when $\alpha\in(0,\frac12)$, so that $\vt$ in \autoref{eq:theta_curve} is certainly strictly increasing, as required for the RLH model to exist in our frameworks and to have a unique (strong) solution by \autoref{thm:rand_well_posed}.

In \autoref{fig:heston_comparison_skewed} we set $\rho=-0.7$ so that price processes are strongly negatively correlated to their volatility, as is usually the case in equity markets. In \autoref{fig:heston_comparison_symmetric} we instead set $\rho=0$, which is more applicable to FX markets. In both cases $\sigma=\kappa=0.2$ and  $\theta=v=0.04$, $\alpha$ is either 0 or $0.2$, and we show maturities ranging from a week ($\tau=1/52$) to two years ($\tau=2$). Using the scheme in \autoref{thm:rlh_convergence}, a separate simulation with $4,096$ paths is run in \texttt{python} for each maturity, with temporal and spatial step sizes of $\tau/512$ and $\theta\tau/512$ respectively. 

We always obtain implied volatilities from put option payoffs $\#(S) := \max\{K-S_T,0\}$, as suggested following \autoref{cor:vol_converge}. This convergence result thus applies with $\Psi=\mathrm{IV}=\mathrm{BS}^{-1}$, and $\mathrm{BS}$ as in \autoref{eq:bs_formula}. Following convention, we present implied volatilities in log-strike $k:=\log(K)$ space, and scale them up by 100. Log-strikes are selected which return a `delta' $\mathrm{N}(-d_+)$ from \autoref{eq:bs_formula} roughly in the interval $(0.005,0.995)$, so that our strike range always roughly captures $99\%$ of simulated prices. We utilise the variance reduction techniques recommended in \cite{McCrickerd_2018}. As reported there, we find that these techniques impart negligible statistical bias on estimated data, so we have not reported these biases here. Finally note that the classical Heston data points are obtained via this model's characteristic function and numerical integration, following \cite{Gatheral_2006}.

A simplified version of the code used to simulate the RLH model via the scheme in \autoref{def:rlh_polygon} is given in the \hyperlink{appendix}{Appendix} for additional clarity, with a price path shown in \autoref{fig:simulation_output}. As discussed at the end of \autoref{sec:simulation}, this code takes 75 ms to run, and we find that the use of 4,096 paths here is sufficient to bring all of the $\alpha=0$ RLH implied volatilities in \autoref{fig:heston_comparison_skewed} and \autoref{fig:heston_comparison_symmetric} within 0.1 of the numerically integrated Heston counterparts.

\vspace{-1mm}

\vspace{4mm}

\vspace{-4mm}

In both \autoref{fig:heston_comparison_skewed} and \autoref{fig:heston_comparison_symmetric}, the RLH implied volatilities clearly coincide with those of the classical Heston model when $\alpha=0$, validating \autoref{thm:heston_ode}. Given that all implied volatilities are similar for the two year maturity, the effect on shorter maturities when increasing $\alpha$ to 0.2 is also clear: in \autoref{fig:heston_comparison_skewed}, we observe increasingly pronounced skews as maturities fall to a week, and in \autoref{fig:heston_comparison_symmetric}, we observe increasingly pronounced curvatures.

Now in \autoref{fig:skews} and \autoref{fig:curvatures} we approximate (at-the-money) skews and curvatures by finite difference, defined for each maturity $\tau$ via the following absolute partial derivatives
\begin{equation}\label{eq:skew_curv}
    \mathrm{skew}(\tau):=\left|\frac{\partial\mathrm{IV}(k,\tau)}{\partial k}\right|_{k=0},\quad \mathrm{curvature}(\tau):=\left|\frac{\partial^2\mathrm{IV}(k,\tau)}{\partial k^2}\right|_{k=0}.
\end{equation}

\vspace{-4mm}

\vspace{-8mm}

In \autoref{fig:skews} and \autoref{fig:curvatures} we hope to observe power law skews and curvatures, like those generated by leading rough volatility models, and considered a `stylised fact' of equity markets. We thus include power laws of type $\tau^{-\alpha}$ and $\tau^{-2\alpha}$ respectively, which are those predicted by the theory of \cite{Alos_2007} and \cite{Alos_2017} when translating the Hurst parameter $H$ there to our fractional derivative $\alpha$ via H\"older regularities, i.e.~$H=0.5-\alpha$. Despite these power laws being (short-time) approximations themselves, similarities between them and our finite difference RLH skews and curvatures are still clearly evident, suggesting that the RLH model indeed behaves like the leading rough volatility models in this respect.

Now in \autoref{fig:rough_heston_comparison}, we simulate implied volatilities using a modified RLH model, which are similar to those from Figures 1 and 2 in \cite{El_Euch_2019b}, i.e.~the rough Heston model. 

\vspace{-4mm}

Contrasting the classical Heston variance process in \autoref{eq:heston_plots}, the rough Heston counterpart in \cite{El_Euch_2019b} is a weak solution of the singular stochastic Volterra equation~
\begin{equation}\label{eq:rough_heston}
    V_t = \xi(t) + \frac{\nu}{\Gamma(H+\frac12)}\int_0^t(t-s)^{H-\frac12}\sqrt{V_s}\dd W^1_s,
\end{equation}
for a forward variance curve $\xi(t)=\EE[V_t]$, $H\in(0,\frac12)$ and $\nu>0$. For a comparison with the RLH model, we thus modify the cumulative variance process $X$ in \autoref{def:RLH_model} to solve
\begin{equation}\label{eq:rough_heston_RLH}
    X'_t = \vt(t) + \nu W^{\frac12-H}_{X_t}\ \  \iff\ \  V_t = \vt(t) + \frac{\nu}{\Gamma(H+\frac12)}\left(\int_0^{\cdot}(\cdot-x)^{H-\frac12}\dd W^1_x\right)_{\int_0^t V_s\dd s},
\end{equation}
where $V:=X'$, we allow $\vt(0)>0$, we have prioritised the rough Heston Hurst parameter $H=\frac12-\alpha$, and have removed the drift component $-X_t$ from \autoref{eq:RLH_model_summary} entirely. We note that this model has a unique (strong) solution by \autoref{thm:rand_well_posed} provided $\vt$ is strictly increasing. However, our martingality result depending on \autoref{cor:gauss_heston_mgf} no longer applies given this drift $-X_t$ is removed. This is practically irrelevant, however, because it \emph{does} apply if a drift $-\ep X_t$ is included in the r.h.s.~of \autoref{eq:rough_heston_RLH} for \emph{any} $\ep>0$, e.g.~$\ep:=2^{-100}$.

There are clear similarities between the rough Heston model in \autoref{eq:rough_heston} and our modified RLH model in \autoref{eq:rough_heston_RLH}. This is validated by \autoref{fig:rough_heston_comparison}, especially because this figure is produced \emph{not} by \emph{calibrating} our parameters $H,\nu,\rho$ to replicate the rough Heston output, but by simply adopting the rough Heston parameters from \cite{El_Euch_2019b}. Note however that the RLH model produces higher implied volatilities in the left tails in \autoref{fig:rough_heston_comparison} in general, so there is still room for improvement through an actual calibration. Recalling \autoref{eq:theta_curve}, it remains to ensure the equivalence $\vt(t)=\xi(t) - \nu\EE[W^{\frac12-H}_{X_t}]$ to reasonable accuracy. We manage this on the fly during the simulation producing \autoref{fig:rough_heston_comparison}, but we find that curves of type $\vt(t)=\vt_0+\vt_1(1-2H)t^{2H}$ also produce reasonable output.

\vspace{2mm}

Through \autoref{fig:skews}, \autoref{fig:curvatures} and \autoref{fig:rough_heston_comparison}, we have thus provided convincing evidence that the RLH model behaves like the leading rough volatility models. However, more theoretical or numerical evidence is required to validate a relationship with the rough Heston model, given that our simulations producing \autoref{fig:rough_heston_comparison} depend on very high fractional derivatives. We clarify some potential future research regarding the RLH model at the end of \autoref{chap:conclusion}.

Finally, recall from the beginning of this chapter that our primary objective when defining the RLH model was to promote understanding for the wider volatility modelling frameworks in \autoref{fig:venn}, via the familiar \emph{classical} Heston model. These apparent \emph{rough} Heston similarities are a bonus. Unlike the rough Heston model, the RLH model exists in a framework where all models can be flexibly modified without compromising their unique strong solution, and have a continuous solution map w.r.t.~uniform convergence over compacts. Conditions for the rough Heston model (and related stochastic Volterra equations) to have a unique strong solution are still not yet known, despite attempts. The rough Heston model has a characteristic function which can be approximated, however, enabling semi-analytic pricing.

% \clearpage

\subsection{Fractional Heston-NIG limits}\label{sec:price_limits}

In this final section the plan is to apply the limiting results from \autoref{chap:solutions}, most notably \autoref{thm:exit_limit}, to the RLH model from \autoref{def:RLH_model}. After doing so, we will demonstrate some surprising classical CIR and Heston limiting results as a special case. The former will establish an entirely new connection between the time-integrated CIR process and the IG L\'evy process, a consequence of which is the weak convergence on Skorokhod's $\uM_1$ topology.

The latter results will strengthen the Heston and NIG relationship discussed at length in the \hyperlink{prologue}{Prologue}. Recall that connections were already established in \cite{Keller_2011} and \cite{Forde_2011} between the Heston process for large times and the NIG \emph{distribution}, and \autoref{thm:mech_pro} from \cite{Mechkov_2015} established the first connection between the marginal distributions of \emph{processes}. The Heston and NIG relationships here are therefore the first \emph{functional} results, illustrating how these processes are related (and not related, as it turns out) for all times simultaneously. These results are not as accessible as the CIR-related ones, with weak convergence being violated on all of Skorokhod's five topologies, for example.

To draw these conclusions on the classical CIR and Heston processes, the relationship from \autoref{thm:heston_ode} between these processes and the RLH model, when setting the fractional derivative $\alpha=0$, will be used. So recall the RLH model for an $\cF_{X_t}$-martingale price process $S=\{S_t\}_{t\in\RR_+}$ and its cumulative variance $X=\{X_t\}_{t\in\RR_+}$, summarised by the equations
\begin{equation}\label{eq:RLH_model_limit}
    X'_t = \sigma W^\alpha_{X_t} + \kappa(\vt(t) - X_t) + v,\quad S_t=\exp(W^\rho_{X_t}-\tfrac12 X_t),
\end{equation}
where we define the usual processes $W^\alpha:=D^\alpha(W^1)$ and $W^\rho:=\rho W^1+\sqrt{1-\rho^2}W^0$. We are now interested in sequences of such models which can be expressed just in terms of sequences of the implicit underlying random field in \autoref{eq:RLH_model_limit}. Specifically, we consider fields
\begin{equation}\label{eq:RLH_field_seq}
    Y^n_{t,x} := \sigma_n W^\alpha_x + \kappa_n(\vt_n(t) - x) + v_n,
\end{equation}
noting that this means the Brownian motion $(W^0,W^1)$ and parameters $\alpha,\rho$ are now fixed. 

When applying \autoref{thm:exit_limit} to such fields, the curves $\kappa_n\vt_n(t)+v_n$ play the same role provided a limit is found uniformly over compacts as $n\to\infty$, so in order to help draw direct conclusions on the classical CIR and Heston processes we will just consider the (classical) cases of $\vt_n(t):=\theta t$ and $v_n:=v$ for $\theta,v>0$. Fields in \autoref{eq:RLH_field_seq} can thus be expressed
\begin{equation}\label{eq:RLH_model_limit2}
    Y^n_{t,x} := \sigma_n W^\alpha_x + \kappa_n(\theta t - x) + v
\end{equation}
and in the results which follow, the term $\theta t$ can be generalised to a limiting curve $\vt_0(t)$.

Recall from \autoref{thm:heston_ode} that when we set the fractional derivative $\alpha=0$ in \autoref{eq:RLH_model_limit2}, the distribution of the random ODE solution $X$ coincides with that of an integrated CIR process, and that of $S$ with a Heston price process. Now depending on how we let $\sigma_n$ and $\kappa_n$ scale with $n$, different, possibly discontinuous, limits will be obtained via \autoref{thm:exit_limit}. 

We are most interested here in limits like those studied in \cite{Mechkov_2015} and summarised in the \hyperlink{prologue}{Prologue}, where $\sigma_n,\kappa_n\xrightarrow{n\to\infty}\infty$ at the same rate, because we know these lead to the most informative and practically useful functional relationships between classical processes. In the \hyperlink{epilogue}{Epilogue}, clarity is provided on alternative limits deriving from the regimes of \cite{Heston_1993} and \cite{Fouque_2011}, also accommodating the case where $Y^n_{0,0} = v_n\xrightarrow{n\to\infty}\infty$.

\vspace{3mm}\textbf{The fast-reversion parameterisation.} In \cite{Mechkov_2015} a particular `fast-reversion' parameterisation of the Heston model is defined which, subject to a relabelling of the parameters $a,b,c>0$, amounts to considering the following It\^o SDEs indexed by any $n>0$~
\begin{equation}\label{eq:fast_reversion_heston}
    \dd V^n_t = na\sqrt{V^n_t}\dd W^1_t + n(b - V^n_t),\quad \dd S^n_t = \sqrt{V^n_t}S^n_t\dd W^\rho_t,\quad (V^n_0,S^n_0)=(c,1).
\end{equation}
The novelty of this parameterisation is due to the linear scaling of \emph{both} the  diffusion and reversion components of the CIR SDE for $V$ with $n$. Through an analysis of characteristic functions (given the system in \autoref{eq:fast_reversion_heston} is affine), the convergence in distribution $S^n_t\xrightarrow{\mathrm{d}}S^0_t$ is obtained as $n\to\infty$ for any $t>0$, where $S^0$ is an exponentiated NIG L\'evy process, with parameters depending only on $a,b$ and $\rho$ and no longer $c$. This result is summarised by \autoref{thm:mech_pro}, will be confirmed by \autoref{cor:heston_fd} then extended in \autoref{cor:heston_hausdorff}. Now we want to parameterise a sequence of RLH models in a similar `fast-reversion' way.

Considering how the RLH model is connected with the classical Heston model, through \autoref{thm:heston_ode}, such a parameterisation like that in \autoref{eq:fast_reversion_heston} is achieved by utilising fields $Y^n$ as in \autoref{eq:RLH_model_limit2} with $\sigma_n:=na$ and $\kappa_n:=n$. This leads simply to the following.

\begin{definition}[Fast-reversion RLH parameterisation]\label{def:fast_RLH}
    In the RLH model from \autoref{def:RLH_model}, set $\sigma=na$, $\kappa=n$, $\vt(t)=b t$ and $v=c$ for some $n,a,b,c>0$, so that the RLH processes $(X^n,S^n)$ are the unique processes over $\RR_+$ which verify the defining equations
    \begin{equation}\label{eq:RLH_model_summary3}
        {X^n_t}' = n\left( a W^\alpha_{X^n_t} + b t - X^n_t\right) + c,\quad S^n_t=\exp\left(W^\rho_{X^n_t}-\tfrac12 X^n_t\right),
    \end{equation}
    for some fixed $\alpha\in(0,\frac12)$ and $\rho\in[-1,1]$. We will say that such an RLH model is in its fast-reversion parameterisation, and will call $n\to\infty$ the RLH model's fast-reversion limit.
\end{definition}

Now our main focus is on establishing a.s.~functional limits of this model via the probability-free results of \autoref{chap:solutions}. Then, by setting $\alpha=0$, these will immediately provide weak limits for the Heston model as in \autoref{eq:fast_reversion_heston}. Although we are interested in a.s.~limits of the RLH model in their own right, notice how the approach here to weak convergence for the classical Heston model contrasts the usual `Prokhorov approach' via finite-dimensional distributions and tightness, as summarised succinctly e.g.~in \cite{Jacod_2003}.

On all topologies from \cite{Skorokhod_1956}, a naive application of Prokhorov's approach to the Heston price process in \autoref{eq:fast_reversion_heston} is doomed, because the functional limits established here go via \autoref{cor:graph_hausdorff}; these are not continuous process limits, not c\`adl\`ag processes, but compact \emph{interval-valued} processes, with paths $\ve(t)=:[\ve_-(t),\ve_+(t)]$ in the set $\uE$ from \autoref{def:excursionary_paths}. Such processes are studied in Chapter 15 of \cite{Whitt_2002}, and their emergence here in finance is not just theoretically fascinating, but practically valuable, given they characterise unexpected behaviour of path-dependent derivatives, discussed in \autoref{chap:conclusion}.

\vspace{3mm}\textbf{Preparatory results.} All of the stochastic process limits of this section derive from the probability-free results from \autoref{sec:bounds_limits} and \autoref{sec:excursionary_limits}. Specifically, we will apply the exit-time and Hausdorff results of \autoref{thm:exit_limit} and \autoref{cor:graph_hausdorff}. So that the application of these results are clear, we first clarify their consequences in the probabilistic setting here.

Recall the set $\bPhi\subset\uD(\RR_+,\RR_+)$ from \autoref{def:setE}, containing the strictly increasing and unbounded c\`adl\`ag paths, and let $d_{\bPhi}$ be the exit-time metric from \autoref{def:exit_metric} satisfying $d_{\bPhi}(\bvp_1,\bvp_2) = \Vert E(\bvp_2) - E(\bvp_1)\Vert_{\RR_+}$. For clarity, $E$ is the exit-time functional from \autoref{def:exit_functional}, and $\Vert\cdot\Vert_{\RR_+}$ is the norm from \autoref{def:uni_metric}, characterising uniform convergence over compacts. Finally recall the set $\uG\subset\uC(\RR_+^2,\RR)$ from \autoref{def:mod_driving_func}, featuring in \autoref{prob:random_ivp}.

The proof of this next result is not given because it is identical to that of point 3.~in \autoref{thm:rand_well_posed}, only replacing the pathwise application of \autoref{thm:wellposed} with that of \autoref{thm:exit_limit}.

\begin{corollary}[Uniform exit-time limits]\label{cor:exit_limit}
    Let $\{Y^n\}_{n\in\NN_0}$ be random fields in $\uG$, let $\{X^n\}_{n\in\NN}$ solve the random IVPs $x'=nY^n_{t,x}$, $x_0=0$ and define $X^0\in\bPhi$ by $X^0_t:=\inf\{x>0:Y^0_{t,x}<0\}$. If $Y^n \xrightarrow[n\to\infty]{\mathrm{a.s.}}Y^0$ uniformly over compacts, then $X^n \xrightarrow[n\to\infty]{\mathrm{a.s.}} X^0$ on the exit-time space $(\bPhi,d_{\bPhi})$.
\end{corollary}

As covered by \autoref{thm:m1_relationship} and \autoref{cor:ae_converge}, recall that this convergence $X^n \to X^0$ on $(\bPhi,d_{\bPhi})$ is stronger than the same on Skorokhod's $\uM_1$ space (defined via the metric in \autoref{eq:m1_metric}) as well as providing the a.s.~pointwise convergence for (Lebesgue) a.e.~$t\in\RR_+$.

Now \autoref{cor:graph_hausdorff} translated into our probabilistic setting provides the following on $(\uE,d_\uE)$, where we recall from \autoref{def:excursionary_paths} that $\uE$ is the set of compact interval-valued paths over $\RR_+$, and $d_\uE$ the Hausdorff metric on this set, defined via the pseudometrics in \autoref{eq:hausdorff_dist}. In the following, we allow processes $\Lambda_X\in\uE$ to return the singleton $\{\Lambda_{X_t}\}$ for each $t\in\RR_+$.

\begin{corollary}[Hausdorff composite limits]\label{cor:hausdorff_random}
    Adopt the assumptions of \autoref{cor:exit_limit}, so that $X^n \xrightarrow[n\to\infty]{\mathrm{a.s.}} X^0$ on $(\bPhi,d_{\bPhi})$, and let $\Lambda=\{\Lambda_x\}_{x\in\RR_+}$ be any process in $\uC(\RR_+,\RR)$. Then the composition processes $\{\Lambda\circ X^n:=\Lambda_{X^n}\}_{n\in\NN}$ verify $\Lambda\circ X^n \xrightarrow[n\to\infty]{\mathrm{a.s.}} \Lambda\bullet X^0$ on $(\uE,d_\uE)$, where
    \begin{equation}
        (\Lambda\bullet X^0)_t := \{\Lambda_x:x\in[X^0_{t_-},X^0_t]\}.
    \end{equation}
\end{corollary}

Recall that the proof of \autoref{cor:graph_hausdorff}, on which \autoref{cor:hausdorff_random} here depends, goes via \autoref{thm:para_uniform}, which is not just a graphical Hausdorff convergence result, but a product convergence result for \emph{specific} parametric representations which \emph{generate} these graphs. The corresponding product statement giving \autoref{cor:hausdorff_random} here is $((X^n)^{-1},\Lambda)\to(E(X^0),\Lambda)$ uniformly over compacts. Although stronger, this does not lead to direct statements on our models, but rather on higher-dimensional representations of them. These representations may be helpful in the future, but for now we prefer to prioritise the likes of \autoref{cor:hausdorff_random}.

Now we are ready to apply these results to understand the fast-reversion limit of the RLH processes $X^n$ and $S^n$ in \autoref{def:fast_RLH}. Indeed, similarities between $\Lambda\bullet X$ in \autoref{cor:hausdorff_random} and our NIG generalisation $S^\bullet$ from \autoref{eq:ex_nig_pro} in the \hyperlink{prologue}{Prologue} should already be evident.

\vspace{3mm}\textbf{Cumulative variance limits.} We now characterise a limit $X^0$ of a sequence of the processes $X^n$ from \autoref{def:fast_RLH} as $n\to\infty$, i.e.~the RLH fast-reversion limit. \autoref{cor:exit_limit} will be applied to establish these, so despite each process $X^n$ being differentiable, the limit $X^0$ will exhibit discontinuities, like the following generalisation of the IG L\'evy process. Recall the Riemann-Liouville (RL) fractional derivative process $W^\alpha$ from \autoref{def:RL_process}.

\begin{definition}[Fractional IG process]\label{def:frac_IG}
    For $a,b>0$ and $\alpha\in(0,\frac12)$, define the RL fractional process $W^\alpha:=D^\alpha(W^1)$ as usual and the process $X^0=\{X^0_t\}_{t\in\RR_+}$ by the exit-time~
    \begin{equation}\label{eq:frac_IG}
        X^0_t :=\inf\big\{x>0 : x - a W^\alpha_x > bt\big\}.
    \end{equation}
    Such a process $X^0$ will be called a fractional IG process of order $\alpha$, with parameters $a,b$.
\end{definition}

By defining $\delta:=a^{-1}b$ and $\gamma:=a^{-1}$, this fractional IG process coincides precisely with the classical IG L\'evy process with parameters $\delta,\gamma>0$ as defined in \cite{Applebaum_2009}, \emph{when} $\alpha=0$ (so when $W^\alpha$ is Brownian motion), which has MGF $e^{\delta(\gamma-\sqrt{\gamma^2-2u})t}$ in general, so $\EE[e^{pX^0_t}] = e^{ba^{-2}(1-\sqrt{1-2a^2u})t}$ in our case. As already discussed, this process could be further generalised via a suitable curve $\vt(t)$ in place of the linear exit-barrier $bt$ in \autoref{eq:frac_IG}.

Like the classical IG process, the fractional IG process has strictly increasing and unbounded c\`adl\`ag paths, but remains finite over $\RR_+$, so is a.s.~in $\bPhi$. This follows e.g.~from \autoref{lem:zeros2}, and \autoref{cor:gauss_heston_mgf}. This latter result actually establishes the MGF existence $M^0_X(p,t)=\EE[e^{pX^0_t}]<\infty$ for all $(p,t)\in\RR\times\RR_+$, from which a.s.~finiteness $X^0_t<\infty$ of course follows. Now applying \autoref{cor:exit_limit} to the RLH model yields the fractional IG process as follows.

\begin{corollary}[Fractional IG limits]\label{cor:frac_IG_limit}
    Let $\{X^n\}_{n\in\NN}$ be a sequence of RLH processes as in \autoref{def:fast_RLH}, and let $X^0$ be the fractional IG process from \autoref{def:frac_IG}, so that
    \begin{equation}\label{eq:frac_ig_sum}
        {X^n_t}' = n\left( a W^\alpha_{X^n_t} + b t - X^n_t\right) + c,\quad X^0_t =\inf\big\{x>0 : x - a W^\alpha_x > bt\big\}.
    \end{equation}
    Then the convergence $d_{\bPhi}(X^n,X^0)=\Vert E(X^0) - (X^n)^{-1}\Vert_{\RR_+}\asc 0$ takes place as $n\to\infty$.
\end{corollary}
\begin{proof}
    The processes $\{X^n\}_{n\in\NN}$ each solve the random IVPs $x'=n Y^n_{t,x}$, $x_0=0$, where
    \begin{equation}
        Y^n_{t,x} := a W^\alpha_x + b t - x + n^{-1}c,
    \end{equation}
    and clearly $Y^n\asc Y^0$ uniformly over compacts of $\RR_+^2$ as $n\to\infty$, where $Y^0_{t,x}:=a W^\alpha_x + b t - x$. All of the RLH fields $\{Y^n\}_{n\in\NN_0}$ are moreover in $\uG$, as confirmed following the MGF existence in \autoref{cor:rlh_mgf_exist}. So the assumptions of \autoref{cor:exit_limit} hold, and we therefore obtain the convergence $X^n \asc X^0$ on $(\bPhi,d_{\bPhi})$, where $X^0\in\bPhi$ is defined by $X^0_t:=\inf\{x>0:Y^0_{t,x}<0\}$. Since this expression for $X^0$ coincides with that in \autoref{eq:frac_ig_sum}, the proof is complete.
\end{proof}

Now recall from \autoref{thm:m1_relationship} that convergence on the exit-time space $(\bPhi,d_{\bPhi})$ is stronger than on Skorokhod's $\uM_1$ space, essentially because the former considers only distances \emph{in time} between paths, rather than both time and space. In turn, like in \autoref{cor:ae_converge}, we get a.s.~(Lebesgue) a.e.~pointwise convergence. That is, for all times $T\in\RR_+$, we a.s.~have
\begin{equation}
    \mathrm{Leb}\left[t\in[0,T]: X^n_t\xrightarrow{n\to\infty}X^0_t\right]=T.
\end{equation}

Finally note that the limiting exit-time process $E(X^0)$ appearing in \autoref{cor:frac_IG_limit} was analysed in \cite{Vellaisamy_2017} in the non-fractional case $\alpha=0$. Using the relationship $E\circ E=M$ noted in \autoref{lem:whitt}, this may be equivalently expressed using the maximal functional from \autoref{eq:maximal_func}, as $E(X^0)=b^{-1}M(\id-aW^\alpha)$, i.e.~$E(X^0)_x=b^{-1}\max_{u\in[0,x]}\{u-aW^\alpha_u\}$. The inverses $\{(X^n)^{-1}\}_{n\in\NN}$ in \autoref{cor:frac_IG_limit} a.s.~find this maximal limit uniformly over compacts, which can be observed on a pathwise basis e.g.~in \autoref{fig:exit_converge}.

\vspace{3mm}\textbf{Classical integrated CIR limits.} Now we clarify what \autoref{cor:frac_IG_limit}, connecting the RLH model with the fractional IG process, means for the classical CIR process in the Heston model from \autoref{eq:fast_reversion_heston}. Given the popularity of the CIR and IG processes in \autoref{cor:IG_limit}, it is surprising that even the 1d reduction of \autoref{cor:IG_limit_fd} is new, despite a large-time connection between a CIR process and IG distribution being known since \cite{Tse_2013}. Accordingly, these results demonstrate our pathwise ODE-based framework's ability to teach us surprising new results about already much-analysed stochastic processes. 

\begin{corollary}[Inverse-Gaussian exit-time limits]\label{cor:IG_limit}
    Let $\{V^n\}_{n\in\NN}$ be a sequence of CIR processes as in \autoref{eq:fast_reversion_heston}, and define $\{X^n\}_{n\in\NN}$ respectively by the time-integrals $X^n_t:=\int_0^t V^n_s\dd s$. Define also the IG process $X^0$ as in \autoref{def:frac_IG} with $\alpha=0$ so, in summary,
    \begin{equation}
        \dd V^n_t = na \sqrt{V^n_t}\dd W^1_t + n(b - V^n_t)\dd t,\quad V^n_0=c,\quad X^0_t =\inf\big\{x>0 : x - a W^1_x > bt\big\}.
    \end{equation}
    Then the weak convergence $X^n\cw X^0$ takes place on the exit-time metric space $(\bPhi,d_{\bPhi})$.
\end{corollary}
\begin{proof}
    To avoid a clash of notation, denote by $\{\tX^n\}_{n\in\NN}$ the processes $\{X^n\}_{n\in\NN_0}$ from \autoref{cor:frac_IG_limit} when setting $\alpha=0$, so we obtain $\tX^n \asc X^0$ on $(\bPhi,d_{\bPhi})$. As shown in \autoref{thm:heston_ode}, we then have the equivalence $\tX^n\ed X^n$ in distribution for every $n\in\NN$. (Given we have adopted the parameterisation in \autoref{def:fast_RLH}, note the parameter relationships there, e.g.~$\sigma=na$.) So from $\tX^n \asc X^0$ on $(\bPhi,d_{\bPhi})$ we obtain $X^n\cw X^0$ as claimed.
\end{proof}

This convergence on $(\bPhi,d_{\bPhi})$ is equivalent to the weak convergence $E(X^n)\cw E(X^0)$ of exit-times w.r.t.~uniform convergence over compacts. Like with \autoref{cor:frac_IG_limit}, convergence on Skorokhod's $\uM_1$ space takes place as a consequence, and now it is natural, and practically relevant, to ask whether we also have convergence of finite-dimensional distributions. This will be denoted $X^n\cfd X^0$ as $n\to\infty$, which means the weak convergence $(X^n_{t_1},\dots,X^n_{t_d})\cw(X^0_{t_1},\dots,X^0_{t_d})$ takes place for any $\{t_k\}_{k=1}^d\subset\RR_+$ of dimension $d\in\NN$.

As demonstrated in Chapter 13 of \cite{Billingsley_1999}, this \emph{does not} follow even from weak convergence on Skorokhod's stronger $\uJ_1$ space, but does if the processes of concern have the property of stochastic continuity, i.e.~$\PP[X_{t_-}= X_t]=1$, where $X_{t_-}:=\lim_{s\uparrow t}X_s$ as usual, and in our setting $X_{0_-}:=0$. This next result shows that $X^n\cfd X^0$ as $n\to\infty$ similarly holds in our setting, provided $X^0$ is stochastically continuous, like any L\'evy process. 

For the proof, recall that $\PP[X_{t_-}= X_t]=1$ provides also $\PP[X_{{t_k}_-}= X_{t_k},k=1,\dots,d]=1$ for any finite $\{t_k\}_{k=1}^d\subset\RR_+$, which can be proved using the basic manipulations in \autoref{eq:full_measure}. Also recall from \autoref{cor:ae_converge} that convergence $\bvp_n\to\bvp_0$ on $(\bPhi,d_{\bPhi})$ provides also the pointwise convergence $\bvp_n(t)\to\bvp_0(t)$ for any point of continuity for $\bvp_0$, which is a.e.~at least. The same was shown in \cite{Skorokhod_1956} to hold for all metrics defined there.

\begin{corollary}[Inverse-Gaussian f.d.~limits]\label{cor:IG_limit_fd}
    Supplementing \autoref{cor:IG_limit}, the convergence $X^n\cfd X^0$ of finite-dimensional distributions over $\RR_+$ also takes place as $n\to\infty$.
\end{corollary}
\begin{proof}
    Let $\{\tX^n\}_{n\in\NN}$ be as in the proof of \autoref{cor:IG_limit}, so that $\tX^n \asc X^0$ on $(\bPhi,d_{\bPhi})$ and $\tX^n\ed X^n$ for each $n\in\NN$. Recall that each $\tX^n$ is differentiable, and $X^0$ is an inverse-Gaussian L\'evy process, thus stochastically continuous. Fixing any finite set $\{t_k\}_{k=1}^d\subset\RR_+$, we therefore have $\PP[X^0_{{t_k}_-}= X^0_{t_k},k=1,\dots,d]=1$.  Since convergence on $(\bPhi,d_{\bPhi})$ provides convergence on $(\RR,|\!\cdot\!|)$ at points of continuity, then from $\tX^n \asc X^0$ on $(\bPhi,d_{\bPhi})$ and the a.s.~continuity $\PP[X^0_{{t_k}_-}= X^0_{t_k},k=1,\dots,d]=1$ we get $(\tX^n_{t_1},\dots,\tX^n_{t_d})\asc(X^0_{t_1},\dots,X^0_{t_d})$ on $(\RR^d,|\cdot|)$. Given $\tX^n\ed X^n$, this provides $(X^n_{t_1},\dots,X^n_{t_d})\cw(X^0_{t_1},\dots,X^0_{t_d})$. So the claim of $X^n\cfd X^0$ is established by definition, given the finite set $\{t_k\}_{k=1}^d\subset\RR_+$ is arbitrary.
\end{proof}

It is of course possible to verify \autoref{cor:IG_limit_fd}, given the integrated CIR and IG processes are affine, so have closed form MGF representations. Actually doing so in the 1d case is similar to the proof via MGFs given for \autoref{lem:int_cir_unbound}. For this let the processes $\{X^n\}_{n\in\NN_0}$ be those in \autoref{cor:IG_limit}, and for $2a^2p<1$ and $t>0$ define the MGFs $M^n_X(p,t):=\EE[e^{pX^n_t}]$. We then obtain $M^n_X(p,t) = e^{\vp^n_0(t)+\vp^n_1(t)c}$, where for $n\in\NN$ and $\lambda:=\sqrt{1-2a^2 p}>0$ we find
\begin{equation}\label{eq:IG_mgf1}
    \vp^n_0(t):= \frac{bt}{a^2}- \frac{2b}{a^2 n}\log\left(\cosh\left(\frac{n \lambda t}{2}\right) + \frac{1}{\lambda}\sinh\left(\frac{n \lambda t}{2}\right)\right),\ \vp^n_1(t):= \frac{2pn^{-1}}{1 + \lambda\coth\left(\frac{n \lambda t}{2}\right)}.
\end{equation}
Subject to redefining parameters, these expressions coincide with those given in \autoref{eq:int_cir_char_func}. From \autoref{eq:IG_mgf1} we can see $\vp_1^n(t)\xrightarrow{n\to\infty}\vp_1^0(t) := 0$ provided $\lambda>0$, which is ensured by $2a^2p<1$. Using similar expansions to those given in \autoref{lem:int_cir_unbound}, we also find
\begin{equation}
    \frac{2}{n}\log\left(\cosh\left(\frac{n \lambda t}{2}\right) + \frac{1}{\lambda}\sinh\left(\frac{n \lambda t}{2}\right)\right)\xrightarrow{n\to\infty}\lambda t.
\end{equation}
So in full we find $\vp_0^n(t)\xrightarrow{n\to\infty}\vp_0^0(t) := ba^{-2}(1 - \lambda)t$, and the resulting MGF limit $M^0_X(p,t) = e^{\vp^0_0(t)+\vp^0_1(t)c} = e^{ba^{-2}(1-\lambda)t} = e^{ba^{-2}(1-\sqrt{1-2a^2 p})t}$ is that of the IG random variable $X^0_t$ from \autoref{cor:IG_limit_fd}, as clarified after \autoref{def:frac_IG}. This reconciles \autoref{cor:IG_limit_fd} in the 1d case, and doing so in higher dimensions is possible via induction, although rather tedious. This contrasts our proof of \autoref{cor:IG_limit_fd}, which can even be visualised like in \autoref{fig:exit_converge}.

Since this result relating the classical integrated CIR and IG processes is just one of the limits arising when applying \autoref{thm:exit_limit}, we clarify in the \hyperlink{epilogue}{Epilogue} all the other limits that can arise when the CIR processes are parameterised differently. These include the \emph{L\'evy} L\'evy process which, as \cite{Applebaum_2009} shows, can be considered a special case of the IG L\'evy process, and both of these L\'evy processes can also arise with random starting points.

\vspace{3mm}\textbf{Price process limits.} In this part we will conduct the same type of analysis as the previous but for the RLH \emph{price} processes $S^n$ in \autoref{def:fast_RLH}. Then in the next part, we will reduce this to consequences for the classical Heston model from \autoref{eq:fast_reversion_heston}, finally strengthening \autoref{thm:mech_pro} as far as we deem possible and thereby answering our questions in the \hyperlink{prologue}{Prologue}. 

As with the fractional IG process from \autoref{def:frac_IG}, we first define \emph{two} candidate limits of the RLH price processes. The senses in which these generalise the classical NIG process, see e.g.~\cite{Barndorff_Nielsen_2001}, \cite{Cont_2003} or \cite{Applebaum_2009}, will be clarified in the next part, when these limits are related to the classical Heston model.

\begin{definition}[Fractional NIG c\`adl\`ag process]\label{def:frac_NIG1}
    Let $X^0$ be the fractional IG process from \autoref{def:frac_IG}, then define the process $S^\circ=\{S^\circ_t\}_{t\in\RR_+}$ as in \autoref{eq:RLH_model_summary3}. So in full,
    \begin{equation}
        X^0_t:=\inf\big\{x>0:x-aW^\alpha_x>bt\big\},\quad S^\circ_t:=\exp\left(W^\rho_{X^0_t}-\tfrac12 X^0_t\right),
    \end{equation}
    for $a,b>0$, $\alpha\in(0,\frac12)$, $\rho\in[-1,1]$, with $W^\alpha:=D^\alpha(W^1)$ and $W^\rho:=\rho W^1+\sqrt{1-\rho^2}W^0$ as usual. $S^\circ$ will be called a fractional NIG \emph{c\`adl\`ag} process of order $\alpha$, with parameters $a,b,\rho$.
\end{definition}

Given $\exp(W^\rho_x-\tfrac12 x)$ is a continuous process over $\RR_+$ and $X^0$ a strictly increasing c\`adl\`ag process over $\RR_+$ with $X^0_0=0$, it is clear that $S^\circ$ is indeed a c\`adl\`ag process over $\RR_+$, with $S^\circ_0=1$. Note that an alternative `Fractional NIG' process is studied in the line of research from \cite{Kumar_2012} to \cite{Wylomanska_2016}, considered for several applications in statistical physics. In this alternative case, fractional Brownian motion is being subordinated, whereas our fractional process $W^\alpha$ is hidden \emph{within} our volatility-related subordinator $X^0$. So the emergence of our candidate limit $S^\circ$ from a sequence of martingales remains plausible. We now define a related \emph{interval-valued} process.

\begin{definition}[Fractional NIG excursion process]\label{def:frac_NIG2}
    Let $a,b,\alpha,\rho,W^0,W^1$ and $X^0$ be as in \autoref{def:frac_NIG1}, but define the real interval-valued process $S^\bullet=\{S^\bullet_t\}_{t\in\RR_+}$ instead using
    \begin{equation}\label{eq:frac_NIG_ex}
        S^{\bullet}_t:=\left\{\exp\left(W^\rho_x-\tfrac12 x\right):x\in[X^0_{t_-},X^0_t]\right\}.
    \end{equation}
    Then $S^\bullet$ will be called a fractional NIG \emph{excursion} process of order $\alpha$, with parameters $a,b,\rho$.
\end{definition}

Again, by the continuity of $\exp(W^\rho_x-\tfrac12 x)$, each $S^\bullet_t$, for $t\in\RR_+$, defines a random \emph{closed subinterval} of $\RR$, not just a random subset. Indeed, we have the equivalent representation
\begin{equation}\label{eq:frac_NIG_ex2}
    S^\bullet_t=\left[\min_{x\in[X^0_{t-},X^0_t]}\exp(W^\rho_x-\tfrac12 x),\max_{x\in[X^0_{t-},X^0_t]}\exp(W^\rho_x-\tfrac12 x)\right]=:[S^-_t,S^+_t].
\end{equation}
It is clear from these expressions that these c\`adl\`ag and excursion processes satisfy $S^\circ_{t_-},S^\circ_t\in S^\bullet_t$ for each $t\in\RR_+$. We should think of $S^\bullet$ as being equivalent to $S^\circ$, only with additional instantaneous excursions attached at the times of the discontinuities of $X^0$ (when $X^0_{t-}<X^0_t$), thus $S^\circ$. The upwards excursions have length $S^+_t-S^\circ_t\ge0$, and the downward excursions $S^\circ_t-S^-_t\ge0$, and both are a.s.~zero at a fixed time, meaning $S^-_t=S^\circ_t=S^+_t$, provided $X^0$ is stochastically continuous. Like the discontinuities of any c\`adl\`ag process such as $X^0$ and $S^\circ$, these excursions are a.s.~countable along a given path, so regardless of stochastic continuity we a.s.~have $S^\bullet_t=\{S^\circ_t\}$ for a.e.~$t\in\RR_+$, meaning again $S^\circ_t=S^-_t=S^+_t$. 

Such an interval-valued process $S^\bullet$, connected with a specific c\`adl\`ag process $S^\circ$, falls beautifully into the setting of Section 15.4 in \cite{Whitt_2002}, arising in queuing theory. Like in \autoref{cor:hausdorff_random}, we will consider $S^\bullet$ as a random element of $\uE$ from \autoref{def:excursionary_paths}. Recall that $\uE$ simply contains \emph{all} real compact interval-valued paths over $\RR_+$, i.e.~not just those of $S^\bullet$ which are connected to the c\`adl\`ag paths of $S^\circ$. It is with respect to the Borel $\sigma$-algebra $\cE$ induced by the excursionary (Hausdorff) metric $d_\uE$ from \autoref{def:excursionary_metric} on $\uE$ that we can consider $S^\bullet$ a bona fide stochastic process, i.e.~a measurable map from $(\Omega,\cF,\PP)$ to $(\uE,\cE)$.

The next two results clarify notions in which a sequence $\{S^n\}_{n\in\NN}$ of RLH price processes from \autoref{def:fast_RLH} converge to the fractional NIG processes $S^\circ$ and $S^\bullet$ respectively. These results constitute straightforward applications of \autoref{cor:ae_composite} and \autoref{cor:hausdorff_random} respectively.

\begin{corollary}[A.e.~fractional NIG limits]\label{cor:price_ae}
    Let $\{S^n\}_{n\in\NN}$ be a sequence of RLH price processes from \autoref{def:fast_RLH}, and $S^\circ$ the fractional NIG c\`adl\`ag process from \autoref{def:frac_NIG1}. Then a.s., the convergence $S^n_t\xrightarrow{n\to\infty}S^\circ_t$ takes place a.e., i.e. for all $T\in\RR_+$, we a.s.~have
    \begin{equation}
        \mathrm{Leb}\left[t\in[0,T]: S^n_t\xrightarrow{n\to\infty}S^\circ_t\right]=T.
    \end{equation}
\end{corollary}
\begin{proof}
    Given $X^n\asc X^0$ on $(\bPhi,d_{\bPhi})$ from \autoref{cor:frac_IG_limit} and that $\Lambda_x:=\exp(W^\rho_x-\frac12 x)$ a.s.~has paths in $\uC(\RR_+,\RR)$, then applying \autoref{cor:ae_composite} on a pathwise basis a.s.~provides 
    \begin{equation}
        \Leb\left[t\in[0,T]:\Lambda_{X^n_t}\xrightarrow{n\to\infty}\Lambda_{X^0_t}\right]=T.
    \end{equation}
    But now just using the definitions of $S^n$, $S^\circ$ and $\Lambda$, we see that this is precisely the claim.
\end{proof}

Although this limiting result is sufficient for some applications, the following is necessary to understand the richer limiting behaviour of path-dependent derivatives. Like in \autoref{cor:hausdorff_random}, we let $S^n$ simultaneously denote the process in $\uE$ returning the singletons $\{S^n_t\}$.

\begin{corollary}[Hausdorff fractional NIG limits]\label{cor:price_hausdorff}
    Let $\{S^n\}_{n\in\NN}$ be a sequence of RLH price processes from \autoref{def:fast_RLH}, and let $S^\bullet$ be the fractional NIG excursion process from \autoref{def:frac_NIG2}. Then the convergence $S^n\asc S^\bullet$ takes place on $(\uE,d_{\uE})$ as $n\to\infty$.
\end{corollary}
\begin{proof}
    Like in \autoref{cor:price_ae}, define the process $\Lambda_x:=\exp(W^\rho_x - \frac12x)$. Then given the convergence $X^n \asc X^0$ on $(\bPhi,d_{\bPhi})$ from \autoref{cor:frac_IG_limit}, \autoref{cor:hausdorff_random} can be applied to obtain $\Lambda\circ X^n \asc \Lambda\bullet X^0$ on $(\uE,d_\uE)$. By the definitions of $S^n$ and $S^\bullet$, this is the claim.
\end{proof}

As noted following \autoref{cor:hausdorff_random}, the graphical Hausdorff result here is actually a consequence of a stronger product convergence result applicable to \emph{explicit} parametric representations of such graphs. In the setting here, we have $((X^n)^{-1},\Lambda)\asc(E(X^0),\Lambda)$ uniformly over compacts. The reduced Hausdorff statement in \autoref{cor:price_hausdorff} is prioritised given it directly applies to the RLH model, rather than a higher-dimensional representation of it.

\vspace{3mm}\textbf{Classical Heston limits.} Now we can set the fractional derivative $\alpha=0$ in the above RLH price process convergence results to establish limits of the classical Heston model from \autoref{eq:fast_reversion_heston}. It will also now become clear how the fractional NIG c\`adl\`ag and excursion processes from \autoref{def:frac_NIG1} and \autoref{def:frac_NIG2} generalise the classical NIG L\'evy process.

First recall e.g.~from \cite{Applebaum_2009} that a NIG L\'evy process $N=\{N_t\}_{t\in\RR_+}$ admits the following `variance-mean mixture' representation in terms of an IG L\'evy subordinator $X$,
\begin{equation}\label{eq:nig_rep}
    X_t:= \inf\big\{x>0:x-a W^1_x > bt\big\},\quad N_t:= \hat\alpha W^0_{X_t} + \hat\beta X_t + \hat\gamma t.
\end{equation}
This is an over-parameterised representation if $\hat\alpha, \hat\beta, \hat\gamma\in\RR$ are not restricted, e.g.~we could simply set $\hat\alpha=1$ here. To draw the clearest comparison with the Heston model, however, these parameters should be restricted like in this next result, which is straightforward but by no means obvious. Notice that the representation of $S^\circ$ in \autoref{lem:NIG_reduc} coincides with that in \autoref{eq:nig_prologue}, and depends only on the \emph{three} parameters $a,b,\rho$, unlike \autoref{eq:nig_rep}.

\begin{lemma}[Fractional NIG reduction]\label{lem:NIG_reduc}
    Let $S^\circ$ be the fractional NIG process from \autoref{def:frac_NIG1}, of fractional order $\alpha=0$. Then $S^\circ$ is an exponentiated NIG process. Specifically,~
    \begin{equation}
        S^\circ_t=\exp\left(\hat\alpha W^0_{X^0_t} + \hat\beta X^0_t + \hat\gamma t\right),\quad \hat\alpha:=\sqrt{1-\rho^2},\quad \hat\beta:=\frac{2\rho-a}{2a},\quad \hat\gamma:=-\frac{\rho b}{a}.
    \end{equation}
\end{lemma}
\begin{proof}
    First separate out the process $W^\rho:=\rho W^1+\sqrt{1-\rho^2}W^0$ in the definition $S^\circ:=\exp(W^\rho_{X^0}-\frac12 X^0)$. Although counter-intuitive, the IG process $X^0$ verifies $X^0_t-aW^1_{X^0_t}=bt$, given its definition $X^0_t:=\inf\{x>0:x-a W^1_x>bt\}$ and continuity of $W^1$. So we can replace the process $aW^1_{X^0_t}$ in $S^\circ$ by $X^0_t-bt$, and doing so we arrive at the claimed representation.
\end{proof}

This next result provides a higher-dimensional generalisation of \autoref{thm:mech_pro}. It is of course possible to verify this using MGFs, like we did in the 1d case following \autoref{cor:IG_limit_fd}. For the 1d case applicable to \emph{price} processes here, \cite{Mechkov_2015} should however be consulted.

\begin{corollary}[Heston f.d.~limits]\label{cor:heston_fd}
    Let $\{S^n\}_{n\in\NN}$ be the sequence of Heston price processes from \autoref{eq:fast_reversion_heston}, and let $S^\circ$ be the process from \autoref{def:frac_NIG1} with $\alpha=0$ (so admitting the exponentiated NIG representation in \autoref{lem:NIG_reduc}). Then $S^n\cfd S^\circ$ over $\RR_+$ as $n\to\infty$.
\end{corollary}
\begin{proof}
    Let $\{\tX^n\}_{n\in\NN}$ be the cumulative variance processes from the proof of \autoref{cor:IG_limit_fd}, so $(\tX^n_{t_1},\dots,\tX^n_{t_d})\asc(X^0_{t_1},\dots,X^0_{t_d})$ on $(\RR^d,|\cdot|)$ for any $\{t_k\}_{k=1}^d\subset\RR_+$ and $\tX^n\ed X^n$ for $n\in\NN$, where $X^n_t:=\int_0^t V_s\dd s$ are the classical Heston processes also from \autoref{cor:IG_limit_fd}. Define the process $\Lambda$ as usual by $\Lambda_x:=\exp(W^\rho_x-\frac12 x)$ and $\tS^n:=\Lambda_{\tX^n}$, recalling that $\tS^n\ed S^n$ by \autoref{thm:heston_ode}. Then by the continuity of $\Lambda$ we obtain $(\tS^n_{t_1},\dots,\tS^n_{t_d}) := (\Lambda_{\tX^n_{t_1}},\dots,\Lambda_{\tX^n_{t_d}})\asc(\Lambda_{X^0_{t_1}},\dots,\Lambda_{X^0_{t_d}}) =: (S^\circ_{t_1},\dots,S^\circ_{t_d})$ on $(\RR^d,|\cdot|)$. Now given that $(\tS^n_{t_1},\dots,\tS^n_{t_d})\ed(S^n_{t_1},\dots,S^n_{t_d})$ for every $n\in\NN$, this provides $(S^n_{t_1},\dots,S^n_{t_d}) \cw(S^\circ_{t_1},\dots,S^\circ_{t_d})$ on $(\RR^d,|\cdot|)$. Since the time points $\{t_k\}_{k=1}^d\subset\RR_+$ are arbitrary, this is equivalent to the claim $S^n\cfd S^\circ$ over $\RR_+$.
\end{proof}

Depending on \autoref{cor:price_hausdorff}, this final result precisely characterises the interval-valued \emph{weak} limit of the classical Heston price process. This limit has paths in the set $\uE$ from \autoref{def:excursionary_paths}, which in itself is extremely surprising. As discussed following \autoref{def:frac_NIG2}, recall that the NIG excursion process $S^\bullet$ here a.s.~returns the singleton $\{S^\circ_t\}$ for a.e.~$t\in\RR_+$, and given we set $\alpha=0$, $S^\circ$ admits the exponentiated NIG representation in \autoref{lem:NIG_reduc}.

\begin{corollary}[Heston Hausdorff limits]\label{cor:heston_hausdorff}
    Let $\{S^n\}_{n\in\NN}$ be the sequence of Heston price processes from \autoref{eq:fast_reversion_heston}, and $S^\bullet$ the process from \autoref{def:frac_NIG2} with $\alpha=0$, so that
    \begin{equation}
        X^0_t:=\inf\big\{x>0:x-aW^1_x>bt\big\},\quad S^{\bullet}_t:=\left\{\exp\left(W^\rho_x-\tfrac12 x\right):x\in[X^0_{t_-},X^0_t]\right\}.
    \end{equation}
    Then the weak convergence $S^n\cw S^\bullet$ takes place on the Hausdorff metric space $(\uE,d_\uE)$.
\end{corollary}
\begin{proof}
    Define processes $\tS^n$ like in \autoref{cor:heston_fd}. Then by \autoref{cor:price_hausdorff}, the convergence $\tS^n\asc S^\bullet$ takes place on $(\uE,d_\uE)$ as $n\to\infty$. Given we have set $\alpha=0$, then \autoref{thm:heston_ode} provides $\tS^n\ed S^n$. So from $\tS^n\asc S^\bullet$ on $(\uE,d_\uE)$ we get the weak claim $S^n\cw S^\bullet$ .
\end{proof}

We consider \autoref{cor:heston_hausdorff} to strengthen \autoref{thm:mech_pro} as much as we deem meaningfully possible. Given this is a consequence of \autoref{cor:price_hausdorff}, which is a consequence of \autoref{cor:hausdorff_random}, which is a consequence of the probability-free \autoref{cor:graph_hausdorff}, then clearly we have not only strengthened \autoref{thm:mech_pro} significantly, but generalised it widely as well. We have therefore achieved our preliminary goal from the \hyperlink{prologue}{Prologue} to strengthen and generalise \autoref{thm:mech_pro}.

Finally recall that \autoref{cor:graph_hausdorff}, on which \autoref{cor:heston_hausdorff} ultimately depends, was demonstrated visually in \autoref{fig:hausdorff_composite}. In the same way we can visualise how the scaled Heston variance processes $n^{-1}V^{n}$ from \autoref{eq:fast_reversion_heston} behave as $n\to\infty$ using \autoref{fig:hausdorff_deriv}. Indeed, \autoref{cor:hausdorff_deriv} can be extended into the probabilistic setting here (like \autoref{cor:graph_hausdorff} was) to show that $n^{-1}V^n$ has an interval-valued limit on $(\uE,d_\uE)$, like $S^n$. This limit a.s.~returns the singleton $\{0\}$ a.e., but still has compact upwards excursions like in \autoref{fig:hausdorff_deriv}, which are dense in $\RR_+$. Given that the processes $V^n$ are not directly tradable, we do not see any practical consequences of this surprising limit, beyond those of \autoref{cor:IG_limit} and \autoref{cor:heston_hausdorff}.

%% file: chapters/5_conclusion.tex
\clearpage
\section{Conclusion}\label{chap:conclusion}

There is inevitably some repetition here of the achievements of this thesis, as summarised in the \hyperlink{abstract}{Abstract} and \autoref{ch:intro}. However, additional clarity is now provided on how these achievements have been met, the value of them, and possible extensions. This value builds upon that covered in the \hyperlink{prologue}{Prologue}, primarily relating to personal motivating experiences and specific, albeit very popular, models. After this clarification, some future directions for research are presented. These range from theoretical generalisations of the new ODE well-posedness results obtained in \autoref{chap:wellposed}, to practical implications of the interval-valued excursion price processes emerging, for the first time in mathematical finance, in \autoref{chap:framework}.

\vspace{3mm}\textbf{The Heston-NIG relationship.} The priority has certainly remained to develop the mathematical theory required to describe how the classical Heston and NIG models are related in the `fast-reversion' limit of \cite{Mechkov_2015}, extending the fixed-time distributive result obtained there, presented here in \autoref{thm:mech_pro}. As set out in the \hyperlink{prologue}{Prologue}, the description had to be sufficiently rich to reveal the class of derivatives whose values converge in this limit, thereby clarifying the applicability and value to practitioners depending on these models. 

The extent of this relationship is now captured by the finite-dimensional limiting result of \autoref{cor:heston_fd}. This indeed reveals a wide class of derivatives with converging prices, namely those depending on the underlying price process \emph{only} through a finite number of fixed time points, sometimes called Bermudan options. But, motivated primarily by generalisations, the novel probability-free and ODE-based approach taken towards these limits in \autoref{chap:solutions} has enabled the stronger, and more informative, Hausdorff limiting result of \autoref{cor:heston_hausdorff}.

This Hausdorff result provides a complete description of the Heston-NIG relationship although, surprisingly, required the introduction of an interval-valued \emph{generalisation} of the classical NIG process. This result is ideal for practical purposes because it not only clarifies the class of continuously-monitored path-dependent derivatives whose prices \emph{will not} converge to those from the classical NIG limit, but also what these prices \emph{will} converge to. 

This particular result should also be of theoretical value to anyone interested in stochastic process limit theorems, in finance or otherwise. In finance, it is the first case of an interval-valued process, like those defined and studied in the context of queuing theory in \cite{Whitt_2002}, emerging. In mathematics more generally, it is the first known example of such a process arising naturally from \emph{continuous} processes through a limit of parameters. This is made all the more surprising given we are not talking about niche continuous processes, but one of the simplest and most popular stochastic volatility models. This is evidenced by \cite{Heston_1993} having 10,000 citations and perhaps more importantly its implementation in numerous financial institutions and commercially available libraries like Numerix.

Over the course of \autoref{sec:price_limits} it was additionally exposed that these Heston-NIG relationships are rooted in a deeper connection between the integrated CIR and IG processes. This connection was established on the new exit-time metric space introduced in \autoref{chap:solutions}, which is stronger than Skorokhod's $\uM_1$ space. It is also easier to understand, given that it does not depend on taking infima over parametric representations, and is homeomorphic to the Polish topology of uniform convergence over compacts on the non-decreasing continuous paths. The \hyperlink{epilogue}{Epilogue} collects several other L\'evy process limits arising from the CIR process under other Heston parameterisations, e.g.~those from \cite{Heston_1993} and \cite{Fouque_2011}.

It is finally worth clarifying that there are immediate multi-dimensional generalisations of these limiting relationships. For example, taking a $d$-dimensional Heston model with a common CIR variance process, an analogous $d$-dimensional exp-NIG L\'evy limit is obtained with a common IG subordinator. This harmoniously connects the popular Heston FX modelling framework of \cite{De_Col_2013} with a less analytically and computationally demanding NIG counterpart, of growing importance as the dimension is raised. In fact this NIG counterpart falls into the tractable FX framework of \cite{Ballotta_2017} built from L\'evy processes, and an implementation is available at \href{https://github.com/ryanmccrickerd/frh-fx}{github.com/ryanmccrickerd/frh-fx}.

\vspace{3mm}\textbf{The wider modelling framework.} All of these classical stochastic process relationships of course originate from the general limiting results for ODE solutions in \autoref{chap:solutions}, most notably \autoref{thm:exit_limit} and \autoref{cor:graph_hausdorff}. These results enabled the fractional generalisations of the Heston-NIG relationship in \autoref{sec:price_limits}, relating to the RLH model from \autoref{sec:RLH_model}.

More generally the RLH model \emph{exemplifies} the general random ODE-based framework from \autoref{sec:base_frame}, existing in both of the sub-frameworks defined in \autoref{sec:heston_frame} and \autoref{sec:martingale}. The route to \autoref{cor:RLH_martingality} brings together several results in these sections. Practically, this result establishes that the RLH price process is a martingale for all parameter combinations, so always generates arbitrage-free derivatives prices. Theoretically, it demonstrates the more general martingale result of  \autoref{them:price_martingale}, which constitutes a novel applications of time-changes and Novikov's martingale condition to random ODE solutions. In turn, necessary integrability requirements depend on the general MGF existence result of \autoref{cor:gauss_heston_mgf}, applicable to a class of generalised Heston models with alternative Gaussian drivers.

Through the recipe in \autoref{def:rlh_polygon} and accompanying \autoref{thm:rlh_convergence} we have demonstrated how such a model can be simulated for this purpose of derivative pricing, and have provided volatility surfaces exhibiting properties associated with promising rough volatility models. So despite the RLH model being defined primarily for illustrative theoretical purposes, this all suggests that a deeper empirical comparison with leading counterparts, such as those of \cite{Bayer_2015} and \cite{El_Euch_2019}, will be worthwhile.

Moving further backwards through the thesis, all models in the framework of \autoref{chap:framework} possess the well-posedness properties from \autoref{chap:wellposed}, perhaps most valuably the uniqueness and continuous dependence robustness captured by \autoref{thm:global_uniqueness} and \autoref{thm:continuity_solution_map}. These ODE results are other firsts, not depending on any spatial regularity properties of the driving functions, such as H\"older regularity, yet still being applicable to maximal solutions. Hence the description of these ODEs throughout as spatially irregular, and their resulting ability to harmoniously accommodate rough volatility models without need for additional well-posedness analysis. This starkly contrasts the ongoing line of theoretical research which aims to accommodate rough volatility within a framework of It\^o-type Volterra integral equations, e.g.~recently studied in \cite{Keller_2018} and \cite{Larsson_2019}.

Practically, this robustness means practitioners will not find that (suitably moderate) adjustments to models lead to counterintuitive consequences, e.g.~on resulting derivative prices. Accordingly, they can safely take advantage of the wide class of models captured by the solution space in \autoref{chap:solutions}, with a relatively low barrier to entry from the outset, given there is strictly no requirement to understand any form of stochastic calculus. Notice that It\^o calculus is introduced sparingly throughout this thesis, and only ever to clarify motivations from, connections with and consequences for other more familiar frameworks.

Of course more research is required until we can fully understand whether our framework built around random ODEs can take centre stage in practice, or whether its primary value will derive from what it can teach us about other frameworks. What we have already is certainly a promising start, having gone all the way from the probability-free well-posedness foundations of \autoref{chap:wellposed} to a specific model in \autoref{chap:framework} which by itself reconciles a popular classical model with rough, discontinuous and even novel excursionary generalisations. 

Finally three directions for future research are provided, which this thesis has made possible.

\vspace{3mm}\textbf{Carath\'eodory ODE extensions.} The discussion leading up to the subset $\uF\subset\uC(\RR^2,\RR)$ of functions from \autoref{def:driving_func} clarifies our main motivation for considering these, given their emergence from the Heston volatility model thus potential for (practical) modelling applications. But, before our maximal uniqueness result of \autoref{thm:global_uniqueness}, we also discussed how this set $\uF$ is (theoretically) related to that in Wend's local uniqueness result in \autoref{thm:wend_uniqueness}. 

Our statement of Wend's theorem is actually a reduced one, applicable only to classical \emph{differentiable} solutions, like the entirety of this thesis and majority of ODE theory. Consulting Theorem 2.6.1 of \cite{Agarwal_1993}, Wend's uniqueness theorem actually holds in the extended setting where functions $f$ are not necessarily in $\uC(\RR^2,\RR)$, but satisfy the weaker `Carath\'eodory' conditions from the existence theorem of \cite{Caratheodory_1927}.

Following Section 2.1 of \cite{Coddington_1955} or \cite{Agarwal_1993}, the Carath\'eodory conditions require each $f(\cdot,x)$ to be \emph{only} measurable, each $f(t,\cdot)$ continuous, and for each compact rectangle $\cX\subset\RR^2$ that there exists a Lebesgue integrable function $m=m_\cX$ such that $|f(t,x)|\le m(t)$ whenever $(t,x)\in\cX$. For any $(\tau,\xi)\in\RR^2$, Carath\'eodory's theorem then provides the existence of an `extended' solution $\vp$ of the ODE $x'=f(t,x)$ over some $(\tau-\ep,\tau+\ep)$ with $\vp(\tau)=\xi$. Given $f(\cdot,x)$ may not be continuous, we define such an extended solution to be absolutely continuous with $\vp'(t)=f(t,\vp(t))$ a.e.~only. The fact that Wend's theorem still holds in this setting ensures a \emph{unique} extended solution over $[\tau,\tau+\ep)$, i.e.~going forwards in time, provided $f(\cdot,x)$ is non-decreasing and $f(t,x)>0$.

Theoretically, it is natural to ask if the functions in $\uF$ can also be relaxed from $f(\cdot,x)$ being continuous to only measurable (other assumptions being equal), without compromising the maximal uniqueness of \autoref{thm:global_uniqueness}. In the generalised Heston framework in \autoref{def:gen_heston_frame}, in which our cumulative variance processes $X$ verify a random ODE of type~
\begin{equation}\label{eq:gen_hest_conc}
    X'_t = \sigma Z_{X_t} + \kappa(\vt(t) - X_t) + v,
\end{equation}
this would allow us to relax the continuity of $\vt$ to e.g.~only right-continuity. This relaxation is however not well-motivated. On the contrary, relaxing the continuous process $Z$ in \autoref{eq:gen_hest_conc} to being only right-continuous would allow us to make use of non-Gaussian L\'evy processes to drive our random ODEs $x'=Y_{t,x}$ thus cumulative variance \emph{extended} solutions $X$. This would provide an alternative to the approach from \cite{Barndorff_Nielsen_2001b}, applied in \cite{Carr_2003}, which instead utilises SDEs driven by non-Gaussian L\'evy processes to obtain cumulative variance processes with dependent, e.g.~reversionary, increments. The L\'evy process $Z$ in these SDEs must have positive increments to ensure that $X$ does, but it is plausible that this could be relaxed in the random ODE of \autoref{eq:gen_hest_conc}.

\vspace{2mm}The main point of this extension to Carath\'eodory ODEs is \emph{not} to widen the set $\Phi$ of possible cumulative variance paths from \autoref{def:solutionset}, since we have already shown in \autoref{cor:construct_limit} how any path in the superset $\bPhi\supset\Phi$ from \autoref{def:setE} can be accommodated as a limit. Rather, we could leverage the \emph{probabilistic} properties of L\'evy processes for analytical and simulation purposes. For example, if in \autoref{eq:gen_hest_conc} $Z$ a L\'evy process, then we can consider optional sampling theory for the evaluation of $\EE[Z_{X_t}]$, thus $\EE[X_t]$, which is e.g.~not possible in the RLH model from \autoref{def:RLH_model} where $Z=W^\alpha$ is a Brownian fractional derivative.

Contrasting the usual Carath\'eodory conditions in which $f(\cdot,x)$ is measurable, e.g.~c\`adl\`ag, and $f(t,\cdot)$ continuous, this motivates the consideration of ODEs depending on the following superset $\uFe\supset\uF$. Note that while paths of L\'evy processes are not necessarily c\`adl\`ag a.s., the stochastic continuity of these processes ensure a `c\`adl\`ag modification', see e.g.~Lemma 1.4.8 in \cite{Applebaum_2009}, meaning that we essentially do not lose generality by assuming this.

\begin{definition}[Set $\uFe$ of functions]\label{def:ext_driving_func}
    Let the set $\uFe$ contain functions $f:\RR^2\to\RR$ with each $f(\cdot,x)$ strictly increasing and continuous, each $f(t,\cdot)$ c\`adl\`ag with \emph{upwards} discontinuities only, so that $f(t,x)-f(t,x_-)\ge0$ for all $(t,x)$, and finally $f(\tau,\xi)>0$ for some $(\tau,\xi)\in\RR^2$.
\end{definition}

\begin{figure}[ht]
    \centering
    \includegraphics[width=0.48\linewidth]{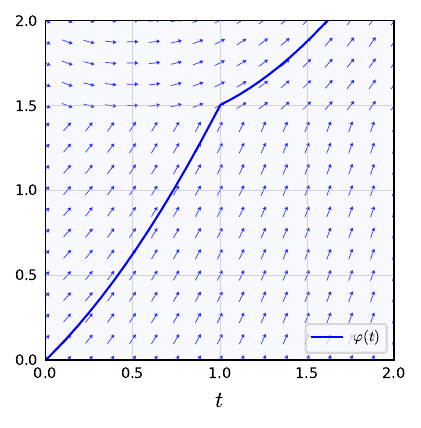}
    \includegraphics[width=0.48\linewidth]{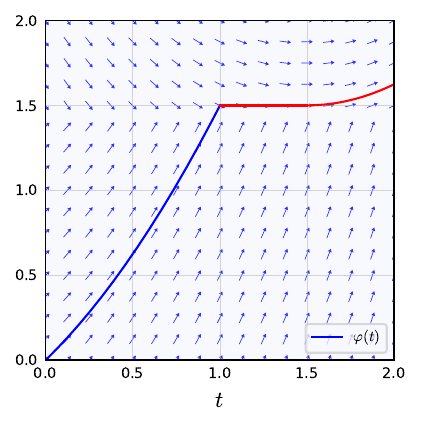}
    \caption{Carath\'eodory extended solutions $\vp$ are shown for the IVP $x'=f(t,x)$, $x(0)=0$, where $f(t,x)=t+z(x)$ and the c\`adl\`ag path $z$ jumps \emph{downwards} at $x=1.5$. If this jump is too large, $\vp$ exists over $[0,1]$ only (right panel), although an a.e.~differentiable uniform limit of forward Euler polygons (red) may still exist.}
    \label{fig:caratheodory}
\end{figure}

It is important to note we are \emph{not} assuming the usual Carath\'eodory conditions in \autoref{def:ext_driving_func}, but rather new ones which clearly represent an inversion of space and time. Like in \cite{Barndorff_Nielsen_2001b}, we assume only upwards discontinuities for now because it is easy to construct IVPs $x'=f(t,x)$, $x(0)=0$ with no maximal extended solutions otherwise, e.g.~of type $f(t,x)=t+z(x)$ in \autoref{fig:caratheodory}. That is, no solution if we do not modify the meaning of an extended solution for our inverted conditions, e.g.~to a path $\vp$ such that the exit-time $E(\vp)$ solves the inverted IVP $x'=1/f(x,t)$, $x(0)=0$ a.e., or more practically a path $\vp$ which forward Euler polygons converge to uniformly over compacts.

Supplementing \autoref{fig:caratheodory} and \cite{Barndorff_Nielsen_2001b}, note that constraints on the downwards jumps of a c\`adl\`ag path peculiarly arise elsewhere, e.g.~the pathwise quadratic variation defined in \cite{Lochowski_2018}. But for now we consider this coincidental. 

The first and most important question towards a non-Gaussian L\'evy-driven random ODE framework is the counterpart to \autoref{thm:global_uniqueness}: are the assumptions $f\in\uFe$ and $f(\tau,\xi)>0$ sufficient for the IVP $x'=f(t,x)$, $x(\tau)=\xi$ to have a unique maximal extended solution?

\vspace{3mm}\textbf{It\^o SDE implications.} The route outlined in \autoref{ch:intro}, from the Heston volatility model in \autoref{eq:heston_intro} to the ODE in \autoref{eq:cir_ivp_well_posed}, is succinctly described by the following arrow
\begin{equation}\label{eq:route_summary}
    \dd V_t = \sigma\sqrt{V_t}\dd W^1_t + \kappa(\theta - V_t)\dd t,\ V_0=v\quad \rightsquigarrow\quad \vp'(t) = \sigma (w\circ\vp)(t) + \kappa(\theta t - \vp(t)) + v.
\end{equation}
The ODE here may be considered a pathwise counterpart of the SDE, with the correspondence $\vp'(t)=V_t(\omega)$, provided we recall that the path $w\in\uC_0(\RR,\RR)$ should not be considered one of $W^1$ but rather the time-changed version $B^1$ from \autoref{eq:tce} or later \autoref{thm:heston_ode}.

Using \autoref{thm:global_uniqueness} we now know that the corresponding IVP $x'=f(t,x)$, $x(0)=0$ has a unique maximal solution $\vp$ when $\sigma,\kappa,\theta,v>0$, for \emph{any} $w\in\uC_0(\RR,\RR)$. So uniqueness holds for the IVP even if the H\"older regularity of $w$ is taken to be far lower than that of Brownian motion, which is a.s.~$\frac12-\ep$ for every $\ep>0$. So it is natural to ask if \autoref{thm:global_uniqueness} can be utilised to establish the pathwise uniqueness of certain SDEs beyond the result of \cite{Yamada_1971}, for which the CIR SDE in \autoref{eq:route_summary} is a well-known boundary case.

This question remains unanswered, even for a simple generalisation of the CIR SDE, say to $\dd V_t = \sigma |V_t|^\alpha\dd W^1_t + \theta\dd t$, $V_0=v$ for some $\alpha\in(0,\frac12)$. There is a partial answer in this case, because when $\theta=0$ this is `Girsanov's SDE', for which pathwise uniqueness fails for $\alpha\in(0,\frac12)$, see Example 1.22 in \cite{Cherny_2005}. But the inclusion of a drift $\theta \dd t$ should not be underestimated, and indeed is what leads to the strictly increasing component $\theta t$ in the ODE of \autoref{eq:route_summary}, important in our proof of \autoref{thm:global_uniqueness}. Popular texts on SDEs display some neglect for such \emph{simple} drifts, because more complicated ones can often be removed by a change of measure, see e.g.~Theorem 5.27.1 in \cite{Rogers_1994}.

The point now is that our new ODE uniqueness result renders this debate somewhat moot, and we demonstrate why with a practical example. Suppose we consider replacing $V_t$ on the r.h.s.~in \autoref{eq:route_summary} with $|V_t|^{2\alpha}$ for some $\alpha\in(0,\frac12)$, which is a natural consideration for a practitioner who has found the Heston model to behave undesirably, compared with their short-time observations. For some $\sigma,\kappa,\theta,v>0$, the SDE under consideration is therefore
\begin{equation}\label{eq:consider_sde}
    \dd V_t = \sigma|V_t|^\alpha\dd W^1_t + \kappa(\theta - |V_t|^{2\alpha})\dd t,\quad V_0=v. 
\end{equation}
A classical result of Skorokhod's guarantees a weak solution of this SDE, see \cite{Skorokhod_1965} or preferably Proposition 1.13 in \cite{Cherny_2005}, the terminology of which we follow here. But this SDE need not have a \emph{unique} weak solution, let alone a unique \emph{strong} solution, i.e.~need not exhibit pathwise uniqueness. See Figure 1.1 in \cite{Cherny_2005} for a succinct reminder of how these properties of SDEs are related. So probabilistic analysis of this SDE is dubious, the applicability for volatility unclear given $V$ may not be non-negative, and the convergence of simulation schemes not guaranteed. Nevertheless, \emph{given} a weak solution $(V,W^1)$, we find, just like \autoref{eq:tce}, $V$ verifies
\begin{equation}\label{eq:new_tce}
    V_t = \sigma B^1_{\int_0^t|V_s|^{2\alpha}\dd s} + \kappa\left(\theta t  - \int_0^t|V_s|^{2\alpha}\dd s\right) + v. 
\end{equation}

Now let $f(t,x):=\sigma w(x) + \kappa(\theta t - x) + v$ be the usual Heston function implicit in \autoref{eq:route_summary} and consider the IVP $x'=f_\alpha(t,x)$, $x(0)=0$ with $f_\alpha\in\uF$ defined by $f_\alpha := \sgn(f)|f|^{2\alpha}$. This IVP is an example of \autoref{prob:ivp}, so by \autoref{thm:global_uniqueness} has a unique maximal solution $\vp_\alpha$ which is strictly increasing. Contrasting the solution $\vp$ in \autoref{eq:route_summary}, $\vp_\alpha$ now verifies
\begin{equation}
    \vp'_\alpha(t) = f_\alpha(t,\vp_\alpha(t)) = |f(t,\vp_\alpha(t))|^{2\alpha} = |\sigma (w\circ\vp_\alpha)(t) + \kappa(\theta t - \vp_\alpha(t)) + v|^{2\alpha}
\end{equation}
where we have neglected the $\sgn(f)$ component of $f_\alpha$ because we know $\vp'_\alpha(t)\ge0$. Applying $|\cdot|^{1/2\alpha}$ to each side, we see $V_t(\omega):=|\vp_\alpha'(t)|^{1/2\alpha}$ verifies the random ODE of \autoref{eq:new_tce} on a pathwise basis, under the identification $w(x) := B^1_x(\omega)$. We have thus constructed a solution of \autoref{eq:new_tce}, which any weak solution of our SDE must verify, using our framework in which all random ODEs have a unique strong solution suitable for volatility modelling.

It is important to see we have not actually claimed the random ODE in \autoref{eq:new_tce} is pathwise unique, because by adding the $\sgn(f)$ component to $f_\alpha$ we conveniently solved a \emph{different} (pathwise unique) random ODE in our framework, the unique solution of which \emph{also} solves \autoref{eq:new_tce}, and is guaranteed non-negative. Our use of $\sgn$ is not necessary; we just need to ensure $f_\alpha(\cdot,x)$ is strictly increasing so $f_\alpha\in\uF$, and the function $|\cdot|^\alpha$ can be replaced throughout by any $\varrho$ bijective from and to $\RR_+$. Then the counterpart of \autoref{eq:consider_sde} is
\begin{equation}\label{eq:consider_sde2}
    \dd V_t = \sigma \varrho(V_t)\dd W^1_t + \kappa(\theta - \varrho^2(V_t))\dd t,\quad V_0=v,
\end{equation}
and the pathwise solution is $(\varrho^2)^{-1}\circ \vp'_\alpha$. Similarly, the curve $\theta t = \int_0^t \theta \dd s$ may be generalised to any strictly increasing $\vt(t)$, and now it is clear that studying \autoref{eq:consider_sde2} in this way provides a fascinating practical perspective on local (volatility of) volatility models stemming from \cite{Dupire_1994}. Indeed, these functions $\vt$ and $\varrho$ could be calibrated like in these models. Alternatively, we could just e.g.~fix $\varrho=\id$, equivalently set $\alpha=1$, to obtain a reversionary extension of the popular SABR model of \cite{Hagan_2002}.

\vspace{2mm}This approach where we map an SDE onto a random ODE that \emph{always} has a unique strong solution (recall from Figure 1.1 in \cite{Cherny_2005} this is `the best possible situation') can be related to alternative time-change and `Doss-Sussman' methods for manipulating SDEs, both covered in \cite{Ikeda_1989} with the latter deriving from \cite{Doss_1977} and \cite{Sussman_1978}. The vital difference is that these alternative methods arrive at ODEs or related integral equations, see Example 2.1 or Theorem 4.3 and its corollary in \cite{Ikeda_1989}, but do not contribute to whether these have a unique solution unless the SDE is already known to. So although one obtains theoretically pleasant relationships between solutions of SDEs and ODEs, rarely can one be used to help the other practically, and certainly nothing can be done should we wish to relax the driving process $B^1$ in \autoref{eq:new_tce} from being Brownian motion. This contrasts our general treatment of random ODEs like \autoref{eq:new_tce} and the wide application to local volatility models just given, because throughout this thesis we have instead prioritised ODEs and answered the question of their well-posedness without any dependence on probability, let alone It\^o SDEs.

\vspace{3mm}\textbf{Empirical testing of models.} Finally we propose specific experiments relating to derivative pricing, which will help to test the models from the martingale framework of \autoref{def:martingale_frame}. We present these in relation to the RLH model from \autoref{def:RLH_model} and extensions, but of course one is free to consider any other in this martingale framework. So recall this model, in which the price process $S$ and its cumulative variance $X = [\log S]$ uniquely verify
\begin{equation}\label{eq:conc_rlh_summary}
    X'_t = \sigma W^\alpha_{X_t} + \kappa(\vt(t) - X_t) + v,\quad S_t=\exp(W^\rho_{X_t}-\tfrac12 X_t).
\end{equation}
In \autoref{sec:vol_surfaces}, implied volatilities generated from this Heston extension are illustrated, and the combination of \autoref{fig:skews}, \autoref{fig:curvatures} and \autoref{fig:rough_heston_comparison} provide convincing empirical evidence that this new model shares important features with the leading (rough) volatility models, namely their short-time skews and curvatures. Specifically, \autoref{fig:rough_heston_comparison} draws comparisons with results in \cite{El_Euch_2019b}, deriving from the alternative \emph{rough} Heston extension.

We have thus shown the RLH model's implied volatilities are sufficiently flexible to justify an independent study of this model's ability to reconcile market data, which only time and space has prohibited here. While a brute-force calibration by simulation is made possible (in reasonable time) by the variance reduction methods of \cite{McCrickerd_2018}, given the promising findings of \cite{Horvath_2021} we suggest also exploring neural network techniques for calibration, utilising our simulation scheme to generate data for training.

Assuming that the RLH model, or an alternative in the framework of \autoref{def:martingale_frame}, performs well enough for financial institutions to consider it in production for the analysis of derivatives on \emph{price} processes, we propose thereafter testing this model's ability to jointly reconcile S\&P 500 and VIX derivative prices. This is known to be a difficult challenge, only recently solved in discrete time by \cite{Guyon_2020}, with \cite{Gatheral_2020} later claiming the first satisfactory model with continuous sample paths; the \emph{quadratic} rough Heston model. Of course similar quadratic RLH models can be considered if required. Analogous to the quadratic model in \cite{Gatheral_2020}, we could define this using the equations
\begin{equation}
    Z'_t = \sigma W^\alpha_{Z_t} + \kappa(\vt(t) - Z_t) + v,\quad X'_t = a(Z'_t - b)^2 + c ,\quad  S_t=\exp(W^\rho_{X_t}-\tfrac12 X_t)
\end{equation}
where $a,b,c>0$. However, this is \emph{not} a model in our martingale framework; note that $X$ is inconveniently adapted to the filtration generated by $W^1_{Z}$, rather than $W^1_{X}$. In our framework it is more natural to replace the implicit RLH random field $Y_{t,x}=\sigma W^\alpha_x + \kappa(\vt(t) - x) + v$ in \autoref{eq:conc_rlh_summary} with a quadratic variant like $\sgn(Y)Y^2$. This idea clearly relates to the use of the function $f_\alpha := \sgn(f)|f|^\alpha$ for solving \autoref{eq:new_tce}, with the difference being that here we will not invert the quadratic transformation by utilising $V_t(\omega):=|\vp_\alpha'(t)|^{1/2\alpha}$ thereafter.

The authors of \cite{Gatheral_2020} repeatedly highlight the importance of the Zumbach effect when treating this joint calibration puzzle, so this effect may as well be tested for the RLH model and its quadratic variant directly, like in \cite{El_Euch_2018b} for the rough Heston model. We can be optimistic about this, given that non-trivial Zumbach effects arise from models which exhibit time reversal \emph{asymmetry}, which our framework indeed exhibits, clarified most plainly by the simple pathwise violation of uniqueness illustrated in \autoref{fig:nonuniqueness2}.

\vspace{4mm} The experiments proposed thus far have a common theme; take a leading volatility model, e.g.~from the conventional frameworks of It\^o or Volterra SDEs, which exhibits desirable features, and show there exists at least one specific model in our random ODE-based framework which competes favourably with it. But of course, besides potentially simplifying and unifying features of more familiar frameworks, a vitally important property of any new theory is the ability of it to make at least one original and experimentally-verifiable prediction.

Towards this, the obvious starting point is to test for the effects of the novel excursion processes which have emerged as fast-reversion limits of models in our framework, and we could again use observed derivative prices to do this. Recall the fractional NIG c\`adl\`ag and excursion processes $S^\circ$ and $S^\bullet$ from \autoref{def:frac_NIG1} and \autoref{def:frac_NIG2}, related to the RLH model, and for simplicity set the fractional derivative $\alpha=0$, so $W^\alpha\!:=\!D^\alpha(W^1)\!=\!W^1$. Then by \autoref{lem:NIG_reduc} $S^\circ$ is the standard exp-NIG L\'evy process limit from the motivating result \autoref{thm:mech_pro}, and $S^\bullet$ is the interval-valued generalisation $S^\bullet_t=:[S^-_t,S^+_t]\ni S^\circ_t$ from \autoref{eq:frac_NIG_ex} which by \autoref{cor:heston_hausdorff} emerges as a weak limit of classical Heston processes. 

Of course, the legal contracts which define financial derivatives do not account for price processes returning intervals of prices over an \emph{infinitesimal} time period, like $S^\bullet$ does. But this is a moot point, because traders who determine prices should be fearful of any excursions which occur over time periods shorter than the duration between their trading activities. Hence recent research into financial excursion risks, such as \cite{Ananova_2020}, depending on It\^o's theory of excursions, reviewed relatively recently in \cite{Watanabe_2010}.

Now to test for such `excursionary' effects in traders' derivative prices, we can first calibrate both price processes $S^\circ$ and $S^\bullet$ to European options. The calibrated parameters of both models will, theoretically, be equivalent, because stochastic continuity ensures the singleton $S^\bullet_T = \{S^\circ_T\}$ is a.s.~returned for any fixed maturity $T>0$, as discussed following \autoref{def:frac_NIG2}. Next we can test which process predicts path-dependent derivative prices better, e.g.~those of barrier options with the same maturity. To be more specific, we could first calibrate the exp-NIG process $S^\circ$ to vanilla put option prices, each interpreted as $\EE[(K - S^\circ_T)_+]$ for maturity $T$ and strike $K$, then determine which of the sets of related put option prices
\begin{equation}\label{eq:barrier_price_bounds}
    \EE\big[(K - \inf_{t\in[0,T]}S^\circ_t)_+\big]\ \le\ \EE\big[(K - \inf_{t\in[0,T]}\inf S^\bullet_t)_+\big] = \EE\big[(K - \inf_{t\in[0,T]}S^-_t)_+\big]
\end{equation}
reconcile observations better. For the ordering here, we have simply used $\inf S^\bullet_t = S^-_t \le S^\circ_t$. It would be striking to find observed barrier option prices are near the upper bound in \autoref{eq:barrier_price_bounds}. Since this price is that predicted by our novel excursion process $S^\bullet$, this finding could be interpreted as confirming excursion risk premia in derivative prices, but more importantly would validate these new excursionary processes for modelling this risk.

%% file: epilogue.tex
\clearpage

\phantomsection\section*{\hypertarget{epilogue}{Epilogue: Integrated CIR-L\'evy relationships}}
\addcontentsline{toc}{section}{Epilogue: Integrated CIR-L\'evy relationships}
\renewcommand\thesubsection{\thesection\Alph{subsection}}

\pagestyle{fancy}
\fancyhf{}
\renewcommand{\headrulewidth}{0pt}
% \rhead{\thepage}
\cfoot{\thepage}
\chead{\nouppercase{\sc Epilogue}}

In \autoref{sec:price_limits}, the limiting results from \autoref{sec:bounds_limits} and \autoref{sec:excursionary_limits} were applied to the RLH model from \autoref{def:RLH_model}, establishing a.s.~limiting connections with the generalised (fractional) IG and NIG processes from \autoref{def:frac_IG}, \autoref{def:frac_NIG1} and \autoref{def:frac_NIG2}. Classical CIR, Heston, IG and NIG weak limit theorems then followed as consequences. 

As mentioned in \autoref{sec:price_limits}, the limits established there depended on the choice to express the RLH model in the specific `fast-reversion' parameterisation in \autoref{def:fast_RLH}, inspired by the classical Heston parameterisation from \cite{Mechkov_2015}, summarised in \autoref{eq:fast_reversion_heston}.

We use this epilogue to simultaneously characterise \emph{all} (eight) of the L\'evy process limits arising from more general fast reverting CIR processes, demonstrating the power of \autoref{thm:exit_limit} in particular. These e.g.~accommodate the parameterisations from \cite{Heston_1993} and \cite{Fouque_2011}. Heston \emph{price} process limits follow from these in exactly the same way that \autoref{cor:heston_fd} and \autoref{cor:heston_hausdorff} did from \autoref{cor:frac_IG_limit}, so are not repeated.

Particularly surprising will be the L\'evy limits arising here which have random starting points. This possibility is accommodated by \autoref{cor:construct_limit}, which enables their \emph{construction}. By covering these, we not only provide reconciliations between classical continuous and jump models of volatility, but also randomised ones, e.g. \cite{Mechkov_2016}, \cite{Jacquier_2016}.

Consider the standard CIR variance process in the classical Heston model of \cite{Heston_1993}
\begin{equation}
    \dd V_t = \sigma\sqrt{V_t}\dd W_t + \kappa(\theta - V_t)\dd t,\quad V_0=v.
\end{equation}
By drastically overparameterising this SDE, we will obtain a variety of limits from it simultaneously, via \autoref{thm:exit_limit}. So let the family $\{V^n\}_{n>0}$ of processes solve the CIR SDEs~
\begin{equation}\label{eq:cir_fr_gen}
    \dd V^n_t = n^\alpha a \sqrt{V^n_t}\dd W_t + n(b - n^{\beta-1} V^n_t)\dd t,\quad V^n_0=n^\gamma c,
\end{equation}
for fixed $a,b,c > 0$ and $\alpha,\beta,\gamma\in(-\infty,1]$. By constraining the exponents $\alpha,\beta,\gamma\le1$, we ensure that the reversionary component $nb$ in \autoref{eq:cir_fr_gen} is never dominated as $n\to\infty$, and so any particular case of $\alpha,\beta,\gamma\in(-\infty,1]$ can be considered a `fast-reversion' regime. The following classical regimes are then recovered when also setting $(n,a,b,c) := (\kappa,\sigma,\theta,v)$
\begin{equation}\label{eq:diff_regimes}
    (\alpha,\beta,\gamma) :=
    \begin{cases}
        \ (0,1,0)         &\text{\cite{Heston_1993}}, \\
        \ (\frac12,1,0)   &\text{\cite{Fouque_2011}}, \\
        \ (1,1,0)         &\text{\cite{Mechkov_2015}}.
    \end{cases}
\end{equation}

We now prepare for the limit of the time-integral processes $X^n_t:=\int_0^t V^n_s\dd s$ deriving from the CIR SDE in \autoref{eq:cir_fr_gen}. As in \autoref{sec:exit_topology}, let $\bPhi\subset\uD(\RR_+,\RR_+)$ contain the strictly increasing and unbounded c\`adl\`ag paths, and $d_{\bPhi}$ be the exit-time metric, satisfying $d_{\bPhi}(\bvp_1,\bvp_2) = \Vert E(\bvp_2) - E(\bvp_1)\Vert_{\RR_+}$. Here, $E$ is the usual exit-time functional, which defines an involutive isometry between the exit-time space $(\bPhi,d_{\bPhi})$ and the set of non-decreasing and unbounded paths in $\uC_0(\RR_+,\RR_+)$ equipped with the `uniform convergence over compacts' norm $\Vert\cdot\Vert_{\RR_+}$.

Recall that convergence on the exit-time metric space $(\bPhi,d_{\bPhi})$ immediately provides convergence on Skorokhod's $\uM_1$ space and pointwise convergence a.e., so also on $\uL_p$ spaces. Recall also that by the weak convergence $X^n\cw X^0$ on a metric space $(\cX,d_\cX)$ we mean the convergence $\EE[\#(X^n)]\xrightarrow{n\to\infty}\EE[\#(X^0)]$ for real, bounded and continuous $\#$ from $(\cX,d_\cX)$. In the following, let the indicator $\mathbbm{1}_x=\mathbbm{1}_{\{1\}}(x)$ return the value 1 if $x=1$ and 0 otherwise.

\begin{theorem}[Integrated CIR fast-reversion limits]\label{thm:cir_fr_limits}
    Let the family $\{V^n\}_{n>0}$ of processes solve the CIR SDEs in \autoref{eq:cir_fr_gen} for fixed $a,b,c > 0$ and $\alpha,\beta,\gamma\in(-\infty,1]$. Define the time-integrals $\{X^n\}_{n>0}$ respectively by $X^n_t:=\int_0^t V^n_s\dd s$. Then the weak convergence $X^n\cw X^0$ takes place on the exit-time space $(\bPhi,d_{\bPhi})$, where $X^0$ is the L\'evy process
    \begin{equation}\label{eq:lim_def}
        X^0_t := \inf\bigg\{ x>0: - \mathbbm{1}_\alpha a W_x + \mathbbm{1}_\beta x > bt + \mathbbm{1}_\gamma c \bigg\}.
    \end{equation}
\end{theorem}
\begin{proof}
    Let $B^n$ be the Brownian motion constructed from $W$ and each $V^n$ as in \autoref{lem:heston_timechange} so that, as in \autoref{thm:heston_ode}, each CIR process $V^n$ equivalently solves the integral equation
    \begin{equation}
        V^n_t = n^\alpha a B^n_{\int_0^t V^n_s\dd s} + n \left(b t - n^{\beta-1}\int_0^t V^n_s\dd s\right) + n^\gamma c.
    \end{equation}
    Then each of the time-integrals $X^n$ solves the random IVP $x'=n Y^n_{t,x}$, $x_0=0$, where
    \begin{equation}\label{eq:random_field_cir}
        Y^n_{t,x} := n^{\alpha-1} a B^n_x + bt - n^{\beta-1}x + n^{\gamma-1}c.
    \end{equation}
    As in \autoref{thm:heston_ode}, this random IVP constitutes an example of \autoref{prob:random_ivp}, with each random field $Y^n$ a.s.~in $\uG$. These are thus well-posed by \autoref{thm:rand_well_posed}. Now let $\tX$ be the unique solution of the random IVP $x'=n \tY^n_{t,x}$, $x_0=0$ where each $\tY^n$ is as in \autoref{eq:random_field_cir} but constructed from the fixed Brownian motion $W$ rather than $B^n$. For clarity, this means~
    \begin{equation}\label{eq:random_field_cir2}
        \tY^n_{t,x} := n^{\alpha-1}a W_x + bt - n^{\beta-1}x + n^{\gamma-1}c,
    \end{equation}
    and this leads to the equivalence $\tX^n\ed X^n$ in distribution. Notice that the convergence $n^{x-1}\xrightarrow{n\to\infty}\mathbbm{1}_x\in\{0,1\}$ takes place for fixed $x\in(-\infty,1]$. Given that $\alpha,\beta,\gamma\in(-\infty,1]$, the convergence $\tY^n\asc\tY^0$ therefore takes place uniformly over compacts as $n\to\infty$, where
    \begin{equation}
        \tY^0_{t,x} := \mathbbm{1}_\alpha a W_x + bt - \mathbbm{1}_\beta x + \mathbbm{1}_\gamma c.
    \end{equation}
    Now define the candidate limit process $X^0\in\bPhi\subset\uD(\RR_+,\RR_+)$ like in \autoref{cor:exit_limit} by
    \begin{equation}
        X^0_t :=\inf\{x>0:\tY^0_{t,x}<0\} = \inf\{ x>0: - \mathbbm{1}_\alpha a W_x + \mathbbm{1}_\beta x > bt + \mathbbm{1}_\gamma c \}.
    \end{equation}
    Then by \autoref{cor:exit_limit} we have the convergence $\tX^n\asc X^0$ on the exit-time space $(\bPhi,d_{\bPhi})$ as $n\to\infty$. Since we have the equivalence $X^n\ed \tX^n$ of distributions for every $n>0$, then this also provides the weak convergence $X^n\cw X^0$ on the space $(\bPhi,d_{\bPhi})$, as claimed.
\end{proof}

\begin{table}[ht]
    \begin{center}
        \begin{tabular}{ c l l }
         $(\mathbbm{1}_\alpha,\mathbbm{1}_\beta,\mathbbm{1}_\gamma)$ & $\ \ \ X^0_t$ & Limit description \\ 
         $(0,0,0)$ & $\quad=\inf\bigg\{ x>0: 0 > bt \bigg\}=\infty$ & Immediate explosion \\ 
         $(0,0,1)$ & $\quad=\inf\bigg\{ x>0: 0 > bt + c \bigg\}=\infty$ & Immediate explosion \\  
         $(0,1,0)$ & $\quad=\inf\bigg\{ x>0: x > bt \bigg\}=bt$ & Deterministic from 0 \\  
         $(0,1,1)$ & $\quad=\inf\bigg\{ x>0: x > bt + c \bigg\}= bt + c$ & Deterministic from $c$\\
         $(1,0,0)$ & $\quad=\inf\bigg\{ x>0: -a W_x > bt \bigg\}$ & L\'evy \\
         $(1,1,0)$ & $\quad=\inf\bigg\{ x>0: -a W_x + x > bt \bigg\}$ & IG\\  
         $(1,0,1)$ & $\quad=\inf\bigg\{ x>0: -a W_x > bt + c \bigg\}$ & L\'evy, random start \\ 
         $(1,1,1)$ & $\quad=\inf\bigg\{ x>0: -a W_x + x > bt + c \bigg\}$ & IG, random start
        \end{tabular}
        \caption{
            Representations and descriptions of the limits arising in \autoref{thm:cir_fr_limits}.
        }\label{tab:cir_limits}
    \end{center}
    \vspace{-4mm}
\end{table}

Given there are eight implicit L\'evy process limits in \autoref{eq:lim_def}, \autoref{tab:cir_limits} describes each of them, mostly found in \cite{Applebaum_2009}. Notice that each case of $(\mathbbm{1}_\alpha,\mathbbm{1}_\beta,\mathbbm{1}_\gamma)$ actually applies to an infinitude of reversionary regimes, except for (1,1,1). E.g., $(\mathbbm{1}_\alpha,\mathbbm{1}_\beta,\mathbbm{1}_\gamma)=(0,1,0)$ corresponds to \emph{any} reversionary regime with $\alpha,\gamma<1$ and $\beta=1$ in \autoref{eq:cir_fr_gen}, like \emph{both} of the \cite{Heston_1993} and \cite{Fouque_2011} regimes defined in \autoref{eq:diff_regimes}.

Some concluding commentary is now provided on all the limits appearing in \autoref{tab:cir_limits} and the approach taken here towards establishing them. These remarks emphasise the surprising ability of the random ODE-based approach taken here to teach us new things about already much studied, and much relied upon, stochastic processes utilised in mathematical finance.

As with \autoref{cor:IG_limit_fd}, which corresponds to the \cite{Mechkov_2015} case of $(1,1,0)$ in \autoref{tab:cir_limits}, the convergence $X^n\cfd X^0$ supplements \autoref{thm:cir_fr_limits} at points where the limit $X^0$ is stochastically continuous. This is all of $\RR_+$, except $\{0\}$ must be removed when the limit violates $X^0_0=0$ a.s., i.e.~in cases 1, 2, 4, 7 and 8 in \autoref{tab:cir_limits}. Defining $X^0_{0_-}:=0$, this can be considered a violation of stochastic continuity. It is clear from their representations that the two random start cases coincide with their deterministic-start counterparts when shifted a distance $-\frac{c}{b}$ backwards in time. By using \autoref{eq:cir_fr_gen}, we see that these random start limits emerge only when the CIR starting points $V^n_0=n^\gamma c=nc$ tend up to $\infty$ as $n$ does.

As demonstrated following \autoref{cor:IG_limit_fd}, all f.d.~convergences can be verified using MGFs, given all processes here are affine. This becomes more difficult if we e.g.~generalise the CIR reversion level $b$ in \autoref{eq:cir_fr_gen} to a c\`adl\`ag path $b(t)$ with strictly increasing and unbounded integral $\int_0^t b(s)\dd s$, but is still possible. As remarked in \autoref{sec:price_limits}, the proof of \autoref{thm:cir_fr_limits} is practically unchanged for this extension and others. E.g., we simply find this integral $\int_0^t b(s)\dd s$ appearing in place of the term $bt=\int_0^t b\dd s$ in the limit in \autoref{eq:lim_def}.

Since the deterministic limits in \autoref{tab:cir_limits} are continuous, convergence in these cases is uniform over compacts. Notice that the reversionary regimes of \emph{both} \cite{Heston_1993} and \cite{Fouque_2011} fall into the third case of (0,1,0), so have the linear limit $X^0_t:=bt$. In these cases, the Heston price process converges weakly to that of Black-Scholes w.r.t.~uniform convergence over compacts. The effect of this was noted in \cite{Fouque_2011}. $V^n$ does not necessarily become deterministic just because  $X^n:=\int_0 V^n_s\dd s$ does, and can in general develop random excursions to $\infty$. This is explained by the limits in \autoref{cor:hausdorff_deriv}, illustrated in \autoref{fig:hausdorff_deriv}.

%% file: notation.tex
\clearpage

\phantomsection\section*{\hypertarget{notation}{Helpful notation}}
\addcontentsline{toc}{section}{Helpful notation}
\renewcommand\thesubsection{\thesection\Alph{subsection}}

\pagestyle{fancy}
\fancyhf{}
\renewcommand{\headrulewidth}{0pt}
\cfoot{\thepage}
\chead{\nouppercase{\sc Notation}}

The following table collects symbols which are used throughout this thesis to denote fixed sets. Descriptions of each set and the page numbers of each symbol's first use are also shown.

\vspace*{6mm}

\begin{table}[ht]
    \begin{center}
        \begin{tabular}{ l l l }
         Symbol & Page & Description\vspace*{4mm} \\ 
         $\NN$ & 25 & Natural numbers, i.e.~$\{1,2,3,\dots\}$ \\
         $\NN_0$ & 35 & Non-negative integers, i.e.~$\{0,1,2,\dots\}=\NN\cup\{0\}$ \\
         $\RR$ & 14 & Real numbers, i.e.~$(-\infty, \infty)$ \\
         $\RR_+$ & 9 & Non-negative real numbers, i.e.~$[0,\infty)$ \\
         $\bRR$ & 41 & Extended real numbers, i.e.~$[-\infty,\infty]=\RR\cup\{\pm\infty\}$ \\
         $\uC$ & 16 & Continuous functions \\
         $\uN$ & 87 & Non-decreasing continuous functions \\
         $\uAC$ & 68 & Absolutely continuous functions \\
         $\uH_\lambda$ & 149 & H\"older continuous functions of order $\lambda\in(0,1)$ \\
         $\uH_\lambda^0$ & 149 & Subset of $\uH_\lambda$ \\
         $\Phi$ & 24 & The solution set of \autoref{prob:ivp2} \\
         $\Phi_\vt$ & 84 & Subset of $\Phi$ \\
         $\Phi'$ & 86 & First derivatives of functions in $\Phi$ \\
         $\Phi^{-1}$ & 88 & Inverses of functions in $\Phi$\\
         $\bPhi$ & 73 & Superset of $\Phi$ \\
         $\uD$ & 16 & C\`adl\`ag functions \\
         $\uE$ & 73 & Excursionary functions \\ 
         $\uF$ & 20 & Subset of continuous functions from $\RR^2$ to $\RR$ \\
         $\uF_\vt$ & 22 & Subset of $\uF$ \\
         $\uG$ & 20 & Subset of continuous functions from $\RR_+^2$ to $\RR$ \\
         $\uG_\vt$ & 25 & Subset of $\uG$
        \end{tabular}
    \end{center}
    \vspace*{-10mm}
\end{table}

%% file: appendix.tex
\clearpage

\phantomsection\section*{\hypertarget{appendix}{Appendix: RLH simulation code}}
\addcontentsline{toc}{section}{Appendix: RLH simulation code}
\renewcommand\thesubsection{\thesection\Alph{subsection}}

\pagestyle{fancy}
\fancyhf{}
\renewcommand{\headrulewidth}{0pt}
\cfoot{\thepage}
\chead{\nouppercase{\sc Appendix}}

\begin{figure}[ht]
    \centering
    \includegraphics[width=1.0\linewidth]{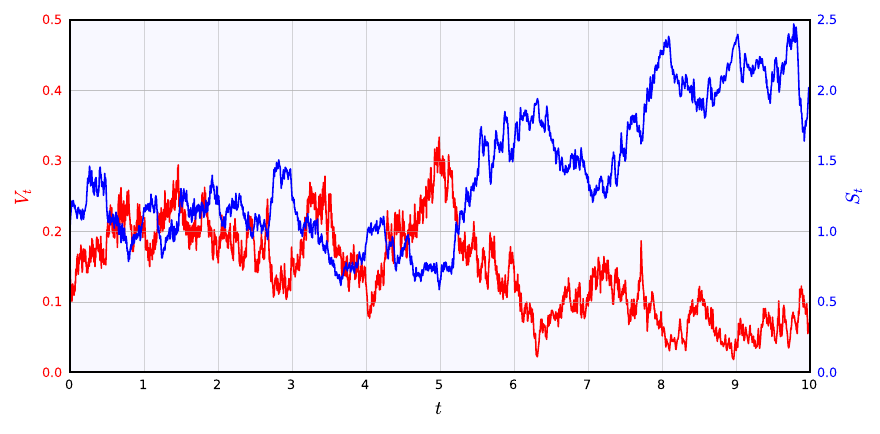}
    \caption{The data in the \texttt{V} and \texttt{S} arrays is shown, after running the \texttt{python} code below with the given seed. This may be compared with Figure 1 in \cite{Gatheral_2020}.}
    \label{fig:simulation_output}
\end{figure}

This appendix provides standalone \texttt{python} code (tested with \texttt{v3.7.3}) which demonstrates how the RLH polygons from \autoref{def:rlh_polygon} may be simulated. The code is self explanatory, except:~we denote $\hat W^\alpha$ and $\hat W^\rho$ by \texttt{Wa} and \texttt{Wr}; the \texttt{kernel} array contains the evaluation points $(x^*_k)^{-\alpha}$ from \autoref{eq:RL_polygon}, and; \texttt{np.convolve} evaluates \emph{all} sums in \autoref{eq:RL_polygon} simultaneously, like in \cite{Bennedsen_2017}. This code takes $75\pm1$ ms to run on a 2.3 GHz Intel Core i5 MacBook Pro, and the arrays \texttt{V} and \texttt{S} from it are illustrated in \autoref{fig:simulation_output}.

\vspace{2mm}

\vspace{-2mm}
\begin{linenumbers}
\texttt{import numpy as np\\
from scipy.special import gamma\\
from scipy.interpolate import interp1d\\
{\color{alt3}\emph{\# Set RLH model parameters, which coincide with Heston's for alpha = 0}}\\
sigma, alpha, kappa, theta, v, rho = 0.1, 0.2, 0.3, 0.4**2, 0.4**2, -0.5\\
{\color{alt3}\emph{\# Set simulation horizons and discretisation steps}}\\
time\_horizon, space\_horizon = 10.0, 1.6\\
time\_steps, space\_steps = 4096, 4096\\
{\color{alt3}\emph{\# Build time and space arrays for forward Euler scheme and random field}}\\
dt = time\_horizon / time\_steps\\
t = np.linspace(0, time\_horizon, time\_steps + 1)\\
dx = space\_horizon / space\_steps\\
x = np.linspace(0, space\_horizon, space\_steps + 1)\\
{\color{alt3}\emph{\# Draw Brownian increments and build Brownian motions}}\\
np.random.seed(1)\\
dW = np.random.normal(size=(space\_steps, 2)) * dx**0.5\\
W = np.zeros((space\_steps + 1, 2))\\
W[1:, :]~= np.cumsum(dW, axis=0)\\
Wr = rho * W[:, 1] + (1 - rho**2)**0.5 * W[:, 0]\\
{\color{alt3}\emph{\# Compute the fractional derivative Wa of W1 using \autoref{eq:RL_polygon}}}\\
Wa = np.zeros(space\_steps + 1)\\
kernel = (x[1:]**(1 - alpha) - x[:-1]**(1 - alpha)) / (1 - alpha) / dx\\
Wa[1:]~= np.convolve(kernel, dW[:, 1])[:space\_steps] / gamma(1 - alpha)\\
{\color{alt3}\emph{\# Build linearly interpolating polygons}}\\
Wa\_polygon = interp1d(x, Wa)\\
Wr\_polygon = interp1d(x, Wr)\\
{\color{alt3}\emph{\# Approximate the RLH random field from \autoref{def:gen_hest_field3}}}\\
def Y(t, x):\\
{\color{white}oooo}return sigma * Wa\_polygon(x) + kappa * (theta * t - x) + v\\
{\color{alt3}\emph{\# Conduct basic forward Euler scheme for cumulative variance X}}\\
V = np.zeros(time\_steps)\\
X = np.zeros(time\_steps + 1)\\
for i in range(time\_steps):\\
{\color{white}oooo}V[i] = Y(t[i], X[i])\\
{\color{white}oooo}X[i + 1] = X[i] + V[i] * dt\\
{\color{alt3}\emph{\# Construct price path}}\\
S = np.exp(Wr\_polygon(X) - 0.5 * X)
}
\end{linenumbers}

%% file: main.bbl
\begin{thebibliography}{139}
\expandafter\ifx\csname natexlab\endcsname\relax\def\natexlab#1{#1}\fi
\expandafter\ifx\csname url\endcsname\relax
  \def\url#1{{\tt #1}}\fi
\expandafter\ifx\csname urlprefix\endcsname\relax\def\urlprefix{DOI: }\fi

\bibitem[{{Abi Jaber}(2019)}]{Abi_Jaber_2019}
{Abi Jaber}, E. (2019).
\newblock {Reconciling rough volatility with jumps}.
\newblock {\em Presentation at the Vienna Congress on Mathematical Finance\/}.

\bibitem[{{Abi Jaber} \& {El Euch}(2019)}]{Abi_Jaber_2019c}
{Abi Jaber}, E. \& {El Euch}, O. (2019).
\newblock {Multifactor Approximation of Rough Volatility Models}.
\newblock {\em {SIAM} Journal on Financial Mathematics\/}, {\em 10\/}(2),
  309--349.
\newline\urlprefix\url{https://doi.org/10.1137/18M1170236}

\bibitem[{Abi~Jaber et~al.(2019)Abi~Jaber, Larsson \& Pulido}]{Larsson_2019}
Abi~Jaber, E., Larsson, M. \& Pulido, S. (2019).
\newblock {Affine Volterra processes}.
\newblock {\em Annals of Applied Probability\/}, {\em 29\/}(5), 3155--3200.
\newline\urlprefix\url{https://doi.org/10.1214/19-aap1477}

\bibitem[{Agarwal \& Lakshmikantham(1993)}]{Agarwal_1993}
Agarwal, R.~P. \& Lakshmikantham, V. (1993).
\newblock {\em Uniqueness and Nonuniqueness Criteria for Ordinary Differential
  Equations\/}.
\newblock World Scientific.
\newline\urlprefix\url{https://doi.org/10.1142/1988}

\bibitem[{Aliprantis(1998)}]{Aliprantis_1998}
Aliprantis, C.~D. (1998).
\newblock {\em Principles of Real Analysis \emph{(3rd ed.)}\/}.
\newblock Academic Press.

\bibitem[{Al{\`{o}}s \& Le{\'{o}}n(2017)}]{Alos_2017}
Al{\`{o}}s, E. \& Le{\'{o}}n, J.~A. (2017).
\newblock On the curvature of the smile in stochastic volatility models.
\newblock {\em SIAM Journal on Financial Mathematics\/}, {\em 8\/}(1),
  373--399.
\newline\urlprefix\url{https://doi.org/10.1137/16m1086315}

\bibitem[{Al{\`{o}}s et~al.(2007)Al{\`{o}}s, Le{\'{o}}n \& Vives}]{Alos_2007}
Al{\`{o}}s, E., Le{\'{o}}n, J.~A. \& Vives, J. (2007).
\newblock On the short-time behavior of the implied volatility for
  jump-diffusion models with stochastic volatility.
\newblock {\em Finance and Stochastics\/}, {\em 11\/}(4), 571--589.
\newline\urlprefix\url{https://doi.org/10.1007/s00780-007-0049-1}

\bibitem[{Ananova et~al.(2020)Ananova, Cont \& Xu}]{Ananova_2020}
Ananova, A., Cont, R. \& Xu, R. (2020).
\newblock {Excursion Risk}.
\newblock {\em arXiv preprint. \\ \emph{URL:
  \texttt{\href{https://arxiv.org/abs/2011.02870}{https://arxiv.org/abs/2011.02870}}}\/}.

\bibitem[{Andersen(2008)}]{Andersen_2007}
Andersen, L. (2008).
\newblock {Simple and efficient simulation of the Heston stochastic volatility
  model}.
\newblock {\em Journal of Computational Finance\/}, {\em 11\/}(3), 1--42.
\newline\urlprefix\url{https://doi.org/10.21314/jcf.2008.189}

\bibitem[{Andersen \& Piterbarg(2010)}]{Andersen_2010}
Andersen, L. \& Piterbarg, V. (2010).
\newblock {\em Interest Rate Modeling\/}.
\newblock Atlantic Financial Press.

\bibitem[{Applebaum(2009)}]{Applebaum_2009}
Applebaum, D. (2009).
\newblock {\em L{\'{e}}vy Processes and Stochastic Calculus \emph{(2nd
  ed.)}\/}.
\newblock Cambridge University Press.
\newline\urlprefix\url{https://doi.org/10.1017/cbo9780511809781}

\bibitem[{Asmussen \& Glynn(2007)}]{Asmussen_2007}
Asmussen, S. \& Glynn, P.~W. (2007).
\newblock {\em Stochastic Simulation: Algorithms and Analysis\/}.
\newblock Springer-Verlag New York.
\newline\urlprefix\url{https://doi.org/10.1007/978-0-387-69033-9}

\bibitem[{Ballotta et~al.(2017)Ballotta, Deelstra \& Rayée}]{Ballotta_2017}
Ballotta, L., Deelstra, G. \& Rayée, G. (2017).
\newblock {Multivariate FX models with jumps: Triangles, Quantos and implied
  correlation}.
\newblock {\em European Journal of Operational Research\/}, {\em 260\/}(3),
  1181--1199.
\newline\urlprefix\url{https://doi.org/10.1016/j.ejor.2017.02.018}

\bibitem[{Barndorff-Nielsen(1997)}]{Barndorff-Nielsen_1997}
Barndorff-Nielsen, O.~E. (1997).
\newblock {Normal Inverse Gaussian Distributions and Stochastic Volatility
  Modelling}.
\newblock {\em Scandinavian Journal of Statistics\/}, {\em 24\/}(1), 1--13.
\newline\urlprefix\url{https://doi.org/10.1111/1467-9469.00045}

\bibitem[{Barndorff-Nielsen et~al.(2018)Barndorff-Nielsen, Benth \&
  Veraart}]{Nielsen_2018}
Barndorff-Nielsen, O.~E., Benth, F.~E. \& Veraart, A. E.~D. (2018).
\newblock {\em Ambit Stochastics\/}.
\newblock Springer International Publishing.
\newline\urlprefix\url{https://doi.org/10.1007/978-3-319-94129-5}

\bibitem[{Barndorff-Nielsen \&
  Shephard(2001{\natexlab{a}})}]{Barndorff_Nielsen_2001}
Barndorff-Nielsen, O.~E. \& Shephard, N. (2001{\natexlab{a}}).
\newblock {Modelling by L{\'{e}}vy Processess for Financial Econometrics}.
\newblock In {\em L{\'{e}}vy Processes\/}. Birkhäuser, Boston, MA.
\newline\urlprefix\url{https://doi.org/10.1007/978-1-4612-0197-7_13}

\bibitem[{Barndorff-Nielsen \&
  Shephard(2001{\natexlab{b}})}]{Barndorff_Nielsen_2001b}
Barndorff-Nielsen, O.~E. \& Shephard, N. (2001{\natexlab{b}}).
\newblock {Non-Gaussian Ornstein–Uhlenbeck-based models and some of their
  uses in financial economics}.
\newblock {\em Journal of the Royal Statistical Society: Series B (Statistical
  Methodology)\/}, {\em 63\/}(2), 167--241.
\newline\urlprefix\url{https://doi.org/10.1111/1467-9868.00282}

\bibitem[{Barndorff-Nielsen \& Shiryaev(2010)}]{Barndorff_Nielsen_2010}
Barndorff-Nielsen, O.~E. \& Shiryaev, A. (2010).
\newblock {\em {Change of Time and Change of Measure}\/}.
\newblock World Scientific.
\newline\urlprefix\url{https://doi.org/10.1142/7928}

\bibitem[{Bartle \& Sherbert(2018)}]{Bartle_2000}
Bartle, R.~G. \& Sherbert, D.~R. (2018).
\newblock {\em Introduction to Real Analysis \emph{(4th ed.)}\/}.
\newblock Wiley.

\bibitem[{Bayer et~al.(2016)Bayer, Friz \& Gatheral}]{Bayer_2015}
Bayer, C., Friz, P. \& Gatheral, J. (2016).
\newblock Pricing under rough volatility.
\newblock {\em Quantitative Finance\/}, {\em 16\/}(6), 887--904.
\newline\urlprefix\url{https://doi.org/10.1080/14697688.2015.1099717}

\bibitem[{Bayer et~al.(2020)Bayer, Harang \& Pigato}]{Bayer_2020}
Bayer, C., Harang, F.~A. \& Pigato, P. (2020).
\newblock {Log-modulated rough stochastic volatility models}.
\newblock {\em arXiv preprint. \\ \emph{URL:
  \texttt{\href{https://arxiv.org/abs/2008.03204}{https://arxiv.org/abs/2008.03204}}}\/}.

\bibitem[{Bennedsen et~al.(2016)Bennedsen, Lunde \& Pakkanen}]{Bennedsen_2016}
Bennedsen, M., Lunde, A. \& Pakkanen, M.~S. (2016).
\newblock {Decoupling the short- and long-term behavior of stochastic
  volatility}.
\newblock {\em arXiv preprint, to appear in Journal of Financial Econometrics.
  \\ \emph{URL:
  \texttt{\href{https://arxiv.org/abs/1610.00332}{https://arxiv.org/abs/1610.00332}}}\/}.

\bibitem[{Bennedsen et~al.(2017)Bennedsen, Lunde \& Pakkanen}]{Bennedsen_2017}
Bennedsen, M., Lunde, A. \& Pakkanen, M.~S. (2017).
\newblock {Hybrid scheme for Brownian semistationary processes}.
\newblock {\em Finance and Stochastics\/}, {\em 21\/}(4), 931--965.
\newline\urlprefix\url{https://doi.org/10.1007/s00780-017-0335-5}

\bibitem[{Bergomi(2016)}]{Bergomi_2016}
Bergomi, L. (2016).
\newblock {\em Stochastic Volatility Modeling\/}.
\newblock Chapman and Hall/{CRC}.
\newline\urlprefix\url{https://doi.org/10.1201/b19649}

\bibitem[{Billingsley(1995)}]{billingsley_1995}
Billingsley, P. (1995).
\newblock {\em Probability and Measure \emph{(3rd ed.)}\/}.
\newblock Wiley.

\bibitem[{Billingsley(1999)}]{Billingsley_1999}
Billingsley, P. (1999).
\newblock {\em Convergence of Probability Measures \emph{(2nd ed.)}\/}.
\newblock Wiley.
\newline\urlprefix\url{https://doi.org/10.1002/9780470316962}

\bibitem[{Blanc et~al.(2017)Blanc, Donier \& Bouchaud}]{Blanc_2017}
Blanc, P., Donier, J. \& Bouchaud, J.-P. (2017).
\newblock {Quadratic Hawkes processes for financial prices}.
\newblock {\em Quantitative Finance\/}, {\em 17\/}(2), 171--188.
\newline\urlprefix\url{https://doi.org/10.1080/14697688.2016.1193215}

\bibitem[{Brigo \& Mercurio(2006)}]{Brigo_2006}
Brigo, D. \& Mercurio, F. (2006).
\newblock {\em Interest Rate Models - Theory and Practice\/}.
\newblock Springer-Verlag Berlin Heidelberg.
\newline\urlprefix\url{https://doi.org/10.1007/978-3-540-34604-3}

\bibitem[{Bru \& Yor(2002)}]{Bru_2002}
Bru, B. \& Yor, M. (2002).
\newblock {Comments on the life and mathematical legacy of Wolfgang Doeblin}.
\newblock {\em Finance and Stochastics\/}, {\em 6\/}(1), 3--47.
\newline\urlprefix\url{https://doi.org/10.1007/s780-002-8399-0}

\bibitem[{Buehler et~al.(2019)Buehler, Gonon, Teichmann \& Wood}]{Buehler_2019}
Buehler, H., Gonon, L., Teichmann, J. \& Wood, B. (2019).
\newblock Deep hedging.
\newblock {\em Quantitative Finance\/}, {\em 19\/}(8), 1271--1291.
\newline\urlprefix\url{https://doi.org/10.1080/14697688.2019.1571683}

\bibitem[{Cantor(1884)}]{Cantor_1884}
Cantor, G. (1884).
\newblock {De la puissance des ensembles parfaits de points: Extrait d’une
  lettre adressée à l’éditeur}.
\newblock {\em Acta Mathematica\/}, {\em 4\/}, 381--392.
\newline\urlprefix\url{https://doi.org/10.1007/BF02418423}

\bibitem[{Carathéodory(1927)}]{Caratheodory_1927}
Carathéodory, C. (1927).
\newblock {\em {Vorlesungen über Reelle Funktionen \emph{(2nd ed.)}}\/}.
\newblock Vieweg+Teubner Verlag Wiesbaden.
\newline\urlprefix\url{https://doi.org/10.1007/978-3-663-15768-7}

\bibitem[{Carr et~al.(2003)Carr, Geman, Madan \& Yor}]{Carr_2003}
Carr, P., Geman, H., Madan, D.~B. \& Yor, M. (2003).
\newblock {Stochastic Volatility for L{\'e}vy Processes}.
\newblock {\em Mathematical Finance\/}, {\em 13\/}(3), 345--382.
\newline\urlprefix\url{https://doi.org/10.1111/1467-9965.00020}

\bibitem[{Carr \& Wu(2004)}]{Carr_2004}
Carr, P. \& Wu, L. (2004).
\newblock {Time-changed Lévy processes and option pricing}.
\newblock {\em Journal of Financial Economics\/}, {\em 71\/}(1), 113--141.
\newline\urlprefix\url{https://doi.org/10.1016/S0304-405X(03)00171-5}

\bibitem[{Cherny \& Engelbert(2005)}]{Cherny_2005}
Cherny, A.~S. \& Engelbert, H.-J. (2005).
\newblock {\em {Singular Stochastic Differential Equations}\/}.
\newblock Springer, Berlin, Heidelberg.
\newline\urlprefix\url{https://doi.org/10.1007/b104187}

\bibitem[{Cid \& Pouso(2009)}]{Cid_2009}
Cid, J.~A. \& Pouso, R.~L. (2009).
\newblock {Does Lipschitz with Respect to $x$ Imply Uniqueness for the
  Differential Equation $y' = f(x, y)$?}
\newblock {\em The American Mathematical Monthly\/}, {\em 116\/}(1), 61--66.
\newline\urlprefix\url{https://doi.org/10.1080/00029890.2009.11920909}

\bibitem[{Ciesielski(1960)}]{Ciesielski_1960}
Ciesielski, Z. (1960).
\newblock {On the Isomorphisms of the Spaces H${}_\alpha$ and $m$}.
\newblock {\em Bulletin of the Polish Academy of Sciences\/}, {\em 8\/}(4),
  217–222.

\bibitem[{Coddington \& Levinson(1955)}]{Coddington_1955}
Coddington, A. \& Levinson, N. (1955).
\newblock {\em Theory of ordinary differential equations\/}.
\newblock McGraw-Hill.

\bibitem[{Cont \& Perkowski(2019)}]{Cont_2019}
Cont, R. \& Perkowski, N. (2019).
\newblock Pathwise integration and change of variable formulas for continuous
  paths with arbitrary regularity.
\newblock {\em Transactions of the American Mathematical Society, Series B\/},
  {\em 6\/}(4), 161--186.
\newline\urlprefix\url{https://doi.org/10.1090/btran/34}

\bibitem[{Cont \& Tankov(2003)}]{Cont_2003}
Cont, R. \& Tankov, P. (2003).
\newblock {\em Financial Modelling with Jump Processes\/}.
\newblock Chapman and Hall/CRC.
\newline\urlprefix\url{https://doi.org/10.1201/9780203485217}

\bibitem[{Cordi et~al.(2020)Cordi, Challet \& Kassibrakis}]{Cordi_2020}
Cordi, M., Challet, D. \& Kassibrakis, S. (2020).
\newblock The market nanostructure origin of asset price time reversal
  asymmetry.
\newblock {\em Quantitative Finance, \emph{forthcoming}\/}.
\newline\urlprefix\url{https://doi.org/10.1080/14697688.2020.1753883}

\bibitem[{Cox et~al.(1985)Cox, Ingersoll \& Ross}]{Cox_1985}
Cox, J.~C., Ingersoll, J.~E. \& Ross, S.~A. (1985).
\newblock {A Theory of the Term Structure of Interest Rates}.
\newblock {\em Econometrica\/}, {\em 53\/}(2), 385--407.
\newline\urlprefix\url{https://doi.org/10.2307/1911242}

\bibitem[{Dambis(1965)}]{Dambis_1965}
Dambis, K.~E. (1965).
\newblock {On the Decomposition of Continuous Submartingales}.
\newblock {\em Theory of Probability {\&} Its Applications\/}, {\em 10\/}(3),
  401--410.
\newline\urlprefix\url{https://doi.org/10.1137/1110048}

\bibitem[{Davis et~al.(2018)Davis, Obłój \& Siorpaes}]{Davis_2018}
Davis, M., Obłój, J. \& Siorpaes, P. (2018).
\newblock Pathwise stochastic calculus with local times.
\newblock {\em Annales de l'Institut Henri Poincaré, Probabilités et
  Statistiques\/}, {\em 54\/}(1), 1--21.
\newline\urlprefix\url{https://doi.org/10.1214/16-aihp792}

\bibitem[{{De Col} et~al.(2013){De Col}, Gnoatto \& Grasselli}]{De_Col_2013}
{De Col}, A., Gnoatto, A. \& Grasselli, M. (2013).
\newblock {Smiles all around: FX joint calibration in a multi-Heston model}.
\newblock {\em Journal of Banking {\&} Finance\/}, {\em 37\/}(10), 3799--3818.
\newline\urlprefix\url{https://doi.org/10.1016/j.jbankfin.2013.05.031}

\bibitem[{Dekking et~al.(2005)Dekking, Kraaikamp, Lopuhaä \&
  Meester}]{Dekking_2005}
Dekking, F.~M., Kraaikamp, C., Lopuhaä, H.~P. \& Meester, L.~E. (2005).
\newblock {\em A Modern Introduction to Probability and Statistics\/}.
\newblock Springer-Verlag London.
\newline\urlprefix\url{https://doi.org/10.1007/1-84628-168-7}

\bibitem[{Doss(1977)}]{Doss_1977}
Doss, H. (1977).
\newblock Liens entre \'equations diff\'erentielles stochastiques et
  ordinaires.
\newblock {\em Annales de l'I.H.P. Probabilit\'es et statistiques\/}, {\em
  13\/}(2), 99--125.

\bibitem[{Dubins \& Schwarz(1965)}]{Dubins_1965}
Dubins, L.~E. \& Schwarz, G. (1965).
\newblock {On Continuous Martingales}.
\newblock {\em Proceedings of the National Academy of Sciences of the United
  States of America\/}, {\em 53\/}(5), 913--916.
\newline\urlprefix\url{https://doi.org/10.1073/pnas.53.5.913}

\bibitem[{Dufresne(2001)}]{Dufresne_2001}
Dufresne, D. (2001).
\newblock The integrated square-root process.
\newblock {\em Minerva Access preprint. \\ \emph{URI:
  \texttt{\href{http://hdl.handle.net/11343/33693}{http://hdl.handle.net/11343/33693}}}\/}.

\bibitem[{Dupire(1994)}]{Dupire_1994}
Dupire, B. (1994).
\newblock Pricing with a smile.
\newblock {\em Risk\/}.

\bibitem[{{El Euch} et~al.(2020){El Euch}, Gatheral, Radoičić \&
  Rosenbaum}]{El_Euch_2018b}
{El Euch}, O., Gatheral, J., Radoičić, R. \& Rosenbaum, M. (2020).
\newblock {The Zumbach effect under rough Heston}.
\newblock {\em Quantitative Finance\/}, {\em 20\/}(2), 235--241.
\newline\urlprefix\url{https://doi.org/10.1080/14697688.2019.1658889}

\bibitem[{{El Euch} et~al.(2019){El Euch}, Gatheral \&
  Rosenbaum}]{El_Euch_2019b}
{El Euch}, O., Gatheral, J. \& Rosenbaum, M. (2019).
\newblock {Roughening Heston}.
\newblock {\em Risk\/}.

\bibitem[{{El Euch} \& Rosenbaum(2019)}]{El_Euch_2019}
{El Euch}, O. \& Rosenbaum, M. (2019).
\newblock {The characteristic function of rough Heston models}.
\newblock {\em Mathematical Finance\/}, {\em 29\/}(1), 3--38.
\newline\urlprefix\url{https://doi.org/10.1111/mafi.12173}

\bibitem[{Feller(1968)}]{Feller_1968}
Feller, W. (1968).
\newblock {\em An Introduction to Probability Theory and Its Applications:
  Volume I\/}.
\newblock Wiley.

\bibitem[{F{\"o}llmer(1981)}]{Follmer_1981}
F{\"o}llmer, H. (1981).
\newblock Calcul d'ito sans probabilites.
\newblock {\em S{\'e}minaire de Probabilit{\'e}s XV 1979/80\emph{,
  143--150}\/}.
\newline\urlprefix\url{https://doi.org/10.1007/bfb0088364}

\bibitem[{Forde \& Jacquier(2011)}]{Forde_2011}
Forde, M. \& Jacquier, A. (2011).
\newblock {The large-maturity smile for the Heston model}.
\newblock {\em Finance and Stochastics\/}, {\em 15\/}, 755--780.
\newline\urlprefix\url{https://doi.org/10.1007/s00780-010-0147-3}

\bibitem[{Fouque et~al.(2011)Fouque, Papanicolaou, Sircar \&
  S{\o}lna}]{Fouque_2011}
Fouque, J.-P., Papanicolaou, G., Sircar, R. \& S{\o}lna, K. (2011).
\newblock {\em Multiscale Stochastic Volatility for Equity, Interest Rate, and
  Credit Derivatives\/}.
\newblock Cambridge University Press.
\newline\urlprefix\url{https://doi.org/10.1017/cbo9781139020534}

\bibitem[{Friz \& Hairer(2014)}]{Friz_2014}
Friz, P. \& Hairer, M. (2014).
\newblock {\em A Course on Rough Paths\/}.
\newblock Springer International Publishing.
\newline\urlprefix\url{https://doi.org/10.1007/978-3-319-08332-2}

\bibitem[{Friz \& Victoir(2010)}]{Friz_2010}
Friz, P.~K. \& Victoir, N.~B. (2010).
\newblock {\em Multidimensional Stochastic Processes as Rough Paths\/}.
\newblock Cambridge University Press.
\newline\urlprefix\url{https://doi.org/10.1017/cbo9780511845079}

\bibitem[{Fukasawa(2011)}]{Fukasawa_2011}
Fukasawa, M. (2011).
\newblock Asymptotic analysis for stochastic volatility: martingale expansion.
\newblock {\em Finance Stoch\/}, {\em 15\/}, 635–654.
\newline\urlprefix\url{https://doi.org/10.1007/s00780-010-0136-6}

\bibitem[{Gatheral(2006)}]{Gatheral_2006}
Gatheral, J. (2006).
\newblock {\em The Volatility Surface: A Practitioner's Guide\/}.
\newblock Wiley.
\newline\urlprefix\url{https://doi.org/10.1002/9781119202073}

\bibitem[{Gatheral et~al.(2018)Gatheral, Jaisson \& Rosenbaum}]{Gatheral_2018}
Gatheral, J., Jaisson, T. \& Rosenbaum, M. (2018).
\newblock Volatility is rough.
\newblock {\em Quantitative Finance\/}, {\em 18\/}(6), 933--949.
\newline\urlprefix\url{https://doi.org/10.1080/14697688.2017.1393551}

\bibitem[{Gatheral et~al.(2020)Gatheral, Jusselin \& Rosenbaum}]{Gatheral_2020}
Gatheral, J., Jusselin, P. \& Rosenbaum, M. (2020).
\newblock {The quadratic rough Heston model and the joint S\&P 500/Vix smile
  calibration problem}.
\newblock {\em Risk\/}.

\bibitem[{Geman et~al.(2001)Geman, Madan \& Yor}]{Geman_2001}
Geman, H., Madan, D.~B. \& Yor, M. (2001).
\newblock {Time Changes for Lévy Processes}.
\newblock {\em Mathematical Finance\/}, {\em 11\/}(1), 79--96.
\newline\urlprefix\url{https://doi.org/10.1111/1467-9965.00108}

\bibitem[{Gerhold et~al.(2016)Gerhold, Gülüm \& Pinter}]{Gerhold_2016}
Gerhold, S., Gülüm, I.~C. \& Pinter, A. (2016).
\newblock {Small-Maturity Asymptotics for the At-The-Money Implied Volatility
  Slope in L{\'{e}}vy Models}.
\newblock {\em Applied Mathematical Finance\/}, {\em 23\/}(2), 135--157.
\newline\urlprefix\url{https://doi.org/10.1080/1350486x.2016.1197041}

\bibitem[{Glasserman(2003)}]{Glasserman_2004}
Glasserman, P. (2003).
\newblock {\em Monte Carlo methods in Financial Engineering\/}.
\newblock Springer-Verlag New York.
\newline\urlprefix\url{https://doi.org/10.1007/978-0-387-21617-1}

\bibitem[{Griffiths \& Higham(2010)}]{Griffiths_2010}
Griffiths, D.~F. \& Higham, D.~J. (2010).
\newblock {\em Numerical Methods for Ordinary Differential Equations\/}.
\newblock Springer-Verlag London.
\newline\urlprefix\url{https://doi.org/10.1007/978-0-85729-148-6}

\bibitem[{Guyon(2020)}]{Guyon_2020}
Guyon, J. (2020).
\newblock {The joint S\&P 500/VIX smile calibration puzzle solved}.
\newblock {\em Risk\/}.

\bibitem[{Guyon \& Henry-Labord\`ere(2013)}]{Guyon_2013}
Guyon, J. \& Henry-Labord\`ere, P. (2013).
\newblock {\em Nonlinear Option Pricing\/}.
\newblock Chapman and Hall/CRC.
\newline\urlprefix\url{https://doi.org/10.1201/b16332}

\bibitem[{Hagan et~al.(2002)Hagan, Kumar, Lesniewski \& Woodward}]{Hagan_2002}
Hagan, P., Kumar, D., Lesniewski, A. \& Woodward, D. (2002).
\newblock {Managing Smile Risk}.
\newblock {\em Wilmott Magazine\/}, {\em 1\/}, 84--108. \\ URL:
  \texttt{\href{https://wilmott.com/managing--smile--risk}{https://wilmott.com/managing\--smile\--risk}}.

\bibitem[{Hamadouche(2000)}]{Hamadouche_2000}
Hamadouche, D. (2000).
\newblock {Invariance principles in Hölder spaces}.
\newblock {\em Portugaliae Mathematica\/}, {\em 57\/}(2), 127--151. \\ URL:
  \texttt{\href{http://eudml.org/doc/48832}{http://eudml.org/doc/48832}}.

\bibitem[{Han \& Kloeden(2017)}]{Han_2017}
Han, X. \& Kloeden, P.~E. (2017).
\newblock {\em Random Ordinary Differential Equations and Their Numerical
  Solution\/}.
\newblock Springer Singapore.
\newline\urlprefix\url{https://doi.org/10.1007/978-981-10-6265-0}

\bibitem[{Hardy \& Littlewood(1932)}]{Hardy_1932}
Hardy, G. \& Littlewood, J. (1932).
\newblock {Some properties of fractional integrals. I.}
\newblock {\em Mathematische Zeitschrift\/}, {\em 27\/}, 565--606.
\newline\urlprefix\url{https://doi.org/10.1007/bf01171116}

\bibitem[{Hardy(1916)}]{Hardy_1916}
Hardy, G.~H. (1916).
\newblock {Weierstrass's Non-Differentiable Function}.
\newblock {\em Transactions of the American Mathematical Society\/}, {\em
  17\/}(3), 301--325.
\newline\urlprefix\url{https://doi.org/10.2307/1989005}

\bibitem[{Hartman(2002)}]{Hartman_2002}
Hartman, P. (2002).
\newblock {\em Ordinary Differential Equations \emph{(2nd ed.)}\/}.
\newblock Society for Industrial and Applied Mathematics.
\newline\urlprefix\url{https://doi.org/10.1137/1.9780898719222}

\bibitem[{Heston(1993)}]{Heston_1993}
Heston, S.~L. (1993).
\newblock {A Closed-Form Solution for Options with Stochastic Volatility with
  Applications to Bond and Currency Options}.
\newblock {\em The Review of Financial Studies\/}, {\em 6\/}(2), 327--343.
\newline\urlprefix\url{https://doi.org/10.1093/rfs/6.2.327}

\bibitem[{Horvath et~al.(2019)Horvath, Jacquier \& Muguruza}]{Horvath_2017}
Horvath, B., Jacquier, A. \& Muguruza, A. (2019).
\newblock {Functional Central Limit Theorems for Rough Volatility}.
\newblock {\em arXiv preprint. \\ \emph{URL:
  \texttt{\href{https://arxiv.org/abs/1711.03078}{https://arxiv.org/abs/1711.03078}}}\/}.

\bibitem[{Horvath et~al.(2021)Horvath, Muguruza \& Tomas}]{Horvath_2021}
Horvath, B., Muguruza, A. \& Tomas, M. (2021).
\newblock Deep learning volatility: a deep neural network perspective on
  pricing and calibration in (rough) volatility models.
\newblock {\em Quantitative Finance\/}, {\em 21\/}(1), 11--27.
\newline\urlprefix\url{https://doi.org/10.1080/14697688.2020.1817974}

\bibitem[{Ikeda \& Watanabe(1992)}]{Ikeda_1989}
Ikeda, N. \& Watanabe, S. (1992).
\newblock {\em Stochastic Differential Equations and Diffusion Processes
  \emph{(2nd ed.)}\/}.
\newblock North Holland.

\bibitem[{Itô(1944)}]{Ito_1944}
Itô, K. (1944).
\newblock Stochastic integral.
\newblock {\em Proceedings of the Imperial Academy\/}, {\em 20\/}(8), 519--524.
\newline\urlprefix\url{https://doi.org/10.3792/pia/1195572786}

\bibitem[{Itô(1951)}]{Ito_1951}
Itô, K. (1951).
\newblock On stochastic differential equations.
\newblock {\em Memoirs of the American Mathematical Society\/}, {\em 4\/},
  1--51.
\newline\urlprefix\url{http://dx.doi.org/10.1090/memo/0004}

\bibitem[{Jacod \& Shiryaev(2003)}]{Jacod_2003}
Jacod, J. \& Shiryaev, A.~N. (2003).
\newblock {\em Limit Theorems for Stochastic Processes \emph{(2nd ed.)}\/}.
\newblock Springer-Verlag Berlin Heidelberg.
\newline\urlprefix\url{https://doi.org/10.1007/978-3-662-05265-5}

\bibitem[{Jacquier et~al.(2018)Jacquier, Pakkanen \& Stone}]{Jacquier_2018}
Jacquier, A., Pakkanen, M.~S. \& Stone, H. (2018).
\newblock Pathwise large deviations for the rough bergomi model.
\newblock {\em Journal of Applied Probability\/}, {\em 55\/}(4), 1078–1092.
\newline\urlprefix\url{https://doi.org/10.1017/jpr.2018.72}

\bibitem[{Jacquier \& Shi(2019)}]{Jacquier_2016}
Jacquier, A. \& Shi, F. (2019).
\newblock {The Randomized Heston Model}.
\newblock {\em SIAM Journal on Financial Mathematics\/}, {\em 10\/}(1),
  89–129.
\newline\urlprefix\url{https://doi.org/10.1137/18m1166420}

\bibitem[{Jin et~al.(2019)Jin, Kremer \& Rüdiger}]{Jin_2019}
Jin, P., Kremer, J. \& Rüdiger, B. (2019).
\newblock {Moments and ergodicity of the jump-diffusion CIR process}.
\newblock {\em Stochastics\/}, {\em 91\/}(7), 974--997.
\newline\urlprefix\url{https://doi.org/10.1080/17442508.2019.1576686}

\bibitem[{Jusselin \& Rosenbaum(2020)}]{Jusselin_2020}
Jusselin, P. \& Rosenbaum, M. (2020).
\newblock No-arbitrage implies power-law market impact and rough volatility.
\newblock {\em Mathematical Finance\/}, {\em 30\/}(4), 1309--1336.
\newline\urlprefix\url{https://doi.org/10.1111/mafi.12254}

\bibitem[{Kaper \& Kwong(1988)}]{Kaper_1988}
Kaper, H.~G. \& Kwong, M.~K. (1988).
\newblock Uniqueness results for some nonlinear initial and boundary value
  problems.
\newblock {\em Archive for Rational Mechanics and Analysis\/}, {\em 102\/}(1),
  45--56.
\newline\urlprefix\url{https://doi.org/10.1007/bf00250923}

\bibitem[{Karatzas \& Shreve(1998)}]{Karatzas_1998}
Karatzas, I. \& Shreve, S.~E. (1998).
\newblock {\em Brownian Motion and Stochastic Calculus\/}.
\newblock Springer-Verlag New York.
\newline\urlprefix\url{https://doi.org/10.1007/978-1-4612-0949-2}

\bibitem[{Keller-Ressel(2011)}]{Keller_2011}
Keller-Ressel, M. (2011).
\newblock Moment explosions and long-term behavior of affine stochastic
  volatility models.
\newblock {\em Mathematical Finance\/}, {\em 21\/}(1), 73--98.
\newline\urlprefix\url{https://doi.org/10.1111/j.1467-9965.2010.00423.x}

\bibitem[{Keller-Ressel et~al.(2018)Keller-Ressel, Larsson \&
  Pulido}]{Keller_2018}
Keller-Ressel, M., Larsson, M. \& Pulido, S. (2018).
\newblock {Affine Rough Models}.
\newblock {\em arXiv preprint. \\ \emph{URL:
  \texttt{\href{https://arxiv.org/abs/1812.08486}{https://arxiv.org/abs/1812.08486}}}\/}.

\bibitem[{Kumar \& Vellaisamy(2012)}]{Kumar_2012}
Kumar, A. \& Vellaisamy, P. (2012).
\newblock {Fractional Normal Inverse Gaussian Process}.
\newblock {\em Methodology and Computing in Applied Probability\/}, {\em 14\/},
  263–283.
\newline\urlprefix\url{https://doi.org/10.1007/s11009-010-9201-z}

\bibitem[{Lakshmikantham \& Leela(1969)}]{Lakshmikantham_1969}
Lakshmikantham, V. \& Leela, S. (1969).
\newblock {\em Differential and Integral Inequalities: Volume I\/}.
\newblock Academic Press.

\bibitem[{Lamperti(1962)}]{Lamperti_1962}
Lamperti, J. (1962).
\newblock On convergence of stochastic processes.
\newblock {\em Transactions of the American Mathematical Society\/}, {\em
  104\/}(3), 430--435.
\newline\urlprefix\url{https://doi.org/10.2307/1993787}

\bibitem[{Lebesgue(1904)}]{Lebesgue_1904}
Lebesgue, H.~L. (1904).
\newblock {\em {Leçons sur l'intégration et la recherche des fonctions
  primitives professées au Collège de France}\/}.
\newblock Cambridge University Press.
\newline\urlprefix\url{https://doi.org/10.1017/cbo9780511701825}

\bibitem[{L\'evy(1953)}]{Levy_1953}
L\'evy, P. (1953).
\newblock {\em Random functions: general theory with special reference to
  Laplacian random functions\/}.
\newblock University of California Press.

\bibitem[{Lipschitz(1876)}]{Lipschitz_1876}
Lipschitz, R. (1876).
\newblock Sur la possibilit\'e d'int\'egrer compl\`etement un syst\`eme donn\'e
  d'\'equations diff\'erentielles.
\newblock {\em Bulletin des Sciences Math\'ematiques et Astronomiques\/}, {\em
  10\/}, 149--159.

\bibitem[{Lochowski et~al.(2018)Lochowski, Perkowski \&
  Prömel}]{Lochowski_2018}
Lochowski, R., Perkowski, N. \& Prömel, D.~J. (2018).
\newblock A superhedging approach to stochastic integration.
\newblock {\em Stochastic Processes and their Applications\/}, {\em 128\/}(12),
  4078--4103.
\newline\urlprefix\url{https://doi.org/10.1016/j.spa.2018.01.009}

\bibitem[{Lusin(1916)}]{Lusin_1916}
Lusin, N. (1916).
\newblock {Int\'egral et s\'erie trigonom\'etrique}.
\newblock {\em Matematicheskii Sbornik\/}, {\em 30\/}(1), 1--242. \\ URL:
  \texttt{\href{http://mi.mathnet.ru/eng/msb6501}{http://mi.mathnet.ru/eng/msb6501}}.

\bibitem[{Mandelbrot \& Van~Ness(1968)}]{Mandlebrot_1968}
Mandelbrot, B.~B. \& Van~Ness, J.~W. (1968).
\newblock {Fractional Brownian Motions, Fractional Noises and Applications}.
\newblock {\em SIAM Review\/}, {\em 10\/}(4), 422--437.
\newline\urlprefix\url{https://doi.org/10.1137/1010093}

\bibitem[{McCrickerd \& Pakkanen(2018)}]{McCrickerd_2018}
McCrickerd, R. \& Pakkanen, M.~S. (2018).
\newblock {Turbocharging Monte Carlo pricing for the rough Bergomi model}.
\newblock {\em Quantitative Finance\/}, {\em 18\/}(11), 1877--1886.
\newline\urlprefix\url{https://doi.org/10.1080/14697688.2018.1459812}

\bibitem[{Mechkov(2015)}]{Mechkov_2015}
Mechkov, S. (2015).
\newblock {Fast-Reversion Limit of the Heston Model}.
\newblock {\em SSRN preprint. \\ \emph{URL:
  \texttt{\href{https://ssrn.com/abstract=2418631}{https://ssrn.com/abstract=2418631}}}\/}.

\bibitem[{Mechkov(2016)}]{Mechkov_2016}
Mechkov, S. (2016).
\newblock {`Hot-start' initialisation of the Heston model}.
\newblock {\em Risk\/}.

\bibitem[{Meerschaert \& Scheffler(2004)}]{Meerschaert_2004}
Meerschaert, M.~M. \& Scheffler, H.-P. (2004).
\newblock Limit theorems for continuous-time random walks with infinite mean
  waiting times.
\newblock {\em Journal of Applied Probability\/}, {\em 41\/}(3), 623--638.
\newline\urlprefix\url{https://doi.org/10.1239/jap/1091543414}

\bibitem[{Muravlev(2011)}]{Muravlev_2011}
Muravlev, A.~A. (2011).
\newblock {Representation of a fractional Brownian motion in terms of an
  infinite-dimensional Ornstein-Uhlenbeck process}.
\newblock {\em Russian Mathematical Surveys\/}, {\em 66\/}(2), 439--441.
\newline\urlprefix\url{https://doi.org/10.1070/rm2011v066n02abeh004746}

\bibitem[{Novikov(1973)}]{Novikov_1972}
Novikov, A.~A. (1973).
\newblock {On an Identity for Stochastic Integrals}.
\newblock {\em Theory of Probability \& Its Applications\/}, {\em 17\/}(4),
  717--720.
\newline\urlprefix\url{https://doi.org/10.1137/1117088}

\bibitem[{Papoulis \& Pillai(2002)}]{Papoulis_2002}
Papoulis, A. \& Pillai, S.~U. (2002).
\newblock {\em Probability, Random Variables, and Stochastic Processes
  \emph{(4th ed.)}\/}.
\newblock McGraw Hill.

\bibitem[{Peano(1890)}]{Peano_1890}
Peano, G. (1890).
\newblock D{\'e}monstration de l'int{\'e}grabilit{\'e} des {\'e}quations
  diff{\'e}rentielles ordinaires.
\newblock {\em Mathematische Annalen\/}, {\em 37\/}, 182--228.
\newline\urlprefix\url{https://doi.org/10.1007/bf01200235}

\bibitem[{Prokhorov(1956)}]{Prokhorov_1956}
Prokhorov, Y.~V. (1956).
\newblock {Convergence of Random Processes and Limit Theorems in Probability
  Theory}.
\newblock {\em Theory of Probability {\&} Its Applications\/}, {\em 1\/}(2),
  157--214.
\newline\urlprefix\url{https://doi.org/10.1137/1101016}

\bibitem[{Puhalskii \& Whitt(1997)}]{Puhalskii_1997}
Puhalskii, A.~A. \& Whitt, W. (1997).
\newblock {Functional large deviation principles for first-passage-time
  processes}.
\newblock {\em The Annals of Applied Probability\/}, {\em 7\/}(2), 362--381.
\newline\urlprefix\url{https://doi.org/10.1214/aoap/1034625336}

\bibitem[{Račkauskas \& Suquet(2004)}]{Rackauskas_2004}
Račkauskas, A. \& Suquet, C. (2004).
\newblock {Necessary and sufficient condition for the Lamperti invariance
  principle}.
\newblock {\em Theory of Probability and Mathematical Statistics\/}, {\em
  68\/}, 127--137.
\newline\urlprefix\url{https://doi.org/10.1090/S0094-9000-04-00601-5}

\bibitem[{Revuz \& Yor(1999)}]{Revuz_1999}
Revuz, D. \& Yor, M. (1999).
\newblock {\em Continuous Martingales and Brownian Motion\/}.
\newblock Springer-Verlag Berlin Heidelberg.
\newline\urlprefix\url{https://doi.org/10.1007/978-3-662-06400-9}

\bibitem[{Rogers \& Williams(2000)}]{Rogers_1994}
Rogers, L. C.~G. \& Williams, D. (2000).
\newblock {\em {Diffusions, Markov Processes, and Martingales \emph{(2nd
  ed.)}}\/}.
\newblock Cambridge University Press. \\ Vol.1 DOI:
  \texttt{\href{https://doi.org/10.1017/cbo9781107590120}{https://doi.org/10.1017/cbo9781107590120}}
  \\ Vol.2 DOI:
  \texttt{\href{https://doi.org/10.1017/cbo9780511805141}{https://doi.org/10.1017/cbo9780511805141}}.

\bibitem[{Royden \& Fitzpatrick(2010)}]{Royden_2010}
Royden, H. \& Fitzpatrick, P. (2010).
\newblock {\em Real Analysis \emph{(4th ed.)}\/}.
\newblock Prentice Hall.

\bibitem[{Rudin(1976)}]{Rudin_1976}
Rudin, W. (1976).
\newblock {\em Principles of Mathematical Analysis \emph{(3rd ed.)}\/}.
\newblock McGraw-Hill.

\bibitem[{Saks(1937)}]{Saks_1937}
Saks, S. (1937).
\newblock {\em {Theory of the Integral \emph{(2nd ed.)}}\/}.
\newblock Hafner Publishing Company New York.

\bibitem[{Samko et~al.(1993)Samko, Kilbas \& Marichev}]{Samko_1993}
Samko, S., Kilbas, A. \& Marichev, O. (1993).
\newblock {\em Fractional Integrals and Derivatives\/}.
\newblock CRC Press.

\bibitem[{Sato(1999)}]{Sato_1999}
Sato, K. (1999).
\newblock {\em {L\'{e}vy Processes and Infinitely Divisible Distributions}\/}.
\newblock Cambridge University Press.

\bibitem[{Skorokhod(1956)}]{Skorokhod_1956}
Skorokhod, A.~V. (1956).
\newblock {Limit Theorems for Stochastic Processes}.
\newblock {\em Theory of Probability {\&} Its Applications\/}, {\em 1\/}(3),
  261--290.
\newline\urlprefix\url{https://doi.org/10.1137/1101022}

\bibitem[{Skorokhod(1965)}]{Skorokhod_1965}
Skorokhod, A.~V. (1965).
\newblock {\em {Studies in the Theory of Random Processes}\/}.
\newblock Addison-Wesley.

\bibitem[{Soong(1973)}]{Soong_1973}
Soong, T. (1973).
\newblock {\em Random Differential Equations in Science and Engineering\/}.
\newblock Academic Press.

\bibitem[{Srinivasan \& Vasudevan(1971)}]{Srinivasan_1971}
Srinivasan, S.~K. \& Vasudevan, R. (1971).
\newblock {\em Introduction to Random Differential Equations and Their
  Applications\/}.
\newblock Elsevier Publishing Company.

\bibitem[{Strand(1968)}]{strand_1968}
Strand, J.~L. (1968).
\newblock {\em Stochastic Ordinary Differential Equations\/}.
\newblock PhD Thesis, University of California (Berkeley).

\bibitem[{Strand(1970)}]{Strand_1970}
Strand, J.~L. (1970).
\newblock Random ordinary differential equations.
\newblock {\em Journal of Differential Equations\/}, {\em 7\/}(3), 538--553.
\newline\urlprefix\url{https://doi.org/10.1016/0022-0396(70)90100-2}

\bibitem[{Sussmann(1978)}]{Sussman_1978}
Sussmann, H.~J. (1978).
\newblock {On the Gap Between Deterministic and Stochastic Ordinary
  Differential Equations}.
\newblock {\em The Annals of Probability\/}, {\em 6\/}(1), 19--41.
\newline\urlprefix\url{https://doi.org/10.1214/aop/1176995608}

\bibitem[{Swishchuk(2016)}]{Swishchuk_2016}
Swishchuk, A. (2016).
\newblock {\em Change of Time Methods in Quantitative Finance\/}.
\newblock Springer International Publishing.
\newline\urlprefix\url{https://doi.org/10.1007/978-3-319-32408-1}

\bibitem[{Tse \& Wan(2013)}]{Tse_2013}
Tse, S.~T. \& Wan, J. W.~L. (2013).
\newblock {Low-bias simulation scheme for the Heston model by Inverse Gaussian
  approximation}.
\newblock {\em Quantitative Finance\/}, {\em 13\/}(6), 919--937.
\newline\urlprefix\url{https://doi.org/10.1080/14697688.2012.696678}

\bibitem[{Vellaisamy \& Kumar(2018)}]{Vellaisamy_2017}
Vellaisamy, P. \& Kumar, A. (2018).
\newblock {First-exit times of an inverse Gaussian process}.
\newblock {\em Stochastics\/}, {\em 90\/}(1), 29--48.
\newline\urlprefix\url{https://doi.org/10.1080/17442508.2017.1311897}

\bibitem[{Vovk(2016)}]{Vovk_2016}
Vovk, V. (2016).
\newblock Purely pathwise probability-free it\^o integral.
\newblock {\em Matematychni Studii\/}, {\em 46\/}(1), 96--110.
\newline\urlprefix\url{https://doi:10.15330/ms.46.1.96-110}

\bibitem[{Watanabe(2010)}]{Watanabe_2010}
Watanabe, S. (2010).
\newblock {Itô’s theory of excursion point processes and its developments}.
\newblock {\em Stochastic Processes and their Applications\/}, {\em 120\/}(5),
  653 -- 677.
\newline\urlprefix\url{https://doi.org/10.1016/j.spa.2010.01.012}

\bibitem[{Wend(1969)}]{Wend_1969}
Wend, D. V.~V. (1969).
\newblock Existence and uniqueness of solutions of ordinary differential
  equations.
\newblock {\em Proceedings of the American Mathematical Society\/}, {\em
  23\/}(1), 27--33.
\newline\urlprefix\url{https://doi.org/10.2307/2037480}

\bibitem[{Whitt(1971)}]{Whitt_1971}
Whitt, W. (1971).
\newblock Weak convergence of first passage time processes.
\newblock {\em Journal of Applied Probability\/}, {\em 8\/}(2), 417--422.
\newline\urlprefix\url{https://doi.org/10.2307/3211913}

\bibitem[{Whitt(1980)}]{Whitt_1980}
Whitt, W. (1980).
\newblock {Some Useful Functions for Functional Limit Theorems}.
\newblock {\em Mathematics of Operations Research\/}, {\em 5\/}(1), 67--85.
\newline\urlprefix\url{https://doi.org/10.1287/moor.5.1.67}

\bibitem[{Whitt(2002)}]{Whitt_2002}
Whitt, W. (2002).
\newblock {\em Stochastic-Process Limits\/}.
\newblock Springer-Verlag New York.
\newline\urlprefix\url{https://doi.org/10.1007/b97479}

\bibitem[{Wintner(1945)}]{Wintner_1945}
Wintner, A. (1945).
\newblock {The Non-Local Existence Problem of Ordinary Differential Equations}.
\newblock {\em American Journal of Mathematics\/}, {\em 67\/}(2), 277--284.
\newline\urlprefix\url{https://doi.org/10.2307/2371729}

\bibitem[{Wyłomańska et~al.(2016)Wyłomańska, Kumar, Połoczański \&
  Vellaisamy}]{Wylomanska_2016}
Wyłomańska, A., Kumar, A., Połoczański, R. \& Vellaisamy, P. (2016).
\newblock {Inverse Gaussian and its inverse process as the subordinators of
  fractional Brownian motion}.
\newblock {\em Physical Review E\/}, {\em 94\/}(4), 21--28.
\newline\urlprefix\url{https://doi.org/10.1103/PhysRevE.94.042128}

\bibitem[{Yamada \& Watanabe(1971)}]{Yamada_1971}
Yamada, T. \& Watanabe, S. (1971).
\newblock On the uniqueness of solutions of stochastic differential equations.
\newblock {\em Journal of Mathematics of Kyoto University\/}, {\em 11\/}(1),
  155--167.
\newline\urlprefix\url{https://doi.org/10.1215/kjm/1250523691}

\bibitem[{Yosie(1925)}]{Yosie_1925}
Yosie, T. (1925).
\newblock {Über die Unität der Lösung der gewöhnlichen
  Differentialgleichungen erster Ordnung}.
\newblock {\em Japanese journal of mathematics\/}, {\em 2\/}, 161--173.
\newline\urlprefix\url{https://doi.org/10.4099/jjm1924.2.0_161}

\bibitem[{Zumbach(2009)}]{Zumbach_2009}
Zumbach, G. (2009).
\newblock Time reversal invariance in finance.
\newblock {\em Quantitative Finance\/}, {\em 9\/}(5), 505--515.
\newline\urlprefix\url{https://doi.org/10.1080/14697680802616712}

\bibitem[{Zygmund(2003)}]{Zygmund_2003}
Zygmund, A. (2003).
\newblock {\em Trigonometric Series \emph{(3rd ed.)}\/}.
\newblock Cambridge University Press.
\newline\urlprefix\url{https://doi.org/10.1017/cbo9781316036587}

\end{thebibliography}
